\documentclass[]{aa} 

\usepackage{graphicx}
\usepackage{txfonts}
\usepackage{natbib}
\usepackage{longtable}
\usepackage{lscape}

\providecommand{\sorthelp}[1]{}

\begin{document}
 
\title{Galactic cold cores
VI. Dust opacity spectral index
\thanks{
{\it Planck} \emph{(http://www.esa.int/Planck)} is a project of the European Space
Agency -- ESA -- with instruments provided by two scientific consortia funded by ESA
member states (in particular the lead countries: France and Italy) with contributions
from NASA (USA), and telescope reflectors provided in a collaboration between ESA and a
scientific consortium led and funded by Denmark.
}
\thanks{{\it Herschel} is an ESA space observatory with science instruments provided
by European-led Principal Investigator consortia and with important
participation from NASA.}
\thanks{
Table 3 is available in electronic form
at the CDS via anonymous ftp to cdsarc.u-strasbg.fr (130.79.128.5)
or via http://cdsweb.u-strasbg.fr/cgi-bin/qcat?J/A+A/
}
}

\author{M.     Juvela\inst{1},
        K.     Demyk\inst{2,3},
        Y.     Doi\inst{5},
        A.     Hughes\inst{3,2,8},
        C.     Lef{\`e}vre\inst{7},
        D.J.   Marshall\inst{9},        
        C.     Meny\inst{2,3},
        J.     Montillaud\inst{4},
        L.     Pagani\inst{7},        
        D.     Paradis\inst{2,3},
        I.     Ristorcelli\inst{2,3},
        J.     Malinen\inst{1},
        L.A.   Montier\inst{2,3},
        R.     Paladini\inst{8},
        V.-M.  Pelkonen\inst{1},
        A.     Rivera-Ingraham\inst{2,10},
        }

\institute{
Department of Physics, P.O.Box 64, FI-00014, University of Helsinki,
Finland, {\em mika.juvela@helsinki.fi}                                
\and
Universit\'e de Toulouse, UPS-OMP, IRAP, F-31028 Toulouse cedex 4, France   
\and
CNRS, IRAP, 9 Av. colonel Roche, BP 44346, F-31028 Toulouse cedex 4, France  
\and
Institut UTINAM, CNRS UMR 6213, OSU THETA, Universit\'e de Franche-Comt\'e, 
41 bis avenue de l'Observatoire, 25000 Besan\c{c}on, France
\and
The University of Tokyo, Komaba 3-8-1, Meguro, Tokyo, 153-8902, Japan
\and
IPAC, Caltech, Pasadena, USA                                          
\and
LERMA, CNRS UMR8112, Observatoire de Paris, 61 avenue de l'observatoire 75014 Paris, France
\and
Max-Planck-Institut für Astronomie, Königstuhl 17, D-69117 Heidelberg, Germany
\and
Laboratoire AIM, IRFU/Service d’Astrophysique - CEA/DSM - CNRS -
Université Paris Diderot, Bât. 709, CEA-Saclay, F-91191,
Gif-sur-Yvette Cedex, France
\and
European Space Astronomy Centre (ESA-ESAC), PO Box 78, 28691,
Villanueva de la Cañada, Madrid, Spain
}

\authorrunning{M. Juvela et al.}

\date{Received September 15, 1996; accepted March 16, 1997}

\abstract { 
The $Galactic$ $Cold$ $Cores$ project has carried out $Herschel$ photometric
observations of 116 fields where the $Planck$ survey has found signs of cold dust
emission. The fields contain sources in different environments and different
phases of star formation. Previous studies have revealed variations in their dust
submillimetre opacity.
} 
{
The aim is to measure the value of dust opacity spectral index and to understand its
variations spatially and with respect to other parameters, such as temperature, column
density, and Galactic location.
}
{
The dust opacity spectral index $\beta$ and the dust colour temperature $T$ are
derived using $Herschel$ and $Planck$ data. The relation between $\beta$ and
$T$ is examined for the whole sample and inside individual fields. 
}
{
Based on $IRAS$ and $Planck$ data, the fields are characterised by a median
colour temperature of 16.1\,K and a median opacity spectral index of $\beta$~=~1.84. 
The values are not correlated with Galactic longitude. 
We observe a clear $T$--$\beta$ anti-correlation. In $Herschel$ observations,
constrained at lower resolution by $Planck$ data, the variations follow the column
density structure and ${\rm \beta_{\rm FIR}}$ can rise to $\sim 2.2$ in individual clumps.
The highest values are found in starless clumps.
The $Planck$ 217\,GHz band shows a systematic excess that is not restricted to
cold clumps and is thus consistent with a general flattening of the dust emission
spectrum at millimetre wavelengths. When fitted separately below and above
700\,$\mu$m, the median spectral index values are ${\rm \beta_{\rm FIR}}\sim 1.91$ and
${\rm \beta(mm) \sim 1.66}$.
}
{The spectral index changes as a function of column density and wavelength. The comparison
of different data sets and the examination of possible error sources show that our
results are robust. However, $\beta$ variations are partly masked by temperature
gradients and the changes in the intrinsic grain properties may be even greater.}


\maketitle

\section{Introduction} \label{sect:intro}

The all-sky survey of the {\it Planck} satellite \citep{Tauber2010} enabled the
identification of cold interstellar dust clouds on a Galactic scale. The high
sensitivity and an angular resolution of better than $5\arcmin$ allowed the detection
and classification of a large number of cold and compact sources that, based on the
low temperature alone, are likely to be associated with prestellar phases of the star
formation process. Analysis of {\it Planck} data led to the creation of the Cold Clump
Catalogue of Planck Objects \citep[C3PO, see][]{planck2011-7.7b}, with more than 10000
sources. The low colour temperatures (mostly $T\la14$\,K) indicate that the fields
have high column density structures that are only partially resolved by the
$\sim5\arcmin$ {\it Planck} beam. Many of the clumps are likely to contain
gravitationally bound cores. 

The {\it Herschel} Open Time Key Programme {\em Galactic Cold Cores} carried out
dust continuum emission observations of 116 fields that were selected based on the {\it
Planck} C3PO catalogue. The fields were mapped with {\it Herschel} PACS and SPIRE
instruments \citep{Pilbratt2010, Poglitsch2010, Griffin2010} at wavelengths of
100--500\,$\mu$m. {\it Herschel} makes it possible to study the {\it Planck} clumps
and their internal structure in detail \citep{GCC-III, GCC-IV, GCC-VI}. The mapped
fields are typically $\sim 40 \arcmin$ in size, enabling investigation of dust
properties on larger scales and, in particular, comparison of dust properties
between 
the dense and cold regions and their lower density environment.
First results
have been presented in \citet{planck2011-7.7b, planck2011-7.7a}, and in
\citet{Juvela2010, Juvela2011, GCC-III} (Papers I, II, and III, respectively).
\citet{GCC-IV} present an analysis of submillimetre clumps and star formation in
these {\it Herschel} fields. Other studies have been carried out in high latitude
fields in \citet{Malinen2014}, \citet{GCC-VI}, and 
Ristorcelli et al. (in preparation).


In this paper we concentrate on the dust opacity spectral index, $\beta$. Our
previous studies indicated significant variations in the dust submillimetre
opacity, $\kappa$, and the average opacity was found to be more than twice
the typical value found in diffuse clouds \citep{GCC-V}. In the densest
clumps -- if nearby and hence well resolved -- $\kappa$ showed additional
increase by more than a factor of two. We therefore might expect to see some
related changes in the spectral index for these sources.

The observed spectral index or emission spectral index, $\beta$, is derived from the
spectral energy distribution (SED) of dust emission assuming that dust is optically
thin, and its emission can be modelled with a modified blackbody. In such a model
$\beta$ describes the asymptotic behaviour of the dust opacity at long wavelengths.
According to physical models describing the optical response of matter with light, such
as the Lorentz model, at far-infrared (FIR) and submillimetre wavelengths $\beta$ is
independent of temperature and may take values in the range 1- 2 depending on dust
composition and structure \citep{Bohren1998}. However, FIR/submillimetre observations
have shown that {\it (i)} $\beta$ values depend on the astrophysical environment, {\it
(ii)} that $\beta$ and the dust temperature are anti-correlated, and {\it (iii)} that
$\beta$ varies with wavelength such that the dust SED flattens for $\lambda \ge$ 500-800
$\mu$m. The $\beta$-T anti-correlation was first observed by \citet{Dupac2003} and
\citet{Desert2008} from PRONAOS and ARCHEOPS observations of various regions of the
interstellar medium (ISM) in the FIR and submillimetre-to-millimetre ranges
respectively. Many studies based on {\em Herschel} and {\em Planck} data find that
$\beta$ and T$_{dust}$ are anti-correlated
\citep{Rodon2010,Veneziani2010,Paradis2010,Etxaluze2011,PlanckII}. These studies have
derived $\beta$ values as low as $\sim1.5$ in warm regions, while $\beta$ values greater
than 2 have been obtained for cold environments.

A flattening of the dust SED at longer wavelengths, first observed by \citet{Reach1995},
is also widely detected by {\em Herschel} and {\em Planck} in various Galactic and
extragalactic environments
\citep{Gordon2010,Galliano2011,Paradis2012,planck2011-6.4b,planck2011-7.0}. In the
literature, this is sometimes referred to as the sub-mm or FIR excess. Rather than a
single value of $\beta$ across the entire FIR-to-millimetre regime, one may thus define
two spectral indices, $\beta_{\rm FIR}$ at $\nu \ge $ 353\,GHz and $\beta_{mm}$ at $\nu
\le $ 353\,GHz. This approach was adopted by {\em Planck} studies of ISM dust at high
galactic latitude \citep{planck2013-XVII} and in the Galactic plane
\citep{planck2013-XIV}, which found that $\beta_{\rm FIR}$ is greater than $\beta_{mm}$
by $\sim0.15$ and $\sim0.3$\,dex, respectively.

The observed $\beta-T$ anti-correlation could stem from the intrinsic properties of
dust grain material, more specifically from its amorphous structure.  Laboratory studies
of interstellar dust analogues show that the dust opacity increases with temperature,
that the spectral index increases when the temperature decreases, and that a single
spectral index is generally not sufficient to describe the experimental data
\citep{Agladze1996,Mennella1998,Boudet2005,Coupeaud2011}. This temperature-dependent
behaviour is related to low energy transitions that are associated with the amorphous
structure of the material. It can be modelled with the TLS model \citep{Meny2007}, which
has also been shown to successfully reproduce observations of ISM dust emission at
FIR/submillimetre wavelengths \citep{Paradis2011,Paradis2014}. Variations in the dust
spectral index could also reflect the dust evolution in the various astrophysical
environments, e.g. grain growth and coagulation in dense clouds \citep{Steinacker2010,
Pagani2010, Koehler2012} or evolution of the carbonaceous dust component
\citep{Jones2013}. It is probable that several processes occur in the ISM and the
interpretation of the T-$\beta$ anti-correlation is therefore not straightforward.
Finally, it is also important to note that $\beta$ and T$_{dust}$ are partially
degenerate. Estimates of $\beta$ are affected by noise in the data, temperature
variations along the line-of-sight and by the method used to fit the observations
\citep{Shetty2009a,Shetty2009b,Juvela2012_bananasplit,Malinen2011,Juvela2012_Tmix,Ysard2012,Juvela2013_TBmethods,
Pagani2015}. The robustness and interpretation of any observed anti-correlation between
$\beta$ and T$_{dust}$ requires the accurate determination and proper treatment of these
biases.

The structure of the paper is the following. The observations are
described in Sect.~\ref{sect:obs}. The main results are presented in
Sect.~\ref{sect:results}, including estimates of the average value of
$\beta$ and its dependence on dust temperature. The results are
discussed in Sect.~\ref{sect:discussion}. We summarise our conclusions
in Sect.~\ref{sect:conclusions}.

\section{Observations and methods}  \label{sect:obs}

\subsection{{\it Herschel} data} \label{sect:Herschel_data}

The target fields for {\it Herschel} observations were selected using the information
that was available from {\em Planck} observations and ancillary data at that time. The
goal was to cover a representative set of {\em Planck} clumps with respect to Galactic
longitude and latitude, estimated dust colour temperature, and clump mass. The
selection of the {\it Herschel} fields is described in \citet{GCC-III} and an overview
of all the maps is given in \citet{GCC-IV} and \citet{GCC-V}. The sample does not
include sources in the Galactic plane, $|b|<1^{\degr}$, which is covered by the Hi-GAL
programme \citep{Molinari2010}. Similarly, regions included in other {\it Herschel}
key programmes like the Gould Belt survey \citep{Andre2010} and HOBYS
\citep{Motte2010} were avoided.

Our observations cover 116 fields. The SPIRE maps correspond to wavelengths
250\,$\mu$m, 350\,$\mu$m and 500\,$\mu$m. They have an average size of
$\sim1800$\,arcmin$^2$. PACS maps were obtained at 100\,$\mu$m and 160\,$\mu$m.
They are smaller, with an average size of $\sim 660$\,arcmin$^2$. The fields are
listed in Table~\ref{table:fields}. The {\it Herschel} observation numbers can be
found in \citet{GCC-IV}.

The SPIRE observations were reduced with the {\it Herschel} Interactive Processing
Environment HIPE v.12.0, using the official pipeline with the iterative destriper
and the extended emission calibration options. The maps were produced with the
naive map-making routine. The PACS data at 100\,$\mu$m and 160\,$\mu$m were
processed with HIPE v. 12.0 up to Level 1 and the final maps were produced with
Scanamorphos v.23 \citep{Roussel2013}. In order of increasing wavelength, the
resolution is approximately 7$\arcsec$, 12$\arcsec$, 18$\arcsec$, 25$\arcsec$, and
37$\arcsec$ for the five bands.
The raw and pipeline reduced data are available via the {\it Herschel} Science
Archive.
The accuracy of the absolute calibration of the SPIRE observations is expected to
be better than 7\% and the relative calibration between the SPIRE bands better than
2\%\footnote{SPIRE Observer's manual, \\ {\em
http://herschel.esac.esa.int/Documentation.shtml}}. For PACS, when combined with
SPIRE data, we adopt an error estimate of 10\%. The value is consistent with the
differences between PACS and Spitzer MIPS measurements of extended emission
\footnote{http://herschel.esac.esa.int/twiki/bin/view/Public/PacsCalibrationWeb}.

{\it Herschel} observations were convolved to the resolution of the 500\,$\mu$m
data using the latest information about the beam
shapes\footnote{http://herschel.esac.esa.int/twiki/bin/view/Public/\\SpirePhotometerBeamProfileAnalysis}.
The effective point spread function (PSF) depends on the source spectrum. For
spectra characteristic of our fields (temperatures $\sim$10--20\,K, spectral index
$\sim$1.5-2.5) the variation of the PSF shape at a given distance from its
centre is at a level of $\sim1$\,\%. The net effect on the final surface brightness
maps is smaller, because of radial averaging and because the signal is never
produced by a single point source.  
The difference with the Gaussian approximation
used in the previous papers \citep[e.g.][]{GCC-III, GCC-IV} is more significant
\citep[see][]{Griffin2013}.
We use beams calculated for a modified blackbody spectrum with fixed parameters
$T=15.0$\,K and $\beta=2.0$. These correspond to the expected parameter values for
cold and compact sources but the beam shape does not vary significantly for the
range of $T$ and $\beta$ values found in our fields. We constructed circular
symmetric convolution kernels that, together with the original PSF, result in an
effective PSF identical (to within a fraction of one per cent) to that of the
500\,$\mu$m band. After convolving data to the common resolution of the 500\,$\mu$m
band, all maps were resampled onto the 14$\arcsec$ pixel grid of the original
500\,$\mu$m maps. This provides adequate sampling of the $\sim37\arcsec$ beam. For
part of the analysis in this paper, the data were further convolved to $1.0\arcmin$
resolution.

\begin{figure}
\includegraphics[width=8.8cm]{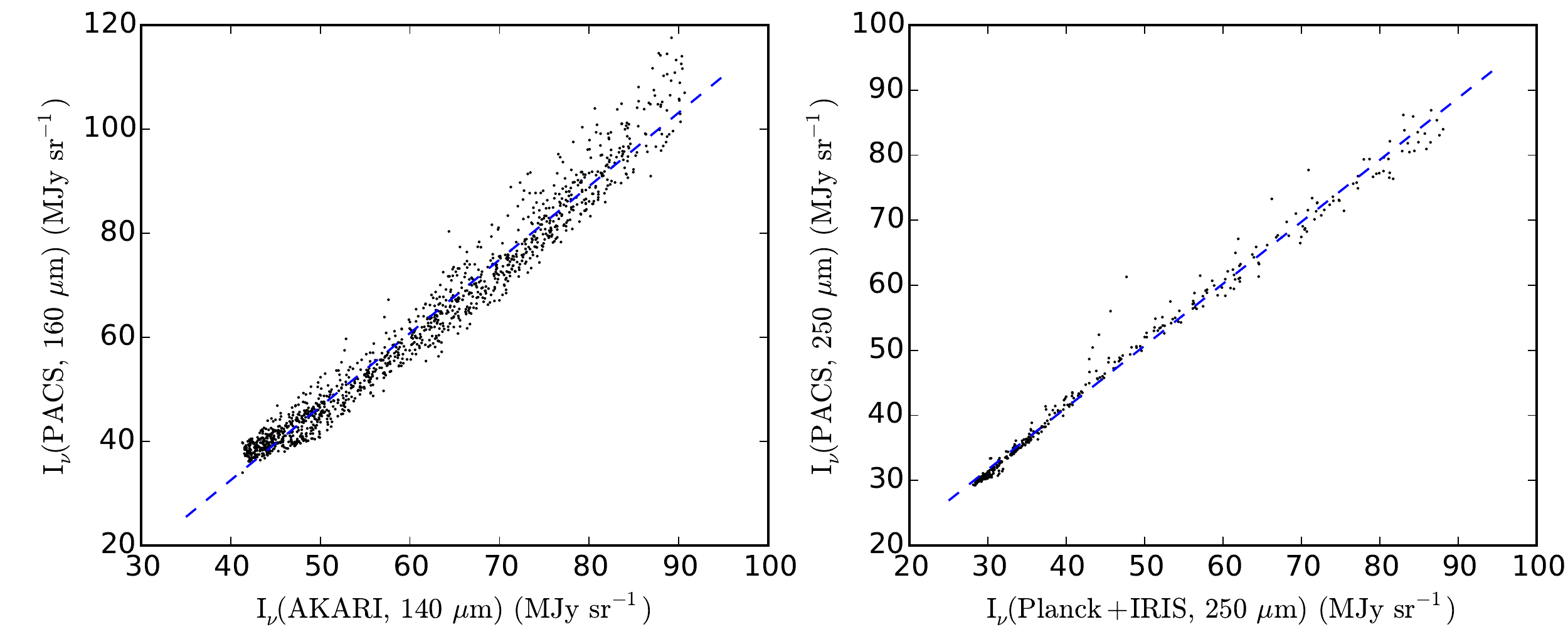}
\caption{
Correlations between {\em Herschel} and reference data in the field G300.86-9.00. PACS
160\,$\mu$m is plotted against AKARI 140\,$\mu$m data (left frame, 1$\arcmin$ resolution)
and SPIRE 250\,$\mu$m data against interpolated IRIS and {\em Planck data} (right frame,
5.0$\arcmin$ resolution). 
}
\label{fig:refcor}
\end{figure}

To determine the absolute zero point of the intensity scale of {\it Herschel} SPIRE
data, we compared the data with {\it Planck} (see Sect.~\ref{sect:Planck_data}) and
IRIS 100\,$\mu$m \citep{MAMD2005} observations. This was done at the
$\sim$4.85$\arcmin$ resolution of {\em Planck} (IRIS and the 857\,GHz {\it Planck}
data were convolved to this resolution; see below Sect.~\ref{sect:Planck_data}).
The detailed procedure of the comparison is described in \citet{GCC-IV}. The {\it
Planck} and IRIS measurements are interpolated to {\it Herschel} wavelengths 
using modified black body curves with $\beta=1.8$. Linear least squares fits
between {\it Herschel} and these reference data provide estimates of the {\em
Herschel} surface brightness zero points and their statistical uncertainty. Unlike
in \citet{GCC-III}, the results are used to correct not only the zero point
but also the gain calibration. At 250\,$\mu$m the estimated zero point errors are
$\sim$1\,MJy\,sr$^{-1}$ or less. The formal uncertainties of the slopes (gain
correction) are below 1\% and typically $\sim$0.5\%. We continue to assume an
uncertainty of 2\% for the calibration of the SPIRE channels.
Figure~\ref{fig:refcor} shows an example of the quality of the correlations between PACS
160\,$\mu$m and AKARI 140\,$\mu$m and between SPIRE 250\,$\mu$m and interpolated IRIS
and {\em Planck} data. Note that at the lowest surface brightness the PACS error
estimates are already dominated by the multiplicative calibration uncertainty, which is
4\,MJy\,sr$^{-1}$ for a surface brightness of 40\,MJy\,sr$^{-1}$ (see below).

For PACS, the linear correlations are often less well defined. In our previous papers
\citep[e.g.][]{Juvela2011, GCC-III}, the zero points were determined from direct
comparison of the average surface brightness values of the {\it Herschel} and
interpolated {\it IRAS} and {\it Planck} data. Because of the similarity of the
wavelength, here we use zero points that are obtained from a comparison with {\it
AKARI} 140\,$\mu$m all-sky maps \citep{Doi2015} with a resolution of $\sim 1 \arcmin$
and with the surface brightness scale tied to DIRBE (resolution $\sim 40 \arcmin$). The
comparison includes colour corrections and extrapolation between 140\,$\mu$m and
160\,$\mu$m. Both are calculated from the same modified blackbody spectra as above,
using $\beta$~=~1.8 fits to SPIRE channels (see below). The 160\,$\mu$m gain calibration
is not changed and we continue to assume a relative uncertainty of 10\%. The comparison
of {\it Herschel} and {\it AKARI} intensities (averaged over the {\it Herschel}
coverage) involves very small statistical errors. The uncertainty of the 160\,$\mu$m
zero point is thus dominated by the uncertainty of the {\it AKARI} zero point, the
relative calibration, and the wavelength extrapolation. We subtracted from AKARI
data a cosmic infrared background (CIRB) level of 0.95\,MJy\,sr$^{-1}$ but the
uncertainty of this component is estimated at only $\sim$0.2\,MJy\,sr$^{-1}$
\citep{Matsuura2011}.
We assume that the  overall zero point accuracy of 160\,$\mu$m is not better than
$\sim$1\,MJy\,sr$^{-1}$, the statistical uncertainty of the  zero points in the
previously used correlations with {\it Planck} and IRIS. The reason for using {\it
Herschel} 160\,$\mu$m data instead of relying directly on {\it AKARI} 140\,$\mu$m band
is that, in addition to higher spatial resolution, the map quality of {\it Herschel}
data is better, especially at small scales.

The zero point and gain corrections discussed above were calculated iteratively
together with the determination of colour corrections. In the case of SPIRE, the
corrections also take the dependence between source spectrum and beam solid
angles into account. The colour corrections were calculated for modified blackbody spectra with
$\beta=1.8$ (the expected average value over the fields), using colour temperatures that
were calculated pixel-by-pixel using SPIRE data and the same fixed value of $\beta$. At
$T=15.0$\,K the difference between $\beta=1.8$ and $\beta=2.2$ colour corrections would
be $\sim$1.0\% for all channels 160\,$\mu$m--500\,$\mu$m (the 100\,$\mu$m band is not
used in this paper). Part of that error would further cancel out when a change in
$\beta$ is compensated by an opposite change in the fitted temperature. Even in the
160\,$\mu$m band and at a low temperature of 12\,K, a difference of $\Delta \beta=0.4$
in the colour correction translates to a change of less than 1\,\% in surface
brightness. Thus, colour corrections with a fixed value of $\beta$ do not bias the
$\beta$ values that we derive in our subsequent analysis.

The SPIRE error estimates mainly consist of the assumed uncertainty of gain calibration
that is tied to {\em Planck} via least squares fits that are performed separately for
each field and band. Thus, these include a significant statistical component, in
addition to the uncertainty of absolute calibration of {\em Planck} (and IRIS). In the
250\,$\mu$m band, additional uncertainty results from interpolation between 100\,$\mu$m
and 350\,$\mu$m. The PACS uncertainties consist of the uncertainty of the original
{\em Herschel} gain calibration, of zero point uncertainties (resulting from comparison with
AKARI), and of mapping artefacts that can be more important than for SPIRE. The
correlations between PACS and interpolated IRAS and {\em Planck} data (not used in this
paper) also suggest that PACS errors are not dominated by systematic errors. In SPIRE
bands systematic errors should be a larger fractional component because of the 
smaller statistical errors.

\subsection{{\it Planck} and IRIS data} \label{sect:Planck_data}

{\em Planck} data from the 857\,GHz, 545\,GHz, 353\,GHz, and 217\,GHz bands are
used to study the dust spectrum at lower resolution, to make use of the longer
wavelength coverage of {\it Planck}. The 857\,GHz and 545\,GHz channels were
already used above to estimate zero points for the {\em Herschel} surface
brightness data. 
The IRIS 100\,$\mu$m data suffer from the problems mentioned earlier (see
Sects.~\ref{sect:intro} and \ref{sect:beta}), the unknown contribution from
stochastically heated very small grains and the sensitivity to line-of-sight temperature
variations. However, because the combination of {\it Planck} and IRIS has already been
studied over the whole sky \citep[e.g.][]{planck2013-p06}, it is useful to investigate
them also in the selected {\it Herschel} fields.

The {\em Planck} data are described in ~\citet{planck2013-p03f}. We use the data from
the official release that is available via the {\em Planck Legacy
Archive}\footnote{http://pla.esac.esa.int/pla/aio/planckProducts.html} and is described
by the {\em Planck} Explanatory
Supplement\footnote{http://www.sciops.esa.int/wikiSI/planckpla}.
We use zodiacal light subtracted maps where the estimated cosmic infrared background has
also already been subtracted. The cosmic microwave background (CMB) anisotropies were
subtracted using the CMB maps, which have been calculated with the SMICA algorithm
\citep{planck2013-p06} and are available in the {\em Planck Legacy Archive}. To set the
Galactic zero levels, the surface brightness values listed in Table~5 of
\citet{planck2013-p03f} were subtracted.
These do not contain the cosmic infrared background and, for consistency, an
estimated CIRB level of 0.80\,MJy\,sr$^{-1}$ was also subtracted from 100\,$\mu$m IRIS
data \citep{Hauser1998, Matsuura2011}.

Several transitions of the CO molecule fall within the {\it Planck} bands, producing
signal above the pure dust emission. The transition $J$~=~2--1 is inside the 217\,GHz band
and the transition $J$~=~3--2 inside the 353\,GHz band. The higher transitions are likely
to be too weak to be a concern because of the steep rise of the dust spectrum towards
shorter wavelengths. In cold clouds and regions of low mass star formation, typical peak
intensities of $J$~=~3--2 CO lines are $\sim$10\,K\,km\,s$^{-1}$
\citep[e.g.][]{JoseGarcia2013}. This corresponds to $\sim$0.5\,MJy\,sr$^{-1}$ in the
353\,GHz band, but dilution by the $\sim 5\arcmin$ beam is likely to make the
contribution smaller (CO peak values are measured at a resolution of $\sim 1\arcmin$ or
less). In the most prominent massive star-forming regions, the line intensities can be
higher by a factor of ten and CO contributions could be locally significant in the
353\,GHz band. At 217\,GHz, the line intensities are of similar order but the conversion
factors from line intensity to {\it Planck} surface brightness is lower by more than a
factor of two \citep{planck2013-p03a}. However, because the dust signal decreases from
353\,GHz to 217\,GHz by a factor of $\sim$5, CO contamination is more important in the
217\,GHz band and, if not taken into account, could bias the $\beta$ estimates
downwards.

We do not have direct multitransition CO observations of our fields. One exception is
LDN~183 (G6.03+36.73), where the average line ratio $T_{\rm A}$(2--1)/$T_{\rm A}$(1--0)
over the central $3\arcmin \times 6\arcmin$ area is $\sim$0.54.
A second example is cloud LDN~1642 (G210.90-36.55) which has been observed with the SEST
telescope by \citet{Russeil2003}. SEST data gives an average ratio of line areas of
$W(2-1)/W(1-0)$~=~0.61. At a resolution of 4.5$\arcmin$, one reaches a value of 0.72 at
the location of the column density peak. 

{\em Planck} data themselves have been used to estimate, with three
different methods, the intensity of the first CO transitions.
The methods used to derive the CO estimates are described in
\citet{planck2013-p03a}. Planck Type 1 CO maps are calculated based on the differences
in the spectral response of bolometer pairs. This makes it possible to estimate the CO
emission of the three first rotational transitions separately, although with relatively
low signal-to-noise ratios (SN). Type 2 estimates of CO(1-0) and CO(2-1) line
intensities result from a multi-channel fitting of the CO emission together with cosmic
microwave background (CMB), dust emission, and free-free emission. Because of the use of
lower frequency channels, Type 2 maps only have a resolution of 15$\arcmin$. Finally,
Type 3 maps are calculated assuming fixed CO line ratios over the whole sky,
CO(2-1)/CO(1-0)=0.595 an CO(3-2)/CO(1-0)=0.297. This results in maps with the best SN
but, of course, no separate information of the line ratios in any given pixel.
The reason why we do not make the correction for CO emission directly
using Type 1 and Type 2 maps is the low signal-to-noise ratio (SN) of Type 1 maps
(at full Planck resolution) and the low spatial resolution of Type 2
maps. The component separations also assume a fixed dust spectrum
($T=17$\,K, $\beta=1.6$), which may lead to errors in regions that are
colder than the average ISM.  According to \citet{planck2013-p03a},
Type 1 maps give at large scales line ratios $T_{\rm A}$(2--1)/$T_{\rm
A}$(1--0)=0.4 and $T_{\rm A}$(3--2)/$T_{\rm A}$(1--0)=0.2. However,
concentrating on CO bright regions, the same paper found average line
ratios of 0.595 and 0.297.

\begin{figure}
\includegraphics[width=8.8cm]{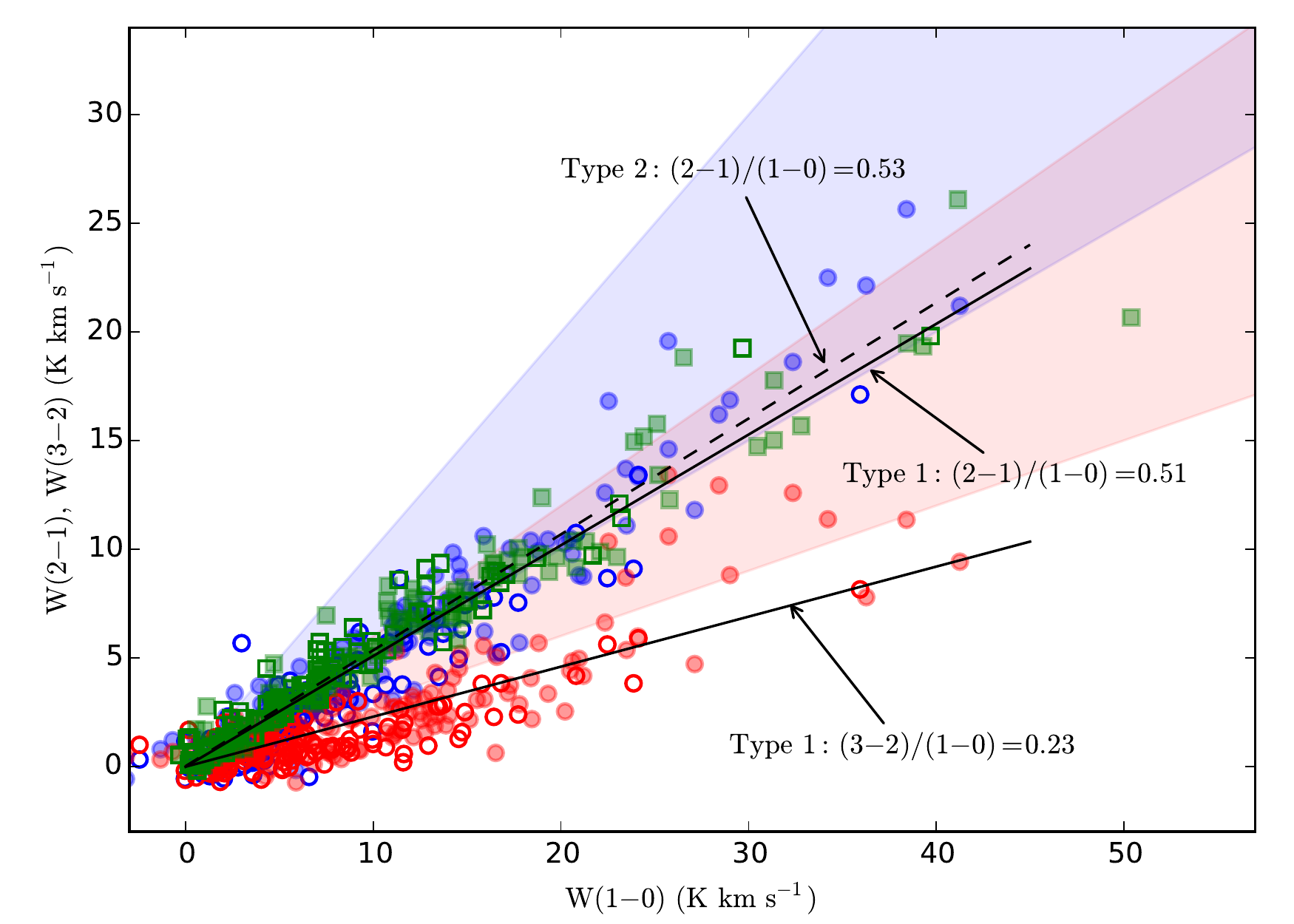}
\caption{
CO line ratios in our fields based on {\em Planck} Type 1 and Type 2 CO maps. Type 1
ratios (2--1)/(1--0) and (3--2)/(1--0) are plotted with blue and red
circles and Type 2 ratios (2--1)/(1--0) with green squares. The open
and the filled symbols are averages over pixels below and above median
column density, respectively. The black lines show least squares fits
to data including both low and high column densities. The shaded
areas indicate the range of line ratios considered in this paper. 
}
\label{fig:CO_ratios}
\end{figure}

Figure~\ref{fig:CO_ratios} shows the line ratios we obtained in our fields for data
averaged over the SPIRE coverage. Type 1 data give values 0.51 and 0.25 for the
(2--1)/(1--0) and (3--2)/(1--0) ratios. Using Type 2 data (with lower resolution but
higher SN) the (2--1)/(1--0) ratio is 0.53. The values are thus in line with the values
quoted above for large area averages and, for (2--1)/(1--0), close to the value observed
in LDN~183. The values are somewhat lower than in CO bright regions in general 
\citep[see above][]{planck2013-p03a}, which may be related to the fact that our fields
are selected based on their low (dust) temperature. In Fig.~\ref{fig:CO_ratios} we have
divided each field to two parts according to its median column density (estimated using
SPIRE data and a constant value of $\beta=1.8$). There are no clear differences in the
line ratios between the two column density samples. However, Type 1 (3--2)/(1--0) ratio
seems to increase in the high column density sample at the highest $W$(1--0) values. A
similar but weaker trend might exist in the (2--1)/(1--0) ratios of Type 1 data but is
not seen in the Type 2 data.

In this paper we correct the 345\,GHz and 217\,GHz data for CO contamination using the
CO(1--0) emission of the {\it Planck} Type 3 CO maps and as a default assuming
line ratios $T_{\rm A}$(2--1)/$T_{\rm A}$(1--0)=0.5 and $T_{\rm A}$(3--2)/$T_{\rm
A}$(1--0)=0.3. The conversion factor between line intensity and its contribution to
surface brightness are given in \citet{planck2013-p03a}. The ratio $T_{\rm
A}$(3--2)/$T_{\rm A}$(1--0)=0.3 is above the average observed ratio (see
Fig.~\ref{fig:CO_ratios}) but is representative of the ratios at the highest column
densities. The correction in the 353\,GHz band is typically $\sim$0.1\,MJy\,sr$^{-1}$,
at most a few per cent of the total signal. The selected ratio $T_{\rm A}$(2--1)/$T_{\rm
A}$(1--0)=0.5 is consistent with the data in Fig.~\ref{fig:CO_ratios}.  The CO
correction in the 217\,GHz band can be more than 10\%.

In \citet{planck2013-p03a} the {\em Planck} Type 3 $^{12}$CO(1-0) maps were compared
with \citet{Dame2001} data of Taurus, Orion, and Polaris. The ratios between the
Planck and Dame et al. line intensities were found to be slightly {\em above} one, with
an average value of 1.085. In case of Type 1 maps, the $J$~=~2--1 estimates were
compared with ground-based data and with estimates derived from COBE FIRAS, both
comparisons showing agreement within $\sim$10\%.
Another comparison of $J$~=~2--1 observations is now possible using the large-scale maps
of Orion A and B clouds presented in \citet{Nishimura2015}. Orion is not representative
of our sample (selected based on the {\em low} dust temperature of the sources) but this
has little effect on the comparison of the CO measurements. Over the area
presented in Fig. 1 of \citet{Nishimura2015}, the average ratio between the {\em Planck}
Type 2 CO estimates and the 1.85\,m Nobeyama radio telescope data is slightly {\em
below} one, 0.91$\pm$0.11. The ratio does not show systematic variations as a function
of column density or dust temperature, although the value is lower, $\sim$0.6, at the
location of Ori\,KL. The mean line value is compatible with the estimated uncertainty of
$\sim$10\% of the intensity calibration of the Nobeyama data. Figure 11 in
\citet{Nishimura2015} also shows that, in spite of large column density changes, the
spatial variations of the (2-1)/(1-0) line ratio are relatively small. In
particular, the ratio remains relatively constant $\sim$0.8 over most of Orion B. This
suggests that the ratio is affected more by the general radiation field of the region
than by the local column density. 

The comparison with ground-based data shows that, within an accuracy of $\sim$10\%, the
combination of {\em Planck} Type 3 and Type 2 data give adequate predictions of the {\em
average} contamination by CO $J$~=~1--0 and $J$~=~2--1 lines \citep[see][]{planck2013-p03a}.
Type 3 data directly trace the CO emission at 6$\arcmin$ resolution. Thus, the main
uncertainty in the {\em local} CO correction of the 217\,GHz band results from possible
variations of line ratios. The examples of LDN~183 and LDN~1642 show that even within
very dense cold clumps the (2--1)/(1--0) ratio does not necessarily rise much above the
mean value of $\sim$0.5. This in spite of the fact that for very high column densities
the line ratio should tend towards one. Higher line ratios ($\ga$1) are possible in hot
cores where the gas kinetic temperature is affected by local star-formation. In Orion,
because of the high-mass stars, the (2--1)/(1--0) ratio is in many places close to one
or even above one \citep[see Fig. 11 in][]{Nishimura2015}. Comparison with dust
temperature maps (based on {\em Herschel} archive SPIRE maps and an assumption of
$\beta=1.8$) shows that the dust temperature of those regions is $\sim$20\,K or above.
At {\em Herschel} resolution, only 19 of our fields contain pixels with $T>20$\,K. These
amount to $\sim$3\% of the total mapped area. Temperatures $T>20$\,K are found mainly in
regions of low column densities and some of the high values result from higher
uncertainties at low intensities and near map borders. Nevertheless, there can be some
denser regions with both strong CO emission and high (2--1)/(1--0) line ratios where our
default CO correction would be clearly too small. The intensity and temperature maps
(see Appendices) can be consulted to identify regions with potential for such errors.

If CO correction is underestimated, it can affect the derived dust spectral index and,
in particular, estimates for $\beta$ variations as a function of wavelength. We 
examine later the effect of applying CO corrections that are twice the default
correction (see Sect.~\ref{sect:wavelength}). This corresponds to an assumption of line
ratios (2--1)/(1--0)=1.0 and (3--2)/(1--0)=0.6. These represent a firm upper limit for
the {\em average} CO contamination. This also is an upper limit for cold clumps where
the CO excitation temperature is below $\sim$15\,K. Larger line ratios are still
possible in regions of very intense radiation field. The shaded regions in
Fig.~\ref{fig:CO_ratios} illustrate the difference between the two assumptions of line
ratios.

The 217\,GHz band could also contain a non-negligible amount of free-free emission. To
estimate and to subtract this contribution, we used the low frequency foreground
component model (available in the {\em Planck Legacy
Archive}\footnote{http://pla.esac.esa.int/pla/aio/planckProducts.html}) and its
predictions at 217\,GHz. The model consists of an estimate for the low frequency
spectral index (Healpix maps with NSIDE=256 and angular resolution of 40$\arcmin$) and
higher resolution estimates of intensity (Healpix maps with NSIDE=2048 and full {\it
Planck} resolution). These result from component separation where the other components
are cosmic microwave background, CO emission, and thermal dust emission
\citep{planck2013-p06}. The component separation cannot take the
local variations of the dust spectral index, dust temperature, relative intensity of
synchrotron and free-free components, and the contribution of anomalous microwave
emission fully into account. 
However, it is a useful indication of the level of non-dust contributions and its
subtraction should reduce whatever small bias this emission might cause. For our
fields, this component ranges from 0.02\% to 2.3\% of the total signal of the
217\,GHz channel. The median value is 0.1\% (for averages over the maps).

For the instrumental noise, we use the error estimates included in {\em Planck} frequency
maps. The accuracy of absolute calibration is estimated to be 10\% for the 857\,GHz
and 545\,GHz channels and $\sim$2\% or better at low frequencies. The error estimates
associated with the CO and low frequency corrections were set conservatively to 50\%
of the correction. For the noise caused by cosmic infrared background (CIB), we use
the estimates given by \citet{planck2013-p06b}. All error components were added
together in quadrature. 
At 857\,GHz and 545\,GHz the error budget is dominated by calibration uncertainties
while at 217\,GHz the uncertainty of the CO correction is often the largest term. The
instrumental noise and CIB fluctuations have significant contribution only at low
column densities, in the most diffuse parts of the fields.

The data between 100$\mu$m and 353\,GHz were fitted pixel-by-pixel using modified
black body spectra with a fixed spectra index of $\beta=1.8$ and the data were colour
corrected for this spectrum. For the range of colour temperatures derived from SPIRE
observations and for spectral index changes of $\pm$0.2, a change in the {\it Planck}
colour corrections could change the surface brightness values by $\sim$1\%. Thus, the
subsequent analysis is only very weakly dependent on the value of $\beta$ assumed in
the colour correction. 

The {\it Planck} beams are characterised in \citet{planck2013-p03c}. The FWHM of
the effective beam is 4.63$\arcmin$, 4.84$\arcmin$, 4.86$\arcmin$, and
5.01$\arcmin$ for the 857\,GHz, 545\,GHz, 353\,GHz, and 217\,GHz bands,
respectively. The beams have some ellipticity that varies across the sky. The
ellipticity is particularly large in the 857\,GHz band, on average $\epsilon\sim
1.39$.  Our calculations are based on the assumption of symmetric beams. In the
zero point determination of {\em Herschel} data, {\it Planck} 857\,GHz maps were
convolved to 4.85$\arcmin$ assuming Gaussian beams. This convolution does little
to mask the effects of beam ellipticity, which thus causes additional noise in
the determination of {\em Herschel} zero points and becomes part of the
estimated statistical error. When {\it Planck} data are used in modified
blackbody fits, all data are convolved to a resolution of 5.0$\arcmin$. We made
some calculations at 8.0$\arcmin$ resolution, where the effects of beam
asymmetries should be small, but did not observe any significant discrepancy with the
5.0$\arcmin$ calculations.

The IRIS versions of the {\it IRAS} 100\,$\mu$m maps \citep{MAMD2005} were convolved
from the original resolution of 4.3$\arcmin$ to 4.85$\arcmin$ (zero point
calculations) or 5.0$\arcmin$ (other analysis), assuming circular symmetric Gaussian
beams. The IRIS data are corrected for the expected contribution of very small grains
(VSGs) which raises the 100\,$\mu$m intensity above that of the pure big grain (BG)
emission. The corrections were derived with the DustEM dust model
\citep{Compiegne2011} as a function of the BG temperature \citep[see][]{GCC-III}. It
accounts for $\sim$10\% at the highest temperatures and rises up to $\sim$20\% for the
coldest fields. The uncertainty of this correction is large but mostly within the
15\% uncertainty adopted for the 100\,$\mu$m band. The VSG correction mainly decreases
the colour temperature estimates and the effect on the spectral index is small.

\subsection{Calculation of dust opacity spectral index} \label{sect:beta}

The spectral index is determined by fitting observed surface brightness values
$I_{\nu}$ with a modified blackbody law
\begin{equation}
I_{\nu} =  I_{0} \times (B_{\nu}(T)/B_{\nu 0}(T)) \times (\nu/\nu_0)^{\beta}.
\label{eq:mbb}
\end{equation}
The fit involves three free parameters: the spectral index $\beta$, the colour
temperature $T$, the intensity $I_0$ at a reference frequency $\nu_0$ and $B_{\nu}$ is
the Planck function. The fit is possible when observations consist of at least three
wavelengths. Observations at long wavelengths (beyond $200$\,$\mu$m) are needed to
constrain the spectral index, while shorter wavelengths ($\lambda\la 200\,\mu$m) are
better for determining colour temperature.
The equation includes the assumption of optically thin emission.

The interpretation of the estimated $\beta(T)$ dependence requires special care, because
errors can produce artificial correlations \citep{Shetty2009b, Juvela2012_bananasplit}.
Even small errors in surface brightness data may lead to significantly higher $T$ and
lower $\beta$ or vice versa.  One must also note that $T$ and $\beta$ only characterise
the shape of the observed spectrum and will differ from the corresponding intrinsic dust
parameters. The colour temperature overestimates the mass-averaged physical dust
temperature and the observed spectral index is smaller than the average opacity spectral
index of the grains \citep{Shetty2009a, Malinen2011, Juvela2012_Tmix}. These differences
are related to a mixture of grain temperatures along the line-of-sight or generally
within the beam. These can be caused by temperature gradients inside the clouds (for
example, cold clumps with warmer envelopes or embedded hot sources) or to the presence
of different grain populations that have different temperatures. Because wavelengths
near and shortward of the peak of the emission spectrum are more sensitive to
temperature, the recovered parameters will depend on the set of wavelengths used. At
$\sim$100\,$\mu$m and below, the contribution of stochastically heated VSGs increases
and the spectrum cannot be approximated as a modified blackbody due to the broader
temperature distribution of VSGs. The PACS 100\,$\mu$m data are therefore not used in
this paper.

The least squares fits of Eq.~\ref{eq:mbb} employ the error estimates given in
Sections~\ref{sect:Herschel_data} and \ref{sect:Planck_data}. The uncertainties
were calculated with direct Monte Carlo simulations or with Markov chain Monte
Carlo (MCMC) methods. 
With MCMC one can calculate the full posterior probability distributions of $I_0$,
$T$, and $\beta$, only using the relative probabilities of the fits performed with
different combinations of the parameter values \citep[e.g.,][]{Veneziani2010,
Juvela2013_TBmethods}. The calculation results in a chain of parameter values and we use
their median as the final parameter estimates.
In selected fits, the zero point uncertainties were included as part of the error
estimates (combined as additional statistical error) or their influence was
examined separately as a systematic error affecting each field individually.

\section{Results} \label{sect:results}

We use different combinations of observational data to seek evidence for spectral index
variations between the fields and between regions characterised by different
temperature and column density. We utilise some internally consistent data sets (e.g.,
a single instrument) and compare the qualitative and quantitative results obtained
with different and even independent data sets (e.g., {\it Herschel} data vs. {\it
Planck} and {\it IRAS} data). 
We start by using Eq.~\ref{eq:mbb} to fit {\it Planck} data in combination with {\it
IRAS} 100\,$\mu$m band. Next we fit {\em Herschel} observations of the three SPIRE
bands, 250\,$\mu$m, 350\,\,$\mu$m, and 500\,$\mu$m. We continue with four bands,
adding the PACS 160\,$\mu$m band to SPIRE data. We finish by using a combination of
{\it Herschel} and {\it Planck} data. In these fits, dust temperature is determined
mainly by {\it Herschel} observations, {\it Planck} providing further constraints on
the spectral index, albeit at a lower spatial resolution.
The different cases are listed in Table~\ref{table:cases}. The table lists the adopted
relative errors to which other sources of uncertainty (CO, CIB, and low frequency
model subtraction) are added.

\begin{table*}
\caption[]{Combinations of observational data used.}
\begin{tabular}{llll}
\hline
Section &  Data  & Resolution & Error estimates \\
\hline
~\ref{sect:PlanckIRIS} & IRIS 100\,$\mu$m, {\it Planck} 857-217\,GHz          &
    5.0$\arcmin$  & 15\%, 10\%, 10\%, 2\%, 2\%   \\
~\ref{sect:SPIRE_only} & {\it Herschel} 250, 350, 500\,$\mu$m                       &
    37$\arcsec$ or 1$\arcmin$ & 2\% + zero point uncertainties, which have mean values of\\
                              & & & 0.39, 0.17, and 0.07\,MJy\,sr$^{-1}$ for the
                              three bands \\
~\ref{sect:with_PACS}  & {\it Herschel} 160, 250, 350, 500\,$\mu$m                       &
    37$\arcsec$ or 1$\arcmin$ & 10\%, 2\%, 2\%, 2\% + a zero point uncertainty, which \\
                              & & & for the 160\,$\mu$m band is assumed to be 1\,MJy\,sr$^{-1}$ \\
~\ref{sect:combined}   & {\it Herschel} 250-500\,$\mu$m and &  $\sim 37 \arcsec$  & {\it Herschel} 3.5\%, \\
& {\it Planck} 857-353\,GHz  &  &  {\it Planck} 10\% (857-545\,GHz) and 2\% (353-217\,GHz) \\
\hline
\end{tabular}
\label{table:cases}
\end{table*}


\begin{figure*}
\includegraphics[width=17.9cm]{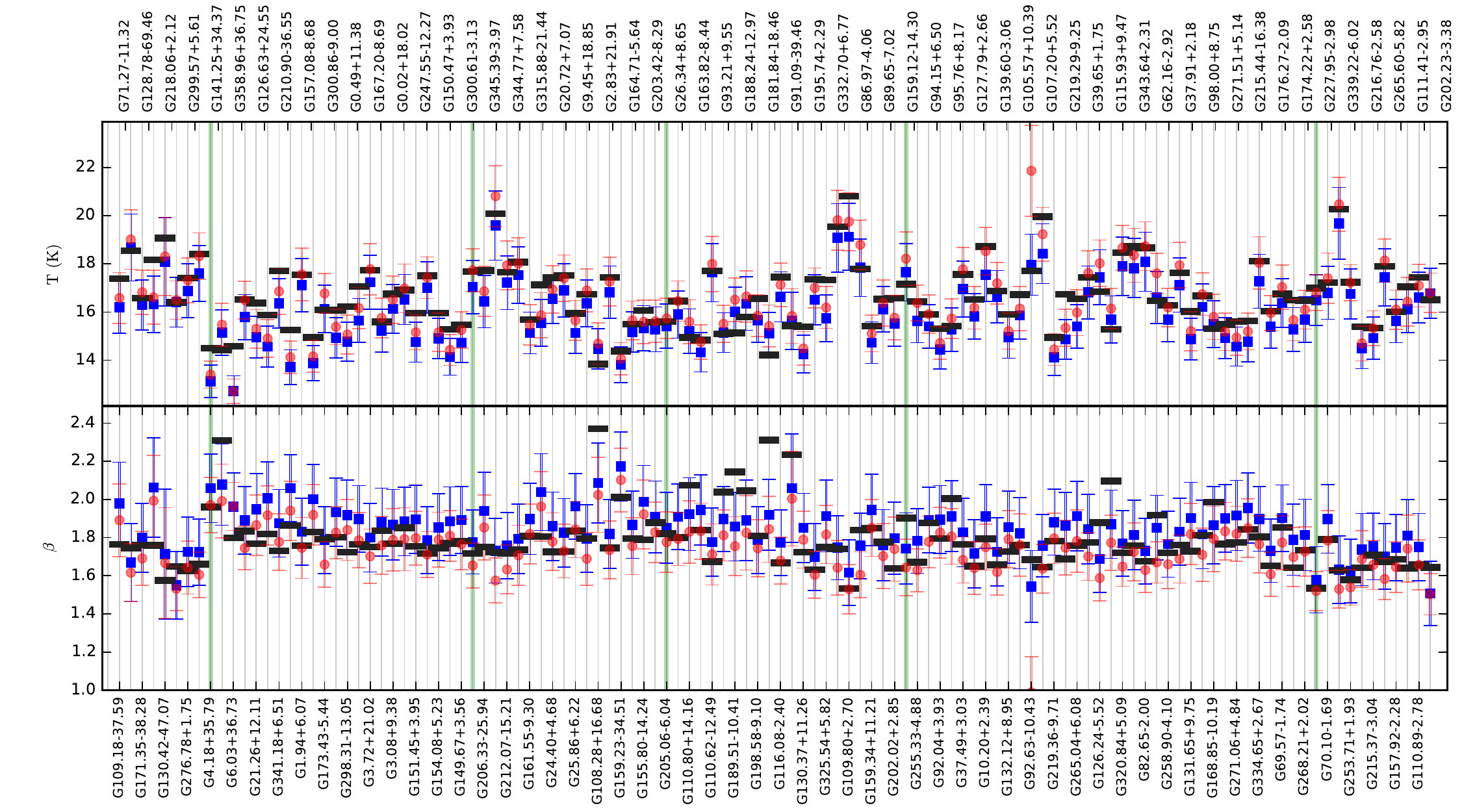}
\caption{
Colour temperature (upper frame) and spectral index values (lower frame) for the 116
fields. For clarity, alternating field names appear above and below the figure. The
values are derived from IRIS and {\it Planck} data, averaged over a single
FWHM=10\,$\arcmin$ beam. Estimates are calculated with (red circles) and without (blue
squares) the 217\,GHz band.
The black horizontal bars show values derived without the 217\,GHz band, using surface
brightness averaged over a larger beam with FWHM=30$\arcmin$. 
The fields are arranged in the order of increasing distance \citep{GCC-IV}. The five
vertical lines indicate the first fields with estimated distances above 0, 200, 400,
800, and 2000\,pc. The first eight fields do not have reliable distance estimates.
}
\label{fig:plot_PI_10arcmin}
\end{figure*}

\subsection{Results from {\it Planck} and IRIS data} \label{sect:PlanckIRIS}

We use IRIS 100$\mu$m and {\it Planck} data in the 857\,GHz, 545\,GHz, 353\,GHz, and
217\,GHz bands to estimate colour temperatures and spectral indices at scales of
5.0$\arcmin$ and larger. 
Unlike in the case of {\em Herschel} data, we need to include the short wavelength IRIS
band to constrain dust temperature. This is expected to be more sensitive to warm dust
and may include contribution from stochastically heated small dust grains. These effects
have been discussed extensively in the literature \citep[e.g.][]{Shetty2009a,
Malinen2011, Juvela2012_Tmix}. Radiative transfer models of Appendix~\ref{app:RT}
suggest that the inclusion of the 100\,$\mu$m band does not make the $\beta$
determination much more sensitive to line-of-sight temperature mixing. For a cloud with
$\tau_{\rm V}=10$\,mag, the combination of IRIS and the three {\em Planck} channels led
to $\beta$ values that were up to 0.07 units lower than for the combination of SPIRE and
the same {\em Planck} bands, the total bias being $\sim$0.10 units.

We start by calculating one SED per field. In each field, we average surface brightness with
a Gaussian beam with FWHM=10.0$\arcmin$, located at the centre of the {\it Herschel} SPIRE
coverage. Because the {\it Herschel} maps are typically 30--40$\arcmin$ in diameter, the
values are representative of most of the field and especially of the higher column density
regions usually found at its centre. SEDs were fitted with modified blackbody functions and
the results are shown in Fig.~\ref{fig:plot_PI_10arcmin}. This figure includes estimates
calculated with and without the 217\,GHz band. The error bars have been estimated with Monte
Carlo simulations, using the surface brightness uncertainties listed in
Sect.~\ref{sect:Planck_data}.

When the lowest frequency is extended from 353\,GHz to 217\,GHz, the mean spectral index is
lower. The average change is 
$\Delta \beta$~=~-0.10 and the mean temperature is higher by $\Delta T$~=~0.52\,K.
The IRIS 100\,$\mu$m data have been corrected for the expected VSG emission (see
Sect.~\ref{sect:Planck_data}). The inclusion of the VSG correction decreases
temperatures by $\Delta T\sim$-0.5\,K and increases spectral index values by $\Delta
\beta \sim 0.05$. With the 217\,GHz band included, the reduced $\chi^2$ values
(normalised by the number of degrees of freedom) are many times higher. Although the
number depends on the relative error estimates of the bands, this already strongly
suggests that a single modified black body curve does not provide a good fit over the
full wavelength range. 

The horizontal black bars show values derived with IRIS and {\it Planck} bands down to
353\,GHz band, using surface brightness measurements averaged over a beam with
FWHM=30$\arcmin$. Compared to the 10$\arcmin$ values (blue points), the temperatures are
higher and spectral index values lower when estimated from the lower resolution data. This
difference is expected because our fields are selected for the presence of cold dust and,
therefore, typically have larger than average column densities. The emission properties are
thus clearly different already between the scales of 10$\arcmin$ and 30$\arcmin$. In
\citet{planck2011-7.7b}, the temperatures observed towards individual {\em Planck} cold clumps
were even lower (the distribution peaking below 14\,K) and the spectral index values were
higher, $\beta \sim 2.1$. However, that analysis concerned individual clumps (sizes below
$\sim$10$\arcmin$) and only the cold emission component that was obtained by subtracting an
estimate of the extended warm background emission.

\begin{figure*}
\sidecaption
\includegraphics[width=12cm]{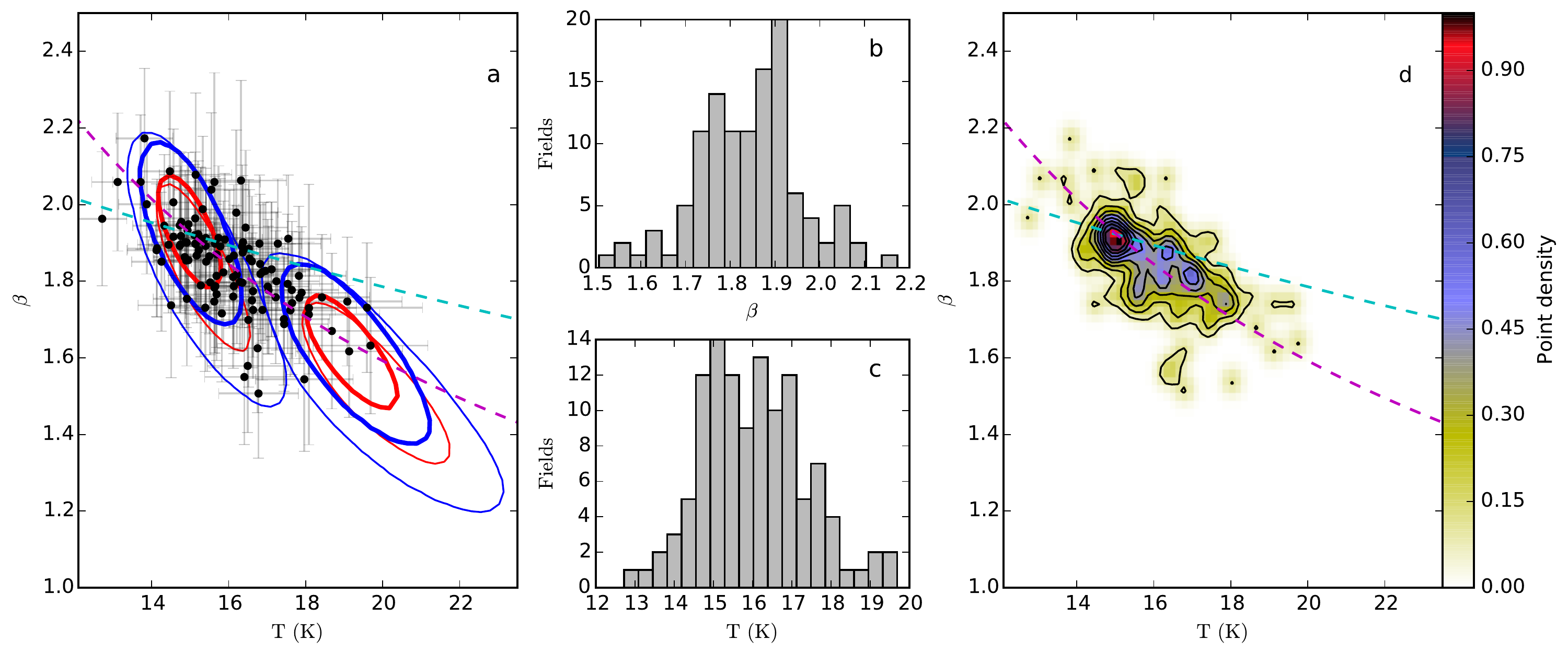}
\caption{
$T$ and $\beta$ values for the 116 fields, calculated from IRIS and {\it Planck} data
($\nu\ge 353$\,GHz) at FWHM=10\,$\arcmin$ resolution. The thin contours show two
examples of the error regions, covering 68\% (red contours) or 95\% (blue contours) of
the distributions. The thick contours correspond to 50\% smaller uncertainties (see
text). The dashed cyan and magenta lines show $T(\beta)$ relation from \citet{Dupac2003}
and \citet{Desert2008}, respectively. Frames b and c show the marginal distributions of
$\beta$ and $T$. Frame d shows the distribution as a colour image (for easier comparison
with Fig.~\ref{fig:ref_PI_TB_pdf}).
}
\label{fig:histo_TB_Planck}
\end{figure*}

Figure~\ref{fig:histo_TB_Planck} shows the correlations between colour temperature and
spectral index for the 10$\arcmin$ beams, for values calculated without the 217\,GHz
band. 
The median values over all fields are $T$~=~15.88\,K and $\beta$~=~1.85.

The distribution of ($T$, $\beta$) points in Fig.~\ref{fig:histo_TB_Planck} shows some
negative correlation between $T$ and $\beta$ that cannot be entirely explained by
measurement errors. The errors scatter ($T$, $\beta$) points along a direction that
has a steeper negative slope than the overall parameter distribution. In the surface
brightness errors, with the possible exception of the 353\,GHz and 217\,GHz bands, the
main component is the uncertainty of the absolute calibration. The band-to-band errors
should be smaller, because calibration errors are likely to be correlated between the
fields. To illustrate the potential importance of this point, we recalculated the
error regions using 50\% smaller uncertainties for the surface brightness measurements
(thick contours in Fig.~\ref{fig:histo_TB_Planck}).

In the figure, we overplot the relation $\beta(T)=(0.4+0.008 \times
T)^{-1}$ derived from PRONAOS observations of cold submillimetre clumps
\citep{Dupac2003} and the corresponding relation $\beta(T)=11.5 \times T^{-0.66}$ from
\citet{Desert2008}. Both relations cross the distribution of our data points.

\begin{figure}
\includegraphics[width=8.8cm]{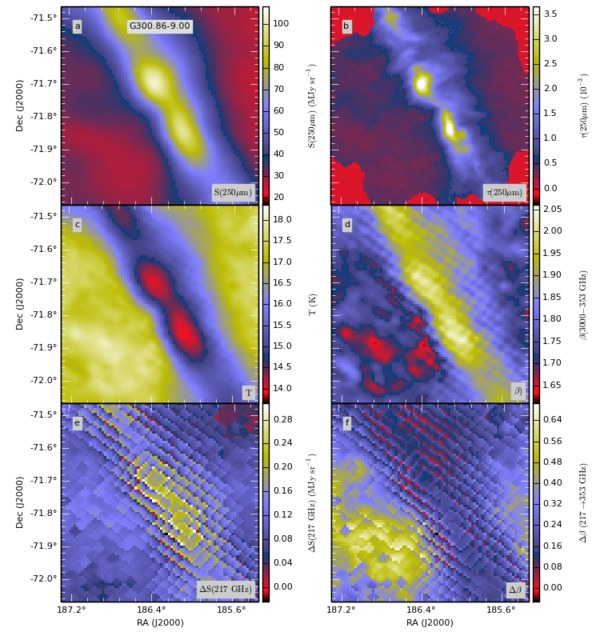}
\caption{
Fit of ($T$, $\beta$) in the field G300.86-9.00 using {\it IRAS} and {\it Planck}
data. The frames $a$, $c$, and $d$ show 250\,$\mu$m surface brightness, colour
temperature, and spectral index fitted to 3000\,GHz ({\it IRAS} 100\,$\mu$m),
857\,GHz, 545\,GHz, and 353\,GHz data, respectively. Frame $b$ shows, for comparison,
the higher resolution {\it Herschel} $\tau(250\,\mu {\rm m})$ map. Frame $e$ is the
surface brightness residual between the 217\,GHz observation and the above fit. The
last frame shows the difference between spectral indices derived with
3000\,GHz-353\,GHz and 3000\,GHz-217\,GHz data. If $\beta$ decreases with wavelength,
the residuals in frames $e$ and $f$ will be positive.
}
\label{fig:PI_FIG_105_main}
\end{figure}

Figure~\ref{fig:PI_FIG_105_main} shows an example of the fits performed
pixel-by-pixel using the {\it IRAS} 100\,$\mu$m and {\it Planck} 857\,GHz, 545\,GHz,
and 353\,GHz data. The field is G300.86-9.00, part of the Musca filament. In frame
$d$, the spectral index is seen to increase towards the filament and especially
towards the two dense clumps. 
The change is from $\beta\sim$1.75 in the background to a peak value of
$\beta$~=~2.03 in both clumps. In Fig. ~\ref{fig:PI_FIG_105_main}, frame $e$ shows
the residuals between {\it Planck} 217\,GHz data and the fit, and frame $f$ shows the
difference between the spectral indices estimated in frequency intervals
3000\,GHz--353\,GHz and 3000\,GHz--217\,GHz. These show that the spectrum is flatter at
the longest wavelengths. In relative terms (frame $f$), the change of spectral index as
a function of wavelength appears to be larger in the diffuse medium. However,
line-of-sight temperature variations tend to decrease the fitted value of $\beta$, thus
reducing the 217\,GHz excess. This may partly explain the spatial pattern in frame $f$.

\begin{figure}
\includegraphics[width=8.8cm]{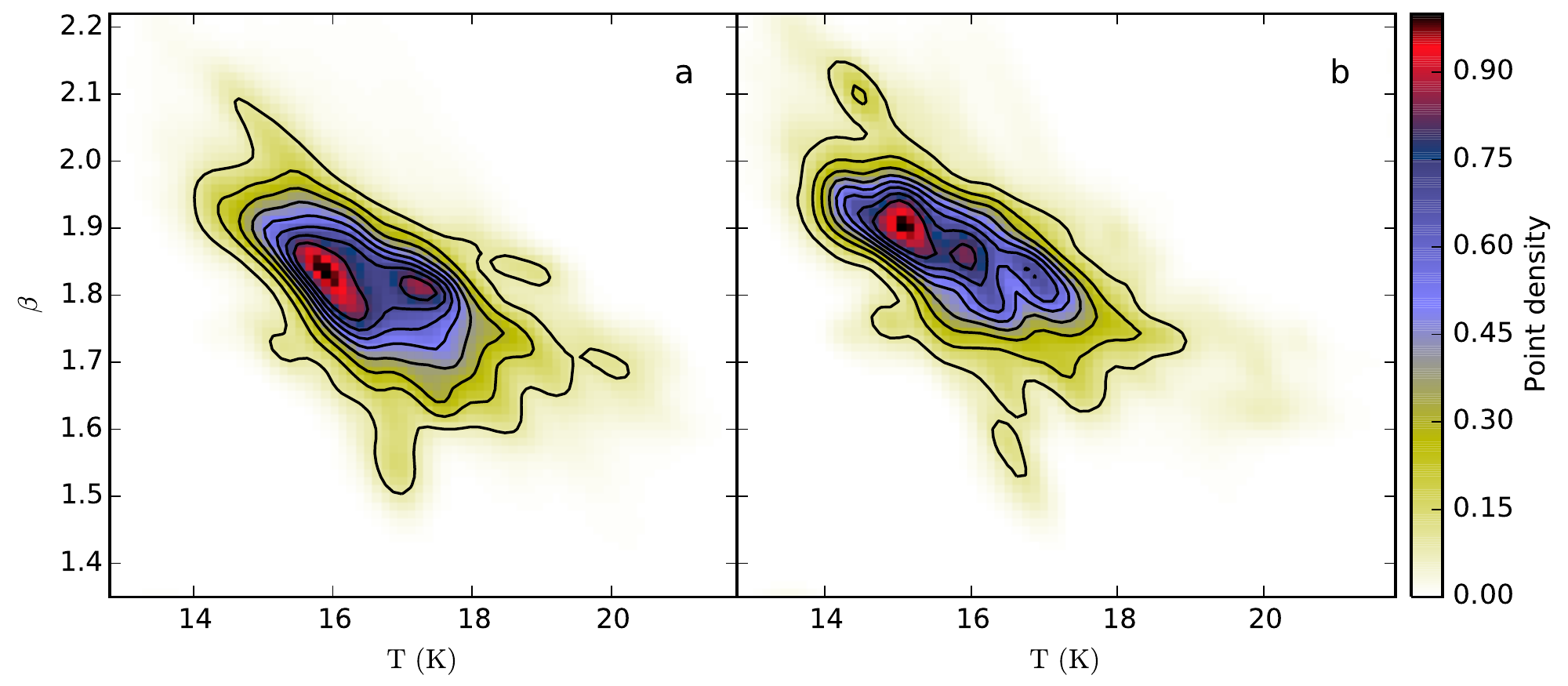}
\caption{
Distribution of $T$ and $\beta$ values estimated with IRIS 100\,$\mu$m data and {\em
Planck} data at frequencies 353\,GHz and higher. The values have been calculated
pixel-by-pixel using maps at a resolution of 5$\arcmin$. Left frame contains all pixels
inside the coverage of SPIRE maps. Right frame contains from each field only 25\% of the
pixels with the highest column densities (fourth quartile). The column densities have
been estimated using the same data and a fixed value of spectral index, $\beta=1.8$.
}
\label{fig:ref_PI_TB_pdf}
\end{figure}

Figure~\ref{fig:ref_PI_TB_pdf} shows correlations between $T$ and $\beta$. These
are calculated pixel-by-pixel, using IRIS and {\em Planck} maps ($\nu\ge353$\,GHz) at
5$\arcmin$ resolution. The figure shows a distribution similar to that of
Fig.~\ref{fig:histo_TB_Planck}. This suggests that spatial averaging from 5$\arcmin$ to
10$\arcmin$ scales has not significantly affected the appearance of
Fig.~\ref{fig:histo_TB_Planck}. In the second frame of Fig.~\ref{fig:ref_PI_TB_pdf} we
use only pixels with estimated column densities in the fourth quartile (25\% of pixels
with the highest column densities). To select the pixels, we use column densities that
are calculated with the same four frequencies but employing a fixed value of the
spectral index, $\beta$~=~1.8. Figure~\ref{fig:ref_PI_TB_pdf} shows that the distribution
is similarly elongated at high column densities but has shifted towards lower
temperatures and higher values of spectral index. 
The median values are $T=16.59$\,K and $\beta=1.80$ for all pixels and 
$T=15.92$\,K and $\beta=1.86$ for pixels with column density in the fourth quartile.

Appendix~\ref{app:pfit} shows the ($T$, $\beta$) fits for all 116 fields.
A 217\,GHz excess (i.e. a flattening of the dust SED at long wavelengths) is present in
most fields. In some fields (e.g. G202.23-3.38 and G253.71+1.93 the situation is less
clear and the residuals are at least partly negative. There are an additional two fields
where, like in G300.86-9.00, the 217\,GHz excess is positive but anti-correlated with
column density. Field G82.65-2.00 contains a long filament that is visible as a region of
lower $\beta$ and smaller 217\,GHz excess. In the densest parts of the filament the excess
is approximately zero. In G6.03+36.73 (LDN~183), the high column density clump is also
associated with a low 217\,GHz excess and slightly lower $\beta$. In these fields, the
temperature variations in the densest clumps may lower the apparent value of the spectral
index so much that the long-wavelength excess disappears.
We return to the question of the long wavelength excess in
Sect.~\ref{sect:wavelength}, in particular regarding the uncertainty of the CO
correction.

In a few fields, the 217\,GHz residuals show a local excess that may be caused by
free-free emission from H{\sc II} regions (e.g., G92.63-10.43, G107.20+5.52). These are
relatively easy to separate from the extended emission where the absolute 217\,GHz
excess follows the column density structure. The extended free-free emission should
already be accounted for by the subtraction of the low frequency model (see
Sect.~\ref{sect:Planck_data}).

\subsection{General results based on SPIRE data} \label{sect:SPIRE_only}

\begin{figure}
\includegraphics[width=8.8cm]{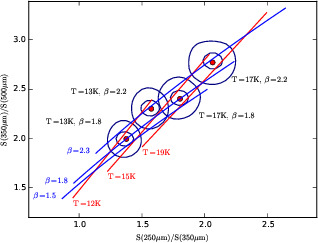}
\caption{
Illustration of the degeneracy between $T$ and $\beta$ in a SPIRE colour-colour
plot. The solid lines correspond to constant values of $\beta$ or $T$. The lengths
of the lines correspond to a temperature range of 10\,K--23\,K or a spectral index
range of 1.0--2.5, respectively. Four distinct parameter pairs are plotted as red
dots. The surrounding contours indicate the FWHM extent of the estimated parameter
distributions in case of 1\% or 5\% uncertainty in all surface brightness data.
}
\label{fig:plot_color_ratios}
\end{figure}

To gain access to smaller spatial scales, we must turn to {\it Herschel} data. The
observations extend only to 500\,$\mu$m, making them less efficient in constraining
the spectral index. We start by examining the colour temperature and spectral index
values determined with SPIRE data only. As shown in Fig.~\ref{fig:plot_color_ratios},
the parameters $T$ and $\beta$ are strongly degenerate, i.e., the colour variations
caused by temperature and spectral index changes follow almost parallel lines.  To
measure $\Delta \beta \sim0.2$ variations with these bands, the surface brightness
data must have an accuracy of almost 1\%. Nevertheless, we analyse these data in an
attempt to obtain results that are mostly independent from the previous {\it IRAS} and
{\it Planck} fits. This is particularly relevant because {\it IRAS} 100\,$\mu$m data
have been used by most studies where a negative correlation between colour temperature
and spectral index has been reported \citep[for example][]{Dupac2003, Desert2008,
planck2013-p06b}. Our only dependence on {\it IRAS} is in the 250\,$\mu$m band where
the zero point and gain correction depend on interpolation between {\it IRAS}
100\,$\mu$m and {\it Planck} 857\,GHz bands. Even there the weight of {\it IRAS} data
is relatively small, thanks to the large distance in wavelength.

\begin{figure}
\includegraphics[width=8.8cm]{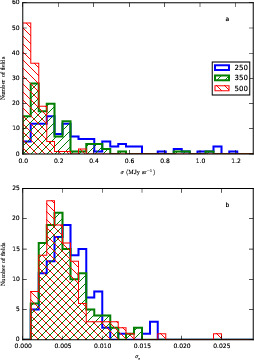}
\caption{
Distributions of estimated statistical errors $\sigma$ in SPIRE surface
brightness zero points for our fields. The values indicated are the
formal errors of linear fits between {\it Herschel} and the
interpolated IRIS and {\it Planck} data. The estimated error in the
250\,$\mu$m zero point is larger than 1.3\,MJy\,sr$^{-1}$ for three
fields. Frame b shows the same distribution as a function of relative
uncertainty, $\sigma$ divided by the median surface brightness of the map.
}
\label{fig:plot_zp_histograms}
\end{figure}

The relative calibration of SPIRE channels is estimated to be better than
2\%\footnote{SPIRE Observer's manual, \\ {\em
http://herschel.esac.esa.int/Documentation.shtml}}. However, when the gain calibration
was adjusted using {\it Planck} and IRIS data, the formal uncertainties were smaller.
The main concern is the accuracy of the zero point correction (see
Sect.~\ref{sect:Herschel_data}), which has a large impact at low column densities. The
distributions of the estimated statistical errors of the intensity zero points are shown
in Fig.~\ref{fig:plot_zp_histograms}. The median values are 0.24, 0.13, and
0.05\,MJy\,sr$^{-1}$ at 250\,$\mu$m, 350\,$\mu$m, and 500\,$\mu$m, respectively. For a
surface brightness of 10\,MJy\,sr$^{-1}$ the statistical errors are thus only at the
level of 1\%. 
Frame b shows the distribution of relative errors calculated using the median
surface brightness of each field. For the lowest column densities, the relative errors
can be larger by a factor of a few.
If the zero point errors are correlated between the SPIRE bands, the dispersion of
the derived $T$ and $\beta$ values would be smaller.
In other words, SPIRE data should be accurate enough to see definite trends, although
the absolute values may be affected by systematic errors.
Because the offset and slope of the linear fits ({\em Herschel} vs reference data)
are anticorrelated and the zero point and calibration errors are added in squares, the
total estimated uncertainty of SPIRE measurements may be overestimated. The effect becomes
negligible outside the faintest regions ($S(250\mu{\rm m})$ above
$\sim$20\,MJy\,sr$^{-1}$) where the errors are dominated by the multiplicative
component.

To study the general behaviour of colour temperature and spectral index values and to
examine the effect of gain and zero point uncertainties, we examined pixel samples
that were selected in each field using a uniform criterion on surface brightness. It
was already known that the highest column densities are associated with the lowest
colour temperatures (estimates based on a fixed value of $\beta$). We selected pixels
where the 350\,$\mu$m surface brightness was within 10\% of a selected surface
brightness level $S_0$. 
The use of 250\,$\mu$m surface brightness would result in a very similar selection
because the ratio between 250\,$\mu$m and 350\,$\mu$m surface brightness changes only
by $\sim$13\% in the temperature range 15$\pm$2\,K (assuming a fixed spectral index).
For a temperature of 15\,K and $\beta=1.8$, the ratio is $I(250\mu{\rm m}/I(350\,\mu{\rm
m})\sim$1.6.
In each field, we then calculated the average surface brightness values of the selected
pixels in the three SPIRE bands, and used this restricted set of measurements to
estimate $T$ and $\beta$. The motivation for this approach is that when surface
brightness is averaged over a large number of pixels, the statistical uncertainty
becomes very small. The errors are then dominated by calibration and zero point errors,
and potentially by other systematic errors that affect entire maps (or parts of them).

\begin{figure}
\includegraphics[width=8.8cm]{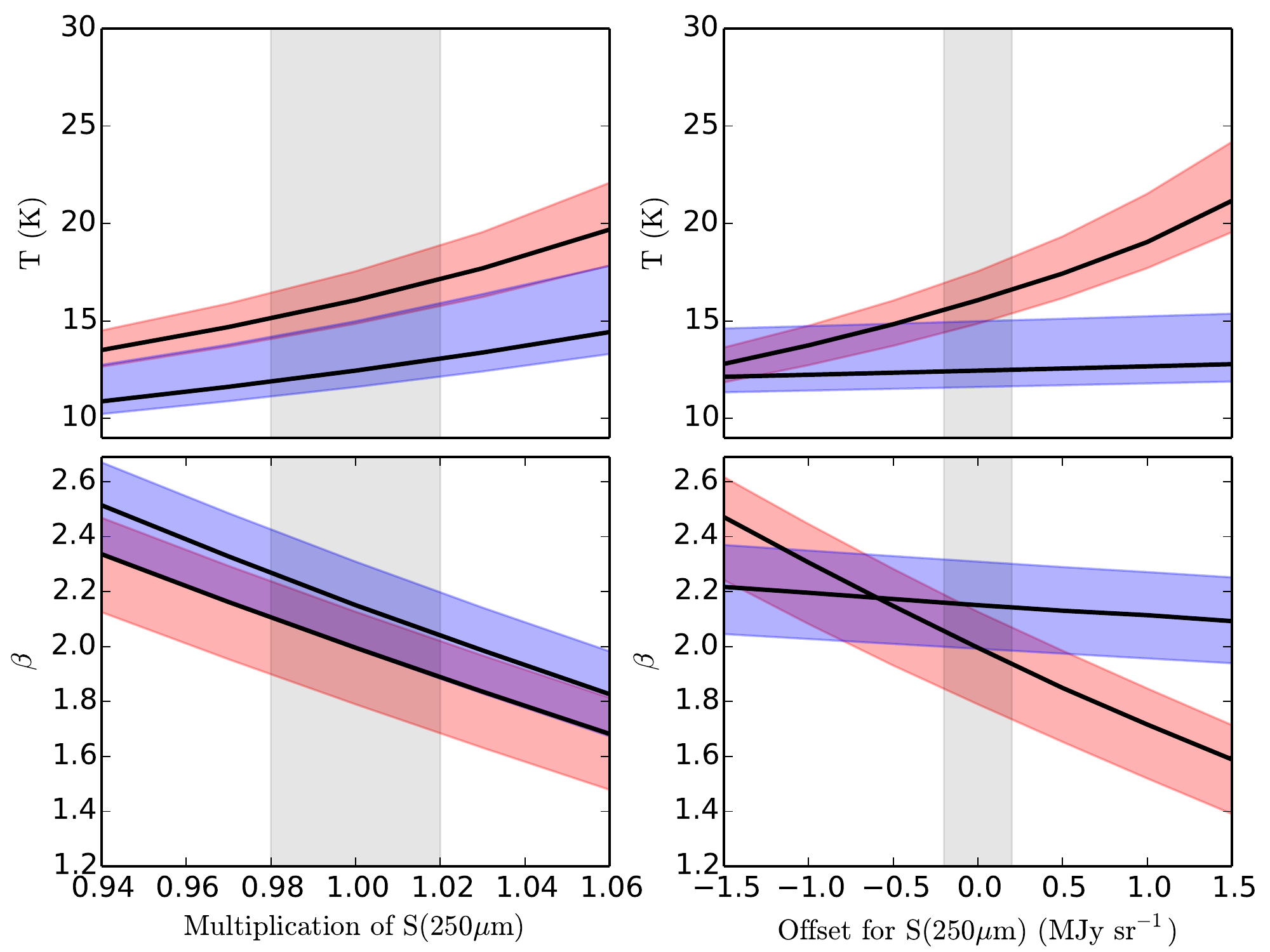}
\caption{
The $T$ and $\beta$ values for pixels with 350\,$\mu$m surface brightness
$\sim$10\,MJy\,sr$^{-1}$ (red band) and $\sim$100\,MJy\,sr$^{-1}$ (blue band). The
values are plotted as a function of multiplicative (left frames) and additive (right
frames) modifications of the measured 250\,$\mu$m surface brightness. The solid curves
are the median values (calculated over all fields) and the shaded region between
the 25\% and 75\% percentiles indicates the field-to-field dispersion. The grey
vertical bands correspond to the probable uncertainties of $\sim$2\% in the relative
gain calibration and $\pm$0.2\,MJy\,sr$^{-1}$ in the surface brightness zero point.
}
\label{fig:try_offsets_250}
\end{figure}

\begin{figure}
\includegraphics[width=8.8cm]{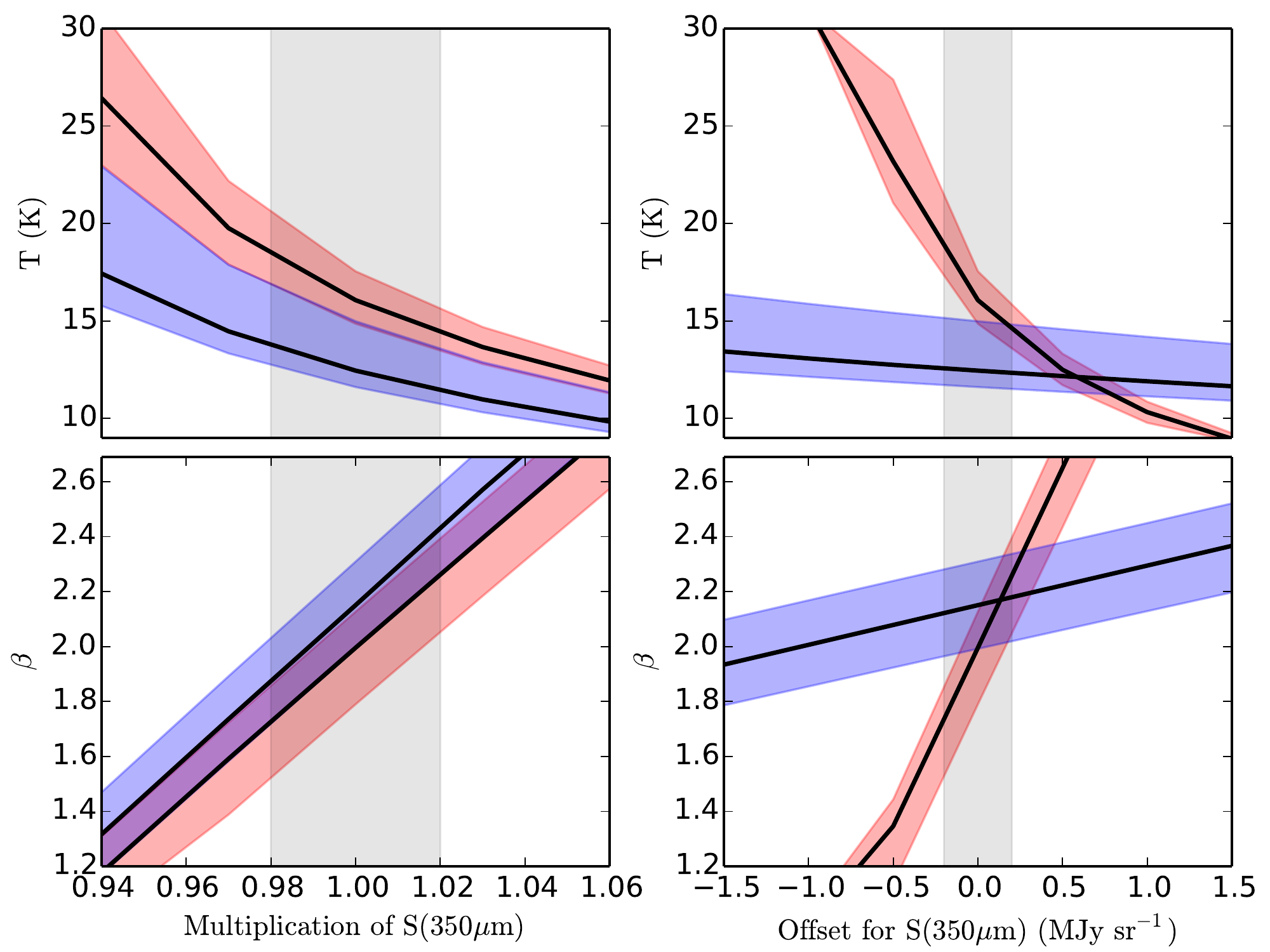}
\caption{
The same as Fig.~\ref{fig:try_offsets_250} but for hypothetical errors in the
350\,$\mu$m surface brightness.
}
\label{fig:try_offsets_350}
\end{figure}

Figure~\ref{fig:try_offsets_250} shows the results as a function of potential errors in
the 250\,$\mu$m data. The diffuse medium with $S(350\mu{\rm m})\sim$10\,MJy\,sr$^{-1}$
(red band) shows higher temperatures and lower spectral index values than the denser
regions with $S(350\mu{\rm m}) \sim $100\,MJy\,sr$^{-1}$ (blue band). The temperature
difference of $\Delta T=+5$\,K corresponds to $\Delta \beta$ of about $-0.2$.
The temperature difference of $\Delta T=+5$\,K corresponds to $\Delta \beta$ of about
$-0.2$. The left panels of Figure~\ref{fig:try_offsets_250} show that errors in the gain
calibration would affect the absolute values of $T$ and $\beta$, but would leave the
difference between the two pixel samples almost unchanged. The right panels, by
contrast, show that zero point errors at a level of 0.2\,MJy\,sr$^{-1}$ would have
little effect on the $S(350\mu{\rm m}) \sim $100\,MJy\,sr$^{-1}$ sample, but a much
larger impact on the $S(350\mu{\rm m}) \sim $10\,MJy\,sr$^{-1}$ sample. Nevertheless,
the difference between the two pixel samples disappears only if we assume that most
fields are affected by a systematic error of $\sim$0.6\,MJy\,sr$^{-1}$. This would be
larger than the typical estimated statistical uncertainties.

The situation changes somewhat if we consider similar uncertainties in the 350\,$\mu$m
band (Fig.~\ref{fig:try_offsets_350}). The multiplicative errors (left panel) again have
little importance but the results are sensitive to zero point errors. A shift of $\Delta
S(350\mu{\rm m})=0.2$\,MJy\,sr$^{-1}$ is enough to make $\beta$ values equal in the two
samples. A larger offset would result in the low surface brightness pixels having larger
$\beta$ values than the high surface brightness pixels. If statistical errors of the
zero points were $\sim$+0.2\,MJy\,sr$^{-1}$, the scatter of $\beta$ values should be
twice as large as what is actually observed. 
For example, for the 100\,MJy\,sr$^{-1}$ sample alone, the apparent $\beta$
values would be expected to range between 1.7 and 2.3 (1 $\sigma$ range; see
Fig.~\ref{fig:try_offsets_350}).
If the errors are statistical and without significant bias,
Fig.~\ref{fig:try_offsets_350} still suggests that $\beta$ is positively correlated with
column density. However, to be confident of this general conclusion, systematic zero
point errors would need to be constrained to a level of $\sim$0.1\,MJy\,sr$^{-1}$.

\begin{figure}
\includegraphics[width=8.8cm]{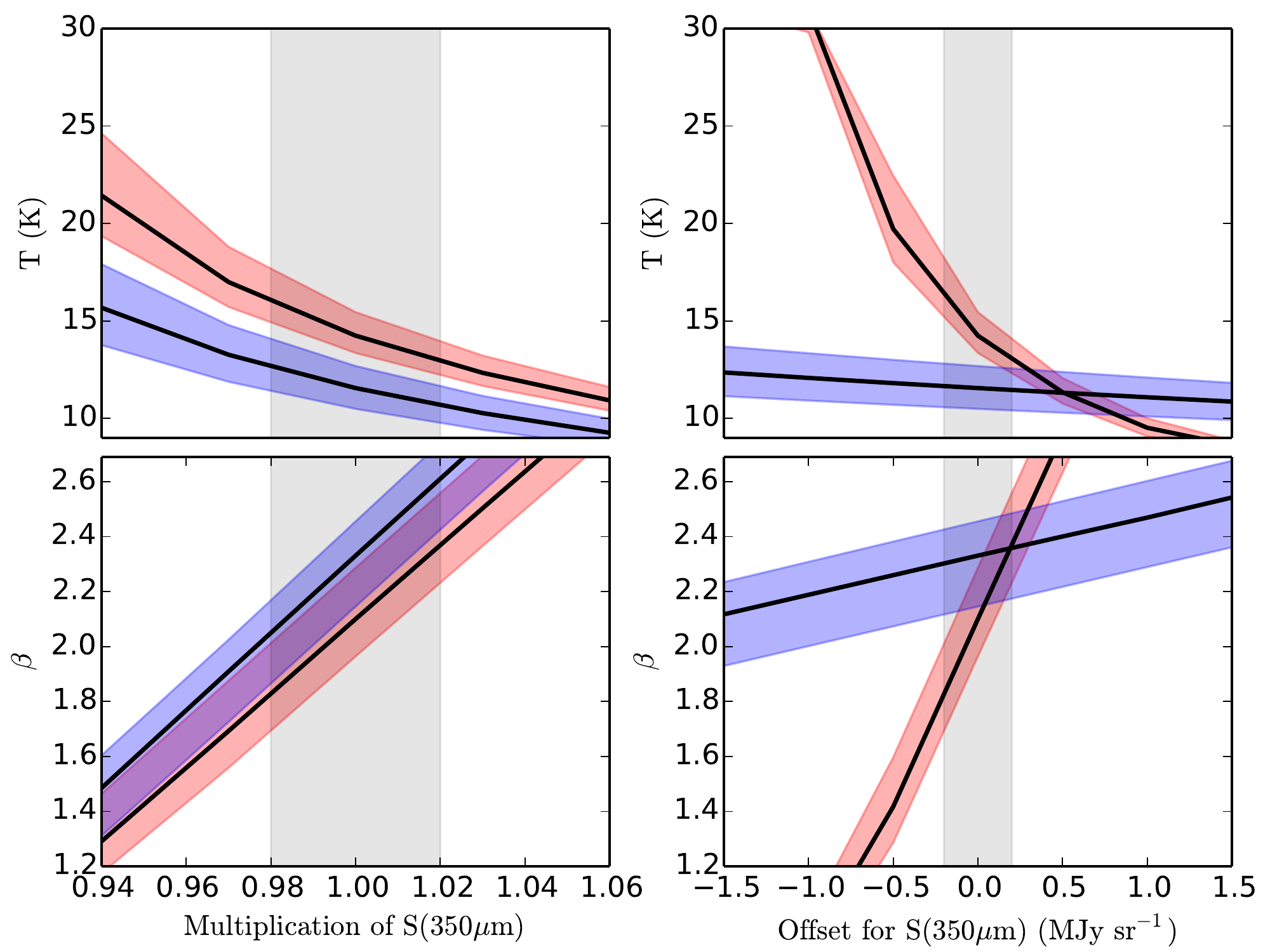}
\caption{
The same as Fig.~\ref{fig:try_offsets_350} but using surface brightness data after 
subtracting the local background.
}
\label{fig:try_offsets_350_bgsub}
\end{figure}

We can derive alternative $T$ and $\beta$ estimates where the zero point errors are
completely independent of those in Fig.~\ref{fig:try_offsets_350}. This is possible by
subtracting the local background. We use the reference regions defined in Table~1
of~\citet{GCC-III}. The results are shown in Fig.~\ref{fig:try_offsets_350_bgsub}. The
selected 350\,$\mu$m surface brightness levels are the same as above, 10\,MJy\,sr$^{-1}$
and 100\,MJy\,sr$^{-1}$. Because these are now background subtracted values, they
correspond to somewhat higher values of absolute surface brightness, the actual
difference depending on the field. Nevertheless, the results are remarkably similar to
Fig.~\ref{fig:try_offsets_350}. The difference between the 100\,MJy\,sr$^{-1}$ and
10\,MJy\,sr$^{-1}$ samples is $\Delta T=-4$\,K and $\Delta \beta=+0.2$.
As before, if the 350\,$\mu$m data contained a systematic error of -0.2\,MJy\,sr$^{-1}$,
the data would be consistent with no spectral index variations correlated with surface
brightness (or temperature). However, this would require the same systematic error in
most fields. While the zero point correction with {\it Planck} and IRIS data could
contain some bias, a similar systematic error is more difficult to understand in the
case of local background subtraction that is carried out independently in each field.

In summary, the SPIRE data gives some indication of spectral index variations, with
higher $\beta$ values being found towards regions of higher column density and lower
temperature. The results are made more plausible by the consistency between the
independent analyses conducted using absolute and relative zero points. Because SPIRE
bands cover a relatively narrow range of wavelength, data must have very high precision
for the absolute values of $T$ and $\beta$ to be measured. 
SPIRE channels may not be enough for accurate determination of the absolute values of
$T$ and $\beta$. Nevertheless, the above experiments show that these data have very
low systematic errors and, therefore, should provide reliable results when combined
with data at other wavelengths.

\subsection{General results from {\em Herschel} 160\,$\mu$m--500\,$\mu$m data} \label{sect:with_PACS}

The inclusion of PACS 160\,$\mu$m data helps to constrain the colour temperature and
should therefore also yield more accurate estimates of spectral index variations. 
This advantage is diminished by (1) the smaller size of the PACS maps, (2) the
uncertainty of the relative calibration between PACS and SPIRE, (3) a possible bias
resulting from the greater sensitivity to the warmest dust along the line-of-sight
\citep{Shetty2009a, Malinen2011, Juvela2012_Tmix, Pagani2015}, and (4) a larger
uncertainty in the determination of the zero point. The last point results from the
smaller size of the maps, and from the lower signal-to-noise ratio at shorter
wavelengths for cold dust emission. Points (3) and (4), together with the contribution
of VSGs, are the reason why we refrain from using the 100\,$\mu$m PACS data in our
analysis.

We continue to use a 2\% uncertainty for the gain calibration in the SPIRE bands
(band-to-band accuracy) and a 10\% uncertainty for the 160\,$\mu$m
measurements\footnote{https://nhscsci.ipac.caltech.edu \\
/sc/index.php/Pacs/AbsoluteCalibration}, the latter number also covering the uncertainty
in the relative calibration of instruments. 

In spite of its problems, the inclusion of 160\,$\mu$m data results in smaller
formal uncertainties of dust temperature, especially when estimated using a constant
value of $\beta$. Thus, in the following we compare data sets that are selected
based on dust temperature or dust column density instead of using surface brightness
thresholds (as in Sect.~\ref{sect:SPIRE_only}).
We start by looking for global trends in $\beta$ in relation to dust temperature. We
selected in each field all pixels where the colour temperature was within 20\% of
$T=14$\,K or $T=$18\,K, based on the fits made with a fixed value of $\beta=1.8$.
Fields that did not contain at least 20 pixels in both categories were excluded.

\begin{figure}
\includegraphics[width=8.8cm]{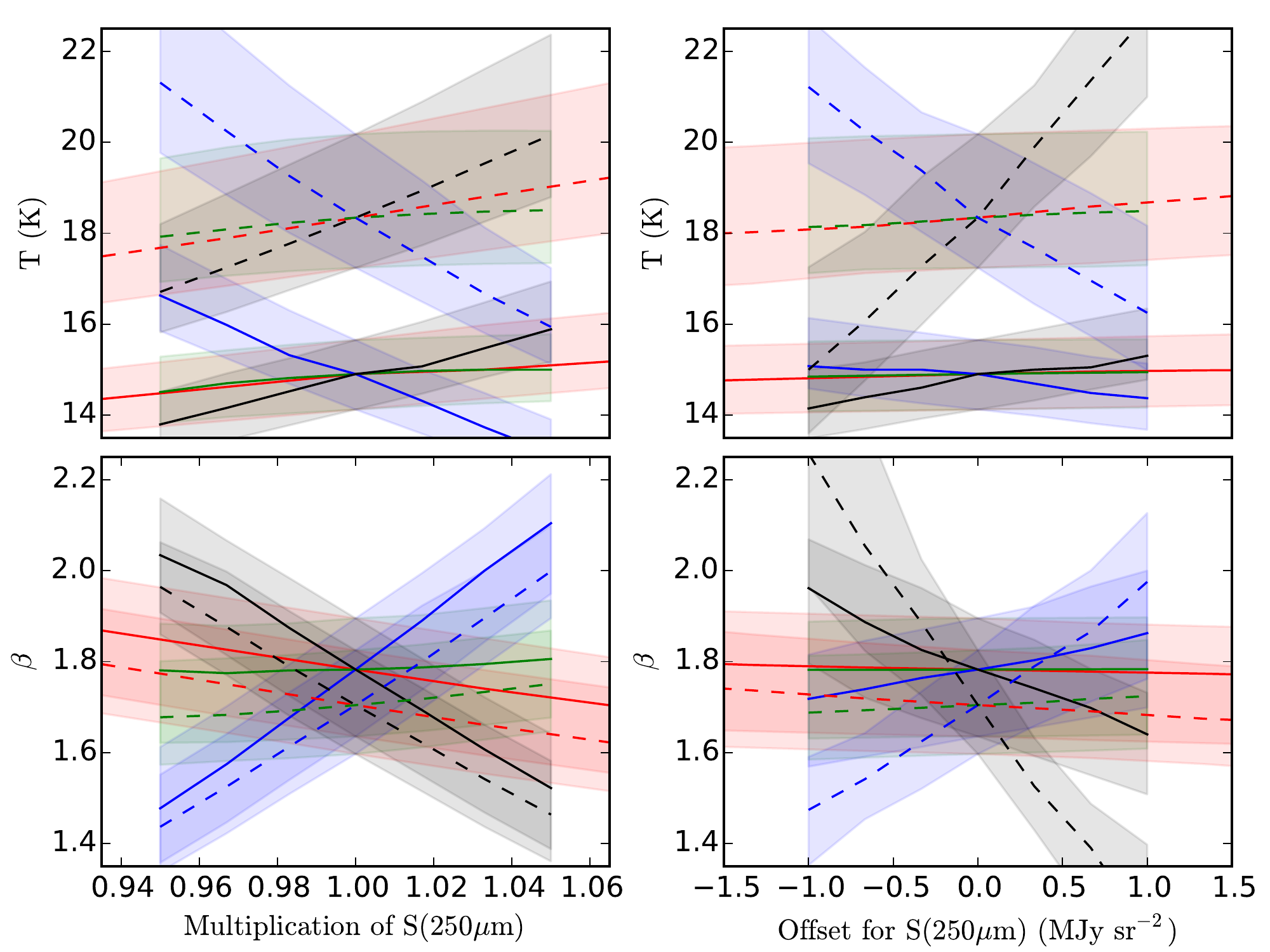}
\caption{
Comparison of median colour temperature and spectral index values derived from {\it
Herschel} 160\,$\mu$m-500\,$\mu$m data for all pixels where modified blackbody fits with
$\beta=1.8$ yield colour temperatures within 20\% of 14\,K (solid lines) or 18\,K
(dashed lines). The lines show the dependence on potential multiplicative or additive
errors in individual bands. The red, green, blue, and black colours correspond to the
four bands in order of increasing wavelength. Assuming no errors (centre of x-axis), the
median $\beta$ is $\sim 0.1$ higher for the sample of cold pixels. The shaded regions
indicate the field-to-field variation (interquartile ranges).
}
\label{fig:quartiles_4_T}
\end{figure}

Figure~\ref{fig:quartiles_4_T} shows the median values of $T$ and $\beta$ and their
sensitivity to surface brightness errors. As seen for the SPIRE data alone, the $\beta$
values are higher for the cold pixel sample, although the field-to-field scatter is
comparable to this difference. Before the 160\,$\mu$m  zero points were adjusted using the
{\it AKARI} comparison, the mean $\beta$ value of the warm pixels was actually $above$
the average $\beta$ of the cold pixels. This illustrates the sensitivity of our analysis
to zero point errors, at least for pixels with low brightness (i.e. at the
10\,MJy\,sr$^{-1}$ level). If we use local background subtraction, the situation remains
essentially the same, with the cold pixels now showing $\beta$ values that have
typically increased by $\sim0.1$. Because of the smaller size of the 160\,$\mu$m maps,
we cannot use the previous reference areas. Therefore, the background was determined by
selecting pixels with surface brightness between the 1\% and 20\% percentiles, and
calculating the average over the subset of those pixels that is common to all four
bands.

\begin{figure}
\includegraphics[width=8.8cm]{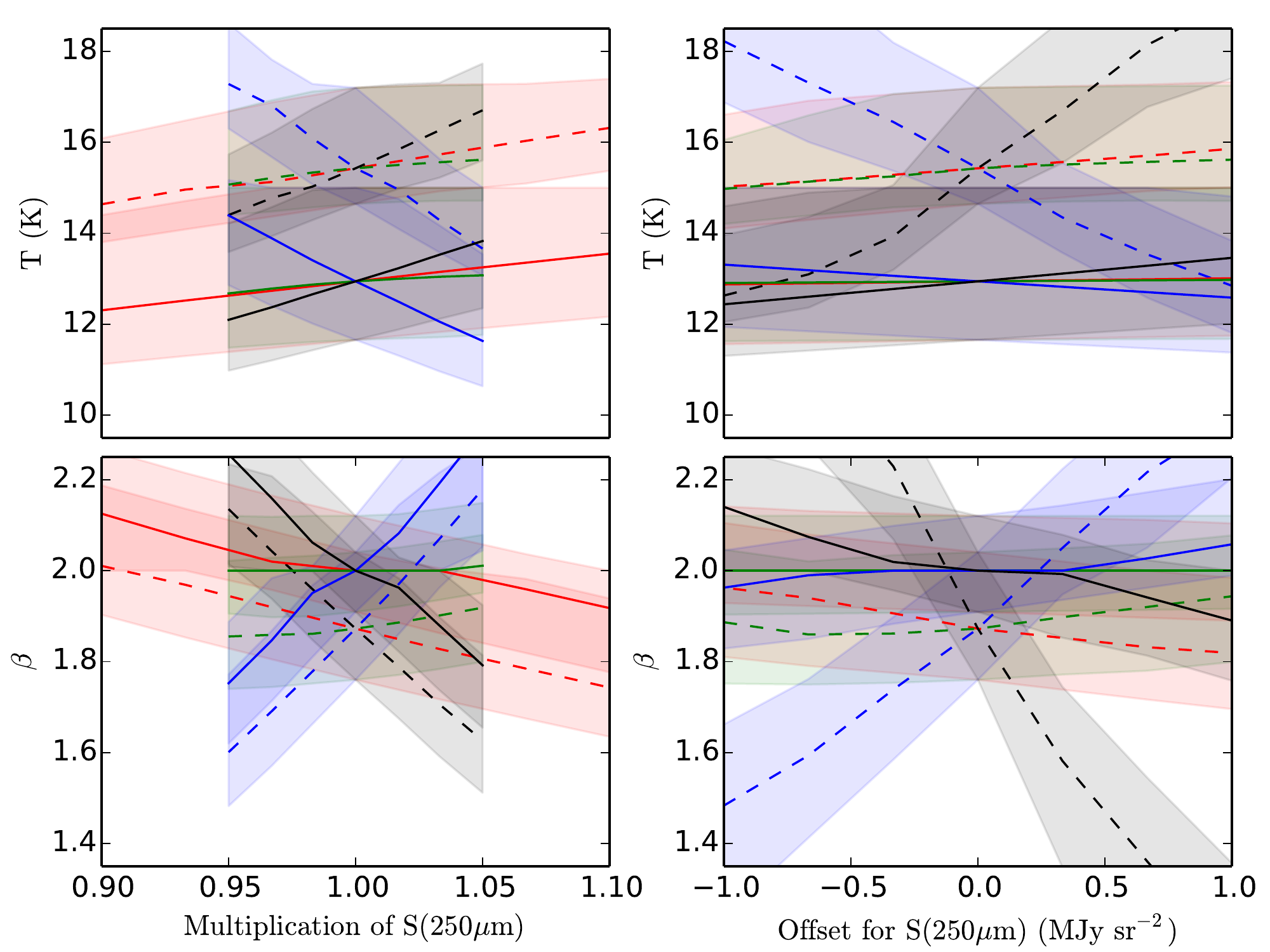}
\caption{
Similar to Fig.~\ref{fig:quartiles_4_T} but comparing pixels with $\tau(250\mu{\rm
m})\sim 1\times 10^{-3}$ (dashed lines) and $\tau(250\mu{\rm m})\sim 4\times 10^{-3}$
(solid lines) and using $T$ and $\beta$ values calculated from surface brightness
after the subtraction of local background.
}
\label{fig:quartiles_4_TAU_BGSUB}
\end{figure}

Systematically higher $\beta$ values are seen also in regions of higher optical depth.
In Fig.~\ref{fig:quartiles_4_TAU_BGSUB}, we compare pixels within 20\% of the values
$\tau(250\mu{\rm m})=1\times 10^{-3}$ and $\tau(250\mu{\rm m})=4\times 10^{-3}$. The
optical depths were derived with a fixed value of $\beta=1.8$ and are thus not affected
by the noise-induced anti-correlation between $T$ and $\beta$ that could lead to some
artificial correlations. Pixels in the higher optical depth interval have $\sim 0.1$
units higher $\beta$ values. The figure shows the results obtained with local background
subtraction. Using the estimated absolute zero points of surface brightness, the results
are similar, the difference of $\beta$ values being $\sim$0.08 instead of $\sim$0.10.

Thus, systematic changes in $\beta$ are seen for pixels selected based on either
temperature or optical depth. Part of the systematic effects may be masked by the
field-by-field variations of the local radiation field. Also, the strongest changes may
be limited in small regions within each field. Therefore, we need to investigate the
$\beta$ variations also directly on maps.

\subsection{{\it Herschel} $T$ and $\beta$ maps}  \label{sect:TBmaps}

The temperature and spectral index maps of all 116 fields are shown in
Appendix~\ref{app:hfit}. These were calculated with a basic MCMC method
\citep{Juvela2013_TBmethods} using either 160--500\,$\mu$m data (four bands) or only
data with 250--500\,$\mu$m (three bands). 
The error estimates are based on the assumption of 2\,\% and 10\,\% relative
uncertainty for the SPIRE and PACS channels, respectively. These estimates (converted
pixel-by-pixel to absolute uncertainties of the surface brightness) and the estimated
absolute uncertainties of the surface brightness zero points were added together in
quadrature. The PACS error estimate covers the uncertainty of the relative calibration
of the two instruments.

The appearance of the $T$ and $\beta$ maps and the dispersion between neighbouring
pixels gives some impression of the purely statistical noise but they are not useful for
estimating systematic effects that zero point errors may cause. In the worst case, a
zero point error can even change the sign of the $T$--$\beta$ correlation.
We calculated 100 realisations, varying the surface brightness zero points within their
estimated uncertainties, each calculation resulting in different $T$ and $\beta$ maps.
The realisations were sorted according to the average values of $\beta$. The
$\beta$ maps correspond to the 16\% and 84\% percentiles of the distribution
($\beta^{\rm min}$ and $\beta^{\rm max}$) were plotted and can be used to indicate how
sensitive the morphology of the $T$ and $\beta$ maps is to zero point errors. 
The $\beta^{\rm min}$ and $\beta^{\rm max}$ maps are estimated through $\chi^2$
minimisation. Because of the a priori constraint on $T$ and $\beta$, the level of the
corresponding MCMC $\beta$ maps can be slightly different.

\begin{figure}
\includegraphics[width=8.8cm]{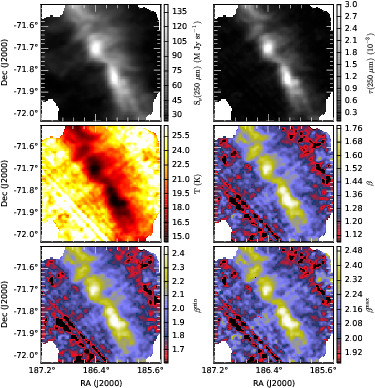}
\caption{
An example of negative correlation between colour temperature and spectral index. The
field is G300.86-9.00 in Musca. The frames, as indicated by their colour scales, show
250\,$\mu$m surface brightness, $\tau(250\,\mu{\rm m})$ optical depth, colour
temperature $T$, and the spectral index $\beta$. The bottom frames show the $\beta$ maps
that correspond to the 16\% and 84\% percentiles in a Monte Carlo study of zero point
errors (see text). The maps are based on SPIRE data and, for plotting only, they have
been convolved to a resolution of 1$\arcmin$.
}
\label{fig:sample_TB_PCC550}
\end{figure}

\begin{figure}
\includegraphics[width=8.8cm]{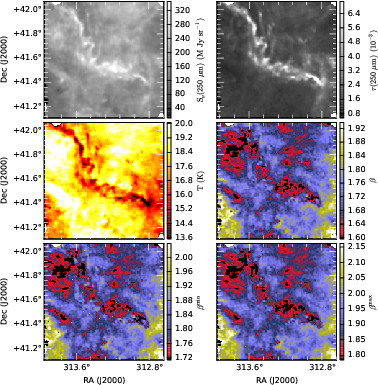}
\caption{
Field G82.65-2.00 as an example of some large scale positive correlation between
colour temperature and spectral index. The frames are as in
Fig.~\ref{fig:sample_TB_PCC550}.
}
\label{fig:sample_TB_G82.65-2.00}
\end{figure}

The maps derived from SPIRE data are shown in Appendix~\ref{app:hfit} and two examples
are included below (Figs.~\ref{fig:sample_TB_PCC550} and
\ref{fig:sample_TB_G82.65-2.00}).
An anti-correlation between $T$ and $\beta$ is seen in most fields. There are some
twenty fields for which the $\beta$ variation is not clearly visible above random
fluctuations or map artefacts. In a few cases (e.g.  G110.89-2.78 and G163.82-8.44), the
observed large-scale anti-correlation is no longer significant when the zero point
uncertainties are taken into account. Nevertheless, an anti-correlation is detected in
nearly 90\% of the fields, showing the ubiquity of this behaviour. However, in a given
field, the effect can be restricted to a small area or even a single core.

There are a few fields where the data show some positive correlations between $T$ and
$\beta$, mainly in small regions near strong point sources (e.g. in fields G343.64-2.31
and G345.39-3.97). In G82.65-2.00, which is shown in
Fig.~\ref{fig:sample_TB_G82.65-2.00}, the correlation between $T$ and $\beta$ is also
partly positive, this being associated with a cold filament that runs diagonally across
the field. At smaller scales, at the level of individual clumps, the parameters are
still preferentially anti-correlated.
In Sect.~\ref{sect:RTeffects}, we argue that a positive correlation could in some cases
be explained by radiative transfer effects that occur when the clouds have high optical
depth for the radiation heating the dust.

When PACS 160\,$\mu$m data are included, the results on the $T$--$\beta$ correlations
are qualitatively similar although the values often differ. The four-wavelength fits
have lower formal error estimates but are affected by the larger uncertainty of the
relative calibration of the two instruments and the different methods used in the zero
point correction of PACS and SPIRE data.
The median values over all fields, calculated over PACS coverage, are $T=16.9$\,K and
$\beta=1.89$ for the 250--500\,$\mu$m fits and $T=18.4$\,K and $\beta=1.66$ for the
160--500\,$\mu$m fits. In the latter case, part of the higher temperature may be
attributed to the fact that shorter wavelengths are more sensitive to line-of-sight
temperature variations and especially to the warmest regions. However, considering the
other sources of uncertainty, this cannot be concluded to be necessarily the main source
of the differences.
Based on the models shown in Appendix~\ref{app:RT}, the inclusion of the
160\,$\mu$m band could be expected to lead to changes $\Delta T \sim$0.2\,K and $\Delta
\beta \sim$-0.04 (in case of the $\tau_{\rm V}=10$\,mag model and compared to
250--500\,$\mu$m data).

\begin{figure*}
\includegraphics[width=18cm]{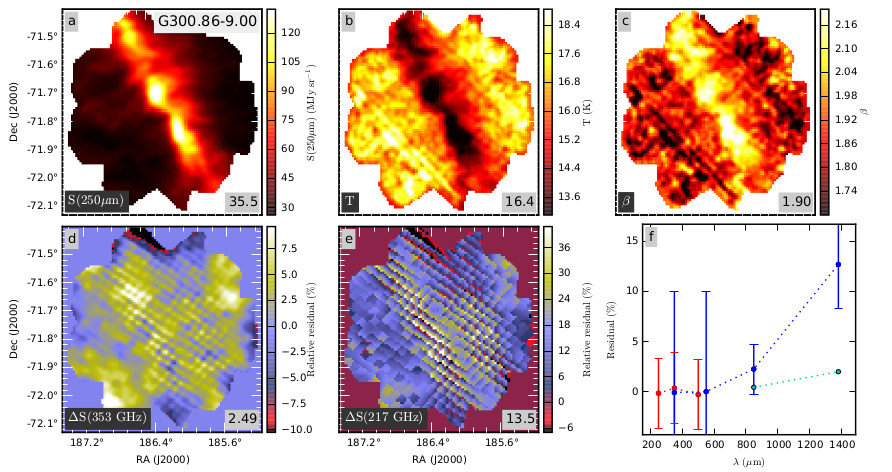}  
\caption{
Modified blackbody fits in field G300.86-9.00 using the combination of {\it
Herschel} and {\it Planck} data. The uppermost frames show the fitted intensity at
250\,$\mu$m, and the colour temperature and spectral index maps. The relative
residuals (observation minus model, divided by model prediction) are shown in frames
$d$ and $f$ for the {\it Planck} bands of 353\,GHz and 217\,GHz. Frame $f$ shows the
median residuals for the three {\em Herschel} bands (red symbols) and the four {\it
Planck} bands (blue symbols). 
The lower dotted line corresponds to twice the default CO correction.
The error bars correspond to the median error estimate
of the surface brightness over the map.
}
\label{fig:BRUTE_RES_main}
\end{figure*}

\subsection{Combined analysis of {\em Herschel} and {\em Planck} data} 
\label{sect:combined}

A final set of $T$ and $\beta$ maps was calculated using {\it Herschel} 250--500\,$\mu$m
data and the four {\it Planck} bands between 857\,GHz to 217\,GHz. The goal is to derive
the most reliable estimates of $\beta$, using the high resolution {\em Herschel}
observations while at the same time making use of {\em Planck} data to constrain $\beta$
at large scales. The {\it Herschel} data are kept at the $\sim 37\arcsec$ resolution
(see Sect.~\ref{sect:Herschel_data}). The 160\,$\mu$m maps are not used, mainly
because of their smaller size. The resulting maps of $T$ and $\beta$ are shown in
Appendix~\ref{app:hpfit}.

In these calculations the observed data could be used at their original resolution
but we continued to use {\em Planck} maps convolved to a $5.0\arcmin$ resolution. The 217\,GHz
data are used at the original 5.01$\arcmin$ resolution, the pixel sampling being visible
in the result maps. The {\it Planck} and IRIS data were colour corrected (see
Sect.~\ref{sect:PlanckIRIS}) but the colour corrections were not updated during the
fitting because their effect on surface brightness values is very small. Furthermore,
if the spectral index is not constant over the full wavelength range, it is not clear
that the joint fit would result in a better description of the SED shape over a given
band.

The model is defined on 28$\arcsec$ pixels and it consists of one intensity, colour
temperature, and spectral index value per pixel. The $\chi^2$ values that are
minimised consist of two components. For {\it Herschel} bands, the model is compared
to the 250--500\,$\mu$m data directly, treating each pixel as an independent
measurement. The second component of the $\chi^2$ value results from the comparison
with {\it Planck} data, requiring the convolution of the model predictions to the
5.0$\arcmin$ resolution. This makes the calculations computationally demanding, partly
because of the convolution operation itself but mainly because of the correlations
that the convolution operation introduces between model pixels. Nevertheless, the
procedure enables us to seek a solution at {\it Herschel} resolution that is still
constrained by the {\it Planck} data on larger scales.

The total probability of a model is a combination of the probabilities of fits to
three {\it Herschel} and four {\it Planck} bands. Considering that each {\it Herschel}
map has been rescaled using the same {\it Planck} and IRIS data, we adopt for SPIRE
channels an error estimate of 3.5\%. This is a compromise between the original 2\%
band-to-band uncertainty and the 7\% uncertainty of the absolute calibration. The zero
point uncertainties are added to these numbers in quadrature (see
Sect.~\ref{sect:TBmaps}). 
For the {\it Planck} data, we keep the 10\% error estimates for the 857\,GHz and 545\,GHz
channels and 2\% at lower frequencies. The 353\,GHz and 217\,GHz error estimates were
further increased with an amount corresponding to 50\% of the CO correction. Because of
this, the relative error estimates are often largest in the 217\,GHz band.

Because {\it Herschel} has many independent measurements over the area covered by a single
{\it Planck} beam, weighting according to the above listed error estimates gives {\it
Herschel} a large weight compared to {\it Planck}. However, at large scales the systematic
errors should be smaller for {\it Planck} because the data are taken from continuous
all-sky maps. {\em Herschel} maps can suffer from minor gradients (e.g. due to map-making)
that cannot be recognised because of the small size of the maps. Therefore, we increased
the relative weight of {\it Planck} data in the $\chi^2$ calculation by an {\em ad hoc}
factor of three. This is not a critical issue because the actual fit only provides a
convenient point of reference, the detailed information being contained in the fit
residuals. 

The calculated $T$ and $\beta$ maps are shown in Appendix~\ref{app:hpfit}. One example
is shown in Fig.~\ref{fig:BRUTE_RES_main}. The $T$ and $\beta$ maps (frames $b$ and $c$)
are relatively similar to the earlier {\em Herschel} fits, again showing clear
anti-correlation between the parameters. The area in Fig.~\ref{fig:BRUTE_RES_main} is
$\sim$80 times the solid angle of a 5$\arcmin$ Gaussian. Thus, considering the potential
effects of noise, the maps still contain a large number of independent samples. Frame
$f$ shows the median residuals for the seven bands used in the fits. The residuals are
small in all {\it Herschel} bands, indicating that these still carry most weight in the
overall fit. The fit is also relatively good in the {\it Planck} 857\,GHz--545\,GHz
bands, the wavelength range that is common with SPIRE. Although {\em Planck} shows
systematically slightly higher surface brightness values, the residuals are at a level
of one per cent and do not show a strong pattern that would be directly related to the
column density structure. 
However, in the 353\,GHz band (850\,$\mu$m) and especially in the 217\,GHz band
(1380\,$\mu$m) the residuals are clearly positive (see frames $d$ and $e$).  In
this case, a single modified blackbody that fits 250\,$\mu$m-550\,$\mu$m data is not a
good description of millimetre emission.
The lower curve corresponds to the larger CO correction that is twice the 
default correction (i.e., assumes line ratios (2-1)/(1-0)=1.0 and (3-2)/(1-0)=0.6, in
units of K\,km/s). In the field G300.86-9.00, this would be sufficient to make residuals
consistent with zero. However, over the area covered by SPIRE maps, {\em Planck} Type 2 CO
maps indicate a line ratio of $\sim$0.4 (a least squares fit gives a slope of 0.39 and
the direct ratio of the two maps gives 0.44 for the average of all pixels and 0.41 for
the brightest pixels in the fourth quartile of $W(1-0)$). The ratio is thus lower than
the assumed default ratio of 0.5. This suggests that the true residuals might be even
slightly larger than indicated by the {\em upper} curve of
Fig.~\ref{fig:BRUTE_RES_main}f.

\begin{table}
\caption{The median values and 1 $\sigma$ scatter of $T$ and $\beta$
calculated for all fields. In each field, the estimates correspond to
spatial averages over a FWHM=10$\arcmin$ beam or, in the case of first
row, FWHM=30$\arcmin$. The values in parentheses indicate the
1\,$\sigma$ scatter between the fields.}
\begin{tabular}{lll}
\hline
Data     &     $T$ (K)   &    $\beta$   \\
\hline
IRIS + Planck, FWHM=30$\arcmin$  &  16.92(1.36)  &  1.71(0.10) \\
IRIS + Planck, no 217\,GHz       &  16.03(1.31)  &  1.84(0.11) \\
IRIS + Planck, with 217GHz       &  16.43(1.59)  &  1.75(0.13) \\
SPIRE                            &  14.89(1.25)  &  2.04(0.16) \\
SPIRE + 160$\mu$m                &  17.46(2.09)  &  1.68(0.15) \\
SPIRE + Planck, with 217\,GHz    &  14.93(1.16)  &  2.03(0.12) \\
\hline
\end{tabular}
\label{table:summary}
\end{table}

\section{Discussion}  \label{sect:discussion}

\subsection{The main results and their robustness} \label{sect:general}

Table~\ref{table:summary} summarises some of the main conclusions of our analysis:
(i) $\beta$ values within the fields are larger than the average value at larger
scales (see Fig.~\ref{fig:plot_PI_10arcmin}), (ii) the $\beta$ values are smaller
when the fits are extended to 217\,GHz, and (iii) the inclusion of PACS 160\,$\mu$m
data results in higher $\beta$ values than an analysis using the SPIRE bands only.
The first two facts are likely to reflect real physical changes, such that high
density regions exhibit with steeper emission spectra. On the other hand, the
difference between the SPIRE and SPIRE+160\,$\mu$m results may indicate the presence
of small systematic errors. 

Figure~\ref{fig:plot_PI_10arcmin} showed that there is a clear difference between
the 10$\arcmin$ and 30$\arcmin$ scales. Because the larger beam also covers diffuse
regions around our target clouds, this already suggests that $\beta$ is correlated
with column density. More direct evidence of dependence between dust temperature,
column density, and spectral index is provided by the maps presented in
Appendices~\ref{app:pfit}-\ref{app:hpfit}. 
In Fig.~\ref{fig:plot_PI_10arcmin}, 
$\beta$ values were $\sim0.1$ units lower when
the 217\,GHz band was included in the fits, the exact value depending on the relative
weight given to the different bands. In Sect.~\ref{sect:combined} and
Appendix~\ref{app:hpfit}, the 217\,GHz residuals are seen to be clearly positive with
only a few exceptions. The wavelength dependence of $\beta$ is examined further in
Sect.~\ref{sect:wavelength}.

The $T$ and $\beta$ maps were estimated with {\em Planck} data
(Appendix~\ref{app:pfit}), with {\em Herschel} data (Appendix~\ref{app:hfit}) data,
and with joint fits of both {\em Planck} and {\em Herschel} data
(Appendix~\ref{app:hpfit}). The anti-correlation between $T$ and $\beta$ cannot be
the result of pure statistical noise because that would not result in spatially
correlated variations in the maps. The only possible explanation would be some kind
of a systematic error that affects both {\em Planck} and {\em Herschel} data in a
similar fashion. This is technically possible because the SPIRE data were scaled and
zero-point corrected using IRIS and {\em Planck} data. However, in
Sect.~\ref{sect:results}, we showed that zero point errors are important only at low
column densities. Furthermore, the SPIRE-only results remained essentially the same
when we used local background subtraction instead of absolute zero points.  We also
demonstrated that gain calibration errors are not able to significantly affect the
$T-\beta$ relation, not even when the data consists of SPIRE bands only
(Sect.~\ref{sect:SPIRE_only}). The {\em Herschel} results were relatively similar to
our previous analysis where the SPIRE gain calibration was not tied to {\em Planck}.
The final possible error source are mapping errors that could produce
artificial $T-\beta$ anti-correlation \citep[e.g.][]{Juvela2013_TBmethods}. 
Mapping artefacts are difficult to detect in surface brightness data even when their
effect on $T$ and $\beta$ estimates should be easily recognisable. The maps in
Appendix~\ref{app:hfit} are thus a good indicator of this kind of errors. Artefacts
would appear as spatial features that are different between the 160-500\,$\mu$m and
250-500\,$\mu$m fits (especially regarding artefacts in PACS data) and/or uncorrelated
with surface brightness. As an example, one can look at the data on the field LDN~134
(G4.18+35.79) in Appendix~\ref{app:hfit}. There are clear features that are following
changes in column density and are similar in the two band combinations. However, the
$\beta_{4}$ map also exhibits a north-south gradient that is correlated with neither the
column density nor the morphology of the $\beta_{3}$ map. This suggests that the PACS
map suffers from a surface brightness gradient (possibly caused by the strong emission
that extends over the southern map boundary and thus complicates map-making). Smaller
scale artefacts are clear, for example, in the south corner of the $T_3$ and $\beta_3$
maps of the field G126.63+23.55. However, in most cases the two band combinations give
comparable results, especially for compact structures, and the observed $\beta$ changes
are well correlated with column density variations. This remains true even in the case
of both G4.18+35.79 and G126.63+23.55.
We therefore conclude that $T-\beta$ anti-correlation is a common physical phenomenon,
which we robustly detect in most of our fields. 

Regarding the change of $\beta$ at wavelengths longer than 500\,$\mu$m, the result
could be affected mainly by errors in the correction for CO emission. However, even
in the joint fits of {\em Herschel} and {\em Planck} data, the 217\,GHz excess is,
after the removal of the expected CO contamination, typically more than 20\% of the
total signal. This is large compared to the expected CO contamination and is thus
very unlikely to be caused by inaccuracy of the CO correction. Furthermore, the
excess is detected clearly also in diffuse regions where the relative importance of
the CO correction should be smaller.

\begin{figure*}
\sidecaption
\includegraphics[width=12cm]{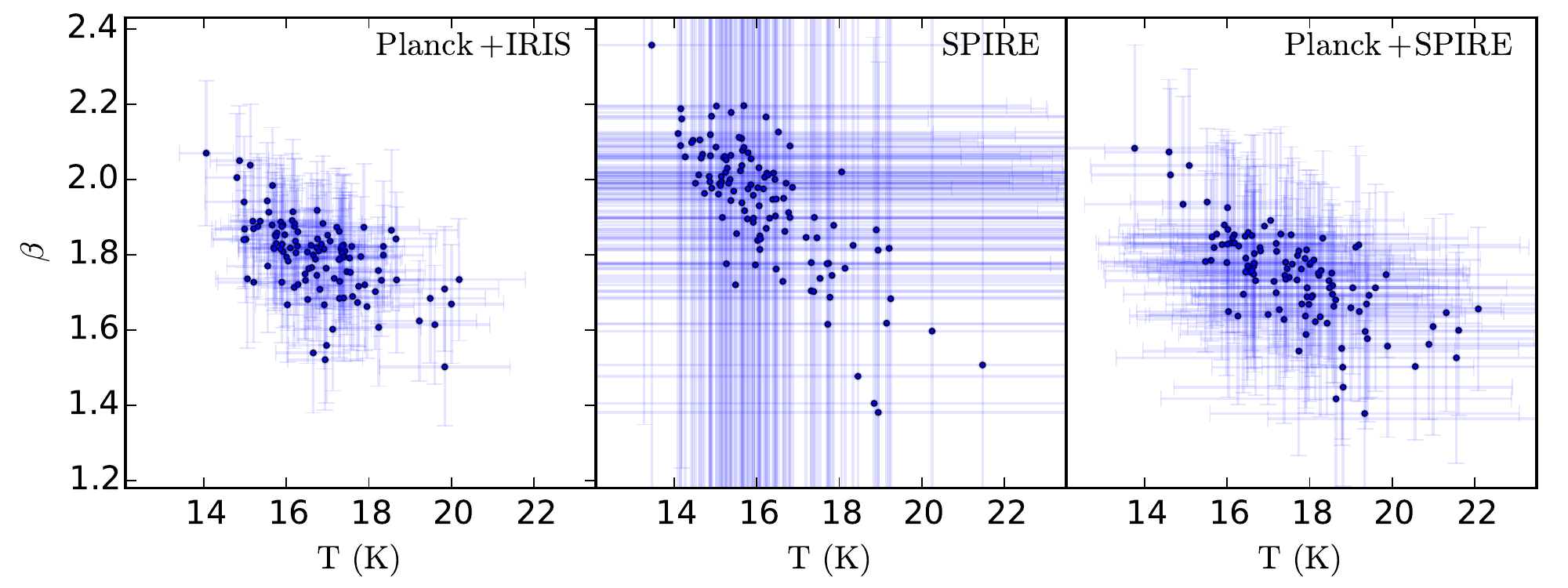}
\caption{
The ($T$, $\beta$) values based on the average surface brightness of the fields.
The values are based on {\it Planck} 857\,GHz-353\,GHz and IRIS 100\,$\mu$m data
(left frame), SPIRE channels (centre frame), or all the previous bands combined
(right frame).
}
\label{fig:stats_TB_1}
\end{figure*}

\subsection{$T$ and $\beta$ correlations between fields} \label{sect:correlations}

\begin{figure}
\includegraphics[width=8.8cm]{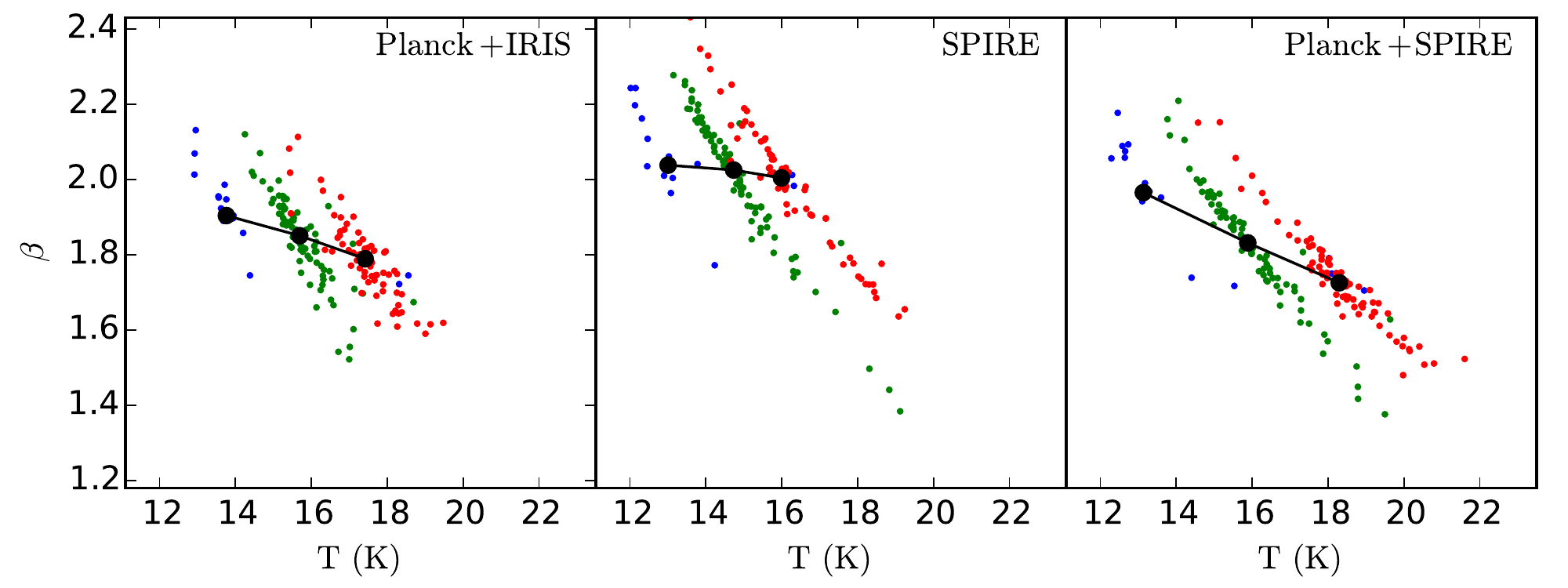}
\caption{
As in Fig.~\ref{fig:stats_TB_1} but using the average surface brightness in three
temperature intervals that are determined with SPIRE data and a fixed value of the
spectral index, $\beta=1.8$. The blue, green, and red symbols correspond to
(14$\pm$0.5)\,K, (16$\pm$0.5)\,K, and (18$\pm$0.5)\,K, respectively. Each point
corresponds to one field. The black lines connect the median values of the three
temperature samples.
}
\label{fig:stats_TB_2}
\end{figure}

\begin{figure}
\includegraphics[width=8.8cm]{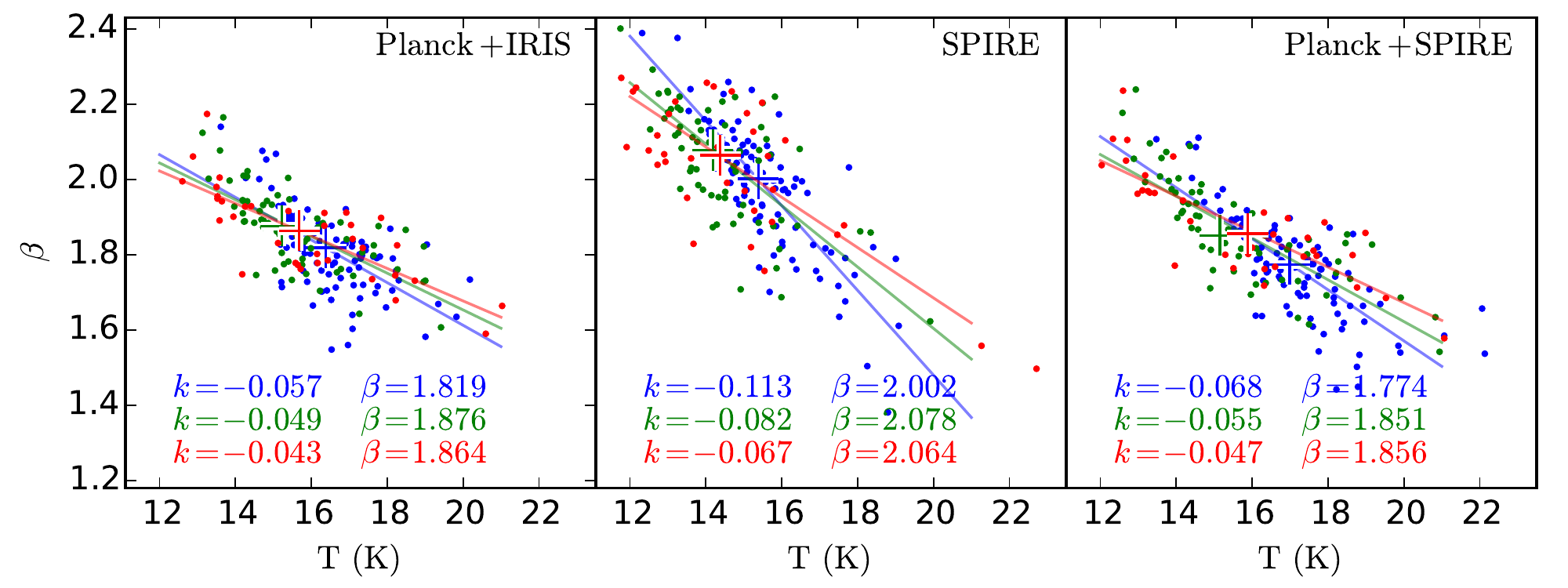}
\caption{
As previous figures but using average surface brightness values in three
$\tau(250\mu{\rm m})$ intervals. The blue, green, and red symbols correspond to
$\tau(250\mu{\rm m})$ equal to $(1\pm 0.5)\times 10^{-3}$, $(2\pm 0.5)\times
10^{-3}$, and $(3\pm 0.5)\times 10^{-3}$ respectively. The optical depths are
derived from a fit to the SPIRE data using a fixed value of $\beta=1.8$. The lines
are least-squares fits to the points of the same colour and the slope values $k$
are listed in the frames (unit K$^{-1}$). The median values are indicated with
crosses (in the middle frame the red and green crosses overlap) and the
median values of $\beta$ are also given.
}
\label{fig:stats_TB_3}
\end{figure}

In Fig.~\ref{fig:stats_TB_1}-\ref{fig:stats_TB_3} we examine the colour temperature
and spectral index values calculated using the average surface brightness of each
field. Fig.~\ref{fig:stats_TB_1} shows the ($T$, $\beta$) relations derived with
(1) IRIS and {\it Planck}, (2) SPIRE, and (3) SPIRE and {\it Planck}. The 217\,GHz
{\it Planck} band was omitted from these fits. The error bars include statistical
and calibration errors, including at 353\,GHz the error associated with the CO
correction. For the SPIRE channels we used an error estimate of 7\%. All three
frames show some degree of $T$-$\beta$ anti-correlation but there are also
differences. For example, the joint fit of {\it Planck} and {\it Herschel} data
results in a wider temperature distribution. When {\it Herschel} data are used, the
error bars strongly overestimate the scatter. This is particularly clear in the
middle frame. Thus, Fig.~\ref{fig:stats_TB_1} shows that the procedure for the
re-scaling of {\it Herschel} data (including the zero points) has been fairly
accurate. This also partly justifies the lower 3.5\% relative uncertainty used for
SPIRE channels in Sect.~\ref{sect:combined}.

Figure~\ref{fig:stats_TB_2} shows similar plots using pixels in given temperature
intervals. The pixel selection is based on $T$ and $\tau(250\mu{\rm m})$ values
calculated with SPIRE data and a constant value of $\beta=1.8$. Therefore, the
selection is not directly affected by the $T$-$\beta$ anti-correlation (whether
physical or noise-produced). In Fig.~\ref{fig:stats_TB_2}, the three temperature
intervals are clearly separated but each forming a narrow distribution that extends
over a wide range of $\beta$ values. The overall dispersion might be caused by noise
but also by the fact that for $\beta=1.8$ a given colour temperature interval
actually contains different $\beta$ values. When $\beta$ was allowed to vary freely,
some points have moved to higher temperature, especially from the lowest $\sim14$\,K
bin. The fact that the orientation of these elongated clouds of points is not the
same as that of the general $T$-$\beta$ anti-correlation does not directly indicate
which of the two effects is stronger, i.e., to what extent the anti-correlation is
caused by noise. However, the comparison of the median values of the three
temperature samples shows the anti-correlation, even when the samples were selected
based on temperatures calculated with a fixed value $\beta=1.8$.

Figure~\ref{fig:stats_TB_3} compares values calculated for three column density
intervals. The ($T$, $\beta$) distributions of the three optical depth intervals
overlap but the median $\beta$ increases with column density. The $\beta$ vs. $T$
slope also becomes flatter with increasing $\tau(250\mu{\rm m})$ for all three
combinations of surface brightness data. Because the x-axis values have been
calculated with a fixed value of $\beta$, the noise-induced anti-correlation should
be reduced. However, higher $\tau(250\,\mu{\rm m})$ values usually correspond to
higher signal-to-noise ratios and the flattening could be partly a noise effect. In
the case of SPIRE data (middle frame), all relations are steeper, this also being
possibly a noise effect.
The $\beta$ values are not quite systematically higher for the
higher column densities, although the lowest column density sample
does correspond to the highest temperatures and the lowest
values of $\beta$. While Fig.~\ref{fig:quartiles_4_TAU_BGSUB} used the
average surface brightness of all pixels within a given optical depth
range, Fig.~\ref{fig:stats_TB_3} examines each field separately. This
increases the risk that result become affected by local artefacts
(e.g, near map boundaries). Depending on the range of column densities
within a map, each plotted symbol may correspond to a very different
number of individual map pixels.

\begin{figure}
\includegraphics[width=8.8cm]{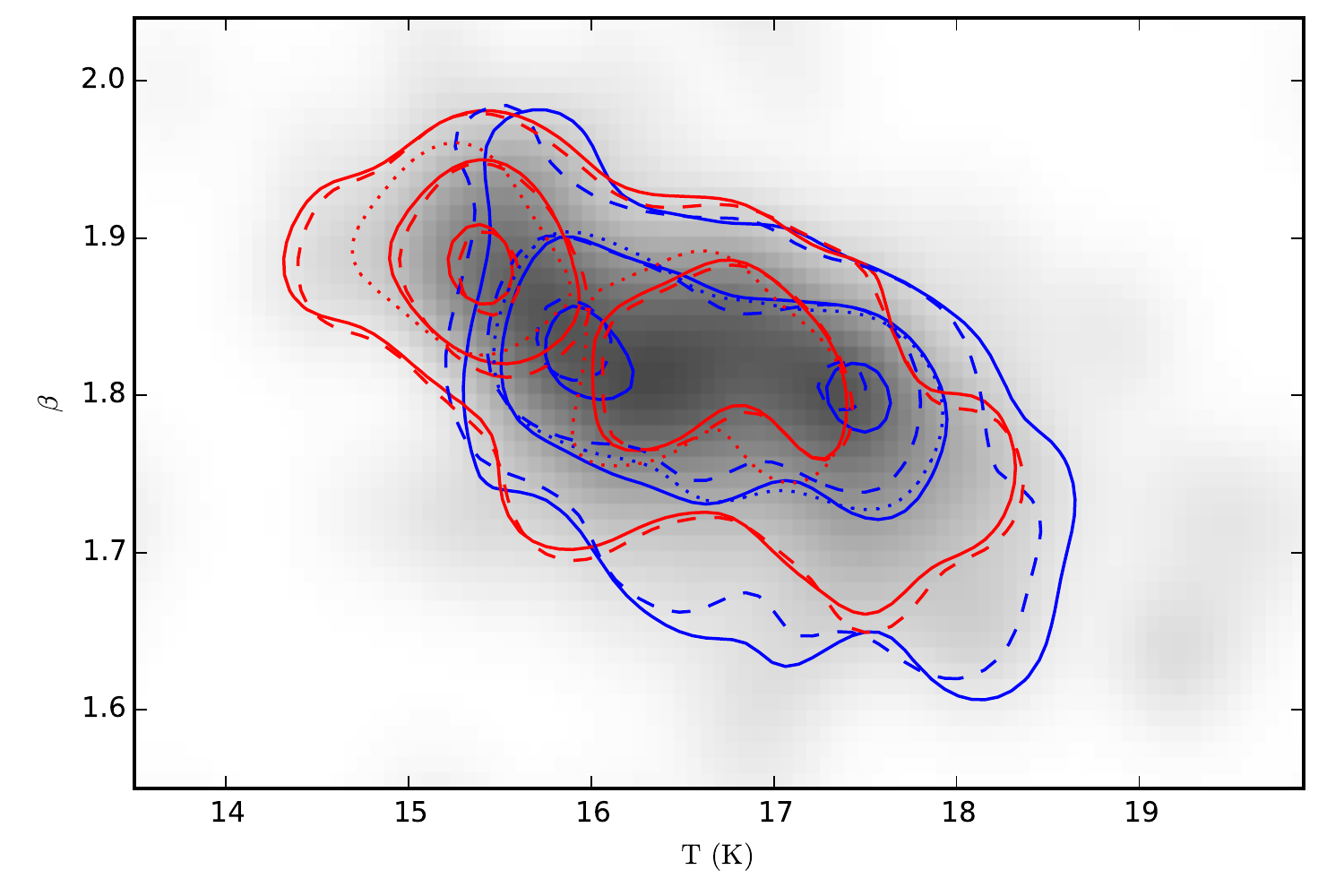}
\caption{
Effect of spatial resolution on field-averaged parameters. The grayscale image shows the
distributions of ($T$, $\beta$) values calculated from IRIS and {\em Planck} data for pixels.
The contours show the distribution separately for pixels below (blue contours) and above
(red contours) the average column density in each field. The contours are drawn at
levels of 20\%, 50\%, and 90\% of the maximum value (data points per area) for 
distributions calculated from surface brightness data at 5$\arcmin$ (solid contours) or
10$\arcmin$ (dashed contours) resolution. The dotted contours (50\% level only)
correspond to a case where the surface brightness values of all pixels below or above
the median value (per field) are averaged before ($T$, $\beta$) calculation.
}
\label{fig:REF_fig}
\end{figure}

Spatial averaging of surface brightness data should increase the bias caused by the
mixing of spectra with different colour temperatures. Figure~\ref{fig:REF_fig} compares
$T$ and $\beta$ estimates derived from IRIS and {\em Planck} surface brightness maps at
a resolution of 5$\arcmin$ and 10$\arcmin$. The data were further divided to low and
high column density parts using column densities calculated using a fixed value
$\beta=1.8$. 
As a third alternative, we averaged the surface brightness of all pixels above and below the
median column density, thus calculating two $T$ and $\beta$ values per field. This last
alternative is expected to maximise the bias and should lead to smaller values of the
spectral index.
The two column density samples are displaced relative to each other, the high column
density pixels extending towards lower temperatures and higher values of the spectral
index. However, the differences between the 5$\arcmin$ and 10$\arcmin$ resolution data
are very small. In particular, there is no systematic shift in $\beta$ values. This
suggests that the effects of temperature mixing are not very pronounced at scales above
5$\arcmin$, up to scales approaching the size of individual fields.

\subsection{Correlation with Galactic location} \label{sect:location}

\begin{figure}
\includegraphics[width=8.8cm]{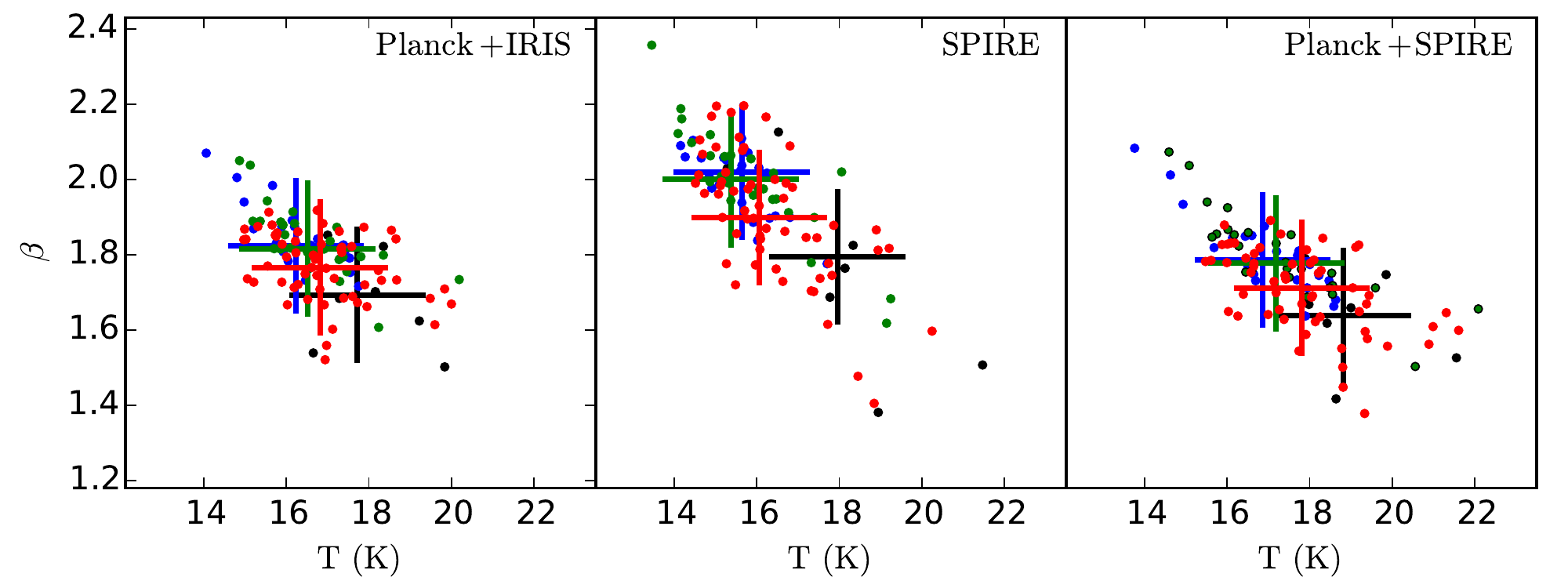}
\caption{
As previous figures but with colours indicating the estimated distance $d$ of
the field. The blue, green, and red symbols correspond to distances $d\le200$\,pc,
$200\,{\rm pc}<d\le 500$\,pc, and $d>500$\,pc, respectively.  The median values are
indicated with the crosses of the same colour. The fields without reliable distance
estimates are marked with black symbols.
}
\label{fig:stats_TB_4}
\end{figure}

Figure~\ref{fig:stats_TB_4} shows ($T$, $\beta$) values for fields in different distance
intervals. The fields with low $T$ and high $\beta$ values are more likely to be located
within 500\,pc. For individual clumps a dependence on distance could be explained by
beam dilution but Fig.~\ref{fig:stats_TB_4} is concerned with the average surface
brightness of entire fields. Nevertheless, in our sample the nearby fields are more
likely to contain a single cold clump that dominates the field-averaged parameters.
Thus, the dependence between $\beta$ and distance is likely to be mainly a selection
effect.

\begin{figure}
\includegraphics[width=8.8cm]{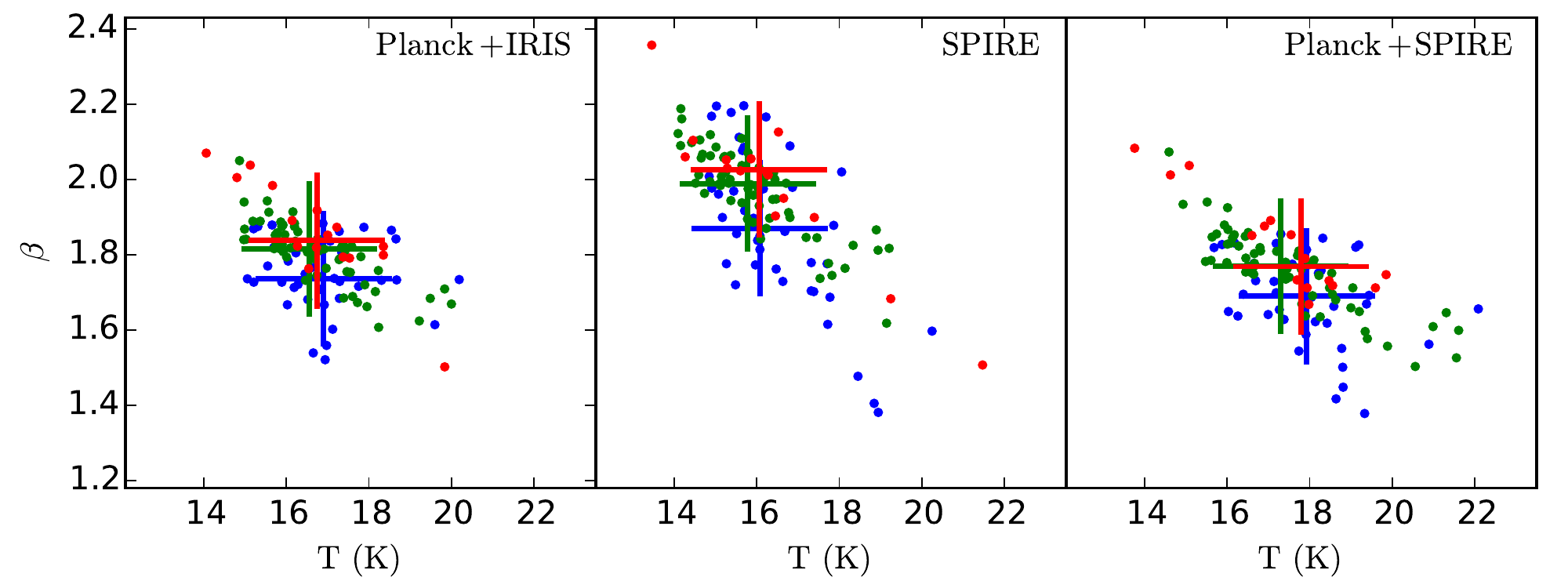}
\caption{
As previous figures but with colours indicating the Galactic latitude. The blue,
green, and red symbols correspond to $|b|<5 \degr$, $5\degr \le |b| <20 \degr$, and
$|b|\ge 20\degr$, respectively. The crosses indicate the corresponding median
values.
}
\label{fig:stats_TB_5}
\end{figure}

Figure~\ref{fig:stats_TB_5} shows the correlation with Galactic latitude. The
scatter is again large but high latitude clouds are preferentially associated with
higher $\beta$ values. The origin of this correlation is partly the same as in
Fig.~\ref{fig:stats_TB_4}, high latitude clouds also being nearby. However, compared
to the case of different optical depth intervals (Fig.~\ref{fig:stats_TB_3}), the
median temperature is less latitude-dependent.

\begin{figure}
\includegraphics[width=8.8cm]{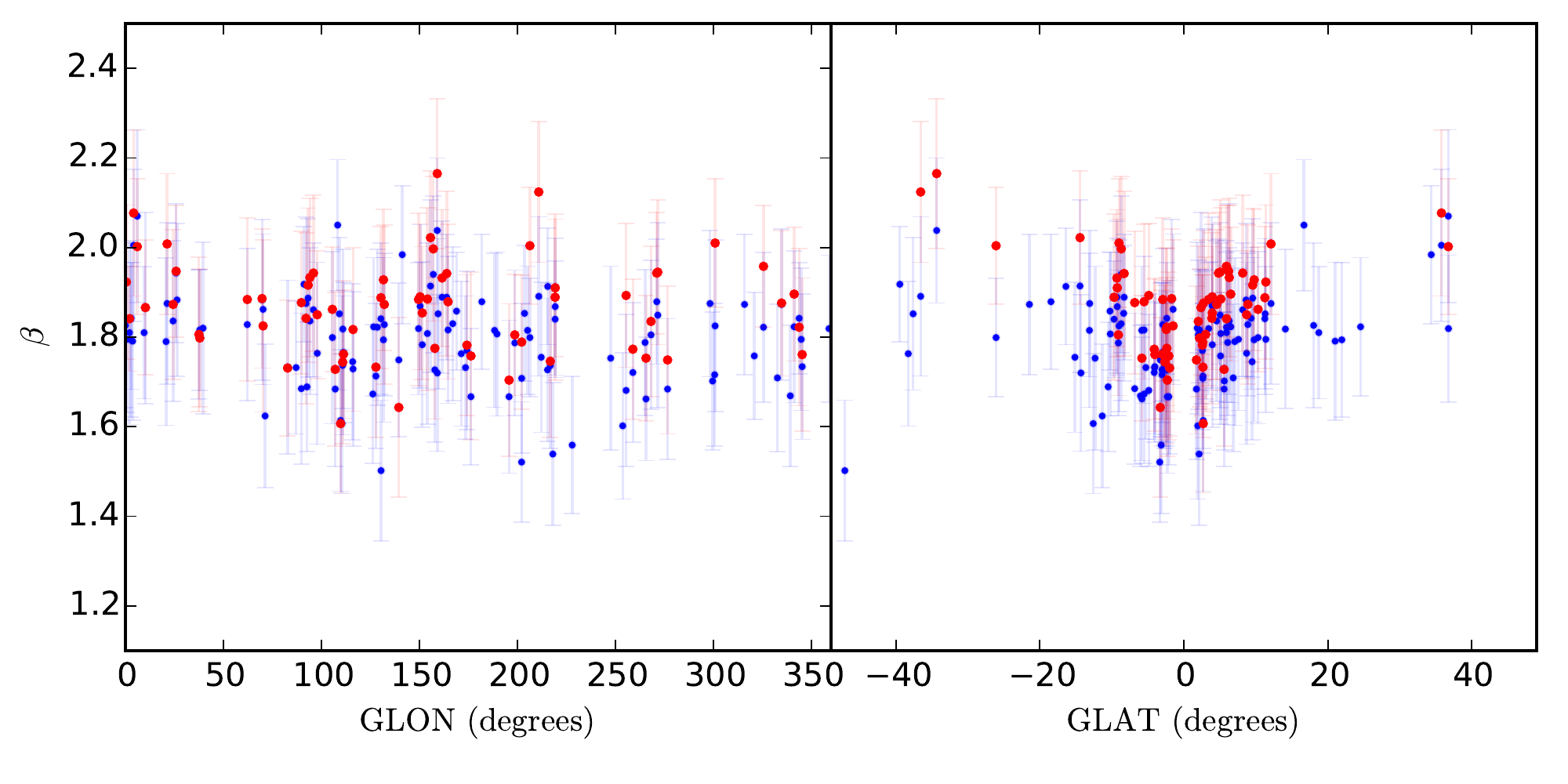}
\caption{
Spectral index values as function of Galactic longitude and latitude.  Values are
calculated using {\it Planck} and IRIS data. The blue symbols correspond to the
average surface brightness in the entire fields. Red symbols correspond to the subset
of pixels with $\tau(250\mu{\rm m})=(2.0\pm 0.5)\times 10^{-3}$.
}
\label{fig:stats_TB_7}
\end{figure}

In Fig.~\ref{fig:stats_TB_7}, the dust parameters are plotted against Galactic
longitude and latitude, the figure showing a general drop of $\beta$ towards low
latitudes. In addition to values averaged over each field, $T$ and $\beta$ are
indicated for pixels $\tau(250\mu{\rm m})=(2.0\pm 0.5)\times 10^{-3}$. Above the
lowest latitudes, these are above the average column density and have higher values
of $\beta$. 
The median values are 1.81 and 1.88 for all the pixels and for the
$\tau(250\mu{\rm})\sim 2\times 10^{-3}$ pixels, respectively. The difference disappears
below $|b|\sim 4\degr$, because the average optical depth rises to and above $2\times
10^{-3}$. The latitude dependence can indicate a change in dust properties. However, at
low latitudes local star-formation and the long lines-of-sight may increase temperature
variations within the beam, thus decreasing the apparent values of $\beta$.

There is no clear dependence on Galactic longitude, although the values for
$l=0-10\degr$ are marginally above the average. We do not see the strong systematic
increase of $\beta$ towards the inner Galaxy that was found in
\citet{planck2013-p06b}.  Partly this is because that effect is strongest in the
Galactic plane, whereas all our fields are at latitudes $|b|>1\degr$ and,
furthermore, within $|l|<30\degr$ all are above $|b|\sim 3\degr$. Secondly, most of
our fields are dominated by emission from within one kiloparsec, too near to probe
physical variations related to, e.g., the Galactic molecular ring.

\begin{figure}
\includegraphics[width=8.8cm]{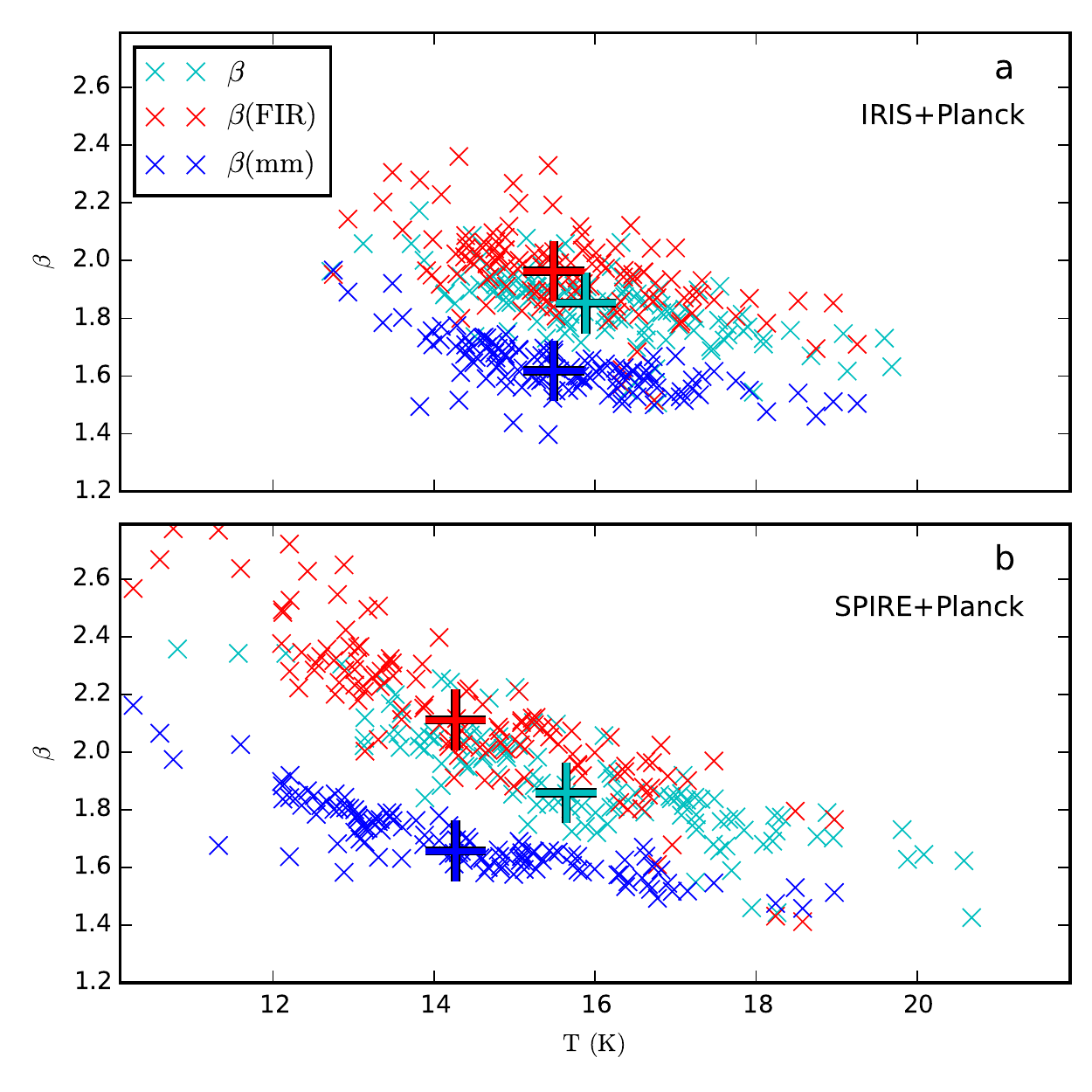}
\caption{
Colour temperature and spectral index fits for surface brightness data averaged over
FWHM=10$\arcmin$ beams centred on each of the 116 fields. The data points correspond
to values obtained from an analysis of the IRIS and {\em Planck} data (frame $a$),
and of the SPIRE and {\em Planck} data (frame $b$). The cyan symbols represent
single component fits that are restricted to frequencies $\nu \ge 353$\,GHz. The
other plot symbols indicate two component fits that employ different spectral
indices below 700\,$\mu$m (red symbols) and above 700\,$\mu$m (blue symbols). The
{\em Planck} 217\,GHz band is included in these two component fits. The plus signs
indicate median values.
}
\label{fig:2B}
\end{figure}

\subsection{Wavelength dependence of $\beta$} \label{sect:wavelength}

In the fits performed with the combination of {\em Herschel} and {\em Planck} data,
the 353\,GHz and 217\,GHz data consistently show an excess relative to the modified
blackbody that describes the emission at wavelengths below $\sim$500\,$\mu$m. 
In Fig.~\ref{fig:PI_FIG_105_main} the relative excess was larger in
diffuse than in the dense medium. This remains true for the joint fit shown in
Fig.~\ref{fig:BRUTE_RES_main}. As mentioned in Sect.~\ref{sect:PlanckIRIS}, this
could be related to line-of-sight temperature variations that decrease the apparent
FIR spectral index of dense regions. However, the quantification of these effects
would require detailed modelling of the individual sources.

We examined the wavelength dependence of $\beta$ using average surface brightness
values towards the centre of each field. As in Fig.~\ref{fig:plot_PI_10arcmin}, the
mean surface brightness is calculated with a Gaussian beam with FWHM equal to
10$\arcmin$. Figure~\ref{fig:2B} shows the single component fits for two data
combinations. These consist of {\em Planck} data with $\nu\ge 353$\,GHz and with
either IRIS 100\,$\mu$m band or the three SPIRE bands. For the latter combination,
the scatter is larger but the median value of $\beta$ is in both cases 1.8--1.9.
Alternative fits were performed using two $\beta$ values, one below and one above
$\lambda_{\rm 0}$~=~700\,$\mu$m. This model thus consists of one intensity parameter,
one colour temperature value, and two spectral index values that are here named
$\beta_{\rm FIR}$ and $\beta_{\rm mm}$. The model is required to be continuous at
700\,$\mu$m.  The {\em Planck} 217\,GHz band is included in these fits. In
Fig.~\ref{fig:2B}, the two component fits show a clear difference between the FIR
part ($\lambda\le 700\,\mu$m) and the millimetre part (217\,GHz--353\,GHz,
$\sim$850--1400\,$\mu$m). 
In the millimetre part the values are ${\rm \beta_{mm}}$~=~1.6--1.7 for both data
combinations. At shorter wavelengths, the median values are significantly higher, ${\rm
\beta_{\rm FIR}}$~=~1.96 for IRIS+{\em Planck} and ${\rm \beta_{\rm FIR}}$~=~2.11 for
SPIRE+{\em Planck}.

\begin{figure*}
\includegraphics[width=17.7cm]{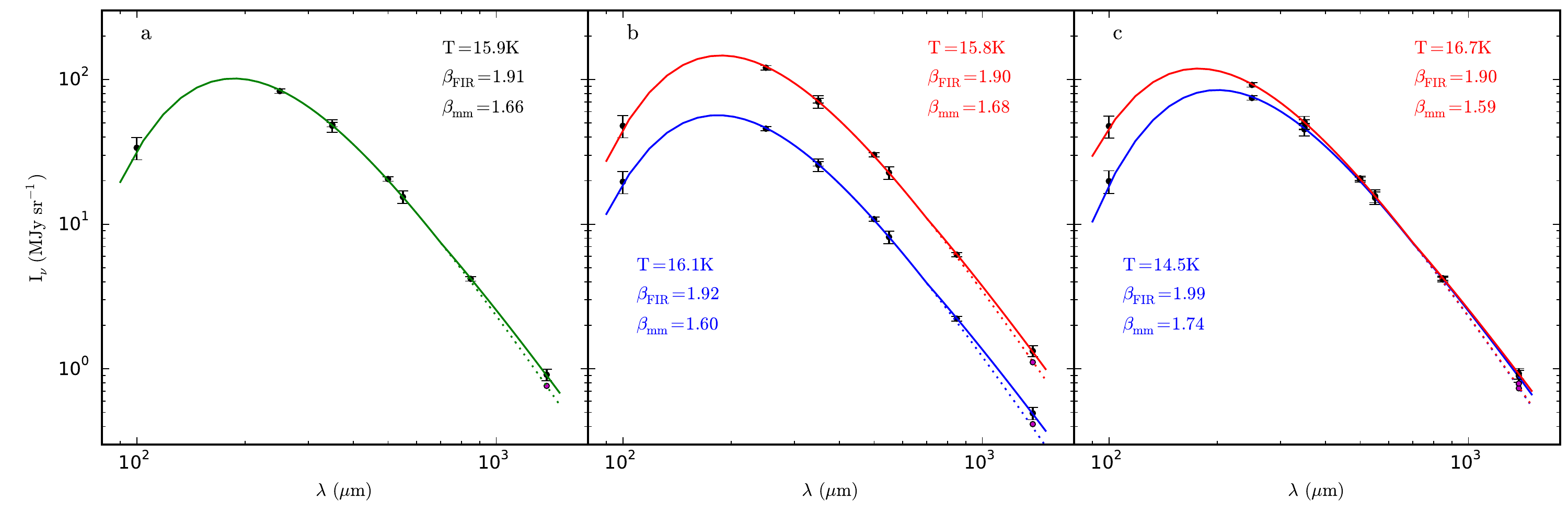}
\caption{
Spectral energy distributions based on surface brightness averaged over fields and over
10$\arcmin$ beams towards the centre of each field. Frame $a$ corresponds to the average
over all fields, frame $b$ to averages over fields below (blue) and above (red) the
median column density, and frame $c$ to averages over fields below and above the median
temperature. The dotted lines are long-wavelength extrapolations using ${\rm
\beta_{FIR}}$. At 217\,GHz the lower points (without error bars) correspond to data that
have been corrected assuming a line ratio of (2-1)/(1-0)=1.0 and thus with twice the
default CO correction.
}
\label{fig:2B_SED}
\end{figure*}

Figure~\ref{fig:2B_SED} shows the SEDs that are estimated from surface brightness values
averaged over a 10$\arcmin$ beam towards each of the fields. The values are further
averaged either over all fields, over fields with column density below or above the
median column density, or over fields with colour temperature below or above the median
colour temperature. The selection of the fields is based on values (${\rm \tau(250\mu
m)}$ and $T$) derived from the analysis of IRIS and {\em Planck} data ($\nu\ge 353$\,GHz) but
all bands are included in the fits shown in Fig.~\ref{fig:2B_SED}. As above, the fits
employ a single dust temperature but two spectral index values. The derived parameter
values are given in the figure. For signal averaged over all fields, the spectral index
values are
${\rm \beta_{\rm FIR}}$~=~1.91 and ${\rm \beta_{mm}}$~=~1.66. Compared to a fit with a
single $\beta$ value (17.0\,K and $\beta=1.72$), the reduced $\chi^2$ value is smaller
by a factor of three. When fields are divided according to column density, higher column
density fields are associated with slightly lower dust temperature, $\Delta T$~=~$0.4$\,K.
The values of ${\rm \beta_{\rm FIR}}$ are practically identical for the two column
density intervals but ${\rm \beta_{mm}}$ is 0.07 units lower for the lower column
density sample.

When the fields are divided based on their colour temperature, differences in the
spectral index are more noticeable. The average signal of cold fields has ${\rm
\beta_{\rm FIR}}$ that is higher by $\sim$0.1. The data are averaged over $10\arcmin$
beams and, for both samples, over almost 60 fields. Thus the noise should be very small
although the results can be affected by systematic errors if these cause scatter between
the $T$ and $\beta$ estimates of individual fields. The zero point errors are one such
possible error source although, by concentrating on high column density central areas,
their effect should be limited. Compared to ${\rm \beta_{\rm FIR}}$, the parameter ${\rm
\beta_{mm}}$ should be more insensitive to $T$ errors but it also shows a clear
difference, ${\rm \beta_{mm}}$ being 0.12 unit higher for the cold field sample. This is
not necessarily very significant, considering the limited wavelength range. On the other
hand, the 217\,GHz point is in every case several $\sigma$ above the spectrum
extrapolated with ${\rm \beta_{\rm FIR}}$.

The CO correction is one of the main uncertainties regarding the magnitude of the
217\,GHz excess. Our default correction assumed a line ratio (2--1)/(1--0)~=~0.5, which is
consistent with the average line ratio obtained from {\em Planck} Type 1 CO maps (see
Fig.~\ref{fig:CO_ratios}). In reality the line ratio is not spatially constant and could
be higher in individual clumps. To examine the possible consequences, we repeated part
of the analysis using CO corrections that were twice as large as the default one. This
corresponds to line ratios $T_{\rm A}$(2--1)/$T_{\rm A}$(1--0)~=~1.0 and $T_{\rm
A}$(3--2)/$T_{\rm A}$(1--0)~=~0.6. According to Fig.~\ref{fig:CO_ratios}, this is a very
conservative estimate of the average (2--1)/(1--0) ratio and also exceeds the largest
observed values of the (3--2)/(2--1) ratio.

\begin{figure}
\includegraphics[width=8.8cm]{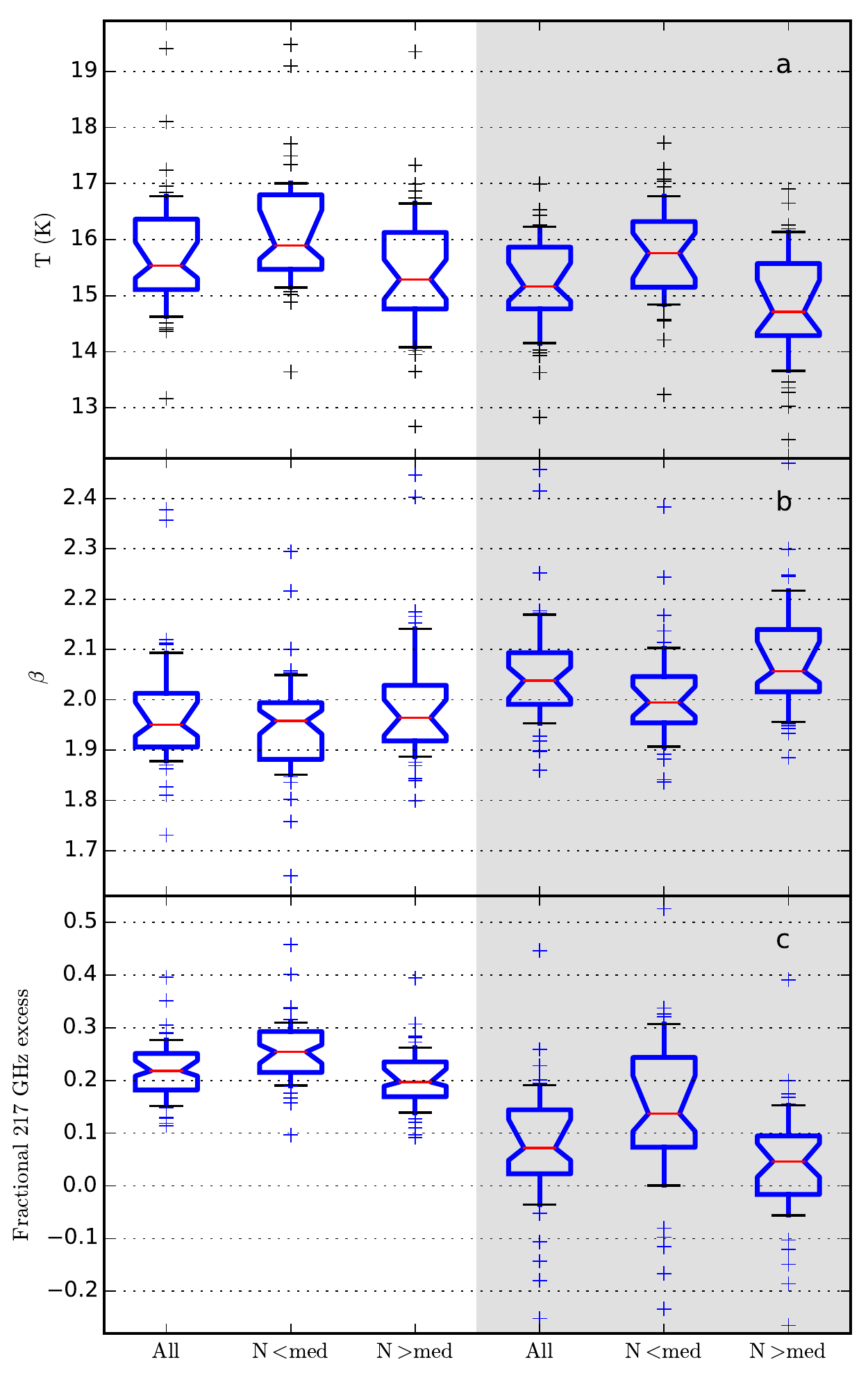}
\caption{
Comparison of fits of {\em Planck} and {\em Herschel} data in case of different 
corrections for CO emission. The boxplots show the distribution of field-averaged values
of $T$, $\beta$, and 217\,GHz excess (relative to fitted modified grey body model, $\sim
5\arcmin$ resolution). Distributions are plotted dividing each field to pixels below
(``$N<{\rm med}$'') and above (``$N>{\rm med}$'') the median column density. The plots
on white background correspond to the default CO correction and the plots on grey
background to a CO correction that is twice as high. Each boxplot shows the median
parameter value (red horizontal line), the interquartile range (box), and the 10\%--90\%
extent of the distribution (``whiskers''). Each individual field outside the [10\%,
90\%] range is plotted with a plus sign.
}
\label{fig:CO_effect_on_217}
\end{figure}

If the CO correction is doubled, the 217\,GHz residuals of Fig.~\ref{fig:2B_SED} remain
positive relative to the extrapolated FIR component with $\beta_{\rm FIR}$. The excess
is $\sim$1\, sigma for the whole sample, more for the low column density sample and
close to zero for the high column density sample. In Fig.~\ref{fig:2B_SED} the
comparison also is affected by spatial averaging that (via temperature mixing) tends to
decrease the value of $\beta_{\rm FIR}$. We therefore also repeated the analysis of
Sect.~\ref{sect:combined} (spatial resolution $\sim 5 \arcmin$ at 217\,GHz) using the
larger CO correction.
Figure~\ref{fig:CO_effect_on_217} shows the resulting $T$ and $\beta$ estimates and the
217\,GHz excess (observed surface brightness in relation to the fitted modified
blackbody with a single-$\beta$ value). The values of $\beta$ again are higher in
regions of higher column density. The difference is larger in case of the larger CO
correction. There also is a clear effect on $T$ and $\beta$ values. The larger
correction leads to almost 0.1 units higher median value of $\beta$. The CO correction
directly reduces the 217\,GHz surface brightness attributed to dust emission and thus
leads to smaller relative 217\,GHz excess. The excess is smaller within the high column
sample, in spite of the higher $\beta$ values in these regions. The observed surface
brightness exceeds the fitted modified blackbody curve in $\sim$75\% of the fields.
Taking into account that the larger CO correction very likely overestimates the average
CO contamination, Fig.~\ref{fig:CO_effect_on_217} strongly suggests that the 217\,GHz
excess is real and ${\rm \beta_{FIR}}$>${\rm \beta_{mm}}$

Although the apparent values of ${\rm \beta_{FIR}}$ are higher, they are expected to
be decreased by line-of-sight variations, which are significant in such high column
density regions (see Sect.~\ref{sect:RTeffects}). This suggests that the intrinsic
dust properties may have an even stronger wavelength dependence. Also, in the two
component fit the exact values of $\beta$ depend on the pivot wavelength. If this is
decreased, the results remain qualitatively similar but values of $\beta_{\rm FIR}$
decrease and values of $\beta_{\rm mm}$ increase. For the $\lambda_{\rm
0}$~=~700\,$\mu$m$\rightarrow$490\,$\mu$m the difference between ${\rm \beta_{FIR}}$ and
${\rm \beta_{mm}}$ decreases by more than half. Detailed characterisation of the
wavelength dependence, including the behaviour at smaller spatial scales, is left for
future papers.

\begin{figure}
\includegraphics[width=8.9cm]{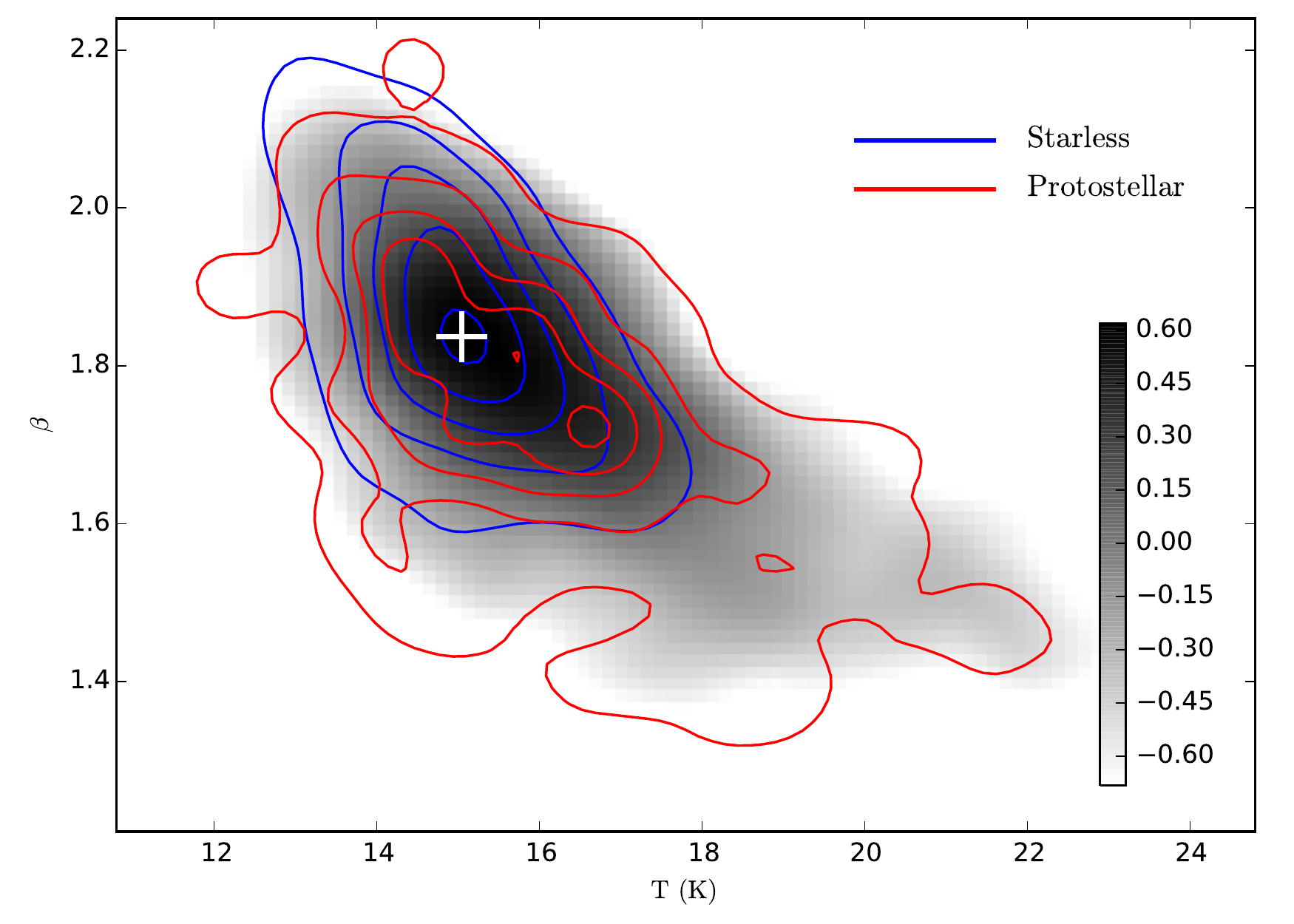}
\caption{
Temperature and spectral index values towards the submillimetre clumps catalogued in
\citet{GCC-IV}. The blue and red contours correspond to starless and protostellar
clumps respectively. There are five equally spaced contours between 15\% and 90\% of
the peak density of points. The background grayscale image and the colour bar show
the distribution of unclassified clumps (logarithm of the number of clumps per bins
of $\Delta T$~=~$0.14$\,K and $\Delta \beta$~=~$0.014$). The white cross shows the peak of
this distribution. The plotted data have been smoothed with a Gaussian with FWHM
equal to 1.0\,K and 0.1 units in $\beta$.
}
\label{fig:YSO_vs_TB}
\end{figure}

\subsection{Effects of temperature variations} \label{sect:RTeffects}

The observed $\beta$ depends on the range of dust temperatures within the beam.  The
observed value will be below the intrinsic opacity spectral index of the dust
grains, the difference increasing with increasing temperature dispersion
\citep{Malinen2011, Ysard2012}. In our sample, we have objects with two types of
temperature gradients: externally heated clumps and cores heated internally by young
stellar objects (YSOs). We examined the effect of YSOs using the catalogue of
submillimetre clumps in \citet{GCC-IV}. The clumps are classified based on WISE
photometric data and submillimetre dust temperature either as protostellar (with YSO
candidates) or starless. For a large fraction of the clumps the classification
remains undetermined \citep[see][for details]{GCC-IV}. In the area covered by both
PACS and SPIRE instruments, there are 204 protostar candidates, 1156 starless
clumps, and 1968 unclassified sources. The distributions of the corresponding ($T$,
$\beta$) values of the clumps are shown in Fig.~\ref{fig:YSO_vs_TB} where the
spectral indices are those derived from 160--500\,$\mu$m data at a spatial
resolution of 60$\arcsec$.

There is significant overlap between the three clump categories and the modes of the
distributions do actually coincide. However, starless clumps are found
preferentially at lower $T$ and higher $\beta$ values while the distribution of
protostellar clumps extends to higher temperatures. The median values of $\beta$ are
1.82 and 1.71 for the starless and the protostellar clumps, respectively. If the
YSOs heat a volume large enough to be significant at long wavelengths (and at the
60$\arcsec$ resolution) the difference could be an indication either of dust
processing (e.g., evaporation of ice mantles or dynamical processing caused by
outflows) or of a direct temperature effect on the grain material \citep{Meny2007}.
However, the observed differences may partly reflect the increased temperature
variations that are caused by the YSOs, without any difference in the intrinsic
$\beta$ of the dust grains \citep[see][]{Shetty2009a, Malinen2011, Juvela2011,
Juvela2012_Tmix}. 

If the temperature dispersion was caused mainly by varying degrees of internal
heating, it could result in the kind of $T-\beta$ anti-correlation as seen in
Fig.~\ref{fig:YSO_vs_TB}. However this is unlikely because the effect of the (mostly
weak) protostellar sources is highly localised and, furthermore, the
anti-correlation is more pronounced for the starless clumps. A complementary
question is whether $\beta$ depends on the grain size distribution. Grains are known
to grow inside dark clouds and their growth is revealed via various phenomena, most
notably the coreshine effect \citep{Steinacker2010, Pagani2010, Steinacker2014,
Lefevre2014}. A variation of $\beta$ due to the grain growth would result in a
segregation of the coreshine cases in the $T-\beta$ plot (such as Figs.
\ref{fig:histo_TB_Planck} and \ref{fig:YSO_vs_TB}). Paladini et al. (in prep.) show
that this is not the case and no preferential region appears in their $T-\beta$
relation.
This gives a clue that the $\beta$ value might not be related to the size of the
grains, or more probably that variations in the dust composition and size, as well
as temperature changes along the line-of-sight do not allow us to see clear trends.
Moreover, for starless cores with the highest values of $\beta$, coreshine should
trace grain growth deep inside the cores while the $T-\beta$ relationship is not
representative of the coldest region \citep{Pagani2015}. For embedded
sources, only a full multiwavelength 3D modelling, including grain property and
temperature variations, can help to remove the degeneracy.

In the case of externally heated clumps, the temperature variations again result in
the observed spectral index values being lower than the intrinsic opacity spectral
index of the dust grains. However, unlike in the case of internal sources, the
temperature variations are likely to favour a positive correlation between $T$ and
$\beta$ \citep{Malinen2011}. The precise effect depends on the strength of the
temperature gradients \citep{Juvela2012_Tmix}. 

We show in Appendix~\ref{app:RT} results from radiative transfer modelling of
optically thick filaments. These suggest that once the visual extinction exceeds
$A_{\rm V}\sim10$\,mag, the observed $\beta$ can be more than $\Delta \beta$~=~$0.1$
below the intrinsic value of the dust grains. The effect is well correlated with
the drop of temperature. When the opacity is increased close to $A_{\rm V} \sim
50$\,mag, the difference between the observed and intrinsic $\beta$ values could be
$\Delta \beta \sim 0.7$. However, such large decreases in $\beta$ values are not
seen in our sample of fields. For example, the main clump of G206.33-25.94 has an
estimated optical depth of $\tau(250\mu{\rm m})$~=~0.0047 (derived with $\beta$~=~1.8
and 160--500\,$\mu$m data at 1$\arcmin$ resolution). Assuming the average
NIR-submillimetre opacity ratio of $\tau(250\mu{\rm m})/\tau_{\rm J}=1.6 \times
10^{-3}$ derived in \citet{GCC-V}, this corresponds to $\tau_{\rm J}\sim 2.9$ and
roughly 10\,mag of visual extinction. The clump is not associated with any dip in
the $\beta$ map and the absolute values of $\beta$ are high, at least
$\beta\sim2.0$ (depending on the data set used). In field G150.47+3.93 the peak
optical depth is $\tau(250\mu{\rm m})\sim 0.007$, which corresponds to $A_{\rm
V}\sim 16$\,mag. Values of $\beta$ are still increasing towards the column density
peak.

Some drop in $\beta$ is observed only towards the centre of G6.03+36.73 (LDN~183) where
the estimated optical depth is $\tau(250\mu{\rm m})=0.017$ (for $\beta=1.8$ and a
resolution of 37$\arcsec$) suggesting $A_{\rm V}\sim50$\,mag. In this source, the
estimated submm dust emissivity is higher than average \citep[see][]{GCC-V}, but for
such high opacities the column density derived from dust emission is still likely to be
underestimated. Based partly on MIR absorption, the estimated maximum extinction exceeds
$A_{\rm V}=100$\,mag, although only in an area with less than 1$\arcmin$ in size
\citep{Pagani2003, Pagani2015}.
At the resolution of 60$\arcsec$, the observed $\beta$ values drop from $\beta\sim 2.0$
in the surrounding regions down to $\beta$~=~1.60 in the central core, the values being
measured using 160--500\,$\mu$m data at a resolution of 1$\arcmin$.  The drop is much
less than would be expected based on the simple models of Appendix~\ref{app:RT}.

In Appendix~\ref{app:RT}, we also tested cases where the dust in the central part
of the filament is associated with higher submillimetre opacity. This leads to 
lower central temperature and thus to larger temperature gradients. As a result,
the observed value of $\beta$ is further decreased towards the centre of the model
cloud. This can happen even when the intrinsic $\beta$ is actually higher in the
centre. 

Compared to the models, the observations do not show such a strong drop of $\beta$.
This could be understood if the clouds had a very clumpy structure (smaller
temperature gradients), the highest column density peaks remained unresolved, or
the increase of the intrinsic $\beta$ values is very large and/or extends clearly
outside the coldest part of the cores. Finally, the filament model may not be a
particularly good description of isolated clumps (such as LDN~183) and this may
lead overestimation of the line-of-sight temperature gradients. Further analysis of
$\beta$ variations would require detailed modelling of all the individual clumps.

\section{Conclusions}  \label{sect:conclusions}

We have used {\em Herschel} and {\em Planck} observations of dust emission to
examine the dust opacity spectral index in a number of fields containing dense
clumps. The fields, typically $\sim 40\arcmin$ in size, were originally selected
based on the signature of cold dust emission but they also cover some more diffuse
areas with higher dust temperatures. Our goal was to measure the variations of the
spectral index and to look for correlations with environmental factors. We have
presented maps of $T$ and $\beta$ derived with different combinations of {\em
Herschel}, {\em IRAS}, and {\em Planck} data. The study led to the following
conclusions:
\begin{itemize}
\item The observed dust opacity spectral index shows clear spatial variations that
are anti-correlated with temperature and correlated with column density. Based on {\em
IRAS} and {\em Planck} observations the median values of field averages are
$T=16.1$\,K and $\beta_{\rm FIR}=1.84$. Using {\em Herschel} data, we find in many
individual cold clumps values up to $\beta\sim 2.2$ ($\nu \le 353\,GHz$).
\item The millimetre emission exhibits a clear excess relative to the
modified blackbody fits that describe the submillimetre data. Two component fits to
the surface brightness data averaged over all fields yield values ${\rm
\beta_{FIR}}$~=~1.91 and ${\rm \beta_{mm}}$~=~1.66. As expected for a sample of high column
density fields, ${\rm \beta_{FIR}}$ is much higher than values reported for diffuse
medium. On the other hand, ${\rm \beta_{mm}}$ is close to the values found in {\em
Planck} studies of diffuse regions at high Galactic latitudes.
\item The amount of CO emission in the Planck bands is one of the main sources of
uncertainty for the $\beta_{\rm mm}$ values. The issue could be resolved by directly
mapping the CO(2-1) emission with ground-based radio telescopes.
\item
The spectral index values are lower close to the Galactic plane, possibly because of the
increased mixing of dust at different temperatures. We do not observe any significant
dependence on Galactic longitude.
\item We have examined how $\beta$ estimates are affected by various error 
sources and the data sets used. We conclude that our results on $\beta$ variations are
robust concerning the detected local spatial variations (vs. dust temperature and column
density) and the variations as a function of wavelength.
\item Because temperature gradients in optically thick
sources lead us to underestimate the intrinsic $\beta$, the true variations of the
spectral index and hence of the dust properties may be even more pronounced.
Quantitative estimates of intrinsic $\beta$ will require detailed radiative transfer
modelling of the observations.
\end{itemize}

\begin{acknowledgements}
This research made use of Montage, funded by the National Aeronautics and Space
Administration's Earth Science Technology Office, Computational Technologies Project,
under Cooperative Agreement Number NCC5-626 between NASA and the California Institute
of Technology. The code is maintained by the NASA/IPAC Infrared Science Archive.
MJ, VMP, and JMa acknowledge the support of the Academy of Finland Grant No.
250741. 
MJ acknowledges the Observatoire Midi-Pyrenees (OMP) in Toulouse for its support for a
2 months stay at IRAP in the frame of the `OMP visitor programme 2014'.
AH acknowledges support from the Centre National d'Etudes Spatiales
(CNES) and funding from the Deutsche Forschungsgemeinschaft (DFG) via
grants SCHI 536/5-1 and SCHI 536/7-1 as part of the priority program
SPP 1573 `ISM-SPP: Physics of the Interstellar Medium'.
AR-I acknowledges support from the ESA Research Fellowship Programme.
\end{acknowledgements}

\bibliography{biblio_with_Planck}

\onecolumn

\begin{landscape}

\begin{longtable}{lcccccccc}
\caption{
The coordinates and sizes of the {\it Herschel} fields. The last columns
give the position and radius of reference regions used in case of
local background subtraction.}\label{table:fields} \\
\hline \hline
Field &  \multicolumn{2}{c}{Centre coordinates} &  Area           &  Distance &  \multicolumn{3}{c}{Reference region} 
      &   Aliases \\
      &   RA(2000.0) & DEC(2000.0)             &  $(\arcmin)^2$ &  (kpc)     &  RA(2000) & DEC(2000) & $r$ ($\arcmin$) 
      &    \\
\hline 
\endfirsthead
\caption{continued.} \\ 
\hline \hline
Field &  \multicolumn{2}{c}{Centre coordinates} &  Area           &  Distance &  \multicolumn{3}{c}{Reference region}  &
Aliases \\
      &   RA(2000.0) & DEC(2000.0)             &  $(\arcmin)^2$ &  (kpc)     &  RA(2000) & DEC(2000) & $r$ ($\arcmin$)  &         \\
\hline      
\endhead      
\hline 
\endfoot

   G0.02+18.02 &    16 40 56.7 & -18 34 59.9 & 1332   & 0.16(0.05)   &   16 40 27.6 & -18 32 16.8   & 2.5 &   - \\
   G0.49+11.38 &    17 04 41.9 & -22 13 51.1 & 1775   & 0.16(+0.20/-0.16)   &   17 04 21.8 & -22 02 20.4   & 2.8 &   LDN15 \\ 
    G1.94+6.07 &    17 28 08.2 & -24 01 53.9 & 2276   & 0.14(0.05)   &   17 28 18.0 & -23 44 56.4   & 3.9 &   LDN69, B77 \\
   G2.83+21.91 &    16 34 28.8 & -14 11 27.8 & 2279   & 0.30(0.30)   &   16 34 27.1 & -13 58 58.8   & 4.7 &   LDN83, MBM135 \\
    G3.08+9.38 &    17 17 51.1 & -21 26 16.4 & 3424   & 0.16(0.05)   &   17 17 59.3 & -21 44 13.2   & 4.2 &    \\ 
   G3.72+21.02 &    16 39 45.9 & -14 02 09.3 & 1776   & 0.16(0.05)   &   16 39 45.4 & -13 51 28.8   & 3.5 &   LDN121 \\
   G4.18+35.79 &    15 53 31.4 & -04 37 55.5 & 1331   & 0.11(0.01)   &   15 52 57.4 & -04 32 27.6   & 2.7 &   LDN134, MBM36, MLB40 \\
   G6.03+36.73 &    15 54 13.9 & -02 51 40.1 & 1332   & 0.11(0.01)   &   15 54 04.6 & -02 42 28.8   & 2.5 &   LDN183, MBM37 \\
   G9.45+18.85 &    17 00 22.2 & -10 51 07.0 & 1777   & 0.28(0.10)   &   17 00 49.7 & -10 44 31.2   & 2.9 &   -  \\
   G10.20+2.39 &    17 59 20.6 & -18 56 22.5 & 1331   & 0.83(0.40)   &   17 59 07.9 & -18 49 37.2   & 2.3 &   -  \\
   G20.72+7.07 &    18 03 38.9 & -07 30 24.5 & 1331   & 0.26(0.26)   &   18 03 01.2 & -07 33 21.6   & 2.8 &   -  \\
  G21.26+12.11 &    17 46 55.7 & -04 36 50.6 & 1331   & 0.12(0.12)   &   17 46 48.2 & -04 26 34.8   & 2.9 &   LDN425, LDN428, LM240 \\ 
   G24.40+4.68 &    18 19 21.5 & -05 32 44.7 & 2284   & 0.26(0.05)   &   18 18 17.3 & -05 36 25.2   & 3.6 &   LDN475, LDN477, LDN470 \\
   G25.86+6.22 &    18 16 20.7 & -03 24 48.9 & 1331   & 0.26(0.05)   &   18 15 39.4 & -03 26 27.6   & 2.5 &   LDN500 \\
   G26.34+8.65 &    18 08 37.9 & -01 51 26.2 & 1331   & 0.40(0.40)   &   18 08 40.3 & -01 42 32.4   & 2.6 &   LDN502, CB112, P61 \\
   G37.49+3.03 &    18 48 57.0 & +05 26 08.7 & 1333   & 0.80(0.60)   &   18 48 31.2 & +05 29 13.2   & 2.4 &   BDN31.48+3.02 \\
   G37.91+2.18 &    18 52 45.4 & +05 26 01.0 & 1331   & 1.06(0.79)   &   18 52 26.4 & +05 31 51.6   & 2.3 &   - \\
   G39.65+1.75 &    18 57 00.1 & +06 47 00.2 & 1774   & 0.99(0.48)   &   18 56 50.6 & +06 41 13.2   & 3.5 &   - \\
   G62.16-2.92 &    19 59 47.3 & +24 16 15.0 & 1329   & 1.02(0.35)   &   19 59 29.3 & +24 22 40.8   & 2.7 &   - \\
   G69.57-1.74 &    20 13 24.7 & +31 16 01.4 & 1772   & 1.78(0.81)   &   20 13 46.1 & +31 08 09.6   & 2.6 &   - \\ 
   G70.10-1.69 &    20 13 58.5 & +31 49 25.7 & 3420   & 2.09(0.83)   &   20 13 39.1 & +32 09 18.0   & 4.3 &   - \\
  G71.27-11.32 &    20 53 14.0 & +26 50 56.4 & 1332   &    -         &   20 53 17.0 & +26 58 04.8   & 3.0 &   - \\
   G82.65-2.00 &    20 53 11.3 & +41 35 16.7 & 4744   & 1.00(+1.00/-0.60)   &   20 52 35.0 & +41 14 27.6   & 5.5 &   LDN914 \\
   G86.97-4.06 &    21 17 20.9 & +43 24 21.6 & 1325   & 0.70(0.10)   &   21 16 53.8 & +43 32 13.2   & 2.3 &   LDN943, LDN944 \\
   G89.65-7.02 &    21 38 23.7 & +43 16 23.6 & 3422   & 0.70(+1.00/-0.70)   &   21 37 30.2 & +43 02 24.0   & 4.5 &   B159, LDN977 \\
  G91.09-39.46 &    23 10 32.6 & +17 06 32.7 & 1330   & 0.52(0.20)   &   23 10 15.1 & +17 12 28.8   & 2.7 &   -  \\
   G92.04+3.93 &    21 03 00.6 & +52 31 46.3 & 2278   & 0.80(0.80)   &   21 01 09.8 & +52 33 50.4   & 3.5 &   LDN1003, ArchG092.03+03.93 \\
  G92.63-10.43 &    22 03 05.7 & +42 14 34.9 & 1330   & 0.90(0.90)   &   22 02 48.7 & +42 07 58.8   & 2.3 &   -  \\
   G93.21+9.55 &    20 37 00.2 & +56 54 46.7 & 1332   & 0.44(+0.46/-0.14)   &   20 37 35.3 & +56 46 12.0   & 2.8 &   LDN1033 \\
   G94.15+6.50 &    20 59 22.8 & +55 35 54.0 & 2279   & 0.80(0.80)   &   20 57 55.4 & +55 37 22.8   & 3.9 &   B357 \\
   G95.76+8.17 &    20 57 24.1 & +58 10 19.9 & 2829   & 0.80(0.80)   &   20 55 23.8 & +58 17 06.0   & 3.5 &   LDN1071, B354  \\
   G98.00+8.75 &    21 03 55.9 & +59 59 36.1 & 1777   & 1.10(0.20)   &   21 03 31.7 & +60 10 22.8   & 4.1 &   ArchG097.82+08.67 \\
 G105.57+10.39 &    21 41 35.6 & +66 33 11.6 & 2829   & 0.90(0.30)   &   21 43 16.8 & +66 25 12.0   & 3.0 &   ArchG105.55+10.45 \\
  G107.20+5.52 &    22 21 14.4 & +63 41 46.0 & 3733   & 0.90(0.30)   &   22 24 24.7 & +63 44 09.6   & 6.3 &   PCC249, LDN1204, S140,  S9 \\
 G108.28+16.68 &    21 09 45.7 & +72 52 58.3 & 1332   & 0.30(+0.20/-0.15)   &   21 08 25.7 & +72 47 13.2   & 2.5 &   -  \\
 G109.18-37.59 &    00 03 50.3 & +24 00 10.8 & 1320   &    -         &   00 04 25.9 & +24 06 32.4   & 4.1 &   -  \\
  G109.80+2.70 &    22 53 30.7 & +62 30 57.7 & 1331   & 0.70(0.10)   &   22 53 20.6 & +62 38 24.0   & 2.8 &   PCC288, S8 \\
 G110.62-12.49 &    23 37 27.7 & +48 29 40.8 & 2280   & 0.44(0.10)   &   23 37 26.2 & +48 43 19.2   & 5.0 &   -  \\
  G110.89-2.78 &    23 18 15.5 & +58 04 01.7 & 2265   & 3.00(1.00)   &   23 17 28.6 & +57 53 52.8   & 4.0 &   -  \\
 G110.80+14.16 &    21 59 02.6 & +72 48 56.3 & 2281   & 0.40(0.10)   &   22 02 28.3 & +72 42 39.6   & 3.8 &   -  \\
  G111.41-2.95 &    23 22 15.4 & +57 50 00.0 & 2264   & 3.00(1.00)   &   23 20 43.4 & +57 55 15.6   & 3.6 &   ArchG111.11-03.01 \\
  G115.93+9.47 &    23 24 05.5 & +71 09 29.0 & 1331   & 1.00(0.50)   &   23 25 43.4 & +71 07 58.8   & 2.8 &   -  \\
  G116.08-2.40 &    23 57 06.7 & +59 42 27.3 & 1332   & 0.50(0.50)   &   23 58 29.0 & +59 41 02.4   & 2.9 &   LDN1257, LDN1257B  \\
  G126.24-5.52 &    01 15 46.6 & +57 12 37.5 & 1327   & 1.00(0.10)   &   01 16 59.3 & +57 08 52.8   & 2.9 &   -  \\
 G126.63+24.55 &    04 23 46.3 & +85 47 06.5 & 1350   & 0.12(0.02)   &   04 28 02.9 & +85 41 20.4   & 3.0 &   S1, LDN1320 \\
  G127.79+2.66 &    01 37 49.1 & +65 05 26.6 & 1344   & 0.80(0.10)   &   01 38 32.9 & +65 12 50.4   & 3.2 &   ArchG127.69+02.65 \\
 G128.78-69.46 &    00 59 22.1 & -06 44 41.2 & 2819   &    -         &   00 59 50.6 & -06 57 57.6   & 4.7 &   -  \\
 G130.37+11.26 &    02 32 07.2 & +72 39 15.1 & 1325   & 0.60(0.10)   &   02 30 49.7 & +72 30 18.0   & 2.4 &   LDN1340 \\
 G130.42-47.07 &    01 12 33.7 & +15 28 59.0 & 1328   &    -         &   01 13 10.1 & +15 34 44.4   & 3.2 &   -  \\
  G131.65+9.75 &    02 39 25.4 & +70 47 11.4 & 1328   & 1.07(+0.52/-0.49)   &   02 41 39.6 & +70 46 01.2   & 2.3 &   S3  \\
  G132.12+8.95 &    02 39 50.5 & +69 48 54.9 & 1762   & 0.85(0.10)   &   02 37 25.0 & +69 52 51.6   & 3.7 &   -  \\ 
  G139.60-3.06 &    02 52 31.6 & +55 39 35.5 & 4074   & 0.85(0.10)   &   02 50 13.4 & +55 24 10.8   & 5.4 &   -  \\
 G141.25+34.37 &    08 48 17.8 & +72 43 16.1 & 1329   & 0.11(0.01)   &   08 46 18.7 & +72 47 38.4   & 3.3 &   MBM27  \\
  G149.67+3.56 &    04 18 07.7 & +55 15 04.8 & 2275   & 0.17(0.05)   &   04 17 25.4 & +55 30 39.6   & 3.6 &   LDN1400, LDN1394 \\
               &               &             &        &              &              &               &     &   B8, B9, MBL71 \\
  G150.47+3.93 &    04 24 37.9 & +55 02 21.6 & 1776   & 0.17(0.05)   &   04 24 30.2 & +55 14 02.4   & 3.5 &   LDN1399, MLB72, MLB74, \\
               &               &             &        &              &              &               &     &   LM17, ArchG150.41+03.91 \\  
  G151.45+3.95 &    04 29 53.9 & +54 14 53.0 & 1331   & 0.17(0.05)   &   04 30 15.6 & +54 06 36.0   & 2.6 &   B12, LDN1407, LDN1400F, \\
               &               &             &        &              &              &               &     &   MLB77, LM25 \\  
  G154.08+5.23 &    04 47 34.4 & +53 05 01.9 & 1329   & 0.17(0.05)   &   04 48 04.8 & +53 13 26.4   & 2.6 &   LDN1426, LM56 \\
 G155.80-14.24 &    03 37 09.4 & +37 42 31.8 & 4765   & 0.35(0.10)   &   03 37 35.0 & +38 03 36.0   & 6.2 &   LDN1434  \\
  G157.08-8.68 &    04 01 39.8 & +41 14 20.4 & 1806   & 0.15(0.15)   &   04 02 37.7 & +41 17 27.6   & 2.9 &   (LDN1443) \\
  G157.92-2.28 &    04 28 51.8 & +45 15 06.3 & 2278   & 2.50(1.00)   &   04 30 20.2 & +45 22 26.4   & 3.9 &   -  \\
 G159.23-34.51 &    02 56 02.5 & +19 39 10.2 & 6527   & 0.33(0.05)   &   02 57 25.2 & +19 11 06.0   & 8.0 &   LDN1457, MBM12, ArchG159.17-34.44  \\
 G159.12-14.30 &    03 50 36.0 & +35 41 58.0 & 1331   & 0.80(0.80)   &   03 50 21.8 & +35 49 44.4   & 3.1 &   -  \\
 G159.34+11.21 &    05 41 17.5 & +52 11 38.8 & 2279   & 0.70(0.70)   &   05 41 30.2 & +51 55 55.2   & 3.6 &   -  \\
  G161.55-9.30 &    04 16 11.2 & +37 46 17.5 & 1358   & 0.25(+0.28/-0.25)   &   04 15 50.4 & +37 39 39.6   & 2.4 &   S7  \\
  G163.82-8.44 &    04 27 13.7 & +36 46 47.2 & 8385   & 0.42(0.10)   &   04 23 21.8 & +36 27 43.2   & 8.6 &   -  \\
  G164.71-5.64 &    04 40 42.8 & +38 09 06.4 & 5002   & 0.33(0.20)   &   04 42 46.1 & +38 20 34.8   & 5.3 &   LDN1481 \\
  G167.20-8.69 &    04 36 34.9 & +34 16 50.4 & 2829   & 0.16(0.16)   &   04 35 08.9 & +34 10 44.4   & 4.2 &   -  \\
 G168.85-10.19 &    04 37 04.3 & +31 49 17.9 & 1330   & 1.30(1.30)   &   04 36 12.7 & +31 49 12.0   & 2.7 &   -  \\
 G171.35-38.28 &    03 18 04.2 & +10 27 01.2 & 1330   &    -         &   03 17 23.3 & +10 25 40.8   & 2.8 &   MBM16  \\
  G173.43-5.44 &    05 08 42.0 & +31 22 38.5 & 3424   & 0.15(0.15)   &   05 08 37.0 & +31 02 13.2   & 4.5 &   -  \\
  G174.22+2.58 &    05 41 42.4 & +35 11 58.2 & 1331   & 1.80(0.20)   &   05 42 26.2 & +35 14 24.0   & 2.3 &   -  \\
  G176.27-2.09 &    05 28 14.3 & +30 57 27.0 & 1356   & 1.57(0.26)   &   05 28 04.3 & +31 06 18.0   & 3.0 &   S6 \\
 G181.84-18.46 &    04 43 56.1 & +16 57 22.9 & 1356   & 0.50(0.50)   &   04 43 53.8 & +16 50 09.6   & 2.8 &   -  \\
 G188.24-12.97 &    05 17 05.1 & +14 54 32.7 & 3422   & 0.45(0.05)   &   05 15 37.4 & +14 55 48.0   & 4.6 &   -  \\ 
 G189.51-10.41 &    05 29 55.3 & +15 27 03.2 & 2831   & 0.45(0.05)   &   05 28 38.6 & +15 25 30.0   & 4.0 &   -  \\
  G195.74-2.29 &    06 10 58.1 & +14 09 56.7 & 1330   & 0.60(0.60)   &   06 11 01.2 & +14 02 02.4   & 2.7 &   ArchG195.73-02.39 \\
  G198.58-9.10 &    05 52 28.8 & +08 19 33.8 & 2279   & 0.45(0.45)   &   05 51 54.7 & +08 16 58.8   & 3.5 &   LDN1598 \\
  G202.23-3.38 &    06 19 33.2 & +08 02 56.7 & 1331   & 3.80(1.00)   &   06 18 49.2 & +08 03 00.0   & 2.7 &   -  \\
  G202.02+2.85 &    06 41 07.4 & +10 47 22.8 & 4079   & 0.76(0.10)   &   06 42 42.7 & +10 42 39.6   & 4.3 &   -  \\
  G203.42-8.29 &    06 04 35.6 & +04 22 31.0 & 2278   & 0.39(0.10)   &   06 03 56.4 & +04 15 57.6   & 3.8 &   -  \\
  G205.06-6.04 &    06 16 27.5 & +04 09 44.0 & 2832   & 0.40(0.10)   &   06 16 40.8 & +04 28 15.6   & 4.6 &   -  \\
 G206.33-25.94 &    05 06 49.1 & -06 14 56.6 & 2276   & 0.21(0.03)   &   05 06 17.0 & -06 00 00.0   & 5.6 &   IC~2118, Witch Head Nebula \\
 G210.90-36.55 &    04 35 07.0 & -14 14 35.3 & 4763   & 0.14(+0.020/-0.028)   &   04 34 50.2 & -14 26 45.6   & 4.3 &   LDN1642, MBM20, IREC305 \\
 G212.07-15.21 &    05 55 57.7 & -06 09 26.4 & 1332   & 0.23(0.10)   &   05 56 21.8 & -06 02 56.4   & 2.8 &   -  \\
  G215.37-3.04 &    06 45 03.3 & -03 32 35.3 & 1329   & 2.40(0.50)   &   06 44 30.5 & -03 30 18.0   & 2.2 &   -  \\
 G215.44-16.38 &    05 57 02.8 & -09 33 26.2 & 1353   & 1.45(1.45)   &   05 57 26.4 & -09 31 15.6   & 2.6 &   S4  \\
  G216.76-2.58 &    06 48 59.8 & -04 36 09.6 & 1330   & 2.40(0.50)   &   06 49 06.2 & -04 24 39.6   & 2.6 &   -  \\
  G218.06+2.12 &    07 08 22.0 & -03 34 59.3 & 1329   &    -         &   07 07 59.3 & -03 38 31.2   & 3.4 &   -  \\
  G219.36-9.71 &    06 28 01.3 & -09 57 53.7 & 2279   & 0.91(0.03)   &   06 28 14.6 & -09 42 54.0   & 4.6 &   LDN1652  \\
  G219.29-9.25 &    06 29 43.0 & -09 48 37.5 & 1331   & 0.91(0.03)   &   06 29 17.0 & -09 41 49.2   & 2.8 &   IREC317  \\
  G227.95-2.98 &    07 07 50.7 & -14 46 23.6 & 2826   & 2.00(0.50)   &   07 07 30.0 & -15 03 57.6   & 4.4 &   -  \\
 G247.55-12.27 &    07 09 26.1 & -36 16 39.4 & 2824   & 0.17(0.17)   &   07 08 39.1 & -36 35 52.8   & 4.0 &   IREC365, DCld 247.5-12.3 \\
  G253.71+1.93 &    08 25 04.9 & -34 26 33.1 & 1318   & 2.26(0.10)   &   08 25 42.2 & -34 23 09.6   & 3.1 &   -  \\
  G255.33-4.88 &    08 01 23.8 & -39 38 07.5 & 1764   & 0.80(0.40)   &   08 00 33.6 & -39 45 36.0   & 2.8 &   DCld 255.4-04.9  \\
  G258.90-4.10 &    08 14 35.7 & -42 08 01.8 & 1323   & 1.04(0.44)   &   08 14 20.9 & -41 59 56.4   & 2.7 &   HMSTG259.1-4.0C  \\
  G265.04+6.08 &    09 17 48.9 & -40 36 11.7 & 1755   & 0.92(0.26)   &   09 18 40.8 & -40 43 26.4   & 2.8 &   -  \\
  G265.60-5.82 &    08 27 59.9 & -48 36 50.8 & 1319   & 2.40(+0.71/-0.85)   &   08 29 07.9 & -48 38 16.8   & 3.2 &   -  \\
  G268.21+2.02 &    09 13 10.0 & -45 36 45.4 & 1756   & 1.87(+0.86/-1.39)   &   09 14 11.3 & -45 33 03.6   & 2.7 &   SDN114, HMSTG268.0+1.8  \\
  G271.06+4.84 &    09 36 23.9 & -45 39 13.0 & 2260   & 1.32(0.50)   &   09 37 00.7 & -45 27 36.0   & 3.9 &   SDN118, HMSTG271.4+4.8  \\
  G271.51+5.14 &    09 39 27.5 & -45 49 41.0 & 3400   & 1.32(0.50)   &   09 41 13.2 & -45 59 20.4   & 4.6 &   HMSTGG271.4+4.8  \\
  G276.78+1.75 &    09 50 16.3 & -51 39 46.6 & 5064   & 2.00(2.00)   &   09 52 58.8 & -51 25 33.6   & 5.3 &   S5, FeSt2-72, DCld 276.9+01.7, \\
               &               &             &        &              &              &               &     &   DCld 276.8+0.1.9 \\
 G298.31-13.05 &    11 38 50.7 & -75 17 27.2 & 1330   & 0.15(0.10)   &   11 40 48.0 & -75 23 24.0   & 3.4 &   SDN138, FeSt2-129, \\
               &               &             &        &              &              &               &     &   HMSTG298.3-13.1, FeSt1-188, \\
               &               &             &        &              &              &               &     &   DCld 298.3-13.1 \\
  G299.57+5.61 &    12 26 45.9 & -57 05 33.6 & 1331   &    -         &   12 25 33.6 & -57 04 44.4   & 3.0 &   HMSTG299.6+5.6   \\
  G300.61-3.13 &    12 28 54.8 & -65 47 40.3 & 1761   & 0.20(0.05)   &   12 27 04.6 & -65 50 34.8   & 2.8 &   HMSTG300.6-3.0 \\
  G300.86-9.00 &    12 25 17.3 & -71 46 05.6 & 1357   & 0.15(0.03)   &   12 26 07.4 & -71 50 38.4   & 3.0 &   PCC550, S10, SDN143, VMF32, \\
               &               &             &        &              &              &               &     &   Musca DN Complex \\
 G315.88-21.44 &    17 19 40.0 & -76 55 16.8 & 1331   & 0.25(0.01)   &   17 16 27.8 & -76 49 22.8   & 2.5 &    -   \\
  G320.84+5.09 &    14 55 10.5 & -53 24 46.6 & 1328   & 1.00(1.00)   &   14 55 15.4 & -53 18 28.8   & 2.7 &   HMSTG320.8+5.1   \\
  G325.54+5.82 &    15 18 44.6 & -50 22 09.3 & 2276   & 0.64(0.44)   &   15 19 04.1 & -50 06 18.0   & 3.7 &   SDN175, HMSTG325.5+5.8, DCld 325.5+05.8 \\
  G332.70+6.77 &    15 49 43.7 & -45 28 32.1 & 1775   & 0.65(0.20)   &   15 50 11.0 & -45 37 19.2   & 2.8 &   -  \\
  G334.65+2.67 &    16 14 40.4 & -47 11 48.8 & 1778   & 1.50(0.50)   &   16 13 45.8 & -47 08 24.0   & 3.7 &   -   \\
  G339.22-6.02 &    17 12 09.7 & -49 34 31.0 & 1334   & 2.09(1.00)   &   17 13 08.9 & -49 37 58.8   & 3.2 &   -   \\
  G341.18+6.51 &    16 25 05.1 & -39 59 10.1 & 1332   & 0.14(0.02)   &   16 25 14.6 & -39 49 40.8   & 3.3 &   SLDN16, HMSTG341.2+6.5, LM170,  \\
               &               &             &        &              &              &               &     &   DCld 341.2+06.5, FeSt1-357 \\
  G343.64-2.31 &    17 10 29.2 & -43 46 28.7 & 2278   & 1.00(0.50)   &   17 10 52.1 & -43 57 43.2   & 3.8 &   HMSTG343.7-2.3, (FeSt1-373) \\
  G344.77+7.58 &    16 33 30.3 & -36 38 59.7 & 1332   & 0.24(0.24)   &   16 33 25.0 & -36 31 48.0   & 2.5 &   -  \\
  G345.39-3.97 &    17 22 50.0 & -43 28 24.7 & 1776   & 0.23(+0.20/-0.10)   &   17 23 42.5 & -43 34 01.2   & 3.8 &   -  \\
 G358.96+36.75 &    15 39 37.9 & -07 13 09.2 & 1355   & 0.11(0.01)   &   15 38 52.6 & -07 10 58.8   & 3.0 &   LDN1780, LDN1778, MBM33 \\
\end{longtable}
\tablefoot{The aliases refer the following catalogues:
Arch = \citet{Desert2008},
B    = \citet{Barnard1919,Barnard1927}, 
BDN  = \citet{Bernes1977},              
CB   = \citet{Clemens1988},             
FeSr = \citet{Feitzinger1984}
HMST = \citet{Hartley1986},             
LDN  = \citet{Lynds1962},               
MBM  = \citet{MBM},                     
MLB  = \citet{MLB1983},                 
SLDN = \citet{Sandqvist1976}.           
Most entries are looked up from \citet{Dutra2002}. Names starting with
PCC and the entries S1-S10 refer to names used in \citet{Juvela2010}
and \citet{planck2011-7.7a} for some of the {\it Herschel} fields.}
\end{landscape}


\clearpage

\appendix

\section{Radiative transfer models} \label{app:RT}

Modelling was used to assess the possible impact that radiative transfer effects
have on the observed spectral index values. The model consists of a cylindrical
filament with a Plummer-like radial density profile
\begin{equation}
   n(r) = \frac{n_0}{1+(r/R_{\rm C})^2},
\end{equation}
where $r$ is the distance from the symmetry axis, $n_0$ is the central density, and
$R_{\rm C}$ determines the extent of the inner flatter part of the density profile.
The filament is illuminated externally by interstellar radiation field
\citep{Mathis1983}. We use the \citet{Ossenkopf1994} dust model for coagulated
grains with thin ice mantles accreted over 10$^5$ years in a density of
$n=10^6$\,cm$^{-3}$. To remove any ambiguity from the comparison of the observed
and the intrinsic $\beta$ values, the model is modified so that $\beta$ is equal to
1.9 at all wavelengths $\lambda>100\mu$m.
Note that this is the intrinsic value of the dust opacity spectral index. Because
of line-of-sight temperature mixing, the apparent value estimated from surface
brightness data can be expected to be lower \citep{Shetty2009a, Juvela2012_Tmix}.

\begin{figure}
\includegraphics[width=8.3cm]{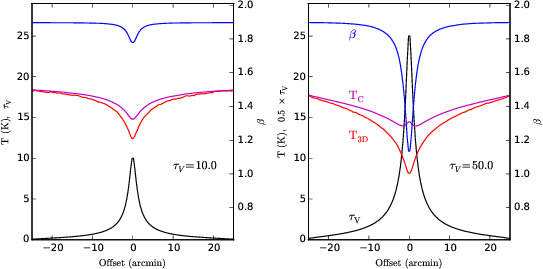}
\caption{
Cross sections of the cylindrical radiative transfer models. The peak optical depths
are $\tau_{\rm V}$~=~10.0 in the left and $\tau_{\rm V}$~=~50.0 in the right panel. Each
frame shows the $\tau_{\rm V}$ profile, the actual dust temperature in the model
midplane ($T_{\rm 3D}$), and the colour temperature $T_{\rm C}$ and spectral index
$\beta$ derived from the simulated surface brightness maps.
}
\label{fig:RT_results}
\end{figure}

The filament is observed in a direction perpendicular to its symmetry axis. The
radiative transfer calculations are used to solve the three-dimensional temperature
structure of the filament and to calculate surface brightness maps at 160, 250, 350,
and 500\,$\mu$m. The surface brightness data are used to derive maps of observed
colour temperature and spectral index. In the calculation we use 10\% and 2\%
relative uncertainties for the PACS channel and the SPIRE channels, the same as in
the case of actual observations. The maps consist of 256$\times$256 pixels
corresponding to the 256$^3$ cell discretisation of the 3D model cloud. The models
are characterised by their total optical depth in V band that, measured through the
filament in the perpendicular direction, is either 10.0 or 50.0. To make the model
more concrete, we assume for the models a linear size of 10\,pc and a distance of
400\,pc. With this scaling, the parameter $R_{\rm C}$ is equal to 0.1\,pc and $n_0$
is determined by the selected optical depths.  Each pixel corresponds to
$\sim$20$\arcsec$ and the final maps are convolved to a resolution of 40.0$\arcsec$
before analysis. Figure~\ref{fig:RT_results} shows cross sections of the modelled
filament in column density (V band optical depth), dust temperature, and the colour
temperature and the spectral index estimated from the surface brightness data.

For the $\tau(V)=10$ model, the apparent spectral index decreases to 1.75, 0.15
units below the intrinsic $\beta$ value of 1.9. For the $\tau(V)=50$ model, the
minimum observed value is only 1.10. This demonstrates the difficulty of making a
direct connection between the apparent spectral indices and the intrinsic dust
properties in the case of optically thick clouds. However, because the line-of-sight
temperature variations always imply a decrease in the apparent spectral index, this
increases the significance of the observed correlation between column density and
spectral index. At least in some regions along the line-of-sight, the actual dust
opacity spectral index should be much higher than the observed values.

We calculated two additional cases where the dust properties were modified in
the central regions with density exceeding 25\% of its maximum value.  The
modification is applied only to wavelengths $\lambda>40\,\mu$m and it is determined
by two criteria. First, the dust opacity is set to twice its original value at
250$\mu$m and, secondly, the long wavelength spectral index $\beta'$ is set to a
value of either 1.7 or 2.2. These are ad hoc models where the dust opacity has
discontinuity at 40\,$\mu$m. This has little practical importance because grains
absorb energy at wavelengths shorter than this limit and correspondingly emit the
energy at much longer wavelengths.

The first result is that the observed (i.e., values derived from surface brightness
data) $\beta$ values towards the filament centre are lower in both cases of modified
dust. We interpret this as an indication that the increased line-of-sight
temperature variations associated with the enhanced submillimetre opacity outweigh
the effect of $\Delta \beta \sim$0.2 changes of the intrinsic spectral index. For
the $\tau_{\rm V}$~=~10 model, the minimum $\beta$ values were 1.41 and 1.67 for the
$\beta'=1.7$ and $\beta'=2.2$ cases, respectively. Both are lower than the previous
apparent value of 1.75. For the $\tau_{\rm V}$~=~50 model, the corresponding minimum
values are 0.57 and 0.67, much lower than the 1.10 obtained before modifications to
the dust model.

\begin{figure}
\includegraphics[width=8.3cm]{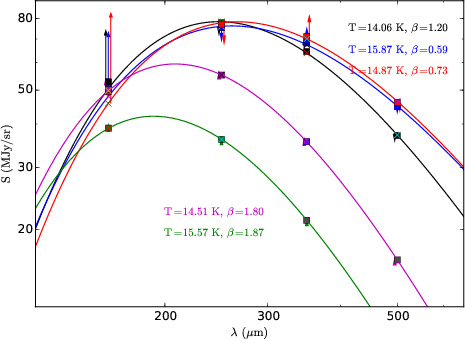}
\caption{
Selected SEDs from $\tau_{\rm V}=50.0$ models. The uppermost three spectra are
towards the centre of the filament, using the original dust model (black line) or
using $\beta=1.7$ dust (blue line) or $\beta=2.2$ dust (red line) in the central
part of the filament. The lower spectra (magenta and green) correspond to larger
offsets (4.4 and 8.7 arcmin) from the symmetry axis that are not directly
affected by the dust modifications. The differences between the measured points and
the modified blackbody curves are indicated with vertical arrows. The length of the
arrow is ten times the difference between the data point and the fitted modified
blackbody curve.
}
\label{fig:RT_spectra}
\end{figure}

Figure~\ref{fig:RT_spectra} shows selected SEDs from the $\tau_{\rm V}$~=~50 models
described above. The uppermost three SEDs are observed towards the filament centre
for the three different dust models. The two lower SEDs are obtained at different
radial offsets, from a region that is unaffected by the $\beta'$ changes that were
applied to the inner part of the filament. We use arrows to indicate the difference
between the fitted modified blackbody curves and the actual data points.  Towards
the optically thick centre, the data do not follow a single modified blackbody
spectrum. The errors are largest at 160\,$\mu$m, because of its smaller weight in
the fit (assuming 10\% uncertainty instead of the 2\% at the longer wavelengths).
However, the fits also systematically overestimate the 250\,$\mu$m intensity and
underestimate the 350\,$\mu$m intensity, this being related to the low values of the
fitted spectral index. The SEDs for both modified dust cases are clearly colder than
the original SED because of the higher opacity at wavelengths where the dust grains
emit most of their energy.

The spectral index was modified at $\lambda>40\mu$m. In the 100--500\,$\mu$m range,
the $\beta$ difference corresponds to a factor of two difference of relative
opacity. Nevertheless, the observed spectra for the $\beta'=1.7$ and $\beta'=2.2$
cases are practically on top of each other over the whole wavelength range probed by
{\it Herschel} observations. After the opacity increase, the central part of the
filament is very cold with dust temperatures close to 5\,K. As a result, even at
500\,$\mu$m the emission per unit dust mass is more than a factor of ten lower than
for the $\sim$17\,K dust on the cloud surface. This means that the $\beta$ changes
at the cloud centre are masked by the warmer, more strongly emitting envelope. In
the model the difference of intrinsic $\beta$ values in the filament centre becomes
evident only at mm wavelengths. In the {\it Herschel} range, the spectral index
determination is clearly very sensitive to temperature variations. If the
160\,$\mu$m data point were omitted from the fits, the spectral index values of
1.16, 0.52, and 0.62 would increase to 1.37, 0.75, and 0.93 respectively, the three
values corresponding to the three models: one with original dust, one with
$\beta'=1.7$ dust in the centre, and one with $\beta'=2.2$ dust in the centre. The
difference between the observed and intrinsic $\beta$ (for given column density)
would also decrease if the column density peak were less steep or if the
observations were averaged over a larger beam. At lower column densities, at larger
offsets from the symmetry axis of the cylindrical filament, the dust SED is
well-described by a single modified blackbody curve, with the observed $\beta$
approaching the intrinsic $\beta$ value.

In spite of such a strong effect in the above model, the observations rarely show
any decrease of $\beta$ towards the dense regions. A weak anti-correlation between
$\beta$ and column density was seen only in a few fields. These include G82.65-2.00
where, based on the observed dust emission, the column density exceeds $N({\rm
H_2})=2\times 10^{22}$\,cm$^{-2}$ in many places \citep{GCC-V}. WISE 12\,$\mu$m data
show clear absorption along the full filament, also suggesting that the visual
extinction is at least $A_{\rm V}\sim 20$\,mag. The average properties of the
filament would therefore be between the two models above but on average closer to
the $\tau_{\rm V}=10.0$ case.  The large scale decrease of $\beta$ towards the
filament is at the level of $\Delta \beta$~=~0.1 and could easily be explained by
temperature variations within the beam. However, even in G82.65-2.00 the $T-\beta$
relation shows an anti-correlation at the smallest scales towards the densest
clumps, contrary to the behaviour in the models.

The temperature mixing in regions like G82.65-2.00 is probably very complex compared
to the simple models above and the net effect on the observed spectral index can be
qualitatively different. The models above showed a maximal effect because we
compared emission between the cold centre and the fully illuminated cloud surface.
In real clouds the dense regions are shielded by diffuse envelopes whose extinction
decreases temperature contrasts at small scales. The drop in the observed $\beta$
would be smaller if the column density profile were less steep or if the data were
smoothed by a larger beam. If the emission region is inhomogeneous, elongated, or
consists of completely separate clumps along the line-of-sight, the radiative
transfer effects would be reduced. Internal heating of some clumps is not able to
explain the general anti-correlation between $T$ and $\beta$ \citep{Malinen2011,
Juvela2012_Tmix}. Therefore, it seems that we are observing real changes in the
intrinsic dust properties that, remarkably, are strong enough to be seen above the
opposite effects caused by temperature variations.

\begin{figure}
\includegraphics[width=8.3cm]{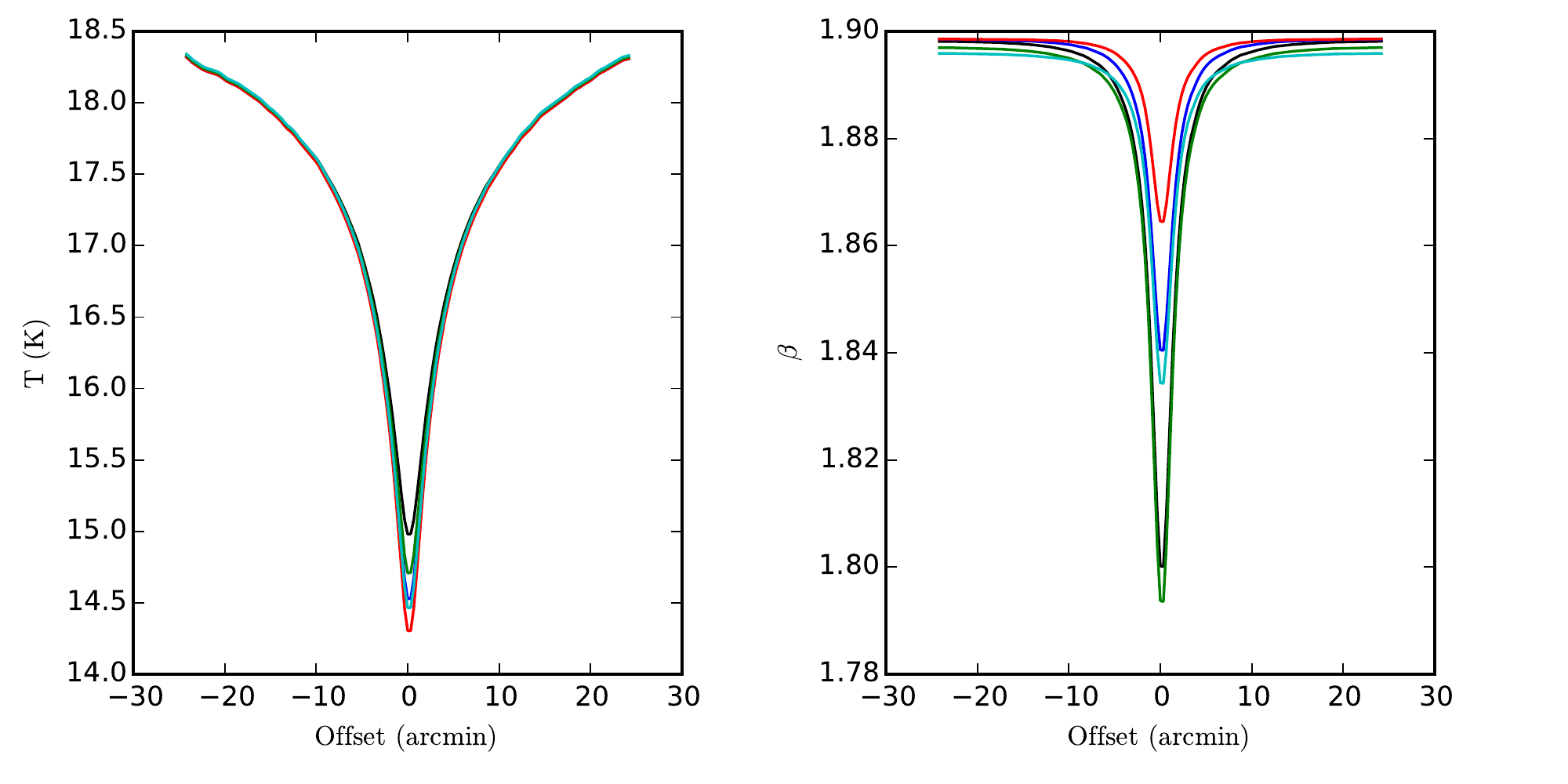}
\caption{
Temperature and spectral index profiles of the $\tau(V)=10$ model observed using
different wavelength combinations. In order of decreasing colour temperature these are:
100, 350, 550, and 850\,$\mu$m (black curves), 160, 250, 350, and 500\,$\mu$m (green
curves), 160, 350, 550, and 850\,$\mu$m (blue curves), 250, 350, and 500\,$\mu$m (cyan
curves), 250, 350, 550, and 850\,$\mu$m (red curves).
}
\label{fig:beta_vs_bands}
\end{figure}

The models can also be used to examine the dependence on the wavelengths used.
Figure~\ref{fig:beta_vs_bands} shows temperature and spectral index profiles of the
$\tau(V)=10$ model that are estimated using five different wavelength combinations. As
expected, $\beta$ values are least biased when long wavelengths, 250--850\,$\mu$m, are
used. However, in the case of this model, even the combination of PACS and SPIRE bands
160--500\,$\mu$m results in values that are lower by no more than $\Delta \beta=0.07$.
This suggests that over most of the observed fields the bias resulting from temperature
variations is not significantly different for the different combinations of frequency
bands.

\clearpage

\section{Fits of $T$, $\beta$ with {\it IRAS} and {\it Planck} data} \label{app:pfit}

The figures \ref{fig:PI_FIG_001}--\ref{fig:PI_FIG_115} show results of modified
blackbody fits of the 100\,$\mu$m {\it IRAS} data and the {\it Planck} 857\,GHz--353\,GHz
observations. The bottom frames show the 217\,GHz residuals (observation minus the
prediction of the fitted model) and the difference in spectral index when the
lowest included frequency is 353\,GHz or 217\,GHz.  For the field G300.86-9.00 the
plot was already shown in Fig.~\ref{fig:PI_FIG_105_main}.

\begin{figure}
\includegraphics[width=8.8cm]{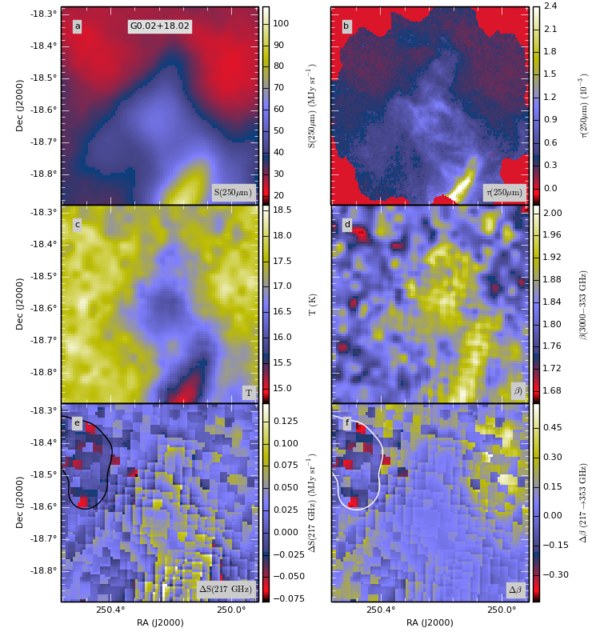}
\includegraphics[width=8.8cm]{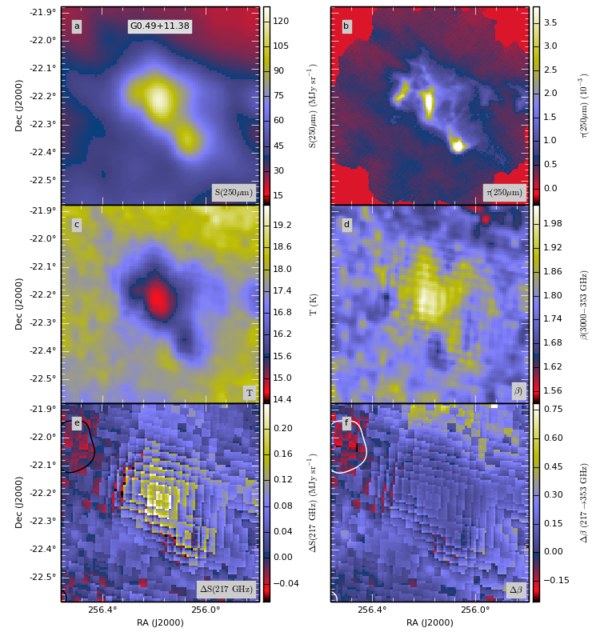}
\caption{
Modified blackbody fits to {\it Planck} and IRIS data in fields G0.02+18.02 and
G0.49+11.38. The frames $a$, $c$, and $d$ show 250\,$\mu$m surface brightness,
colour temperature, and spectral index fitted to 3000\,GHz ({\it IRAS}
100\,$\mu$m), 857\,GHz, 545\,GHz, and 353\,GHz data, respectively. Frame $b$ shows
the higher resolution {\it Herschel} $\tau(250\,\mu {\rm m})$ map. Frame $e$ is the
surface brightness residual between the 217\,GHz observation and the above fit. The
last frame shows the difference between spectral indices derived with fits to
3000\,GHz-353\,GHz and to 3000\,GHz-217\,GHz data. The fields in this and the
following figures are in order of increasing galactic longitude.
}
\label{fig:PI_FIG_001}
\end{figure}

\begin{figure}
\includegraphics[width=8.8cm]{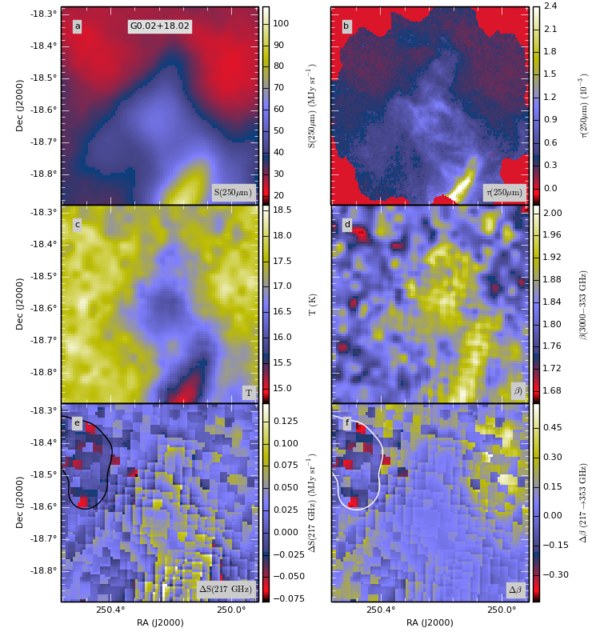}
\includegraphics[width=8.8cm]{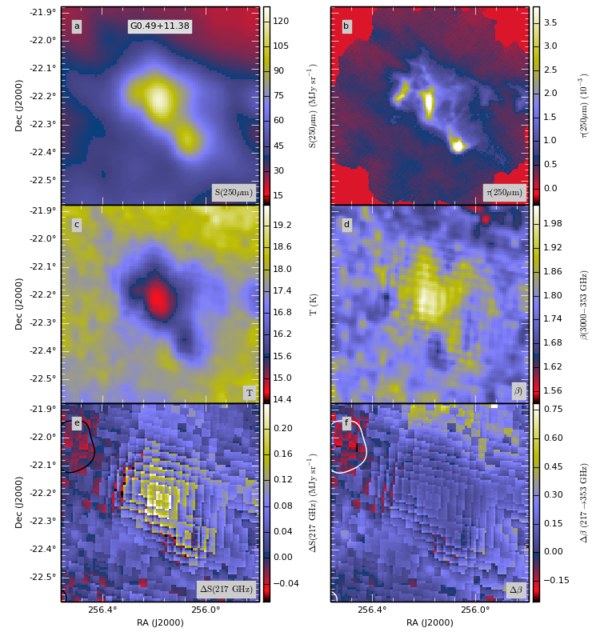}
\caption{Continued{\ldots} Fields G0.02+18.02 and G0.49+11.38}
\end{figure}
\begin{figure}
\includegraphics[width=8.8cm]{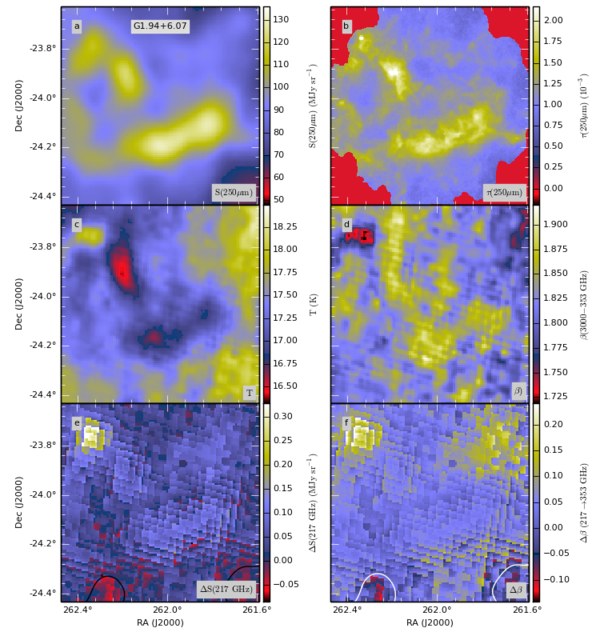}
\includegraphics[width=8.8cm]{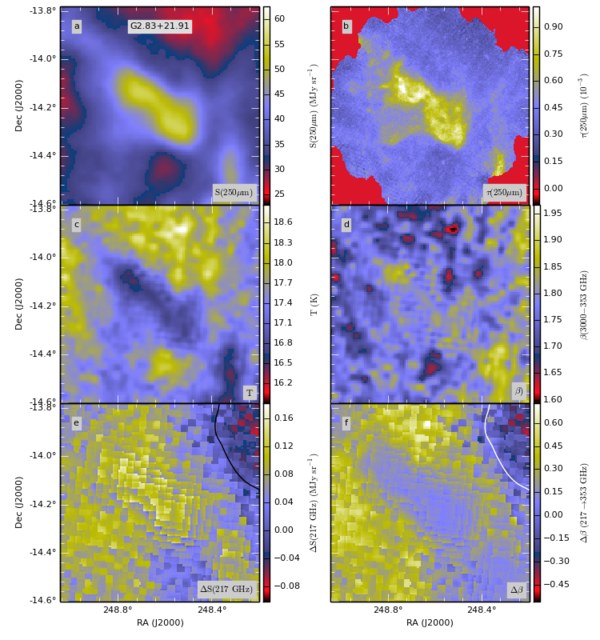}
\caption{Continued{\ldots} Fields G1.94+6.07 and G2.83+21.91}
\end{figure}
\begin{figure}
\includegraphics[width=8.8cm]{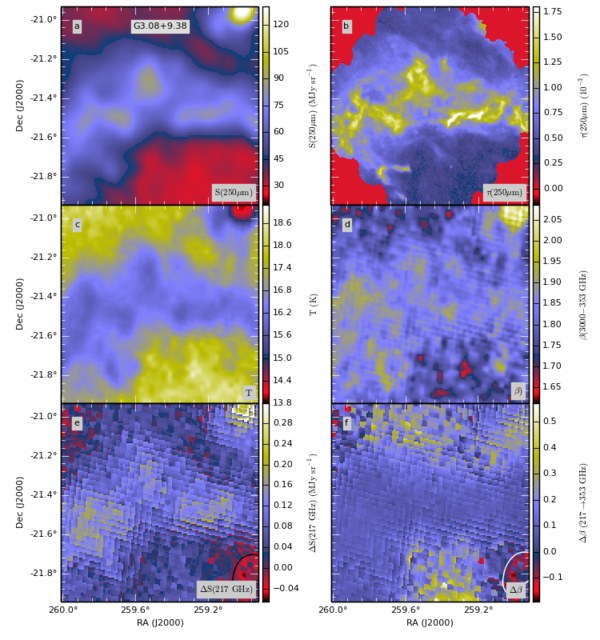}
\includegraphics[width=8.8cm]{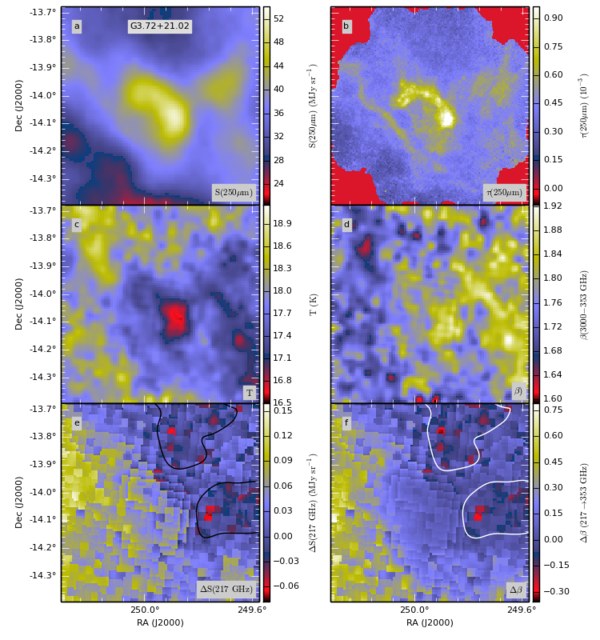}
\caption{Continued{\ldots} Fields G3.08+9.38 and G3.72+21.02}
\end{figure}
\begin{figure}
\includegraphics[width=8.8cm]{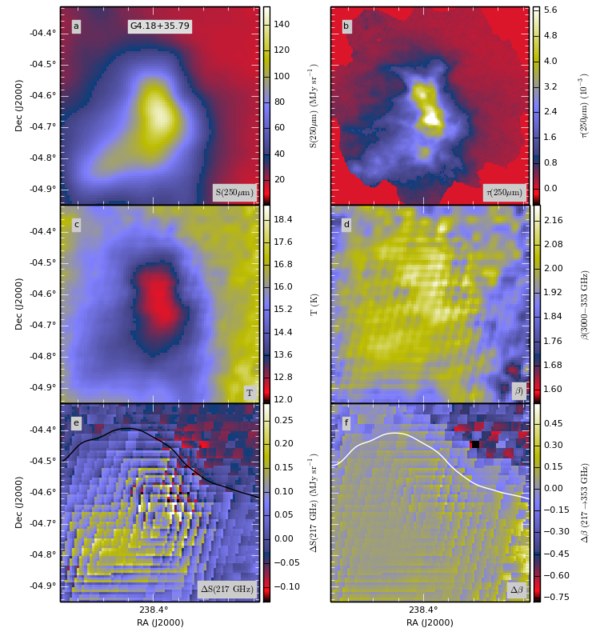}
\includegraphics[width=8.8cm]{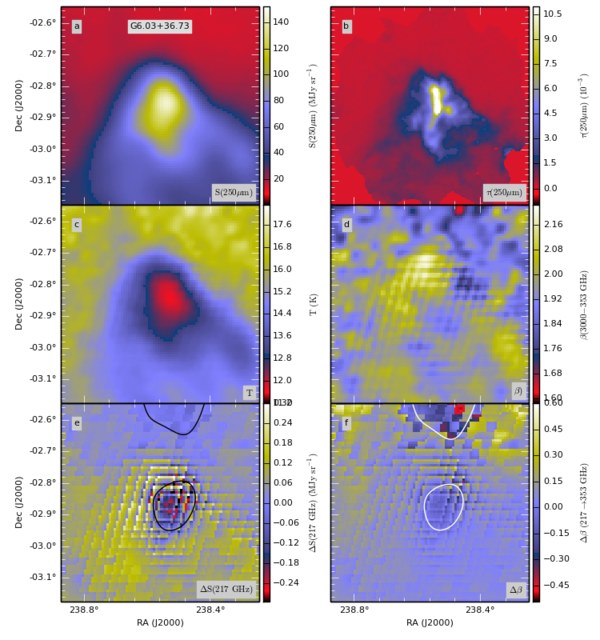}
\caption{Continued{\ldots} Fields G4.18+35.79 and G6.03+36.73}
\end{figure}
\begin{figure}
\includegraphics[width=8.8cm]{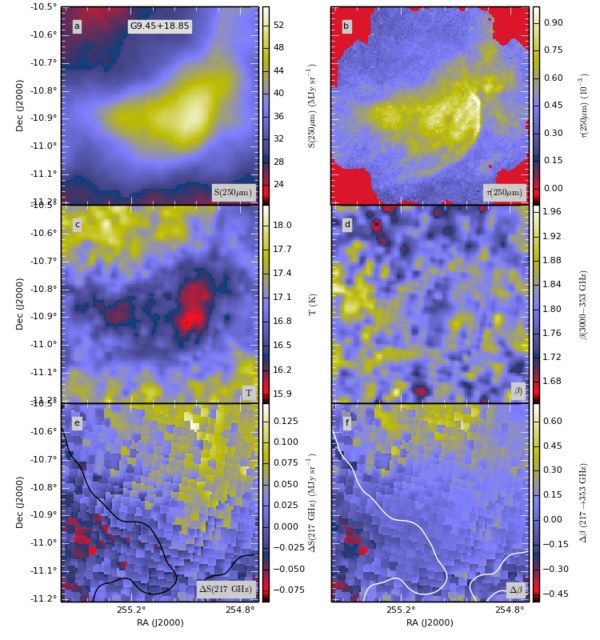}
\includegraphics[width=8.8cm]{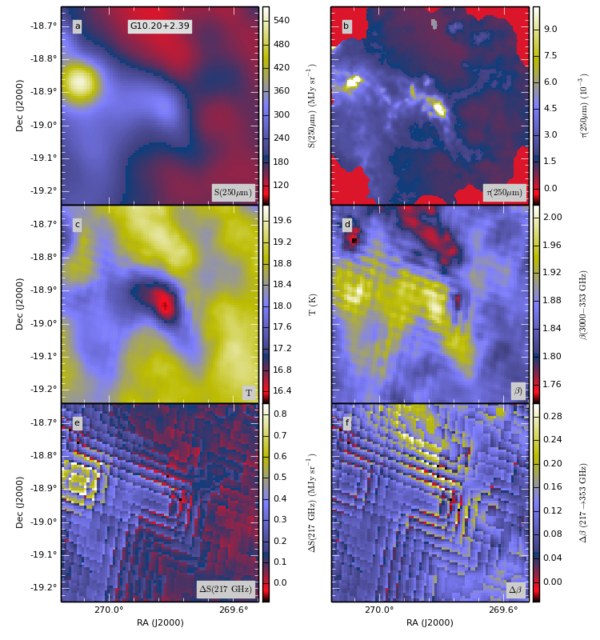}
\caption{Continued{\ldots} Fields G9.45+18.85 and G10.20+2.39}
\end{figure}
\begin{figure}
\includegraphics[width=8.8cm]{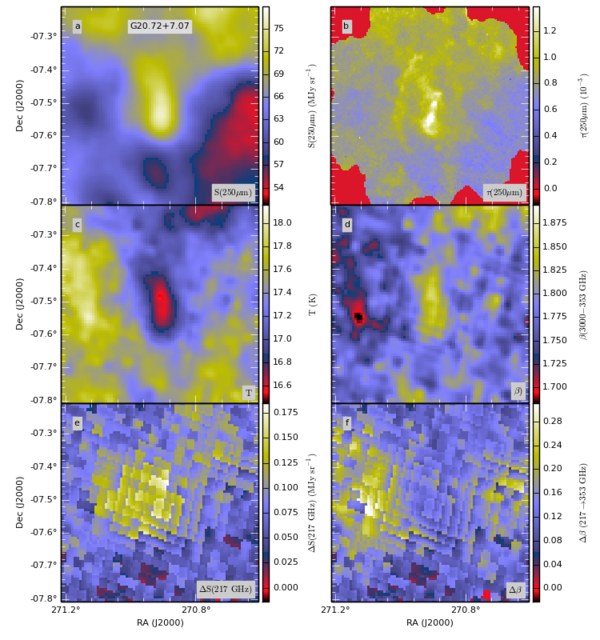}
\includegraphics[width=8.8cm]{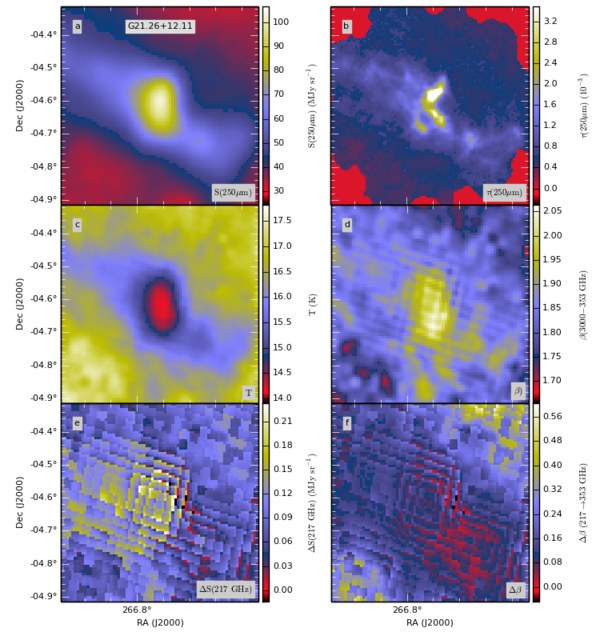}
\caption{Continued{\ldots} Fields G20.72+7.07 and G21.26+12.11}
\end{figure}
\begin{figure}
\includegraphics[width=8.8cm]{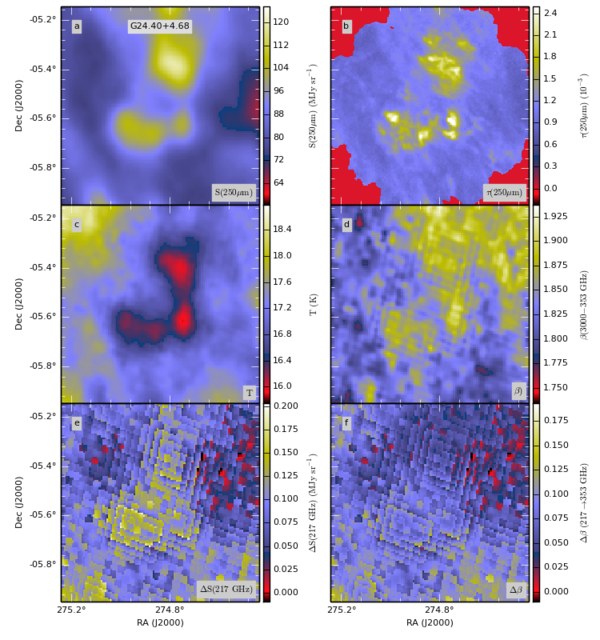}
\includegraphics[width=8.8cm]{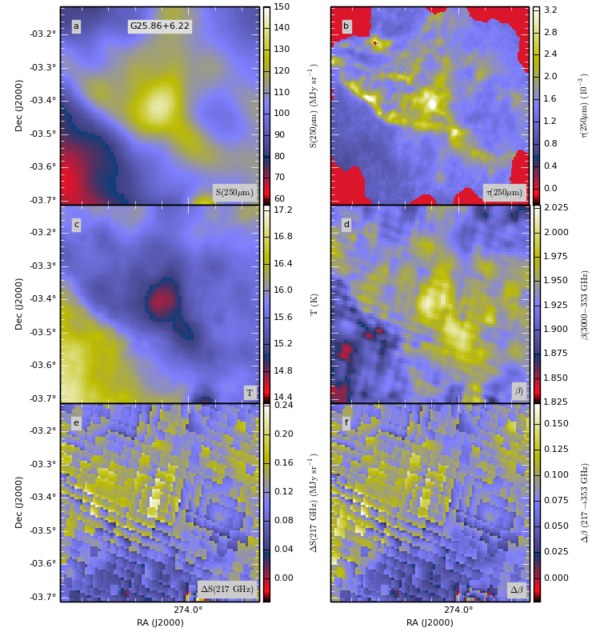}
\caption{Continued{\ldots} Fields G24.40+4.68 and G25.86+6.22}
\end{figure}
\begin{figure}
\includegraphics[width=8.8cm]{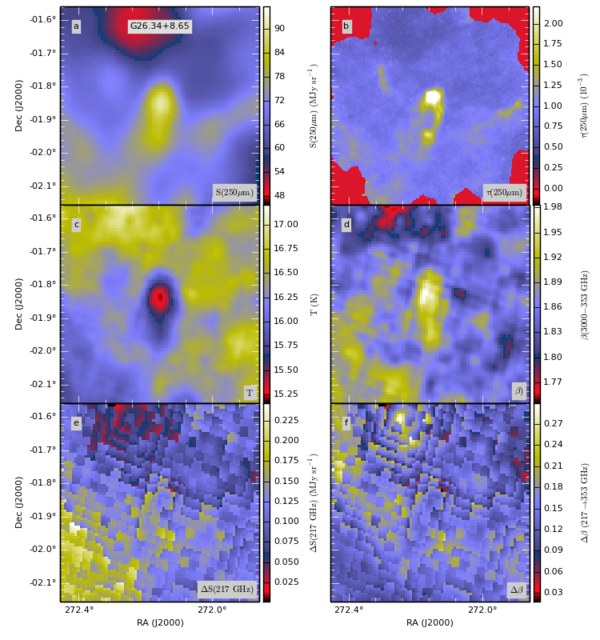}
\includegraphics[width=8.8cm]{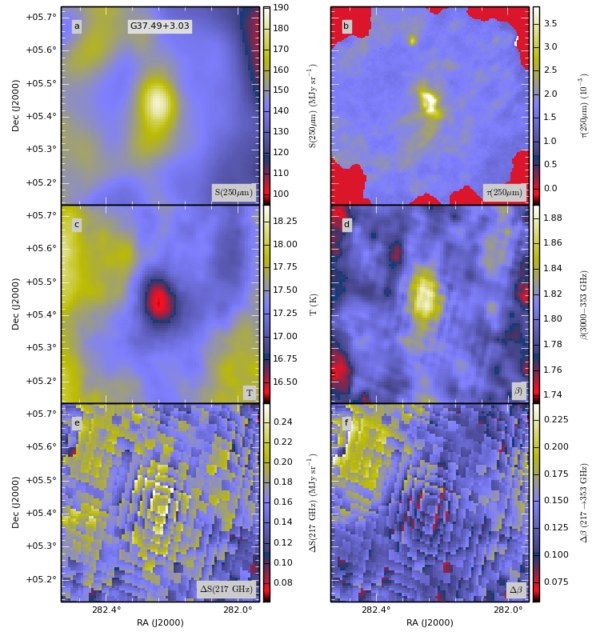}
\caption{Continued{\ldots} Fields G26.34+8.65 and G37.49+3.03}
\end{figure}
\begin{figure}
\includegraphics[width=8.8cm]{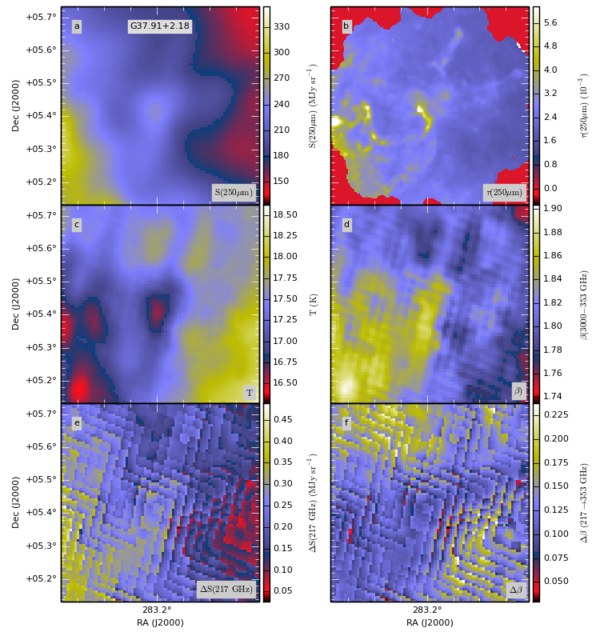}
\includegraphics[width=8.8cm]{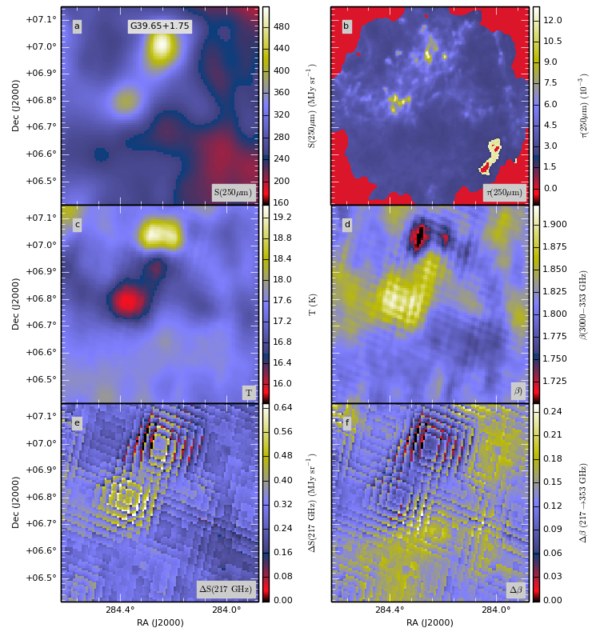}
\caption{Continued{\ldots} Fields G37.91+2.18 and G39.65+1.75}
\end{figure}

\clearpage

\begin{figure}
\includegraphics[width=8.8cm]{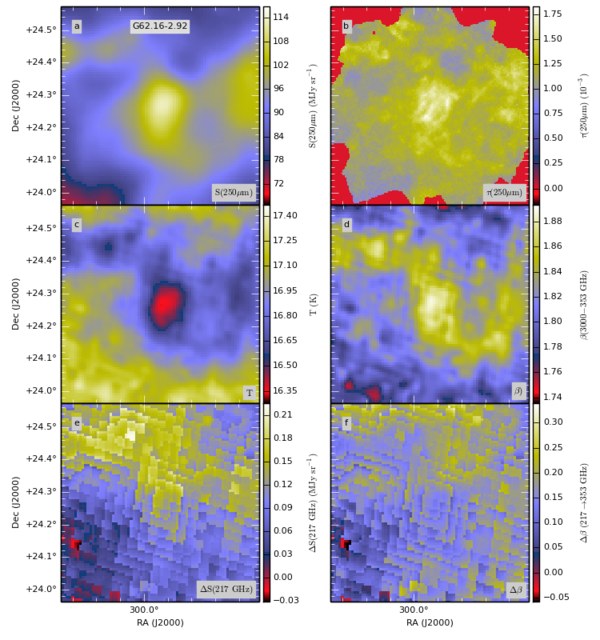}
\includegraphics[width=8.8cm]{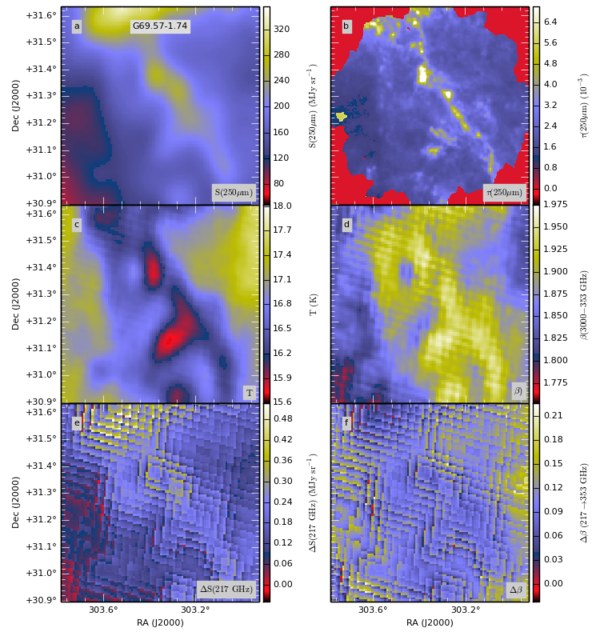}
\caption{Continued{\ldots} Fields G62.16-2.92 and G69.57-1.74}
\end{figure}
\begin{figure}
\includegraphics[width=8.8cm]{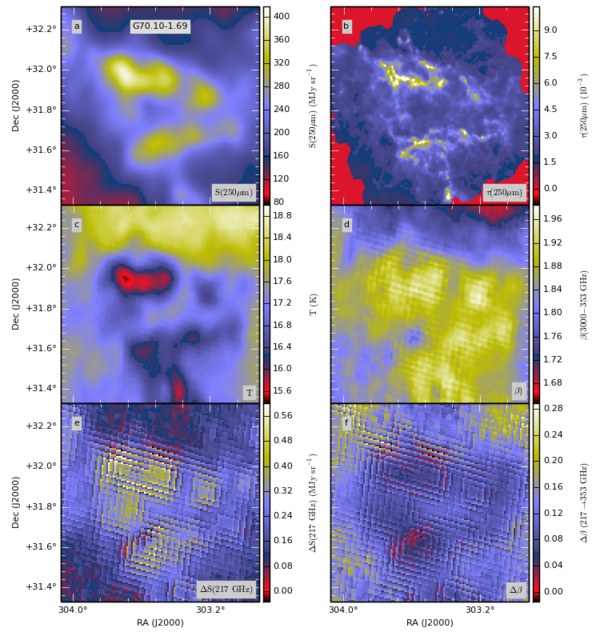}
\includegraphics[width=8.8cm]{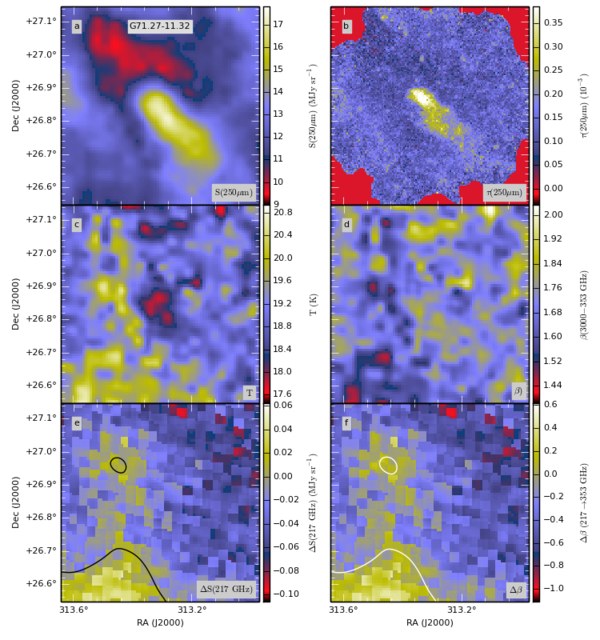}
\caption{Continued{\ldots} Fields G70.10-1.69 and G71.27-11.32}
\end{figure}
\begin{figure}
\includegraphics[width=8.8cm]{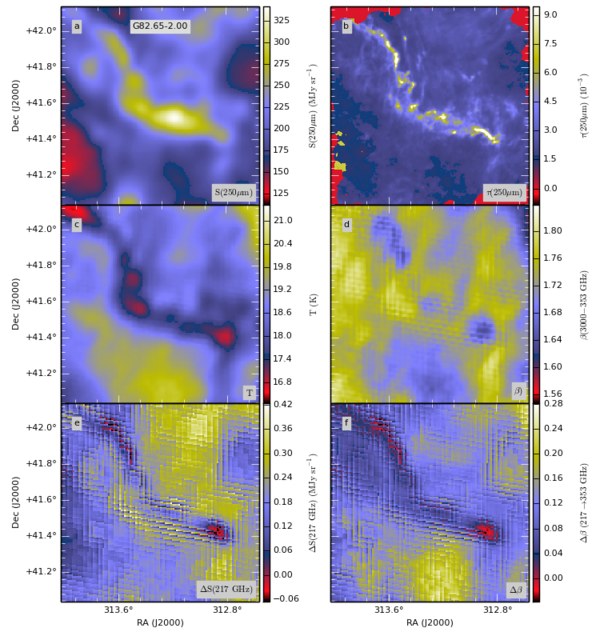}
\includegraphics[width=8.8cm]{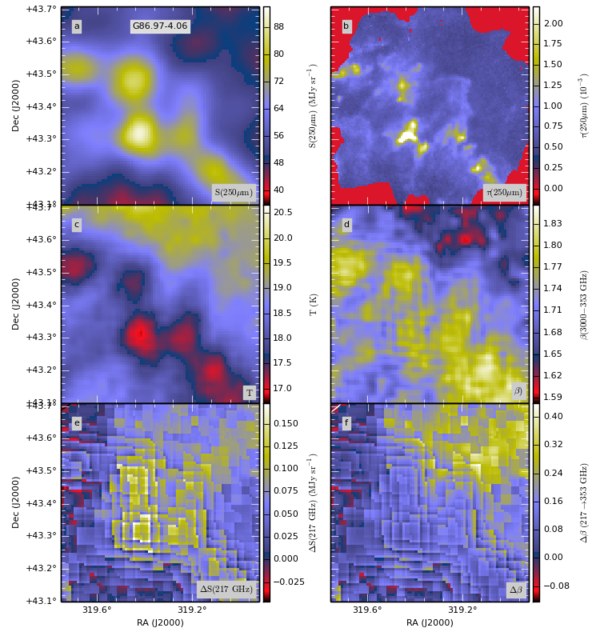}
\caption{Continued{\ldots} Fields G82.65-2.00 and G86.97-4.06}
\end{figure}
\begin{figure}
\includegraphics[width=8.8cm]{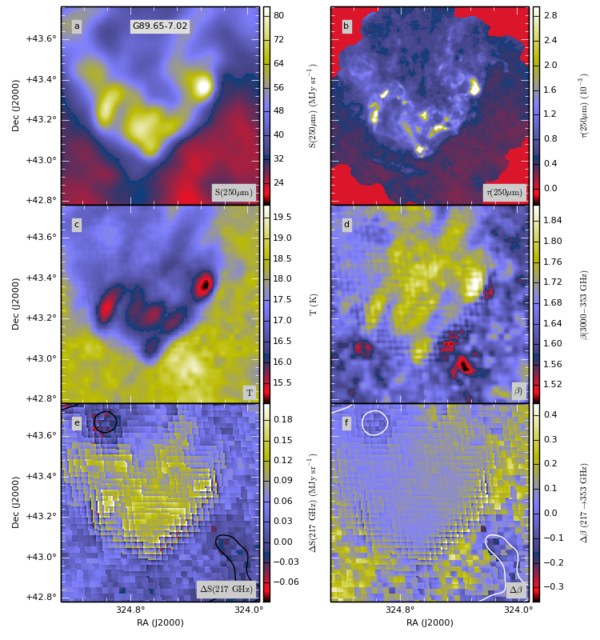}
\includegraphics[width=8.8cm]{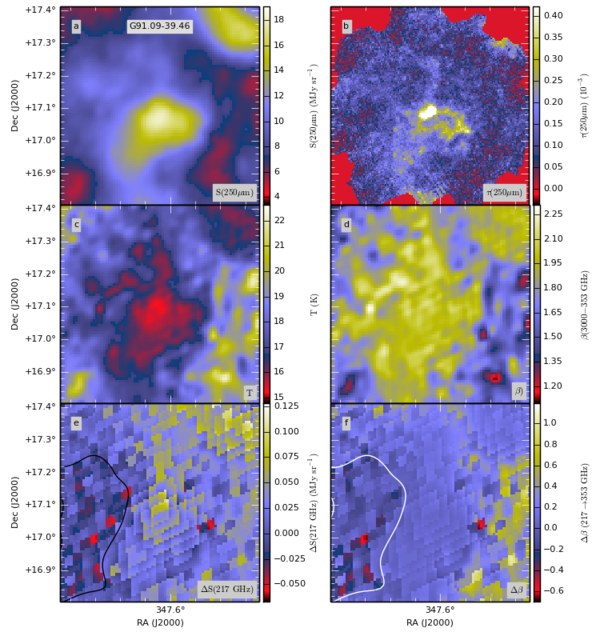}
\caption{Continued{\ldots} Fields G89.65-7.02 and G91.09-39.46}
\end{figure}
\begin{figure}
\includegraphics[width=8.8cm]{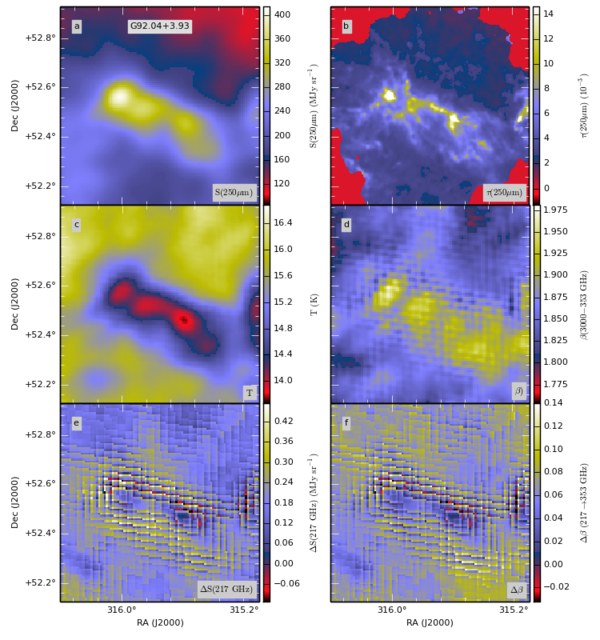}
\includegraphics[width=8.8cm]{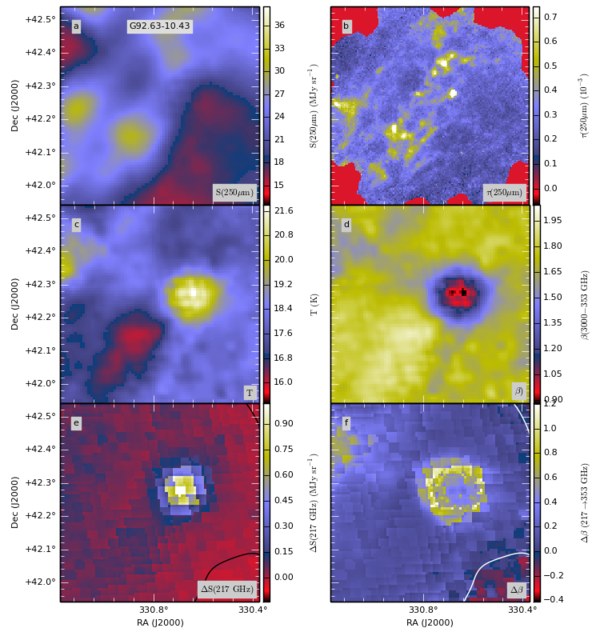}
\caption{Continued{\ldots} Fields G92.04+3.93 and G92.63-10.43}
\end{figure}
\begin{figure}
\includegraphics[width=8.8cm]{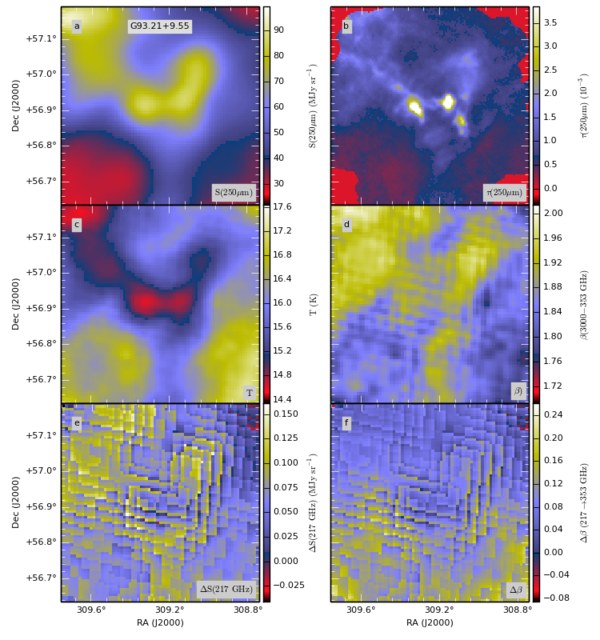}
\includegraphics[width=8.8cm]{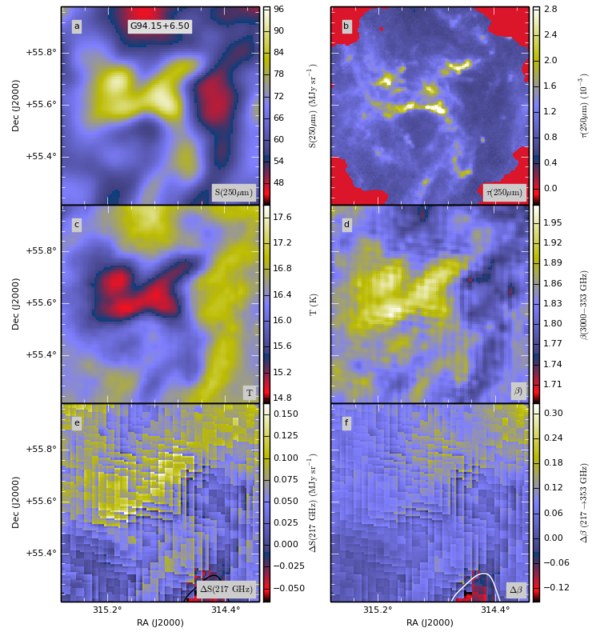}
\caption{Continued{\ldots} Fields G93.21+9.55 and G94.15+6.50}
\end{figure}
\begin{figure}
\includegraphics[width=8.8cm]{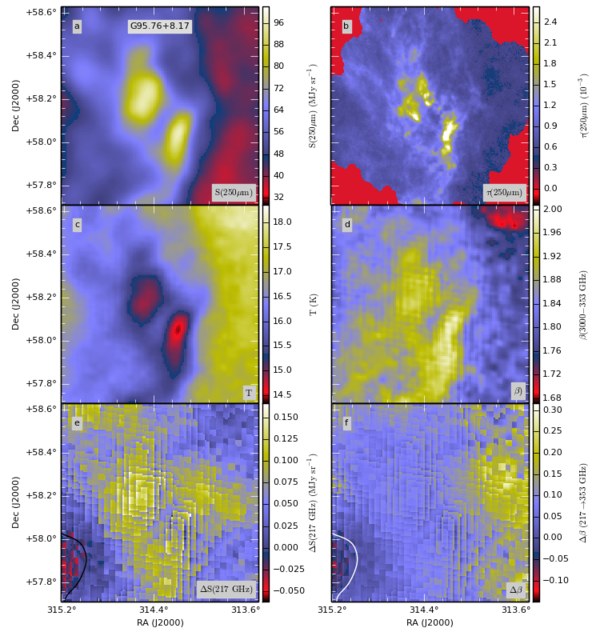}
\includegraphics[width=8.8cm]{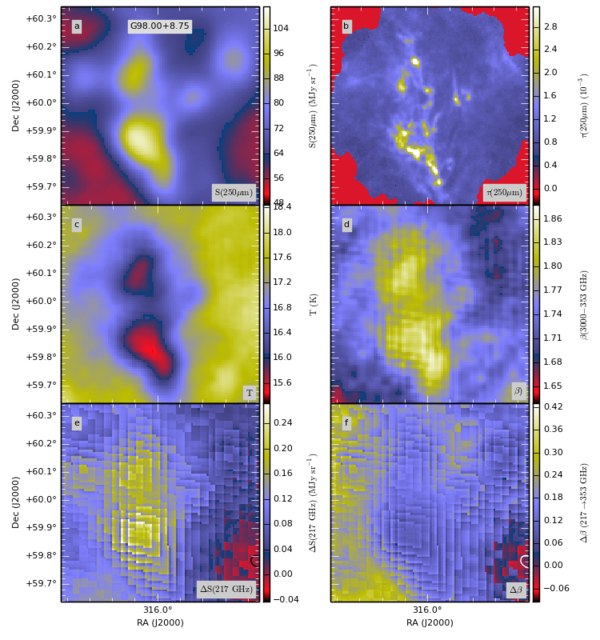}
\caption{Continued{\ldots} Fields G95.76+8.17 and G98.00+8.75}
\end{figure}
\begin{figure}
\includegraphics[width=8.8cm]{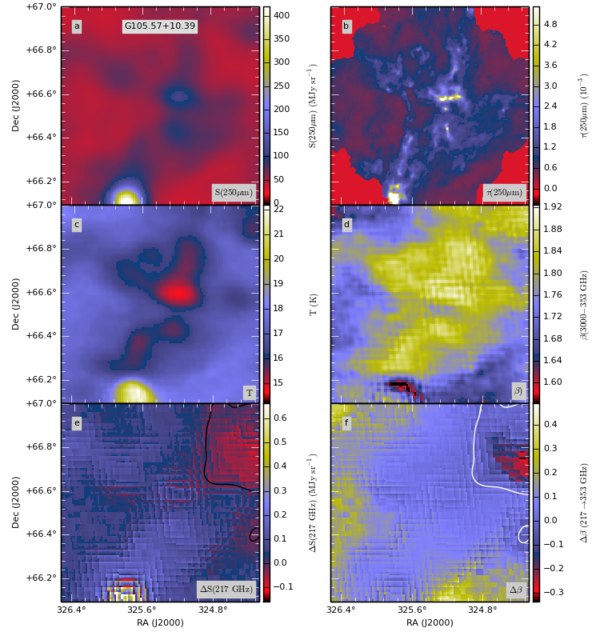}
\includegraphics[width=8.8cm]{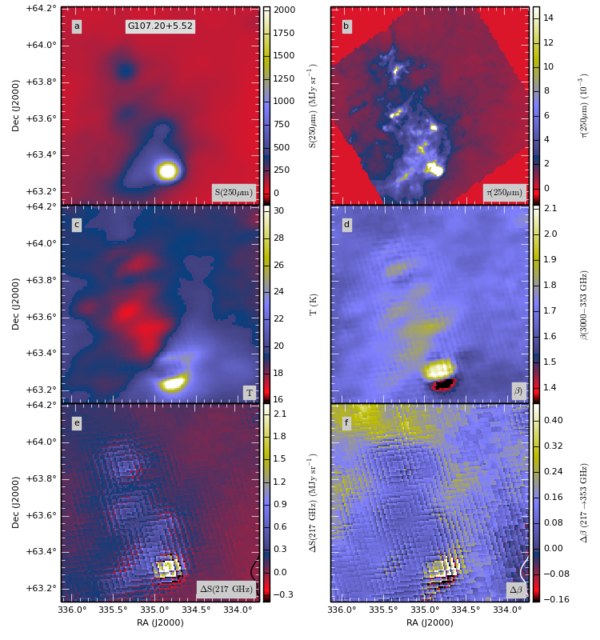}
\caption{Continued{\ldots} Fields G105.57+10.39 and G107.20+5.52}
\end{figure}
\begin{figure}
\includegraphics[width=8.8cm]{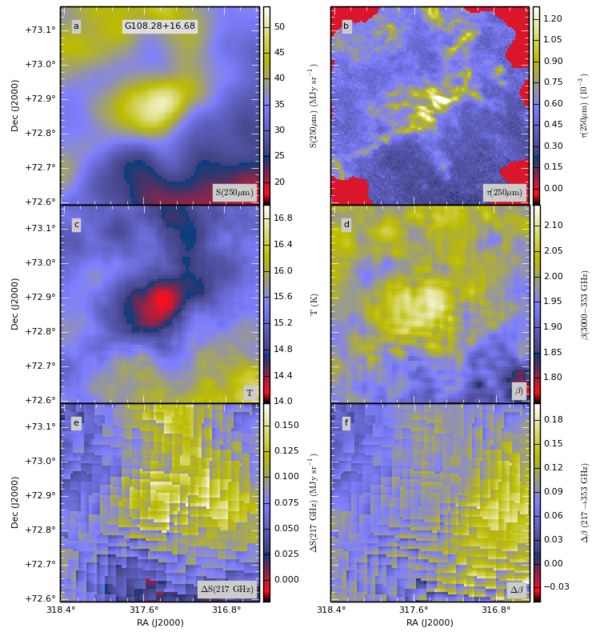}
\includegraphics[width=8.8cm]{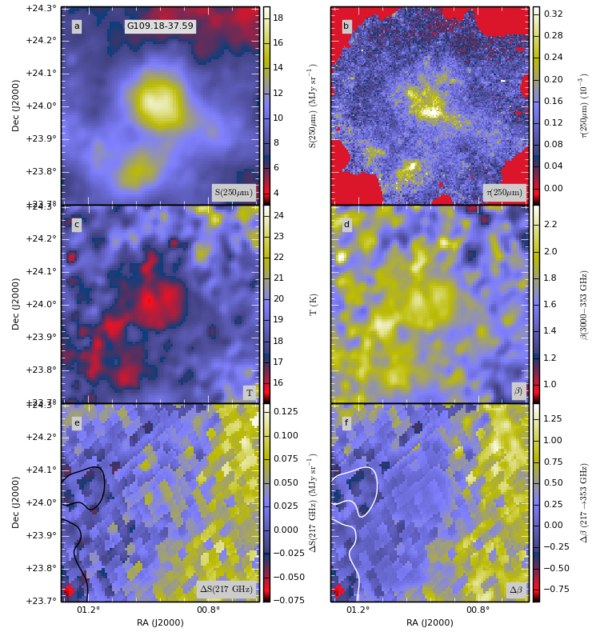}
\caption{Continued{\ldots} Fields G108.28+16.68 and G109.18-37.59}
\end{figure}
\begin{figure}
\includegraphics[width=8.8cm]{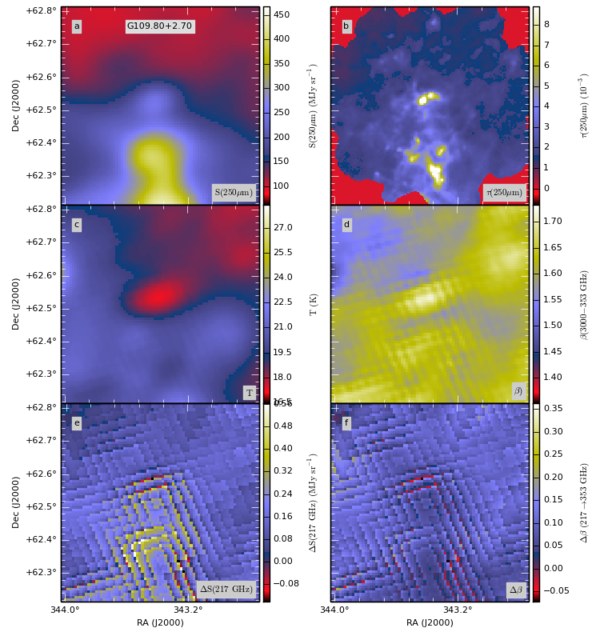}
\includegraphics[width=8.8cm]{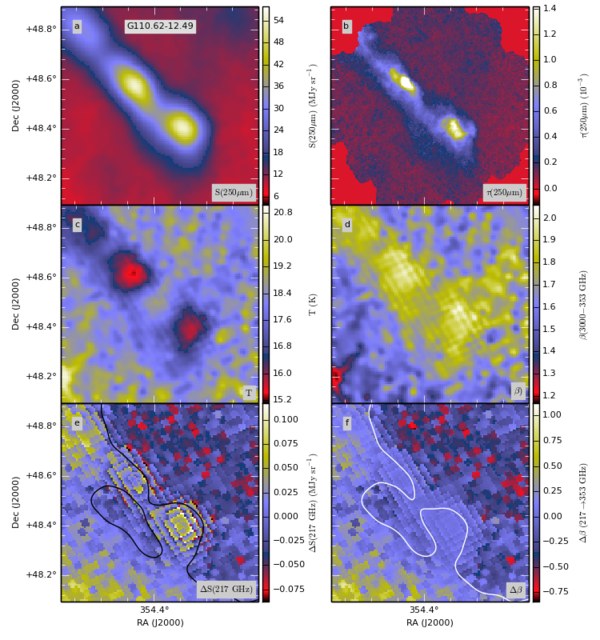}
\caption{Continued{\ldots} Fields G109.80+2.70 and G110.62-12.49}
\end{figure}

\clearpage

\begin{figure}
\includegraphics[width=8.8cm]{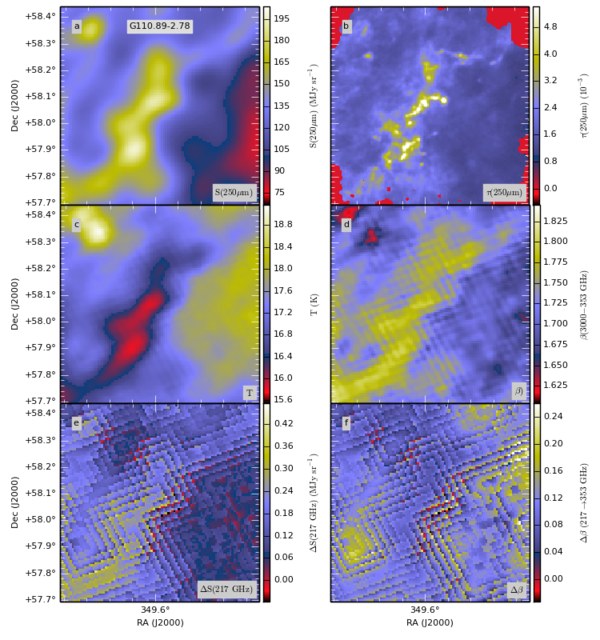}
\includegraphics[width=8.8cm]{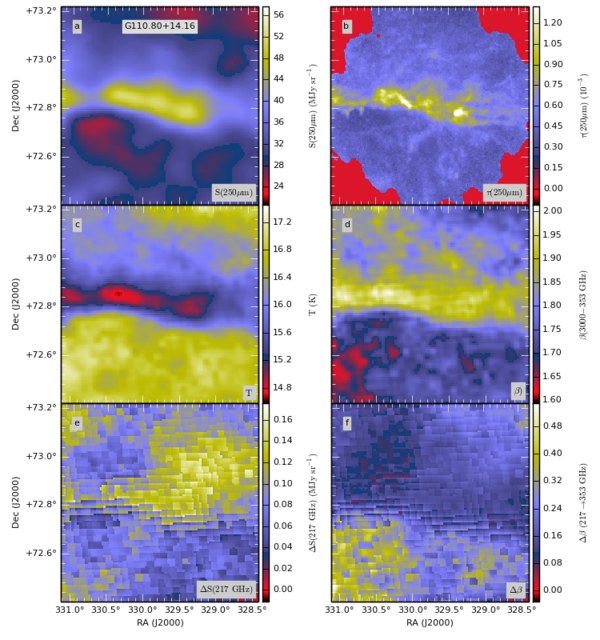}
\caption{Continued{\ldots} Fields G110.89-2.78 and G110.80+14.16}
\end{figure}
\begin{figure}
\includegraphics[width=8.8cm]{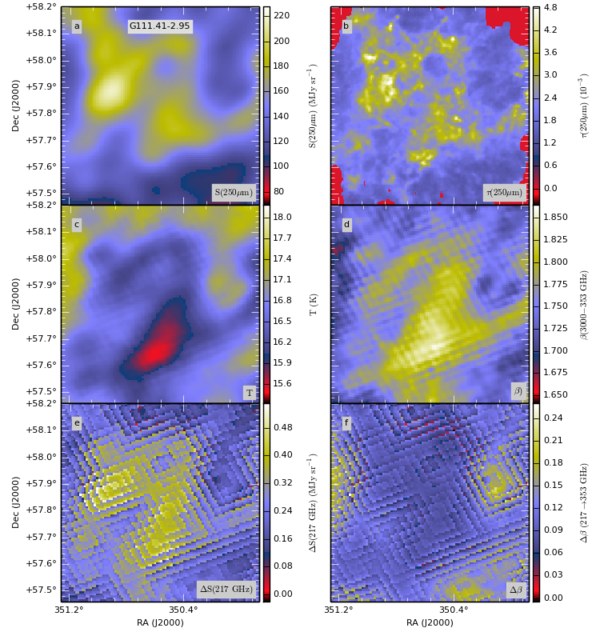}
\includegraphics[width=8.8cm]{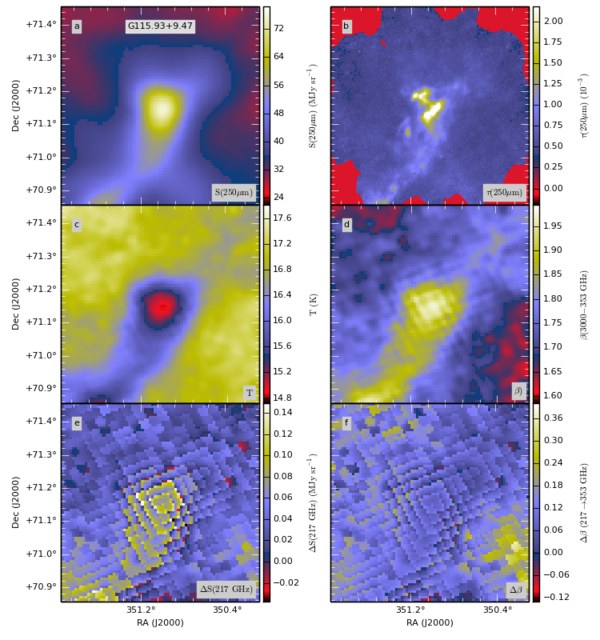}
\caption{Continued{\ldots} Fields G111.41-2.95 and G115.93+9.47}
\end{figure}
\begin{figure}
\includegraphics[width=8.8cm]{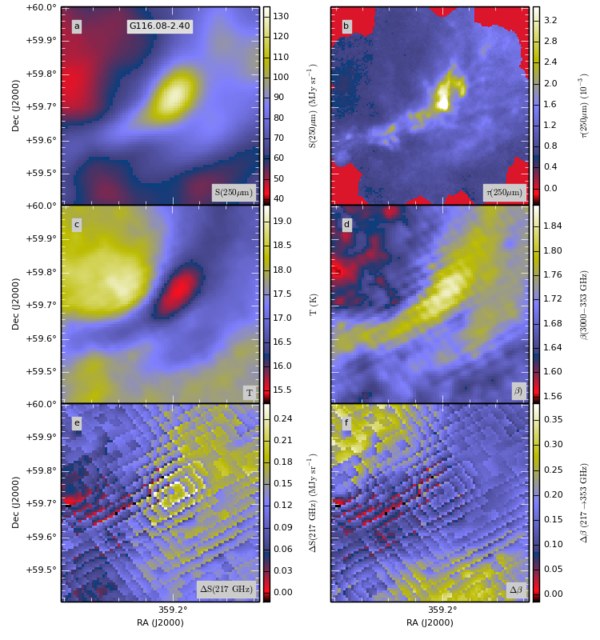}
\includegraphics[width=8.8cm]{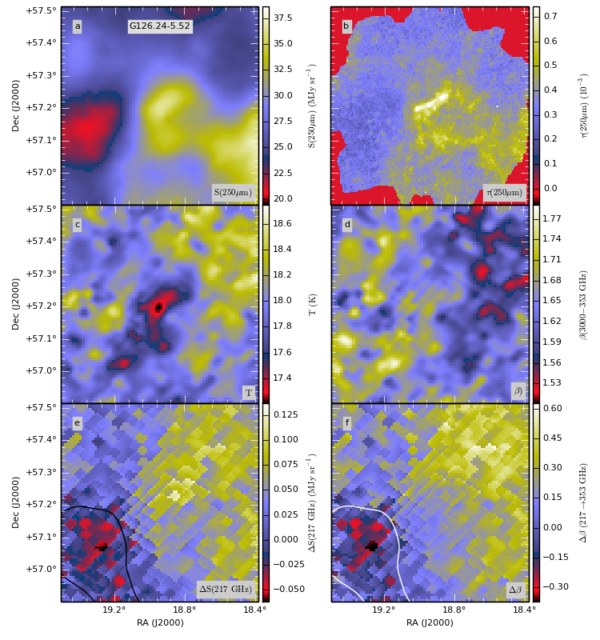}
\caption{Continued{\ldots} Fields G116.08-2.40 and G126.24-5.52}
\end{figure}
\begin{figure}
\includegraphics[width=8.8cm]{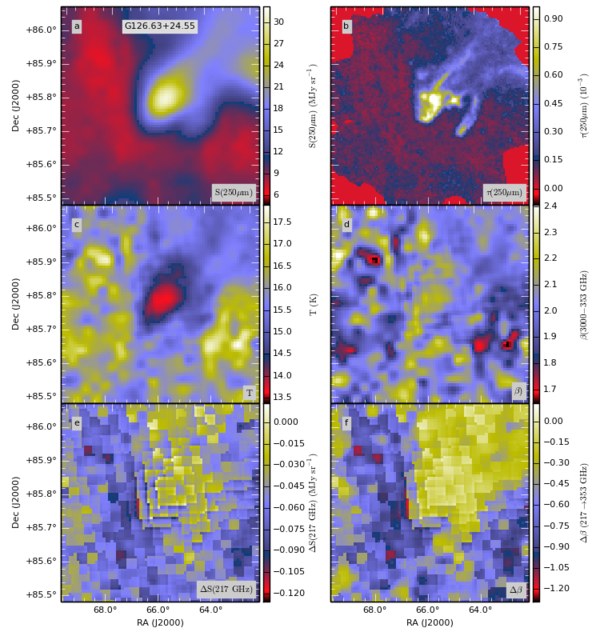}
\includegraphics[width=8.8cm]{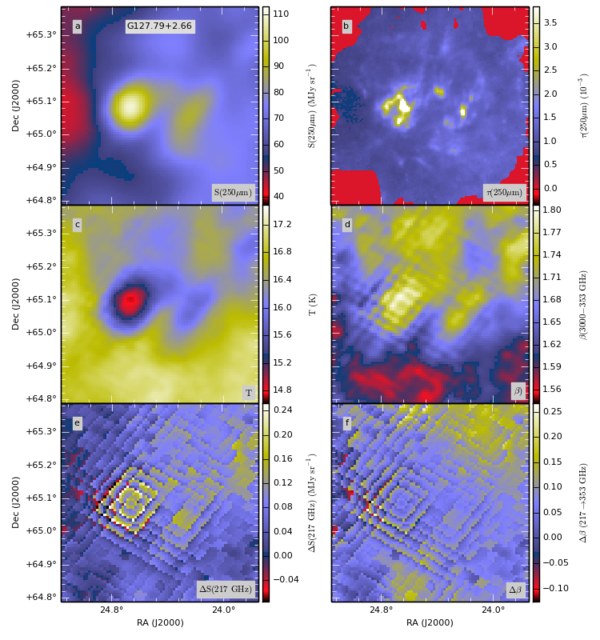}
\caption{Continued{\ldots} Fields G126.63+24.55 and G127.79+2.66}
\end{figure}
\begin{figure}
\includegraphics[width=8.8cm]{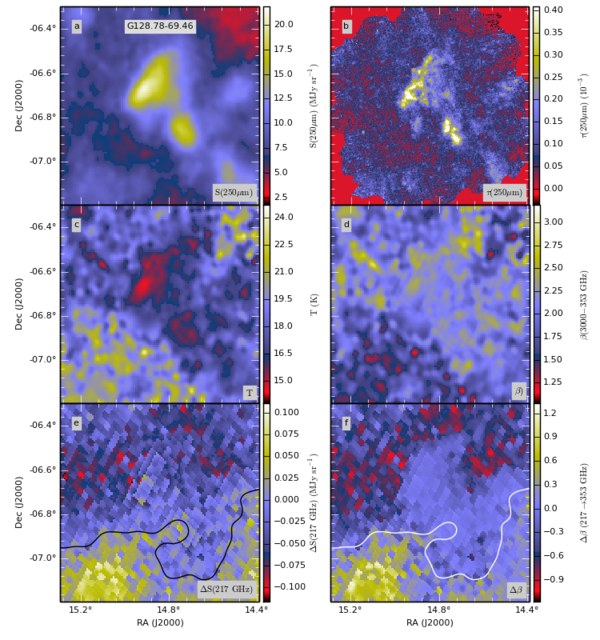}
\includegraphics[width=8.8cm]{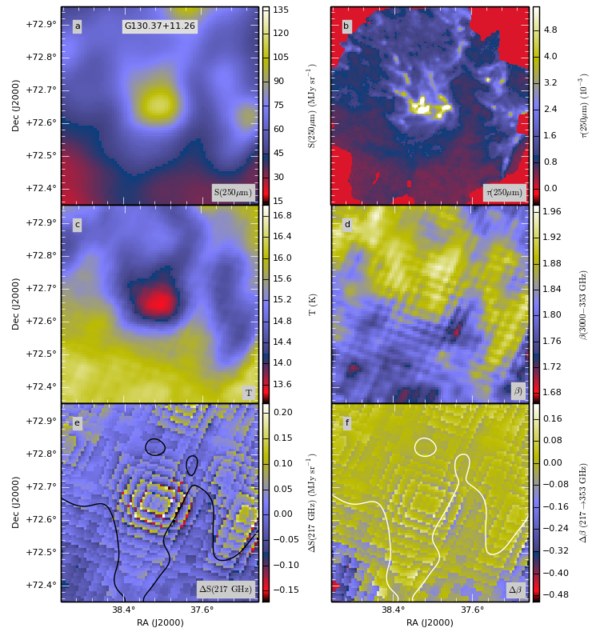}
\caption{Continued{\ldots} Fields G128.78-69.46 and G130.37+11.26}
\end{figure}
\begin{figure}
\includegraphics[width=8.8cm]{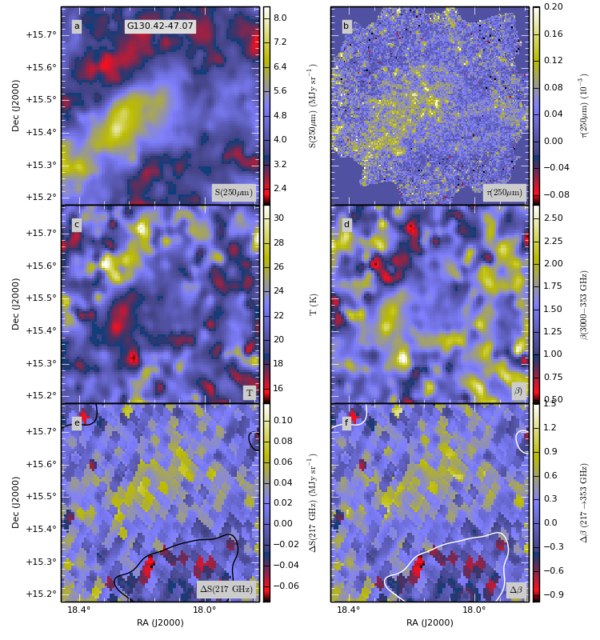}
\includegraphics[width=8.8cm]{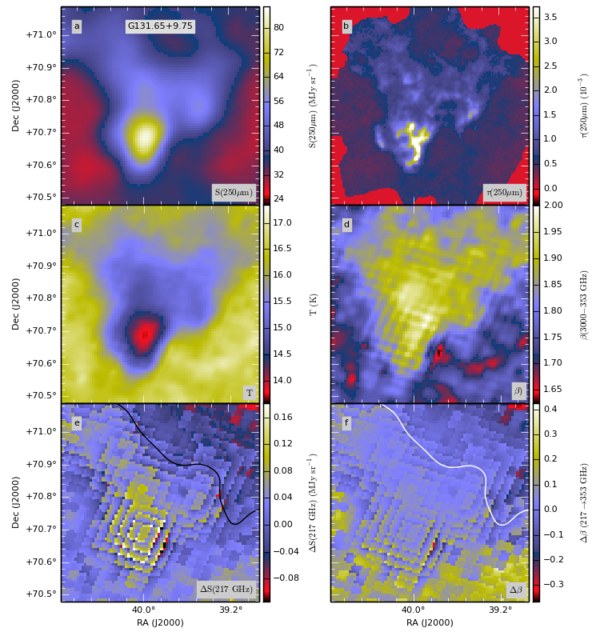}
\caption{Continued{\ldots} Fields G130.42-47.07 and G131.65+9.75}
\end{figure}
\begin{figure}
\includegraphics[width=8.8cm]{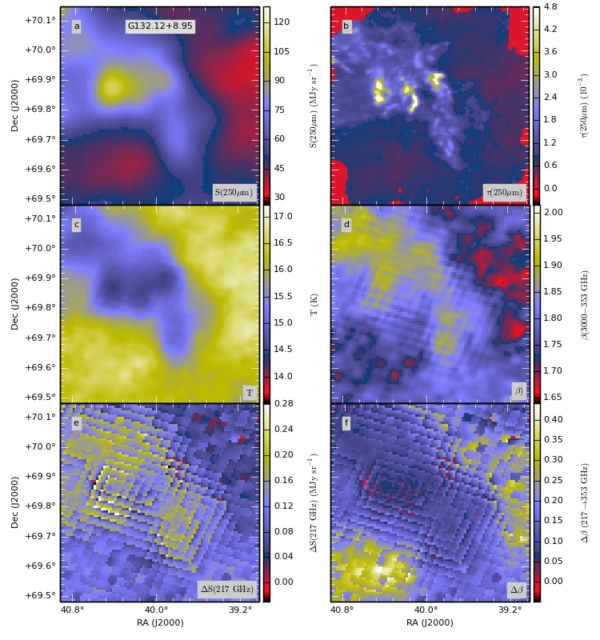}
\includegraphics[width=8.8cm]{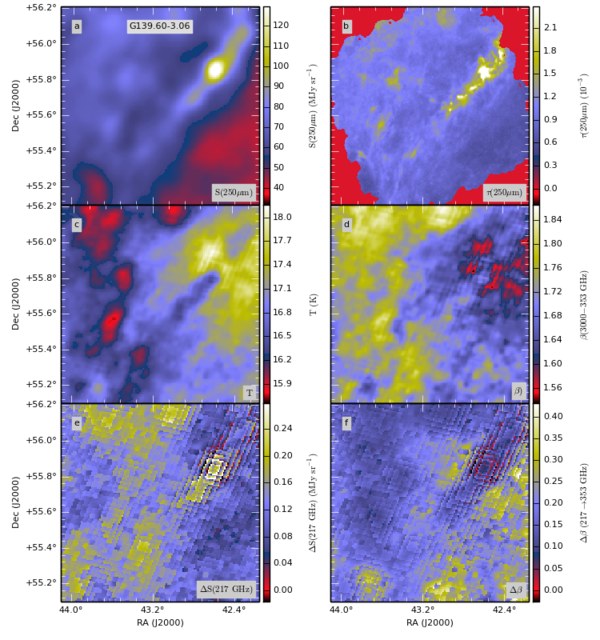}
\caption{Continued{\ldots} Fields G132.12+8.95 and G139.60-3.06}
\end{figure}
\begin{figure}
\includegraphics[width=8.8cm]{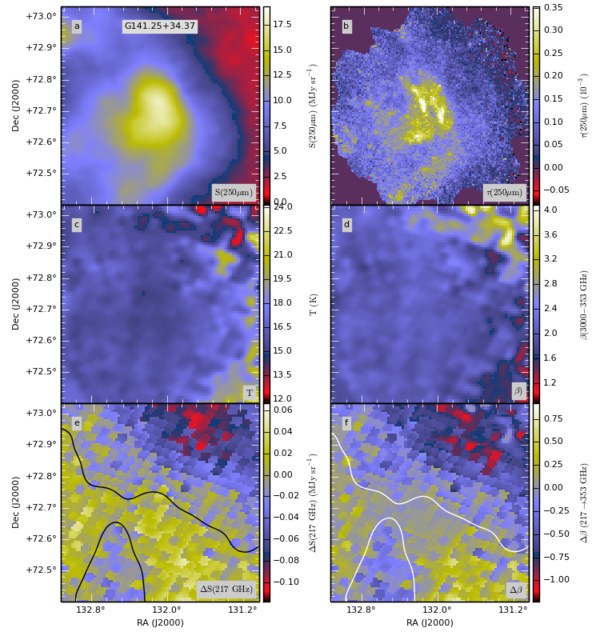}
\includegraphics[width=8.8cm]{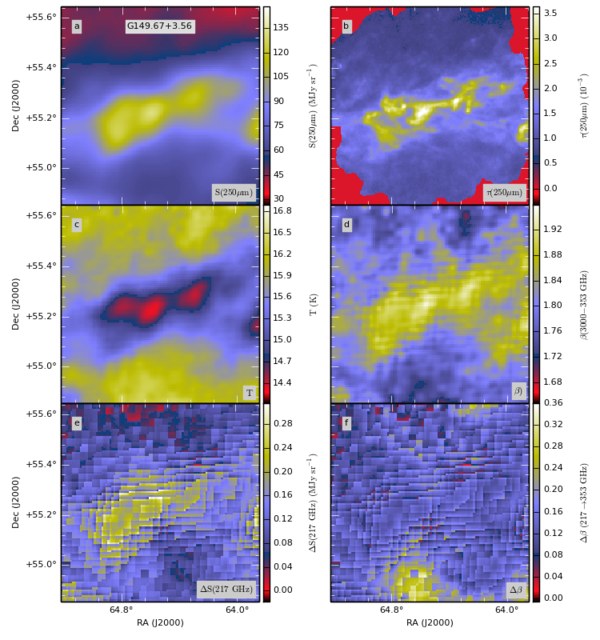}
\caption{Continued{\ldots} Fields G141.25+34.37 and G149.67+3.56}
\end{figure}
\begin{figure}
\includegraphics[width=8.8cm]{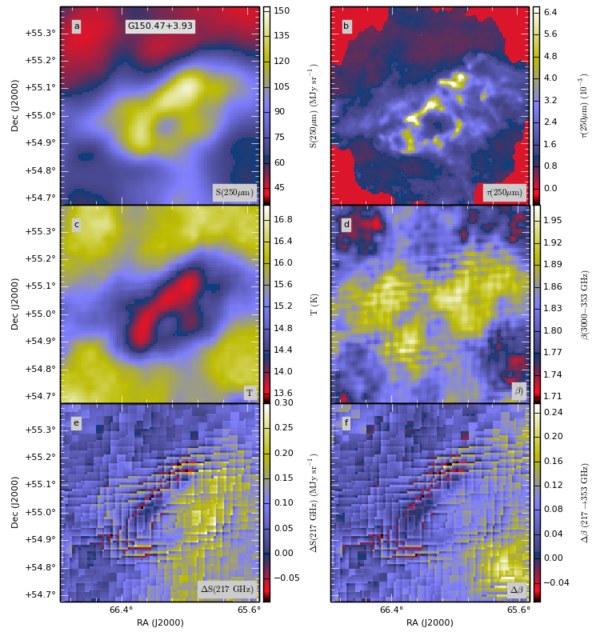}
\includegraphics[width=8.8cm]{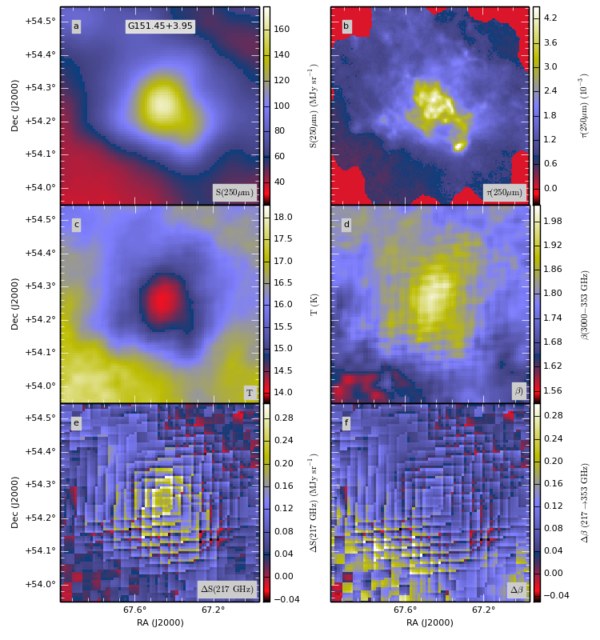}
\caption{Continued{\ldots} Fields G150.47+3.93 and G151.45+3.95}
\end{figure}
\begin{figure}
\includegraphics[width=8.8cm]{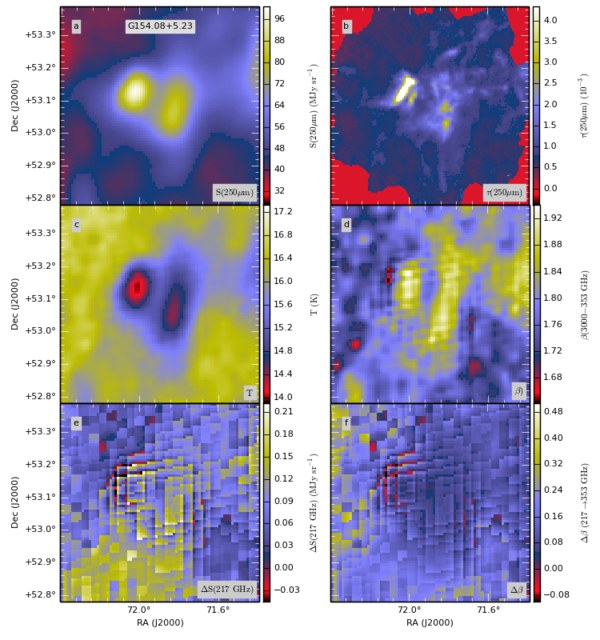}
\includegraphics[width=8.8cm]{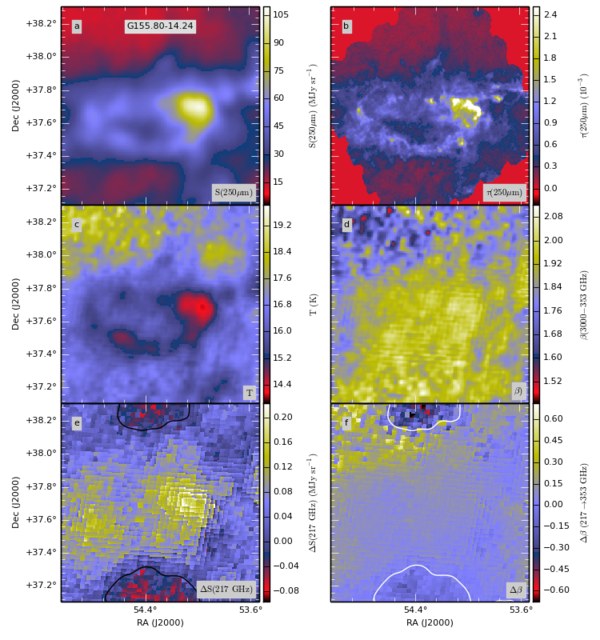}
\caption{Continued{\ldots} Fields G154.08+5.23 and G155.80-14.24}
\end{figure}

\clearpage

\begin{figure}
\includegraphics[width=8.8cm]{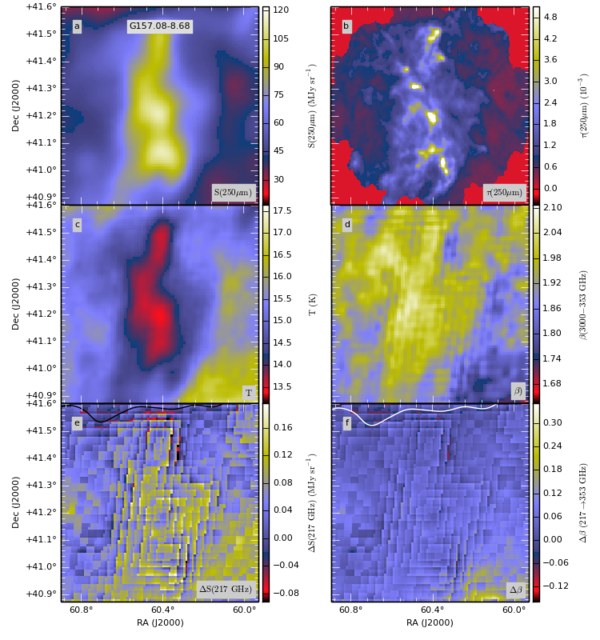}
\includegraphics[width=8.8cm]{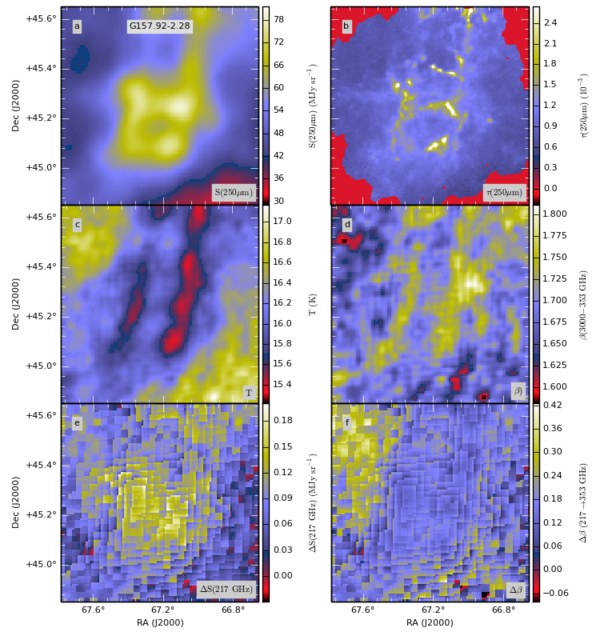}
\caption{Continued{\ldots} Fields G157.08-8.68 and G157.92-2.28}
\end{figure}
\begin{figure}
\includegraphics[width=8.8cm]{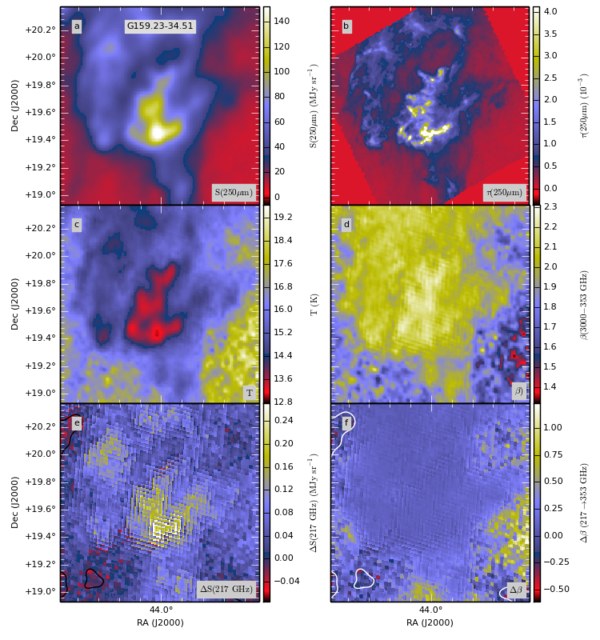}
\includegraphics[width=8.8cm]{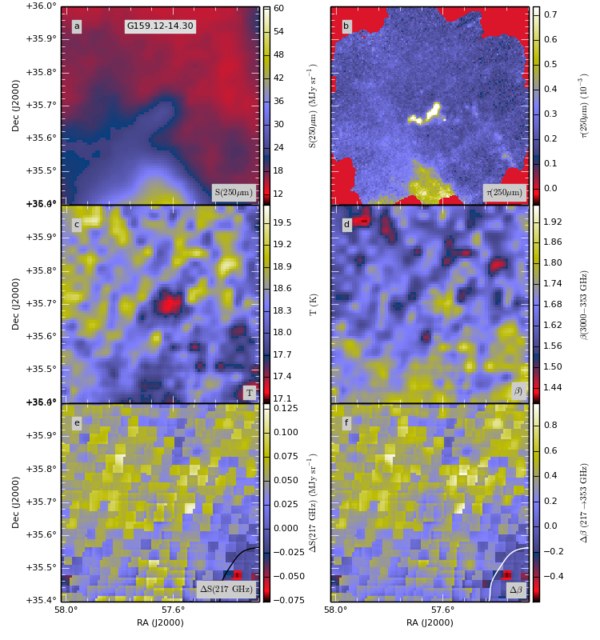}
\caption{Continued{\ldots} Fields G159.23-34.51 and G159.12-14.30}
\end{figure}
\begin{figure}
\includegraphics[width=8.8cm]{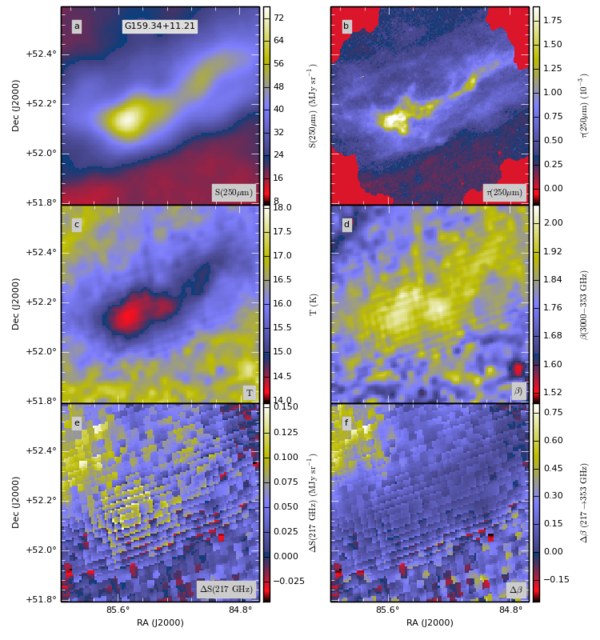}
\includegraphics[width=8.8cm]{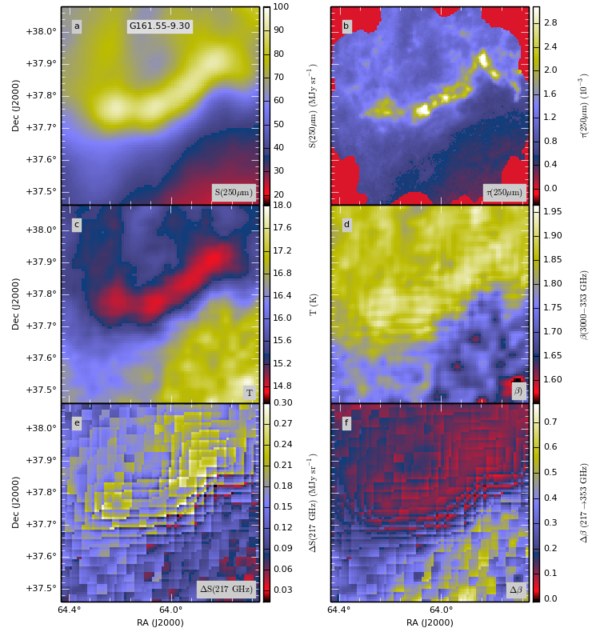}
\caption{Continued{\ldots} Fields G159.34+11.21 and G161.55-9.30}
\end{figure}
\begin{figure}
\includegraphics[width=8.8cm]{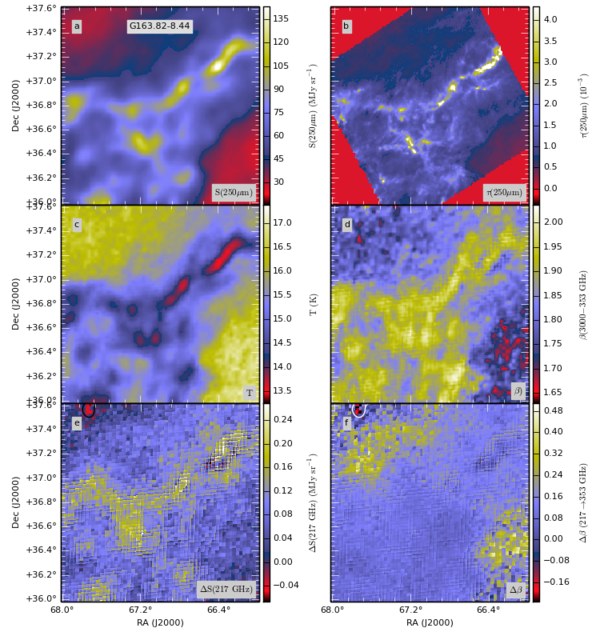}
\includegraphics[width=8.8cm]{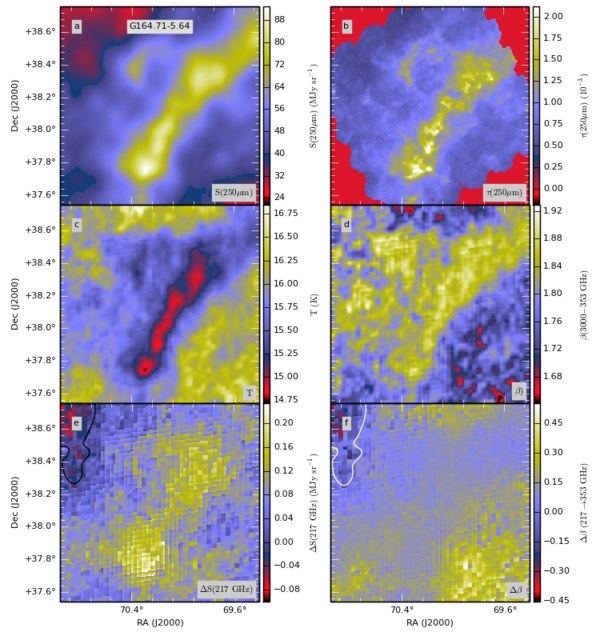}
\caption{Continued{\ldots} Fields G163.82-8.44 and G164.71-5.64}
\end{figure}
\begin{figure}
\includegraphics[width=8.8cm]{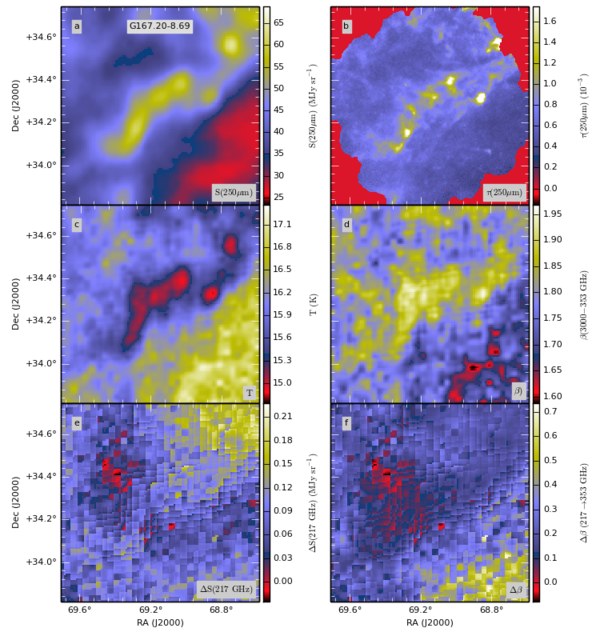}
\includegraphics[width=8.8cm]{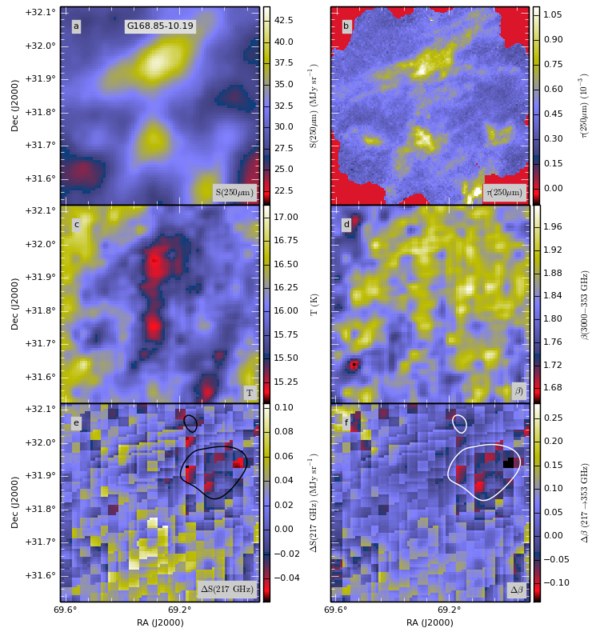}
\caption{Continued{\ldots} Fields G167.20-8.69 and G168.85-10.19}
\end{figure}
\begin{figure}
\includegraphics[width=8.8cm]{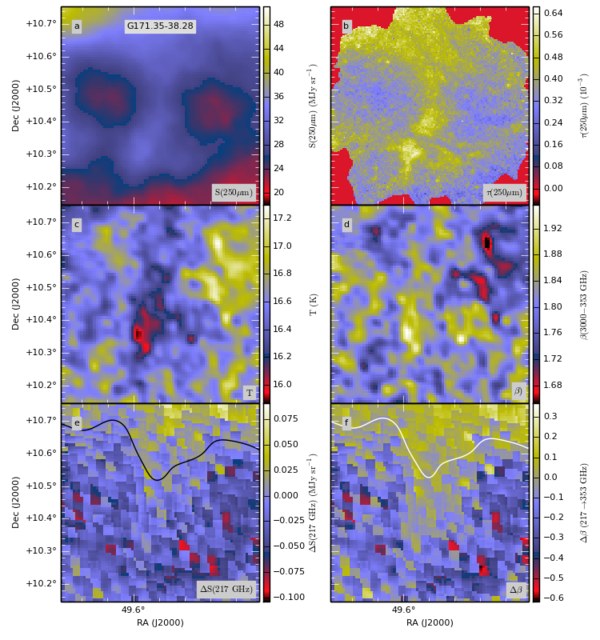}
\includegraphics[width=8.8cm]{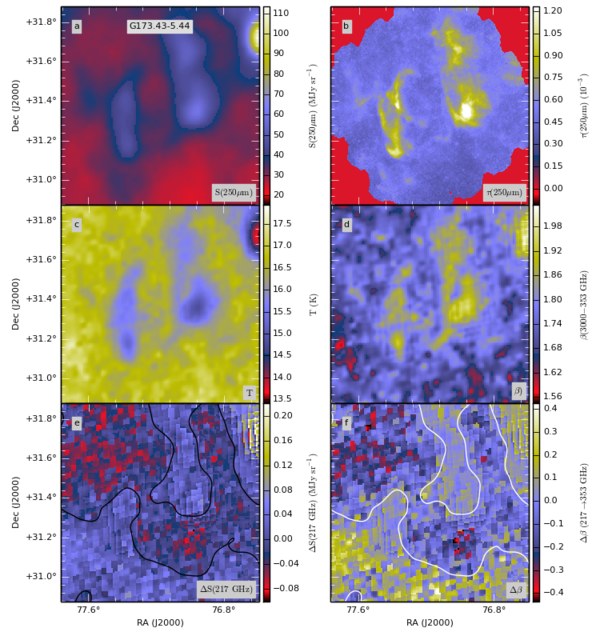}
\caption{Continued{\ldots} Fields G171.35-38.28 and G173.43-5.44}
\end{figure}
\begin{figure}
\includegraphics[width=8.8cm]{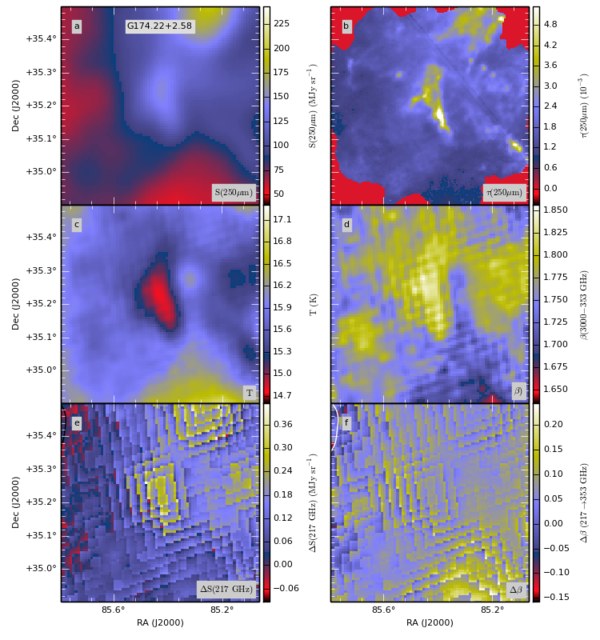}
\includegraphics[width=8.8cm]{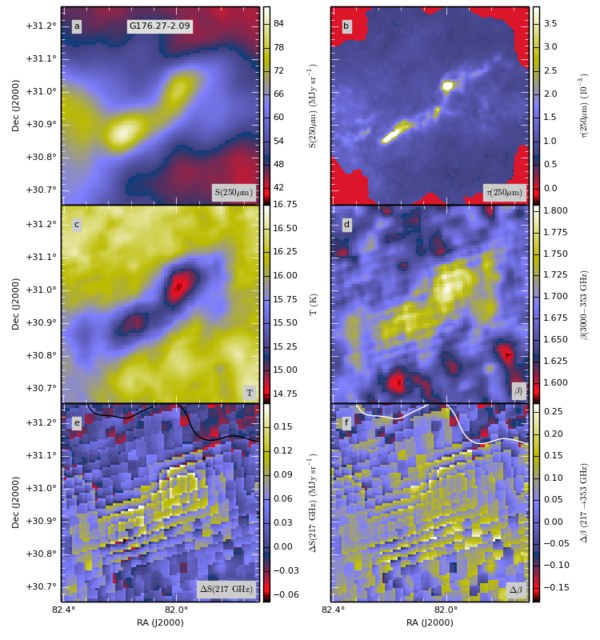}
\caption{Continued{\ldots} Fields G174.22+2.58 and G176.27-2.09}
\end{figure}
\begin{figure}
\includegraphics[width=8.8cm]{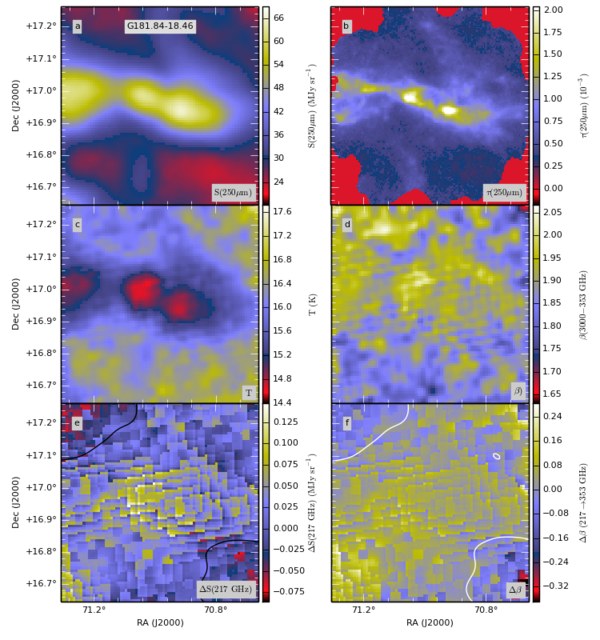}
\includegraphics[width=8.8cm]{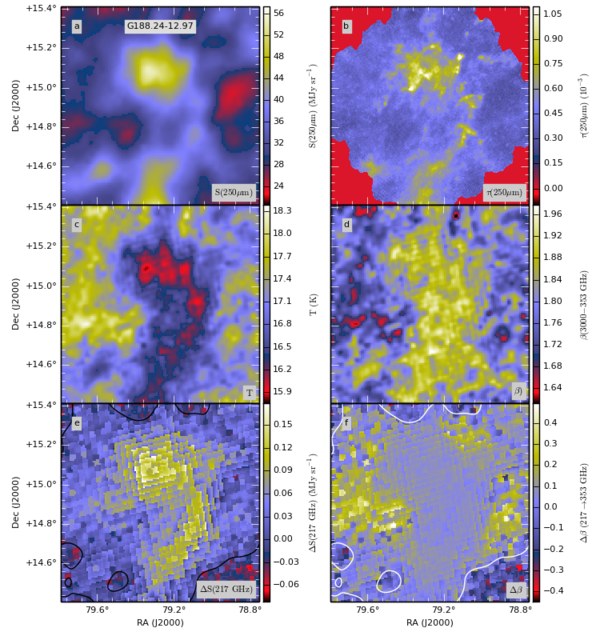}
\caption{Continued{\ldots} Fields G181.84-18.46 and G188.24-12.97}
\end{figure}
\begin{figure}
\includegraphics[width=8.8cm]{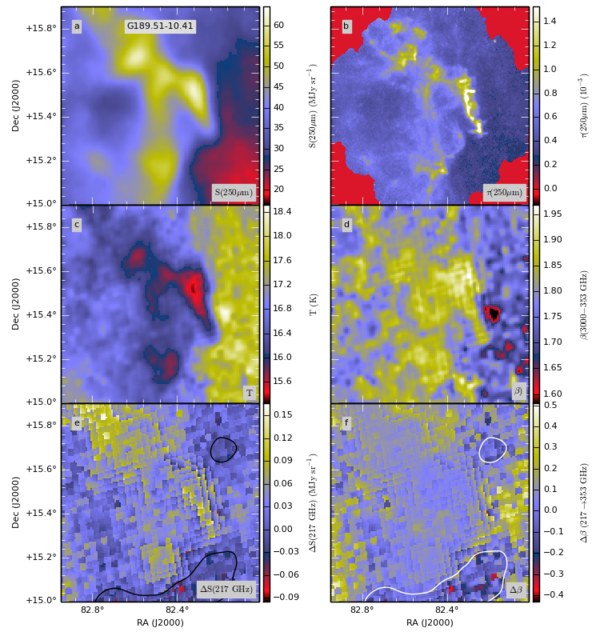}
\includegraphics[width=8.8cm]{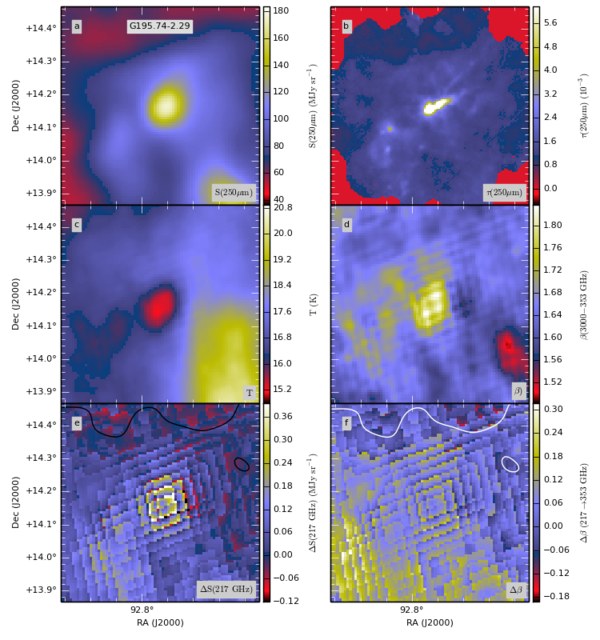}
\caption{Continued{\ldots} Fields G189.51-10.41 and G195.74-2.29}
\end{figure}
\begin{figure}
\includegraphics[width=8.8cm]{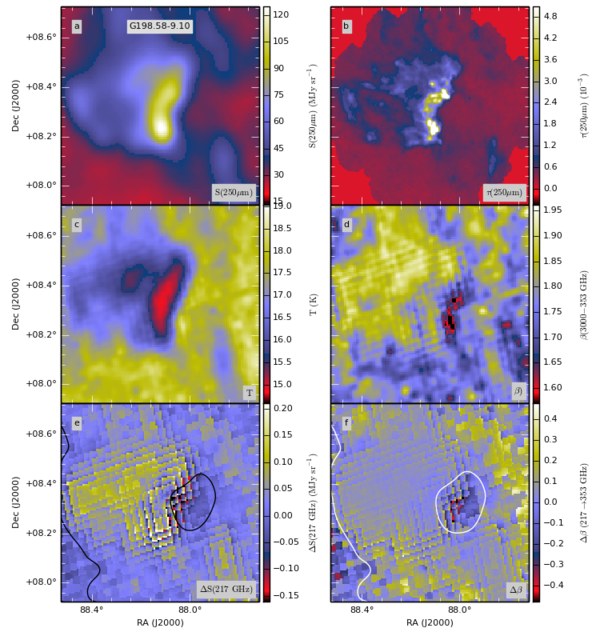}
\includegraphics[width=8.8cm]{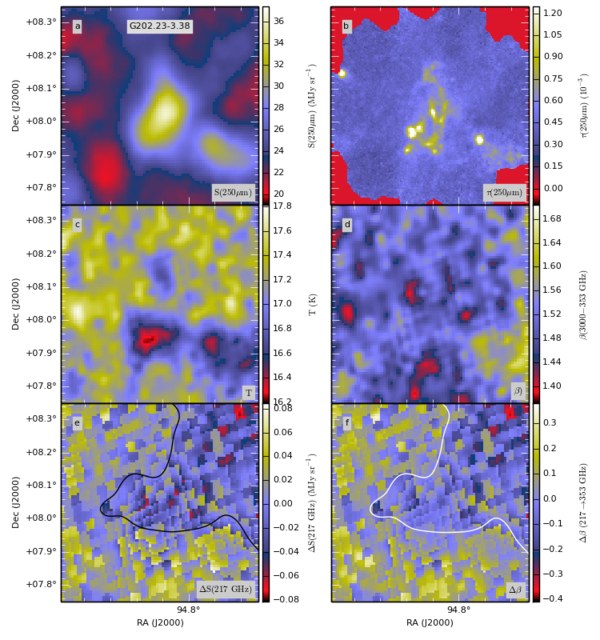}
\caption{Continued{\ldots} Fields G198.58-9.10 and G202.23-3.38}
\end{figure}

\clearpage

\begin{figure}
\includegraphics[width=8.8cm]{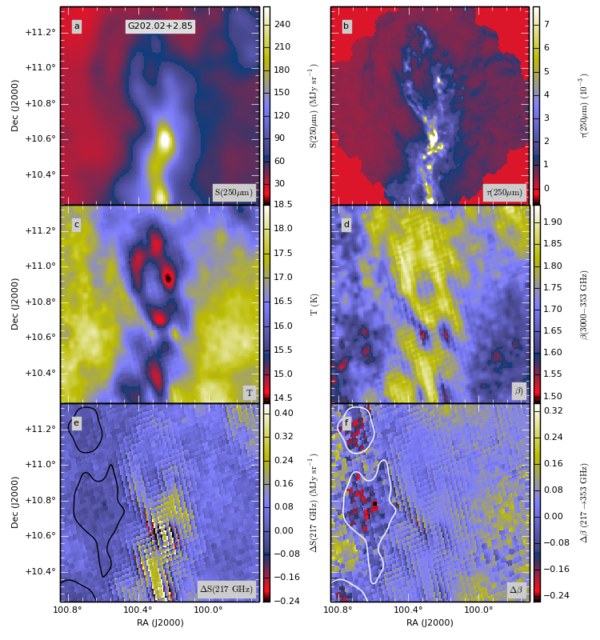}
\includegraphics[width=8.8cm]{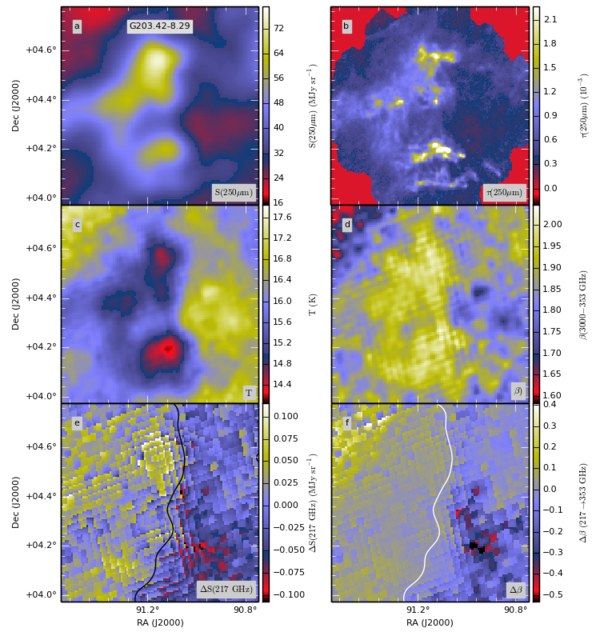}
\caption{Continued{\ldots} Fields G202.02+2.85 and G203.42-8.29}
\end{figure}
\begin{figure}
\includegraphics[width=8.8cm]{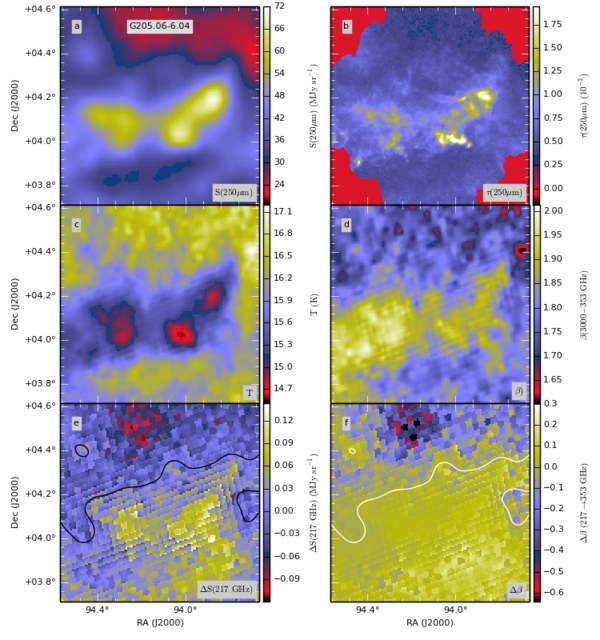}
\includegraphics[width=8.8cm]{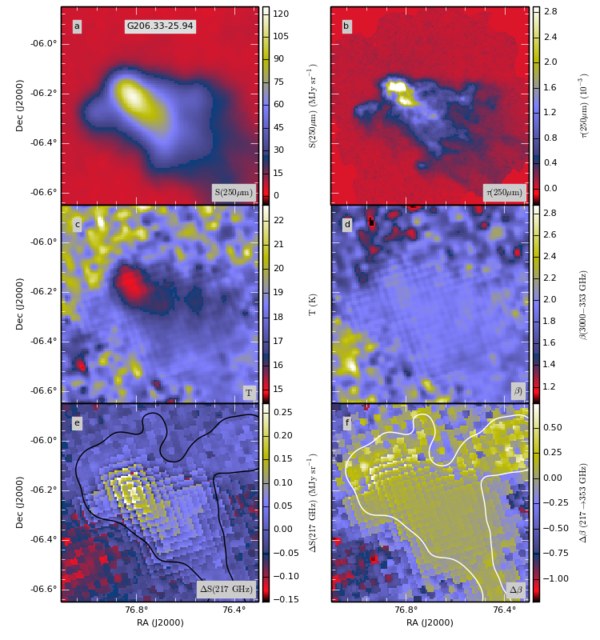}
\caption{Continued{\ldots} Fields G205.06-6.04 and G206.33-25.94}
\end{figure}
\begin{figure}
\includegraphics[width=8.8cm]{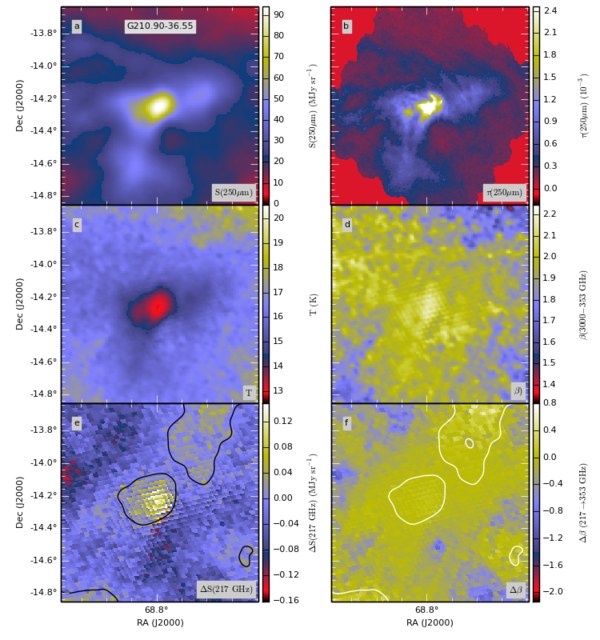}
\includegraphics[width=8.8cm]{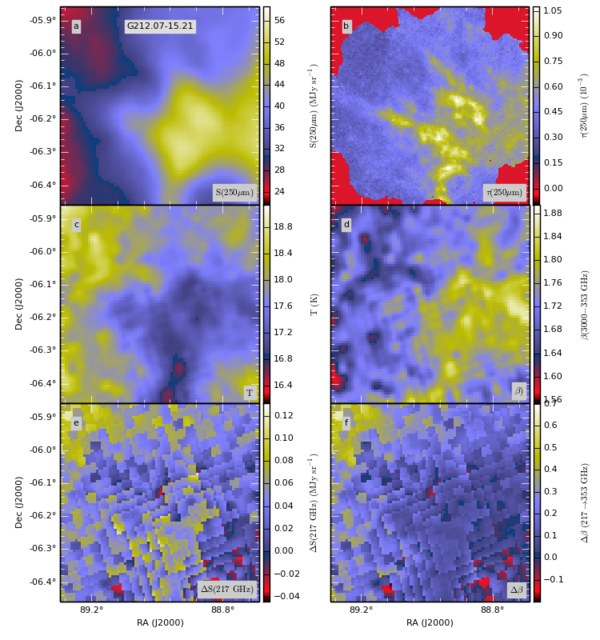}
\caption{Continued{\ldots} Fields G210.90-36.55 and G212.07-15.21}
\end{figure}
\begin{figure}
\includegraphics[width=8.8cm]{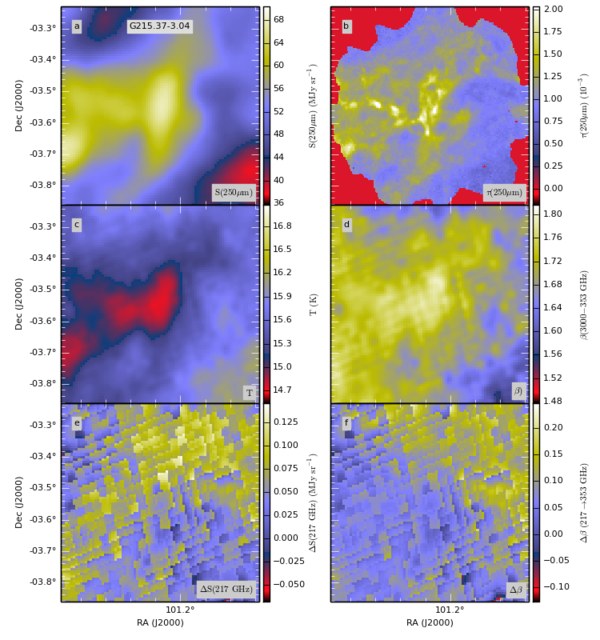}
\includegraphics[width=8.8cm]{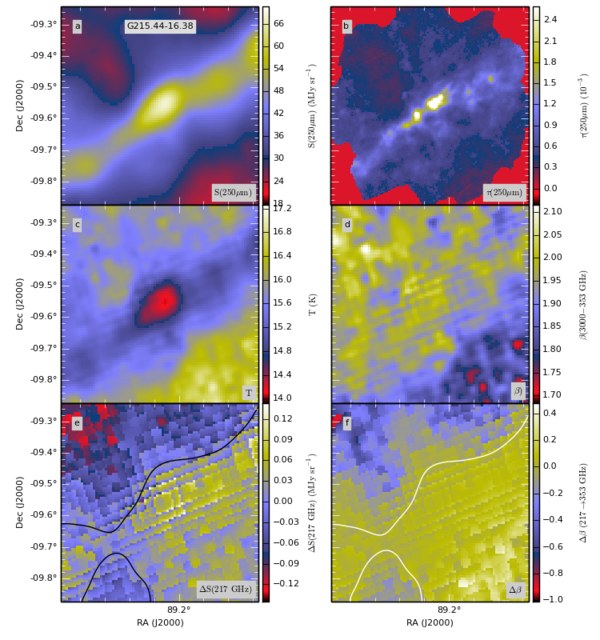}
\caption{Continued{\ldots} Fields G215.37-3.04 and G215.44-16.38}
\end{figure}
\begin{figure}
\includegraphics[width=8.8cm]{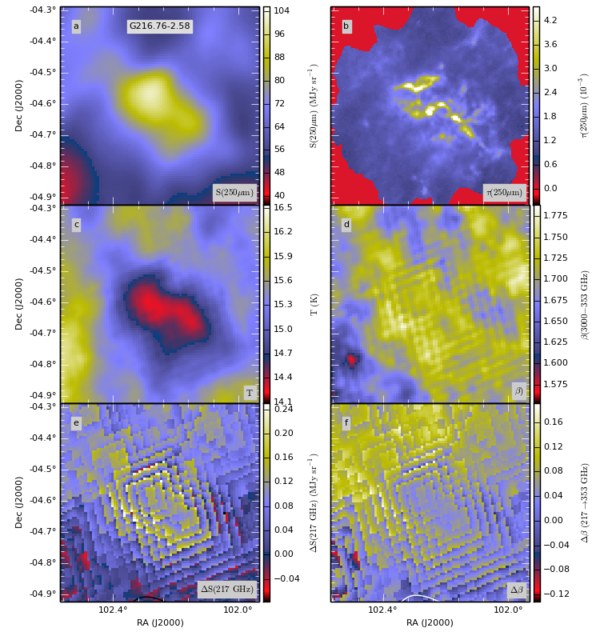}
\includegraphics[width=8.8cm]{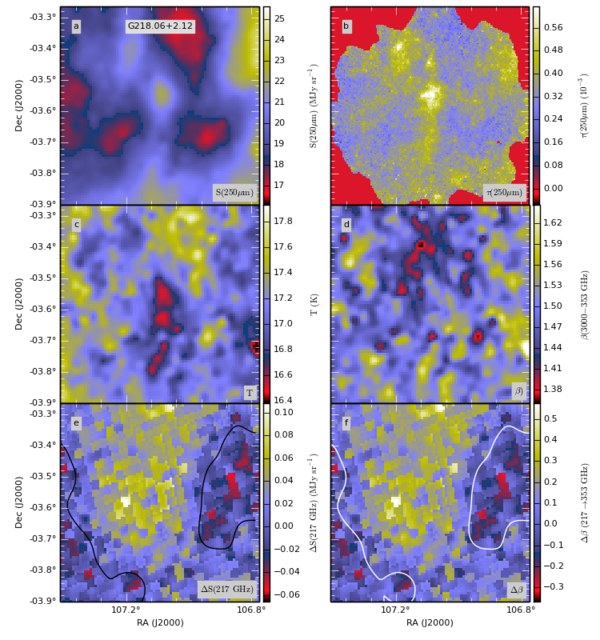}
\caption{Continued{\ldots} Fields G216.76-2.58 and G218.06+2.12}
\end{figure}
\begin{figure}
\includegraphics[width=8.8cm]{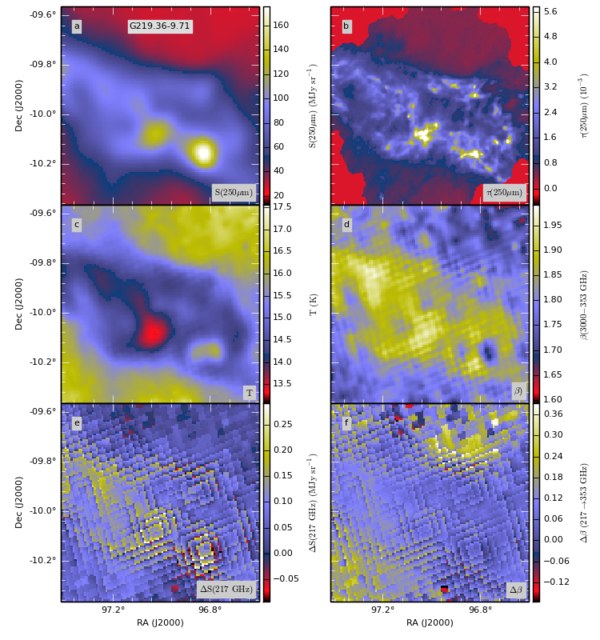}
\includegraphics[width=8.8cm]{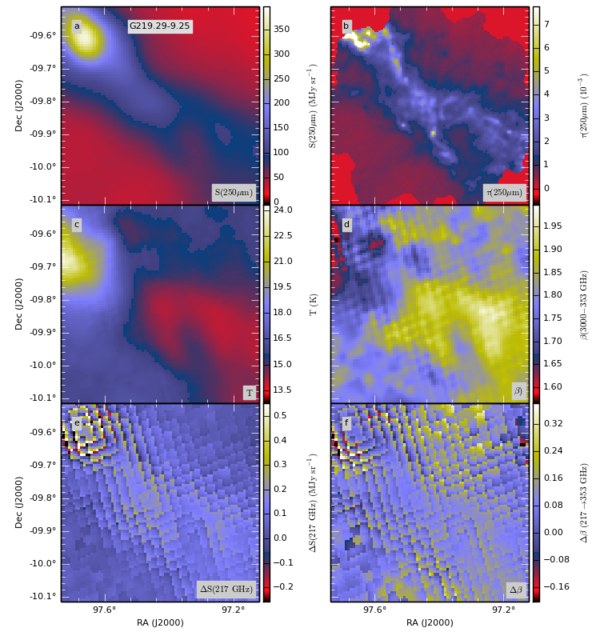}
\caption{Continued{\ldots} Fields G219.36-9.71 and G219.29-9.25}
\end{figure}
\begin{figure}
\includegraphics[width=8.8cm]{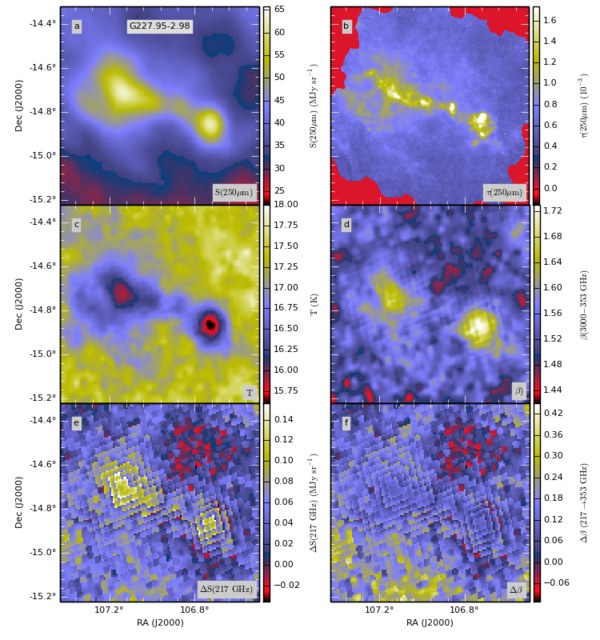}
\includegraphics[width=8.8cm]{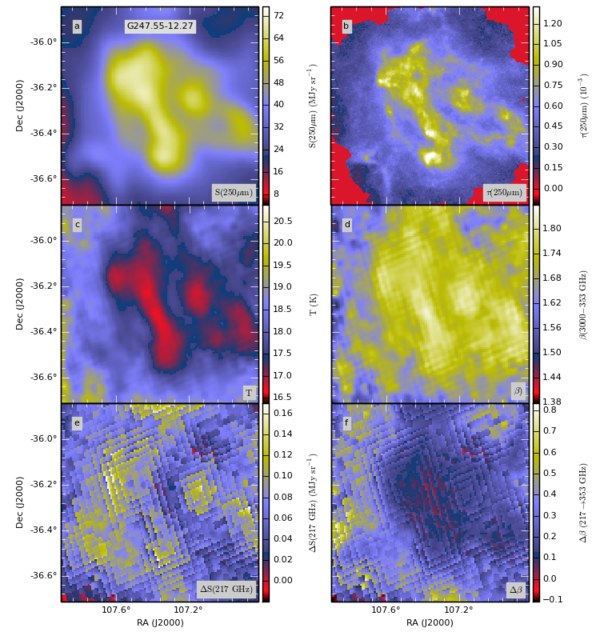}
\caption{Continued{\ldots} Fields G227.95-2.98 and G247.55-12.27}
\end{figure}
\begin{figure}
\includegraphics[width=8.8cm]{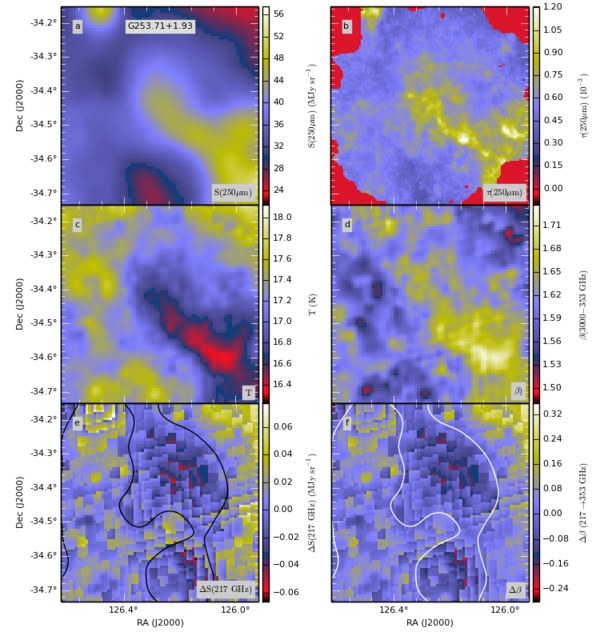}
\includegraphics[width=8.8cm]{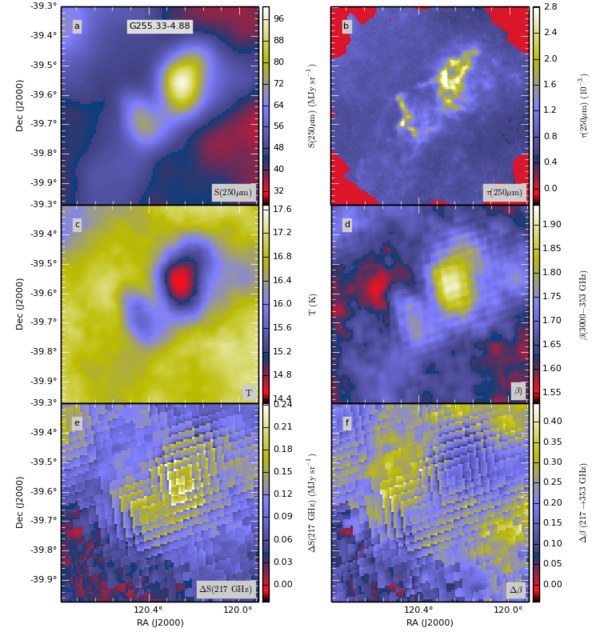}
\caption{Continued{\ldots} Fields G253.71+1.93 and G255.33-4.88}
\end{figure}
\begin{figure}
\includegraphics[width=8.8cm]{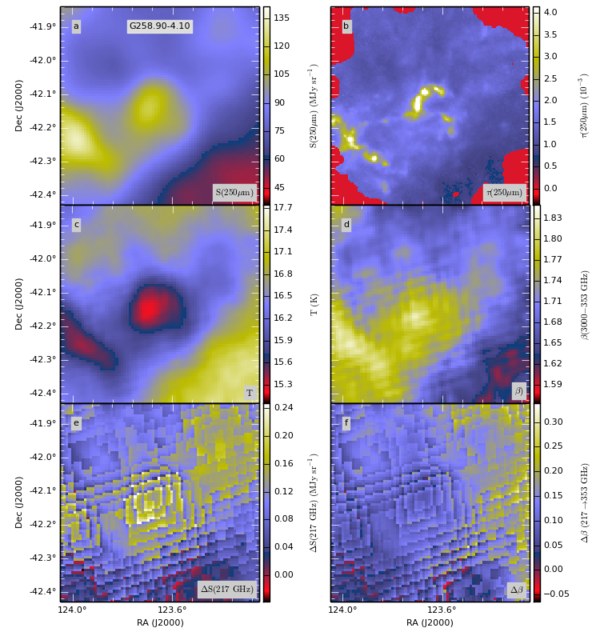}
\includegraphics[width=8.8cm]{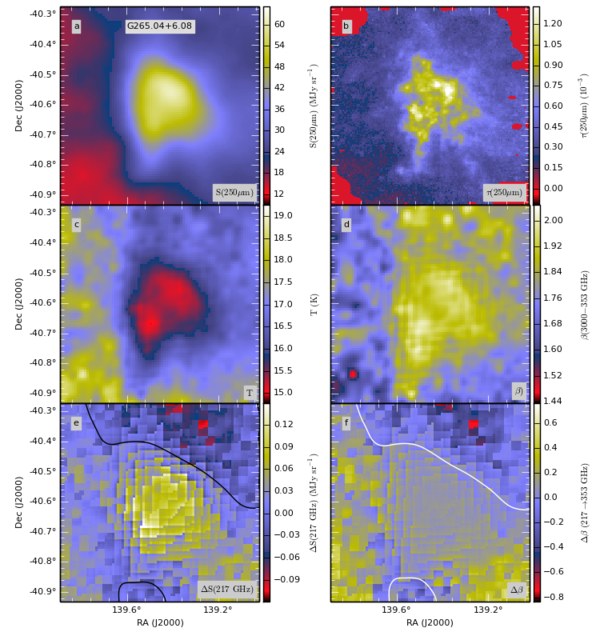}
\caption{Continued{\ldots} Fields G258.90-4.10 and G265.04+6.08}
\end{figure}
\begin{figure}
\includegraphics[width=8.8cm]{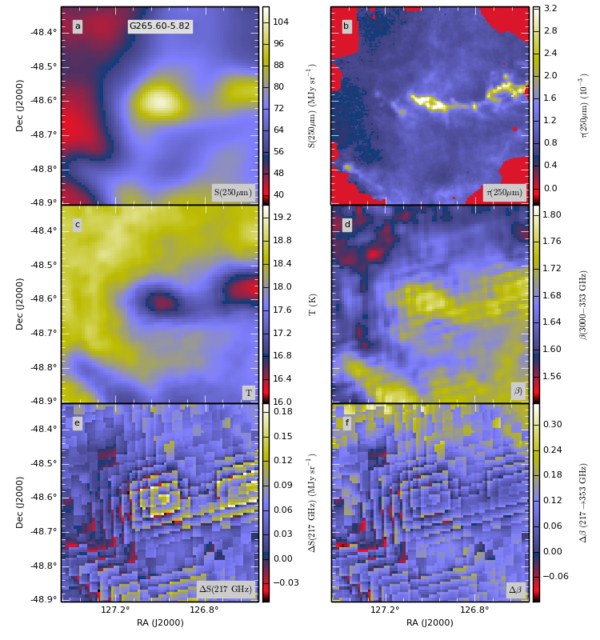}
\includegraphics[width=8.8cm]{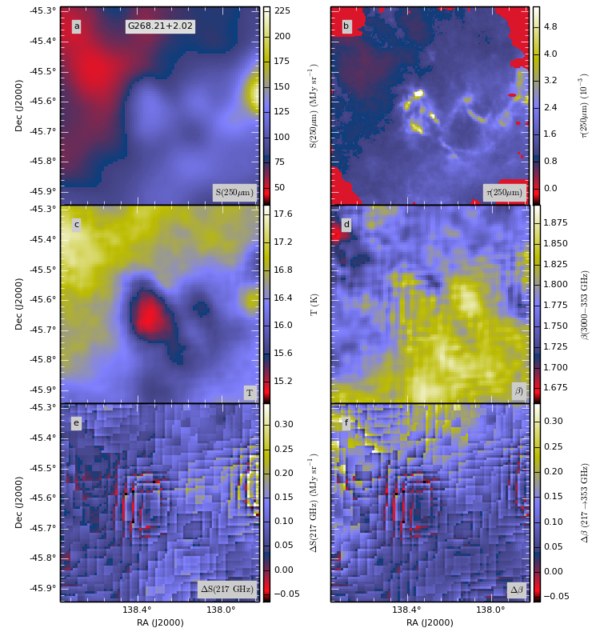}
\caption{Continued{\ldots} Fields G265.60-5.82 and G268.21+2.02}
\end{figure}

\clearpage

\begin{figure}
\includegraphics[width=8.8cm]{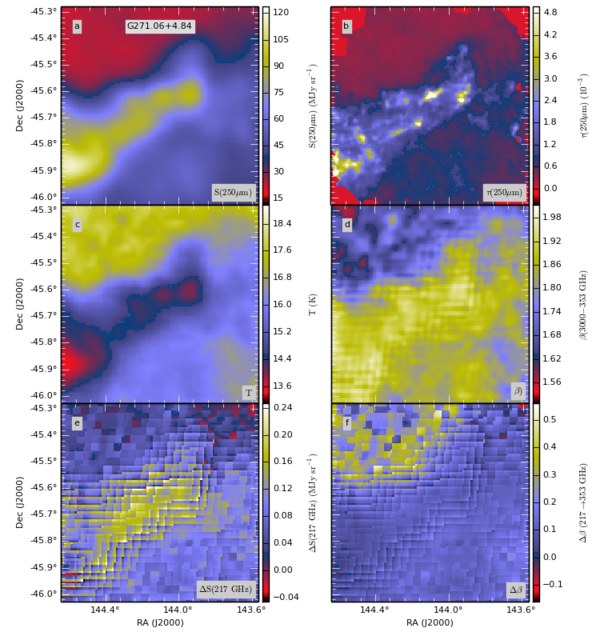}
\includegraphics[width=8.8cm]{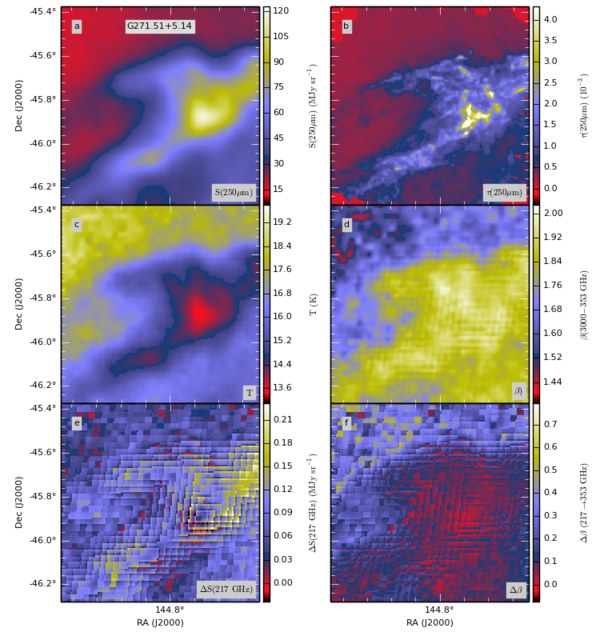}
\caption{Continued{\ldots} Fields G271.06+4.84 and G271.51+5.14}
\end{figure}
\begin{figure}
\includegraphics[width=8.8cm]{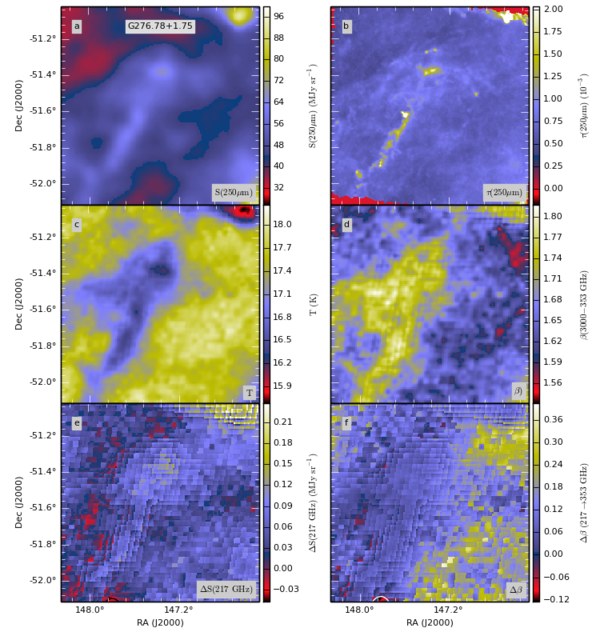}
\includegraphics[width=8.8cm]{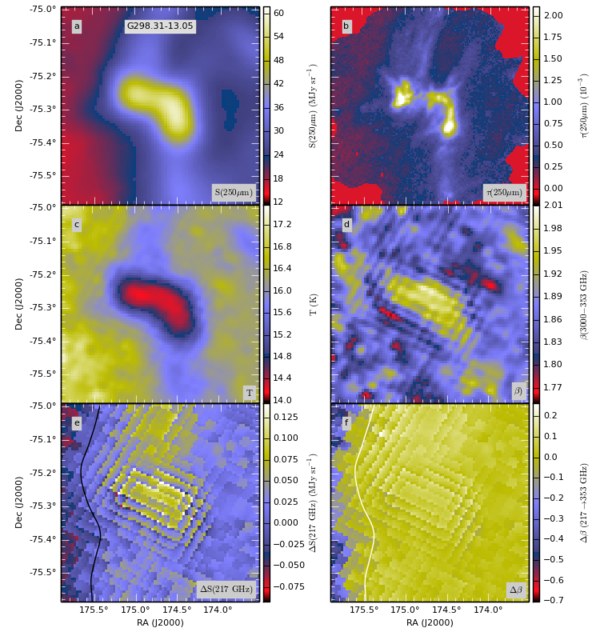}
\caption{Continued{\ldots} Fields G276.78+1.75 and G298.31-13.05}
\end{figure}
\begin{figure}
\includegraphics[width=8.8cm]{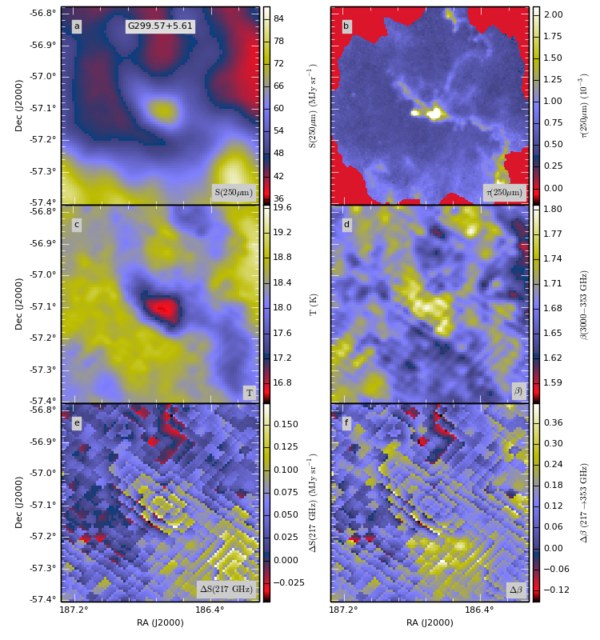}
\includegraphics[width=8.8cm]{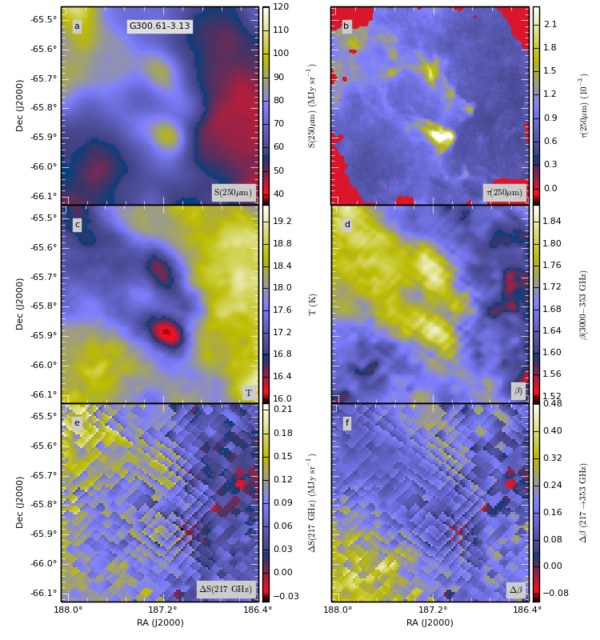}
\caption{Continued{\ldots} Fields G299.57+5.61 and G300.61-3.13}
\end{figure}
\begin{figure}
\includegraphics[width=8.8cm]{PI_FIG_105_VSGC_LFISUB_5am.jpg}
\includegraphics[width=8.8cm]{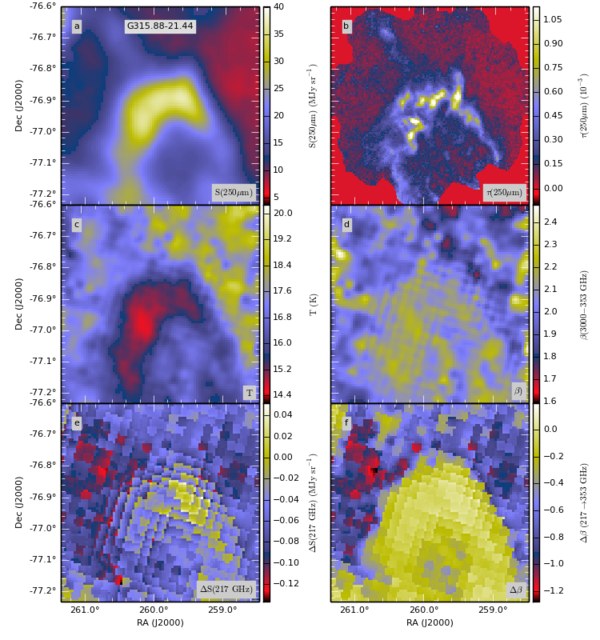}
\caption{Continued{\ldots} Fields G300.86-9.00 and G315.88-21.44}
\end{figure}
\begin{figure}
\includegraphics[width=8.8cm]{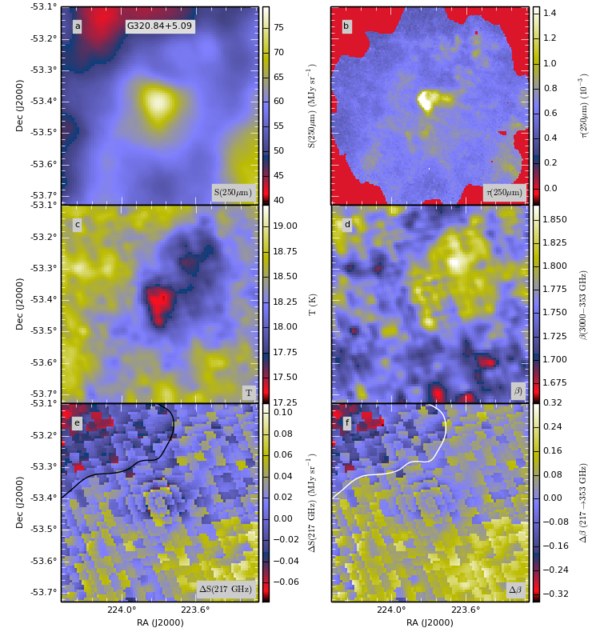}
\includegraphics[width=8.8cm]{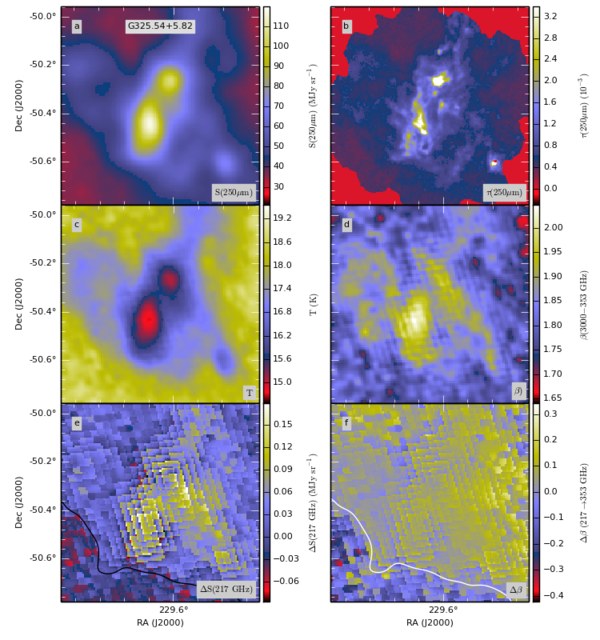}
\caption{Continued{\ldots} Fields G320.84+5.09 and G325.54+5.82}
\end{figure}
\begin{figure}
\includegraphics[width=8.8cm]{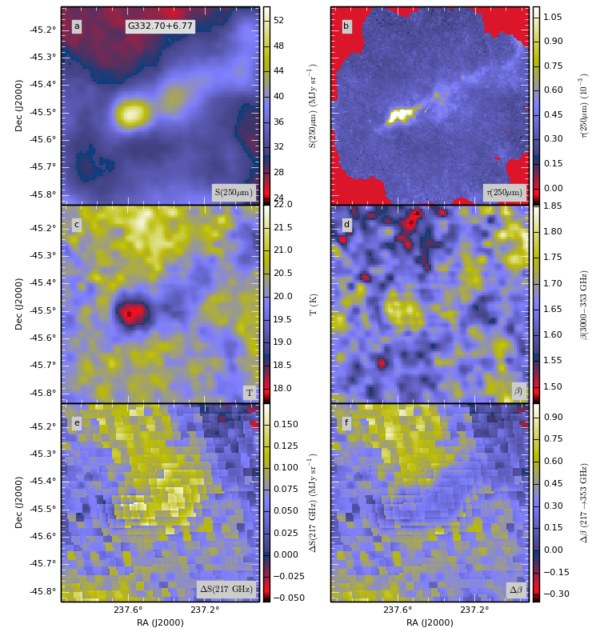}
\includegraphics[width=8.8cm]{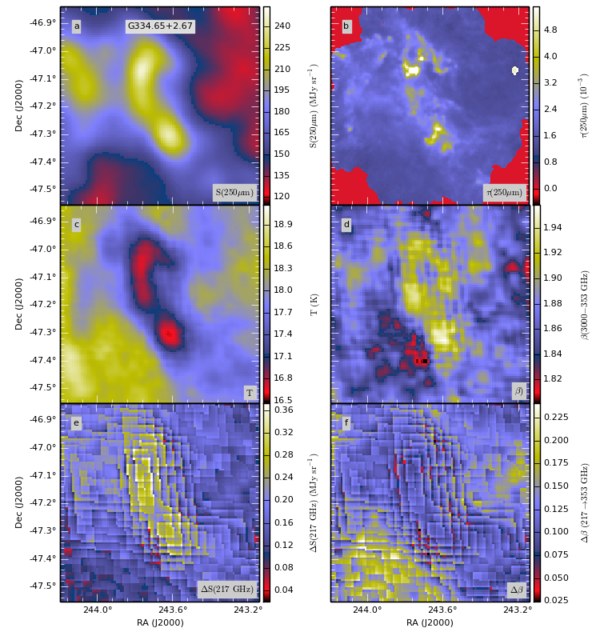}
\caption{Continued{\ldots} Fields G332.70+6.77 and G334.65+2.67}
\end{figure}
\begin{figure}
\includegraphics[width=8.8cm]{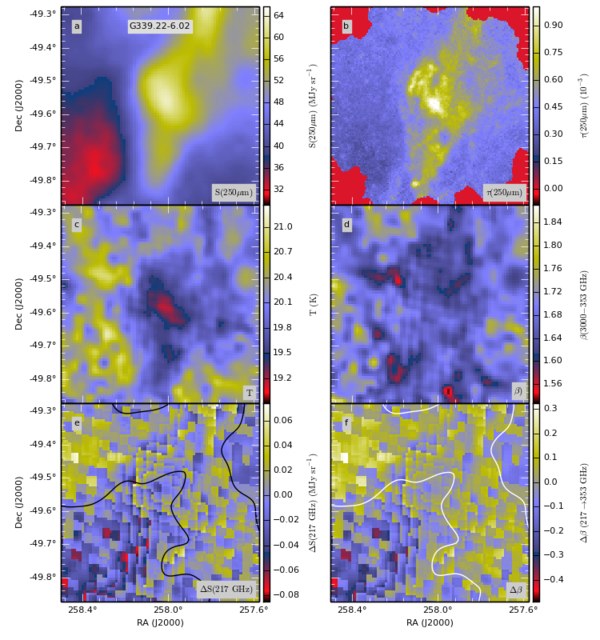}
\includegraphics[width=8.8cm]{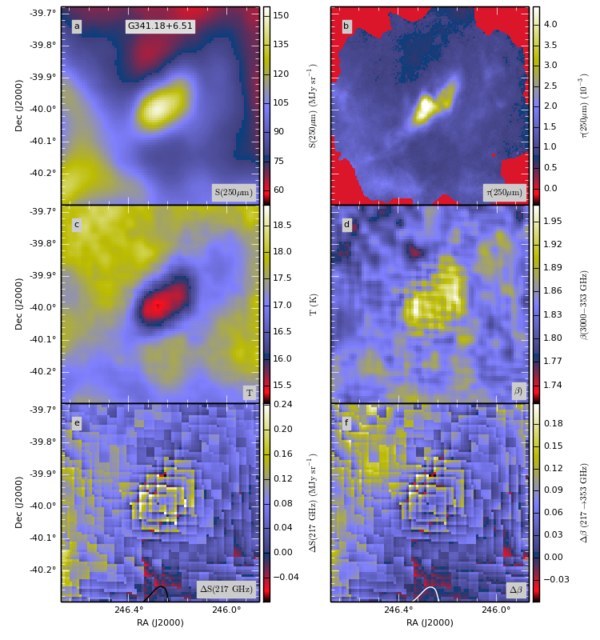}
\caption{Continued{\ldots} Fields G339.22-6.02 and G341.18+6.51}
\end{figure}
\begin{figure}
\includegraphics[width=8.8cm]{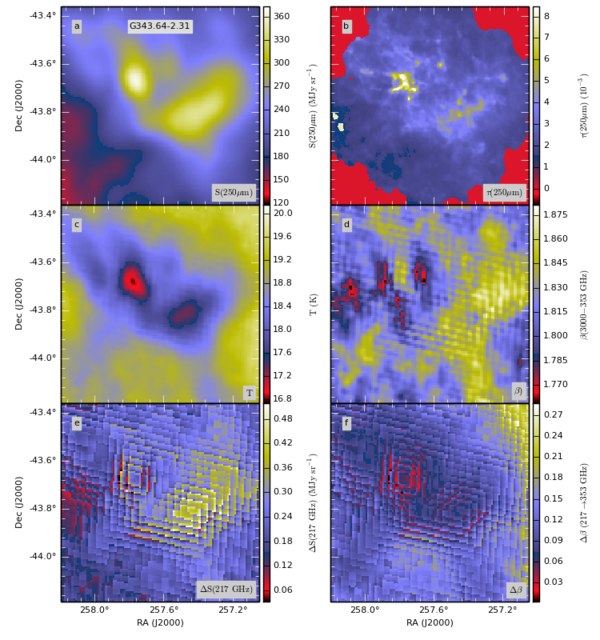}
\includegraphics[width=8.8cm]{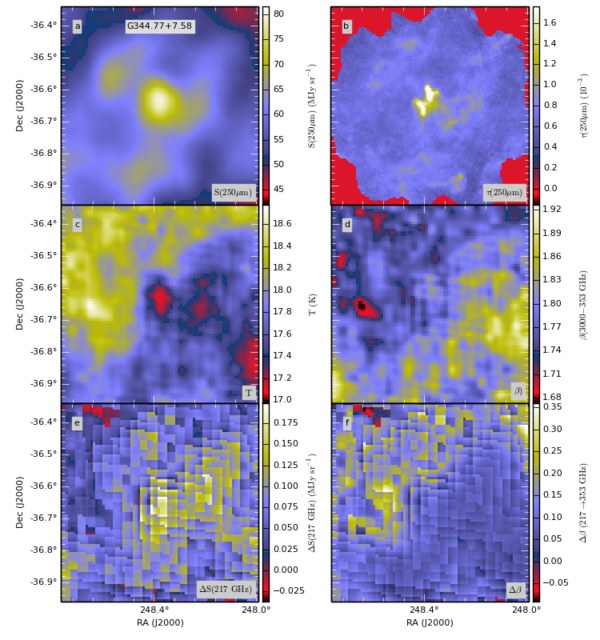}
\caption{Continued{\ldots} Fields G343.64-2.31 and G344.77+7.58}
\end{figure}
\begin{figure}
\includegraphics[width=8.8cm]{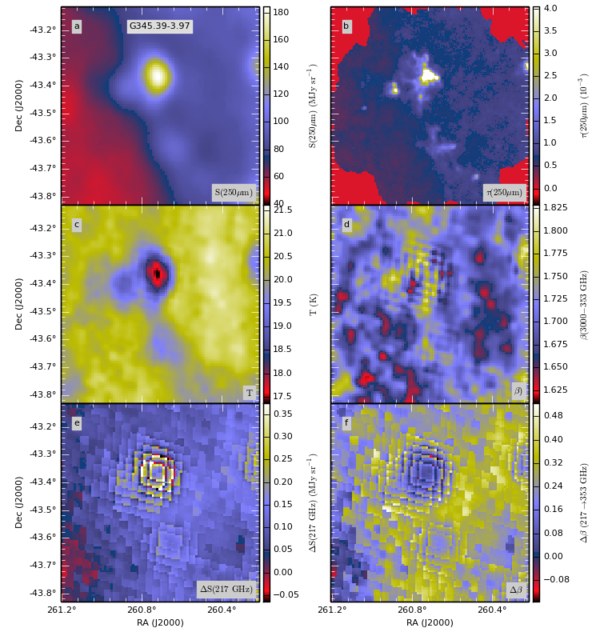}
\includegraphics[width=8.8cm]{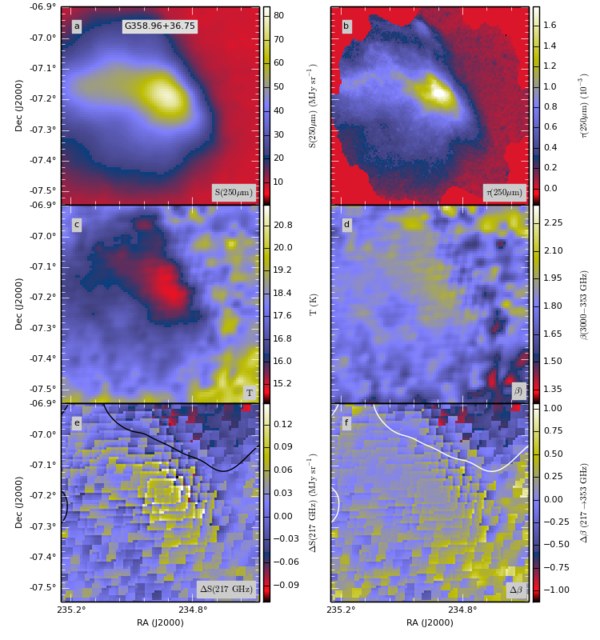}
\caption{Continued{\ldots} Fields G345.39-3.97 and G358.96+36.75}
\label{fig:PI_FIG_115}
\end{figure}

\clearpage

\section{Maps of ($T$, $\beta$) with {\it Herschel} data} \label{app:hfit}

The figures~\ref{fig:TB_PLWBEAM_001}--\ref{fig:TB_PLWBEAM_115} show the colour
temperature and spectral index maps calculated with {\it Herschel} data. The figures
include results derived with both 160--500\,$\mu$m and 250--500\,$\mu$m data.

\begin{figure*}
\includegraphics[width=16cm]{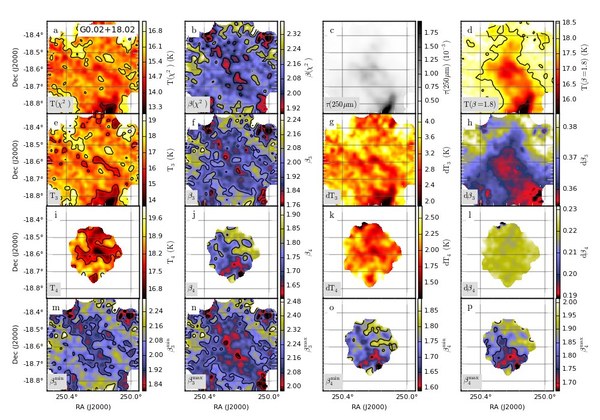}
\caption{
Temperature and spectral index fits of field G0.02+18.02 using {\it Herschel} data.
Frames are: colour temperature and spectral index with SPIRE data and $\chi^2$
minimisation (frames $a-b$), 250$\mu$m optical depth and colour temperature for SPIRE
data and $\beta=$1.8 (frames $c$-$d$), colour temperature and spectral index with SPIRE
data and MCMC calculations (frame $e$-$f$) and the corresponding error maps (frames
$g$--$h$), MCMC results for 160--500\,$\mu$m fits (frames $i$--$j$) with corresponding
error estimates (frames $k$-$l$). The last frames show the effect of zero point
uncertainty in the three-band fits (frames $m$-$n$) and four-band fits (frame $o$-$p$)
($\chi^2$ fits) based on Monte Carlo simulation using $\chi^2$ fits. 
Temperature maps have contours drawn at intervals of 2.0\,K, starting at 10.0\,K.
Spectral index maps have contours at intervals of 0.2, starting at 1.0.
}
\label{fig:TB_PLWBEAM_001}
\end{figure*}

\begin{figure*}
\includegraphics[width=16cm]{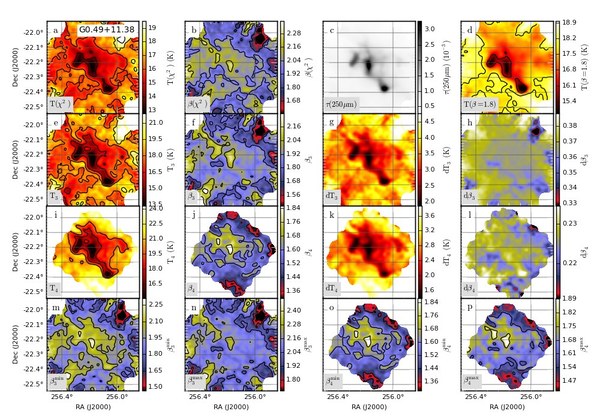}
\caption{Continued{\ldots} Field G0.49+11.38}
\end{figure*}
\begin{figure*}
\includegraphics[width=16cm]{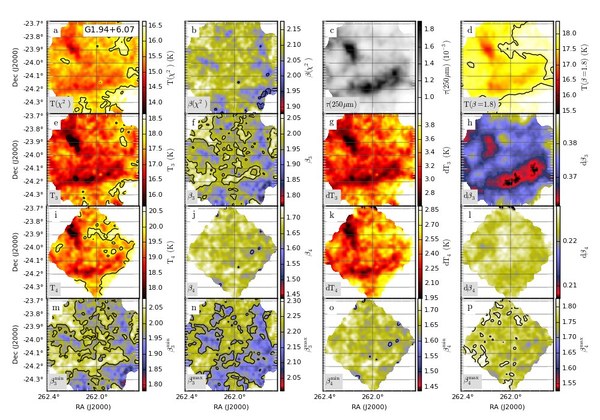}
\caption{Continued{\ldots} Field G1.94+6.07}
\end{figure*}
\begin{figure*}
\includegraphics[width=16cm]{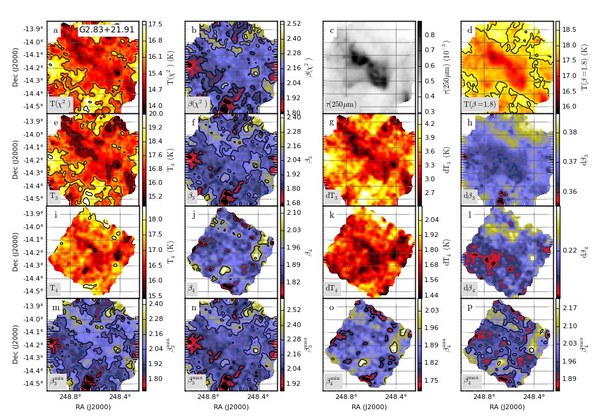}
\caption{Continued{\ldots} Field G2.83+21.91}
\end{figure*}
\begin{figure*}
\includegraphics[width=16cm]{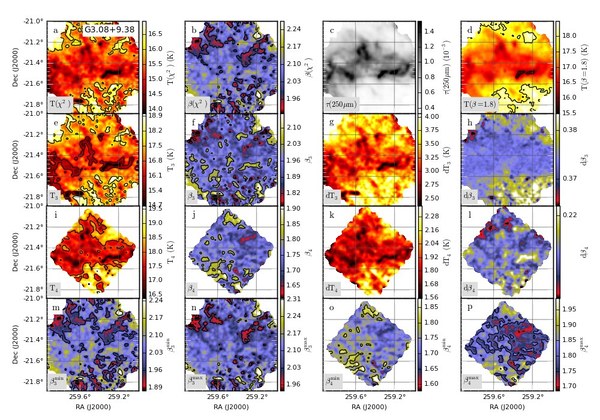}
\caption{Continued{\ldots} Field G3.08+9.38}
\end{figure*}
\begin{figure*}
\includegraphics[width=16cm]{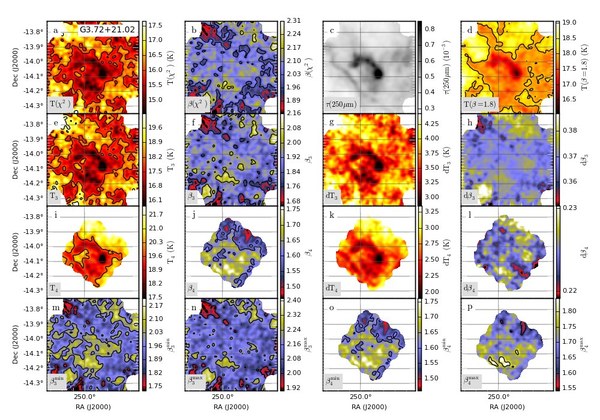}
\caption{Continued{\ldots} Field G3.72+21.02}
\end{figure*}
\begin{figure*}
\includegraphics[width=16cm]{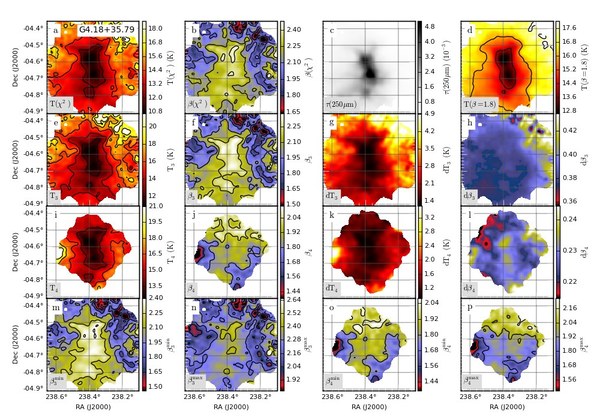}
\caption{Continued{\ldots} Field G4.18+35.79}
\end{figure*}
\begin{figure*}
\includegraphics[width=16cm]{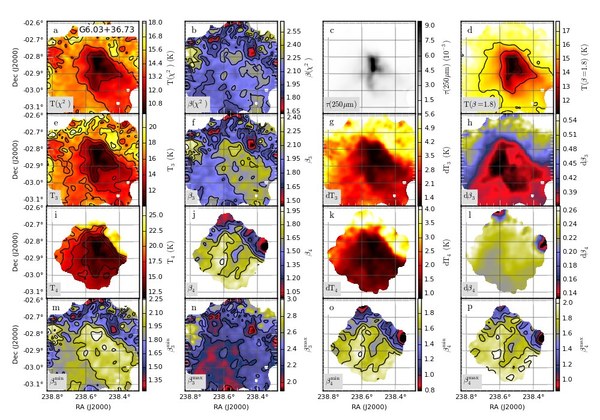}
\caption{Continued{\ldots} Field G6.03+36.73}
\end{figure*}
\begin{figure*}
\includegraphics[width=16cm]{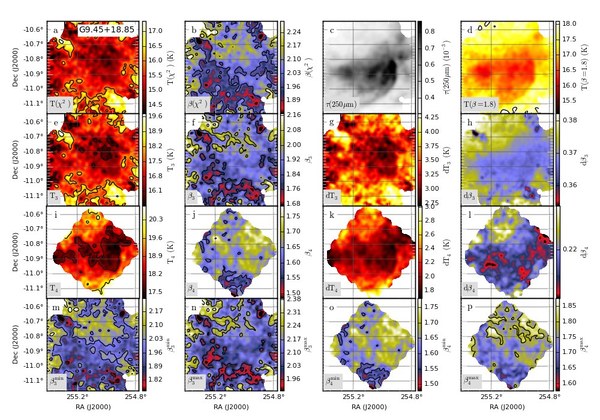}
\caption{Continued{\ldots} Field G9.45+18.85}
\end{figure*}
\clearpage
\begin{figure*}
\includegraphics[width=16cm]{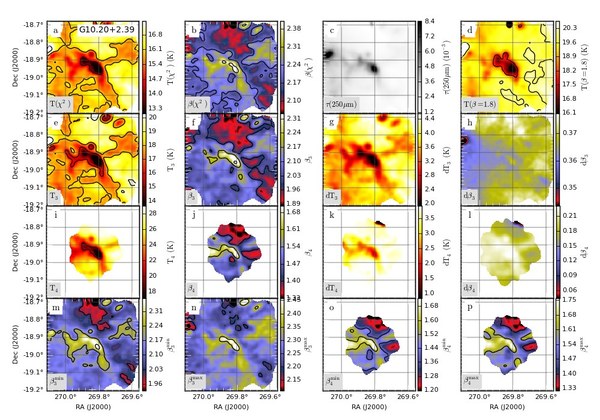}
\caption{Continued{\ldots} Field G10.20+2.39}
\end{figure*}
\begin{figure*}
\includegraphics[width=16cm]{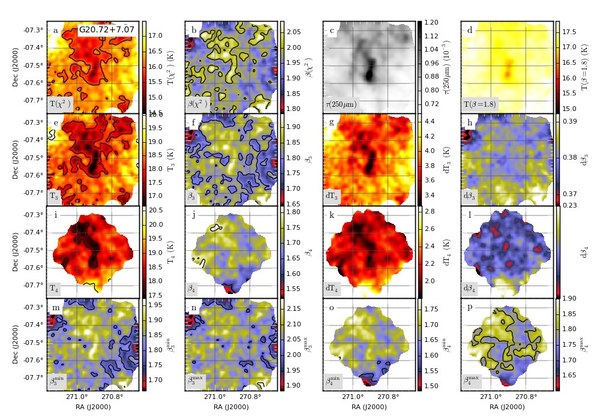}
\caption{Continued{\ldots} Field G20.72+7.07}
\end{figure*}
\begin{figure*}
\includegraphics[width=16cm]{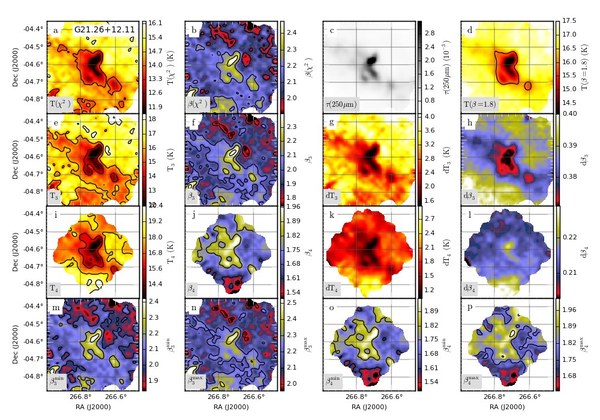}
\caption{Continued{\ldots} Field G21.26+12.11}
\end{figure*}
\begin{figure*}
\includegraphics[width=16cm]{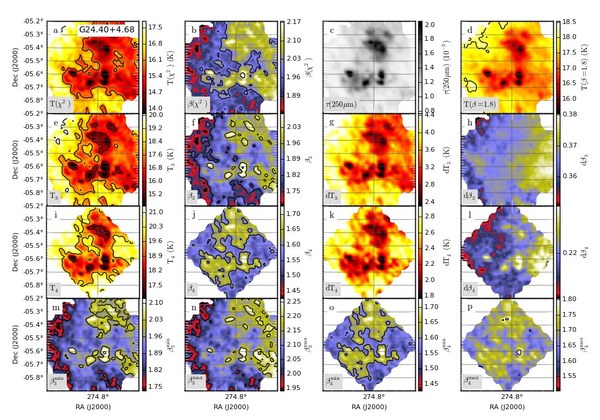}
\caption{Continued{\ldots} Field G24.40+4.68}
\end{figure*}
\begin{figure*}
\includegraphics[width=16cm]{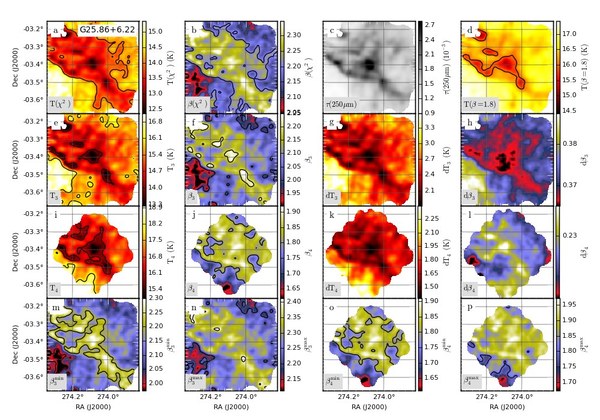}
\caption{Continued{\ldots} Field G25.86+6.22}
\end{figure*}
\begin{figure*}
\includegraphics[width=16cm]{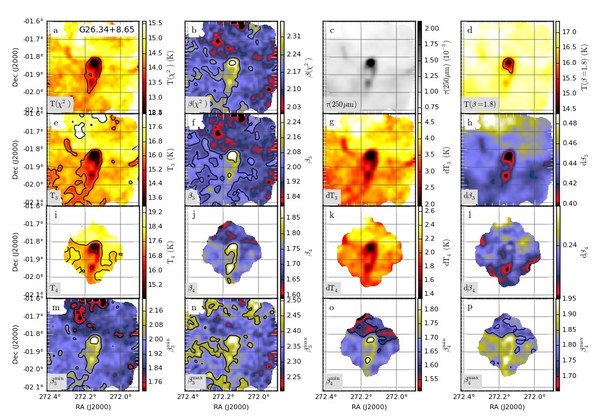}
\caption{Continued{\ldots} Field G26.34+8.65}
\end{figure*}
\begin{figure*}
\includegraphics[width=16cm]{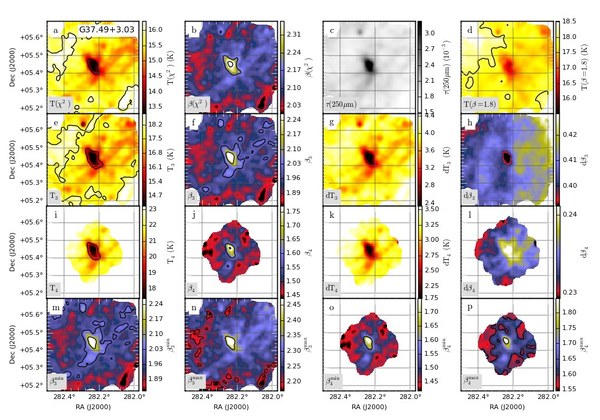}
\caption{Continued{\ldots} Field G37.49+3.03}
\end{figure*}
\begin{figure*}
\includegraphics[width=16cm]{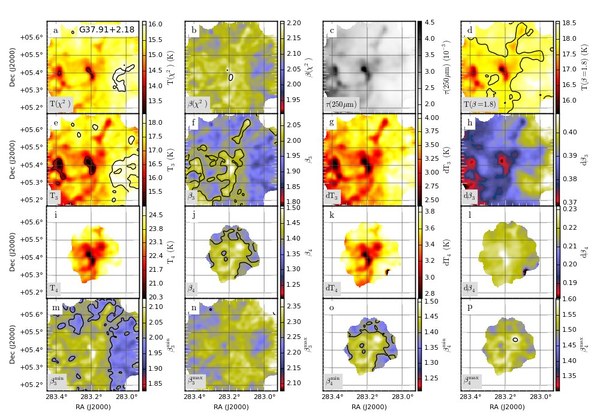}
\caption{Continued{\ldots} Field G37.91+2.18}
\end{figure*}
\begin{figure*}
\includegraphics[width=16cm]{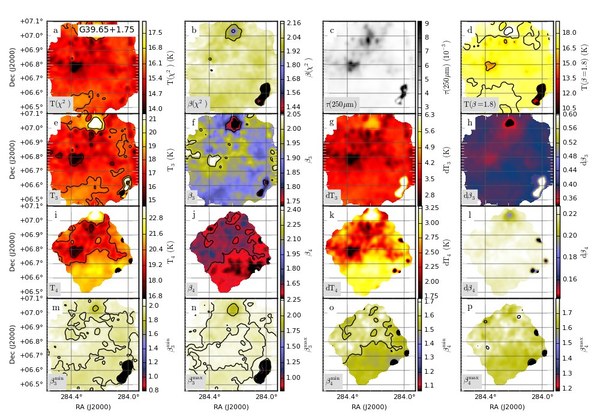}
\caption{Continued{\ldots} Field G39.65+1.75}
\end{figure*}
\begin{figure*}
\includegraphics[width=16cm]{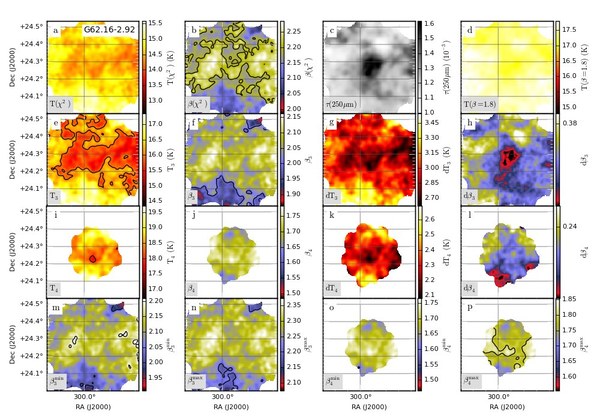}
\caption{Continued{\ldots} Field G62.16-2.92}
\end{figure*}
\clearpage
\begin{figure*}
\includegraphics[width=16cm]{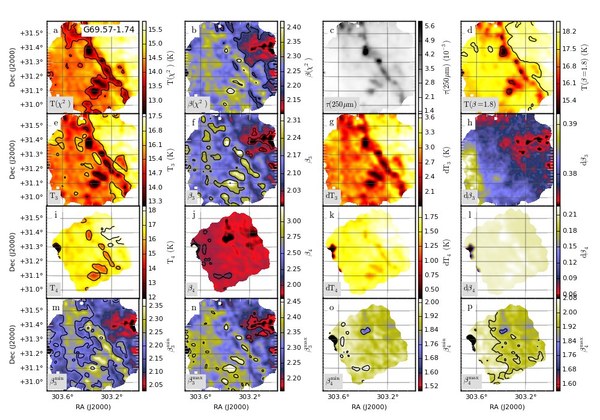}
\caption{Continued{\ldots} Field G69.57-1.74}
\end{figure*}
\begin{figure*}
\includegraphics[width=16cm]{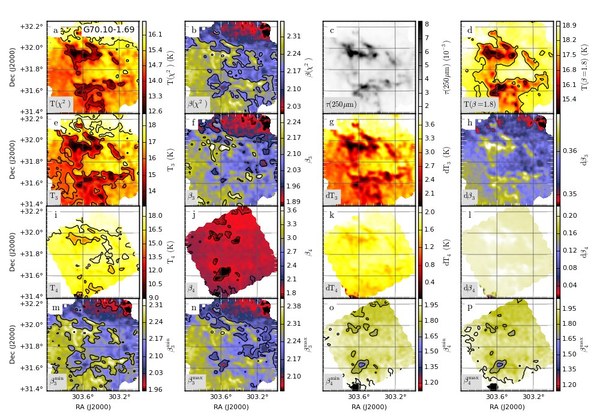}
\caption{Continued{\ldots} Field G70.10-1.69}
\end{figure*}
\begin{figure*}
\includegraphics[width=16cm]{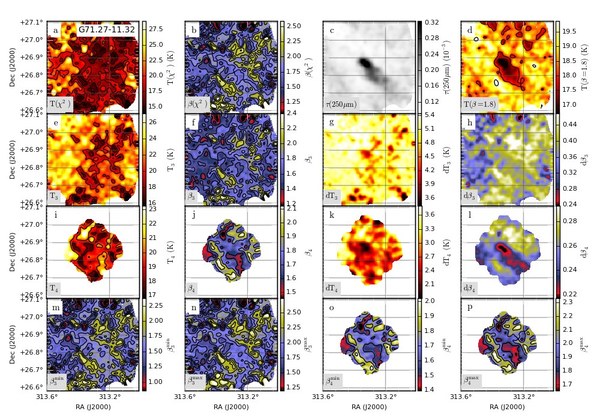}
\caption{Continued{\ldots} Field G71.27-11.32}
\end{figure*}
\begin{figure*}
\includegraphics[width=16cm]{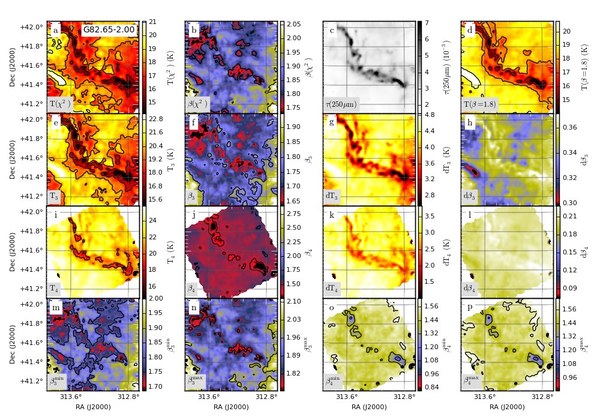}
\caption{Continued{\ldots} Field G82.65-2.00}
\end{figure*}
\begin{figure*}
\includegraphics[width=16cm]{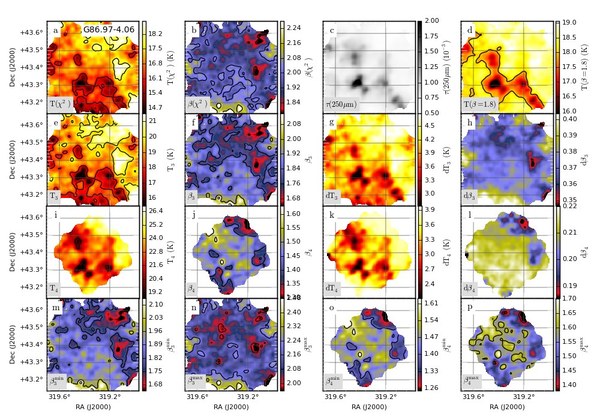}
\caption{Continued{\ldots} Field G86.97-4.06}
\end{figure*}
\begin{figure*}
\includegraphics[width=16cm]{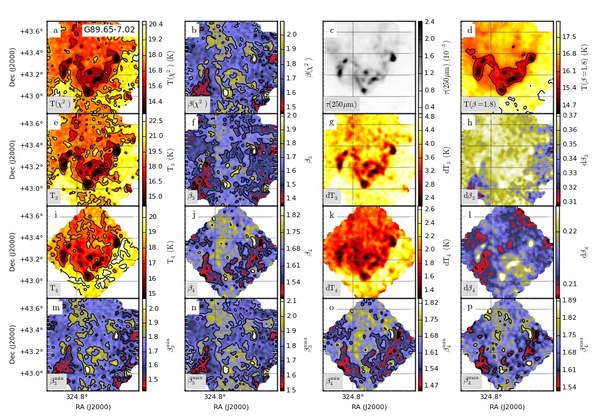}
\caption{Continued{\ldots} Field G89.65-7.02}
\end{figure*}
\begin{figure*}
\includegraphics[width=16cm]{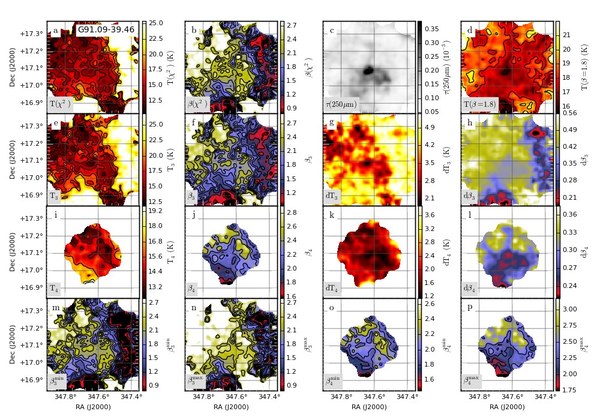}
\caption{Continued{\ldots} Field G91.09-39.46}
\end{figure*}
\begin{figure*}
\includegraphics[width=16cm]{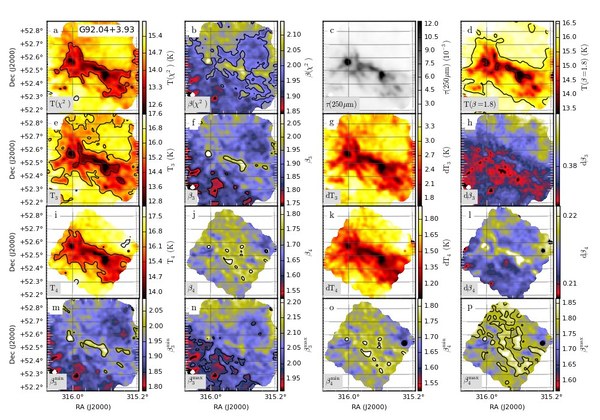}
\caption{Continued{\ldots} Field G92.04+3.93}
\end{figure*}
\begin{figure*}
\includegraphics[width=16cm]{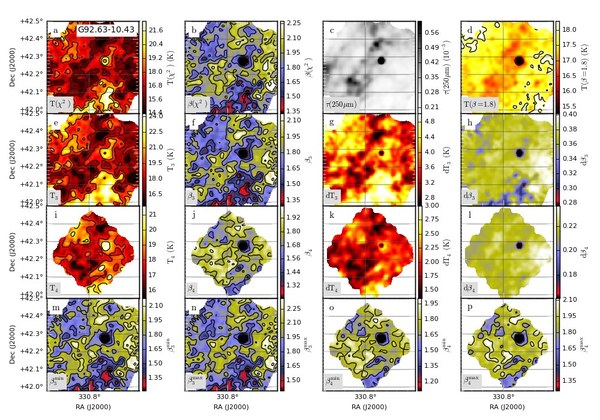}
\caption{Continued{\ldots} Field G92.63-10.43}
\end{figure*}
\begin{figure*}
\includegraphics[width=16cm]{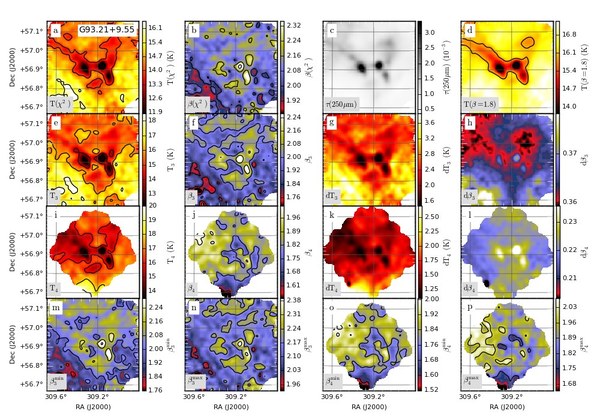}
\caption{Continued{\ldots} Field G93.21+9.55}
\end{figure*}
\clearpage
\begin{figure*}
\includegraphics[width=16cm]{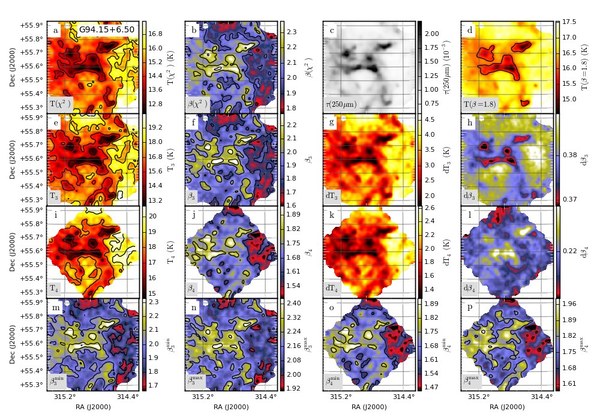}
\caption{Continued{\ldots} Field G94.15+6.50}
\end{figure*}
\begin{figure*}
\includegraphics[width=16cm]{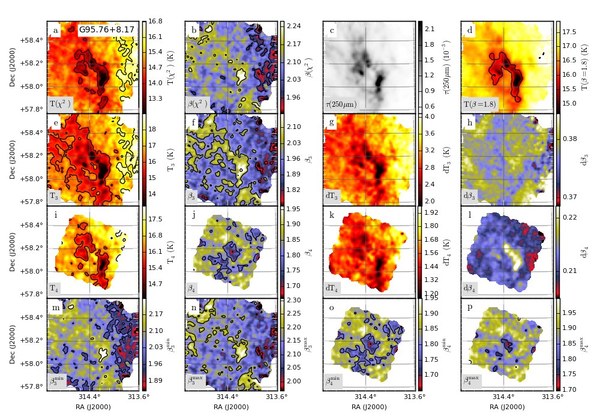}
\caption{Continued{\ldots} Field G95.76+8.17}
\end{figure*}
\begin{figure*}
\includegraphics[width=16cm]{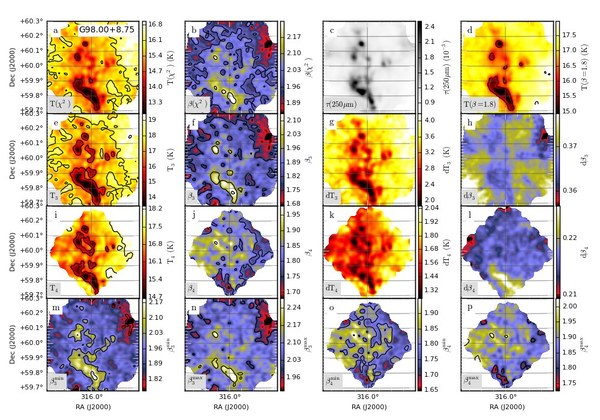}
\caption{Continued{\ldots} Field G98.00+8.75}
\end{figure*}
\begin{figure*}
\includegraphics[width=16cm]{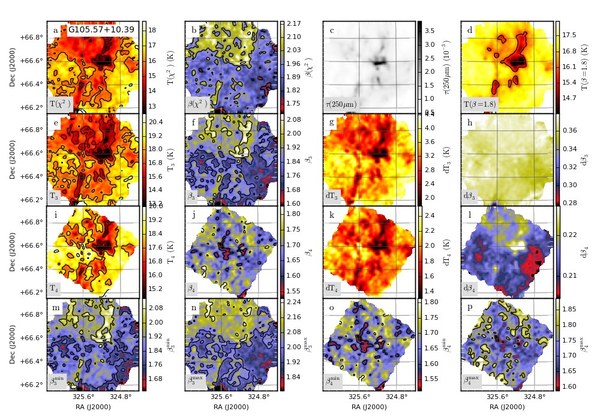}
\caption{Continued{\ldots} Field G105.57+10.39}
\end{figure*}
\begin{figure*}
\includegraphics[width=16cm]{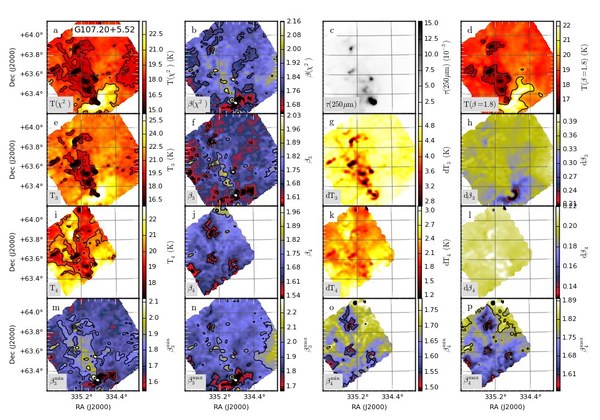}
\caption{Continued{\ldots} Field G107.20+5.52}
\end{figure*}
\begin{figure*}
\includegraphics[width=16cm]{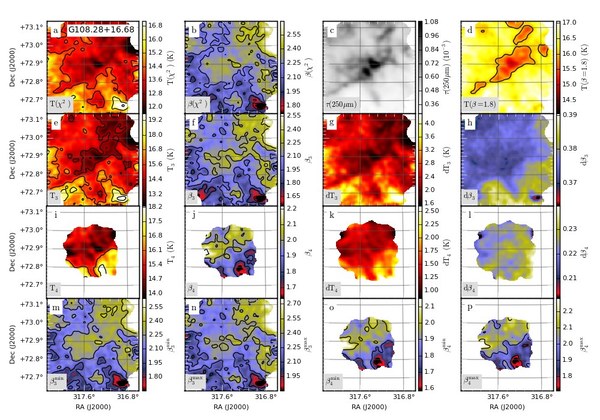}
\caption{Continued{\ldots} Field G108.28+16.68}
\end{figure*}
\begin{figure*}
\includegraphics[width=16cm]{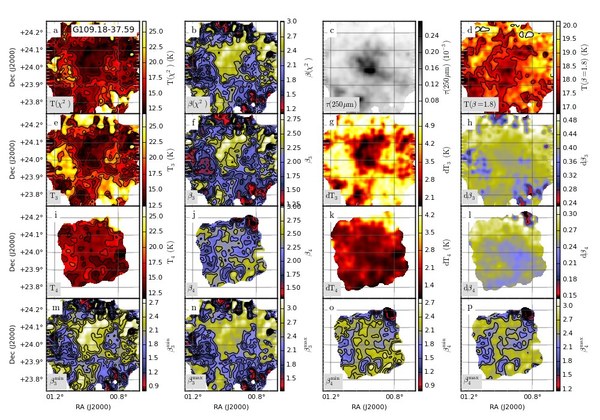}
\caption{Continued{\ldots} Field G109.18-37.59}
\end{figure*}
\begin{figure*}
\includegraphics[width=16cm]{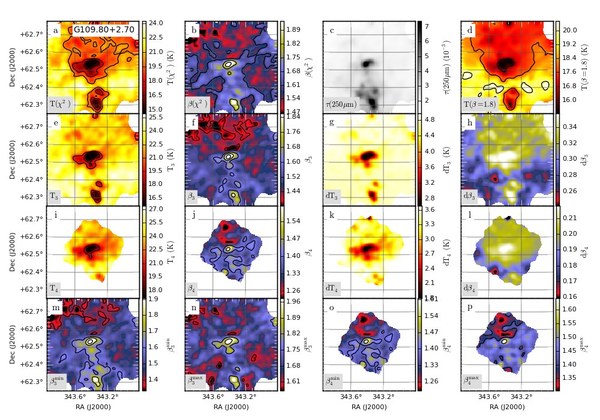}
\caption{Continued{\ldots} Field G109.80+2.70}
\end{figure*}
\begin{figure*}
\includegraphics[width=16cm]{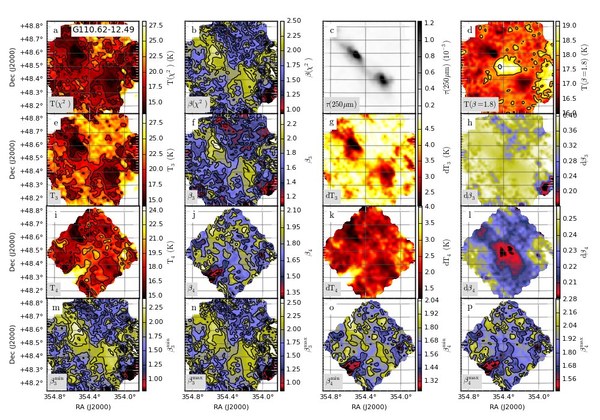}
\caption{Continued{\ldots} Field G110.62-12.49}
\end{figure*}
\begin{figure*}
\includegraphics[width=16cm]{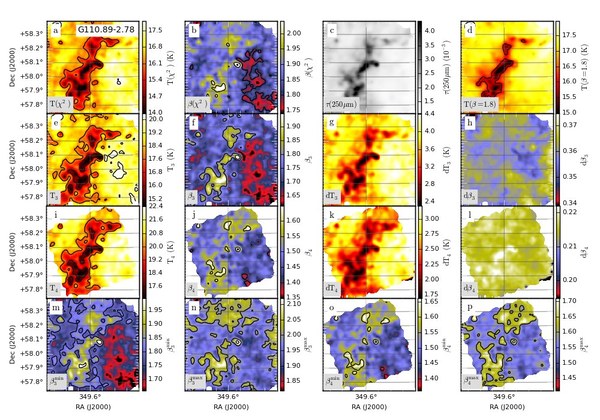}
\caption{Continued{\ldots} Field G110.89-2.78}
\end{figure*}

\clearpage

\begin{figure*}
\includegraphics[width=16cm]{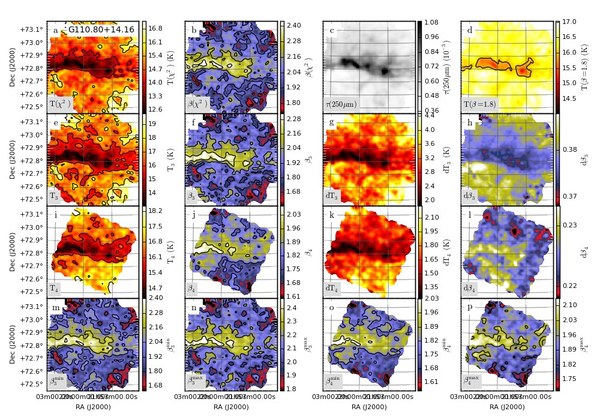}
\caption{Continued{\ldots} Field G110.80+14.16}
\end{figure*}
\begin{figure*}
\includegraphics[width=16cm]{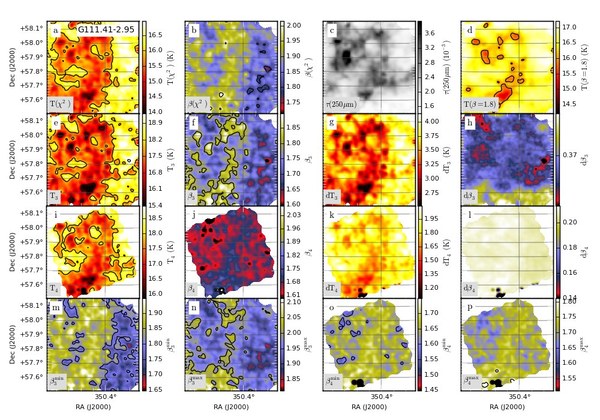}
\caption{Continued{\ldots} Field G111.41-2.95}
\end{figure*}
\begin{figure*}
\includegraphics[width=16cm]{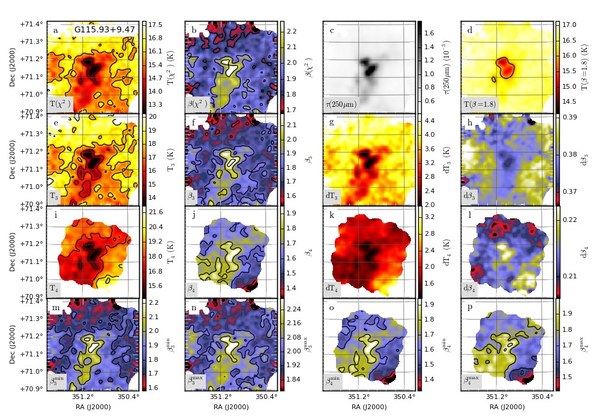}
\caption{Continued{\ldots} Field G115.93+9.47}
\end{figure*}
\begin{figure*}
\includegraphics[width=16cm]{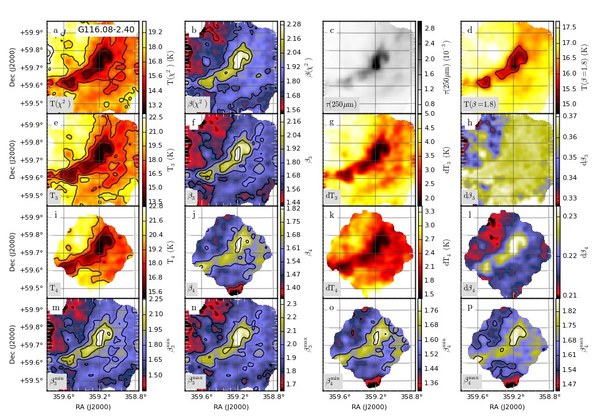}
\caption{Continued{\ldots} Field G116.08-2.40}
\end{figure*}
\begin{figure*}
\includegraphics[width=16cm]{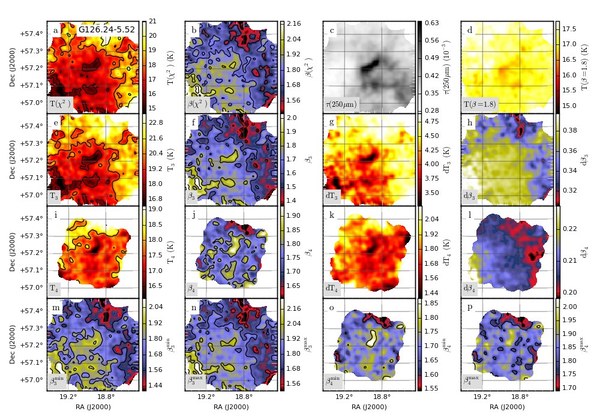}
\caption{Continued{\ldots} Field G126.24-5.52}
\end{figure*}
\begin{figure*}
\includegraphics[width=16cm]{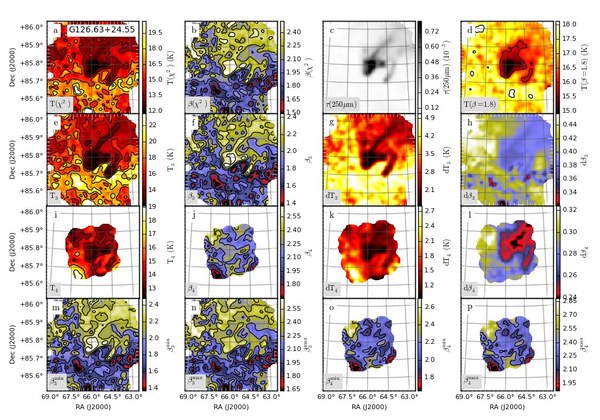}
\caption{Continued{\ldots} Field G126.63+24.55}
\end{figure*}
\begin{figure*}
\includegraphics[width=16cm]{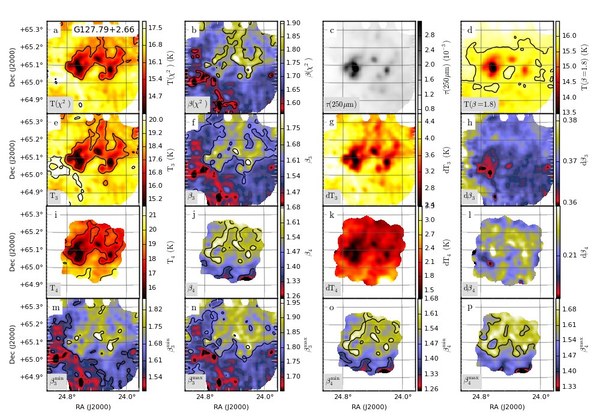}
\caption{Continued{\ldots} Field G127.79+2.66}
\end{figure*}
\begin{figure*}
\includegraphics[width=16cm]{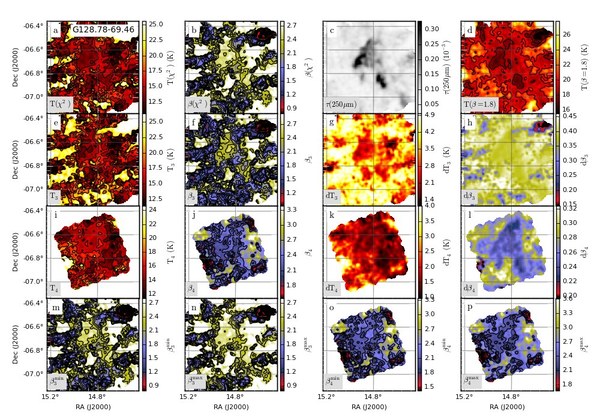}
\caption{Continued{\ldots} Field G128.78-69.46}
\end{figure*}
\begin{figure*}
\includegraphics[width=16cm]{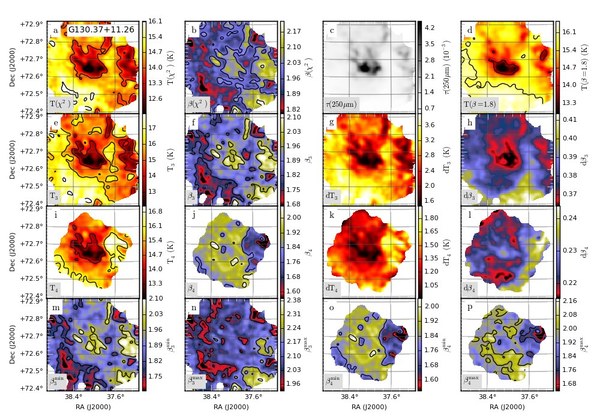}
\caption{Continued{\ldots} Field G130.37+11.26}
\end{figure*}
\begin{figure*}
\includegraphics[width=16cm]{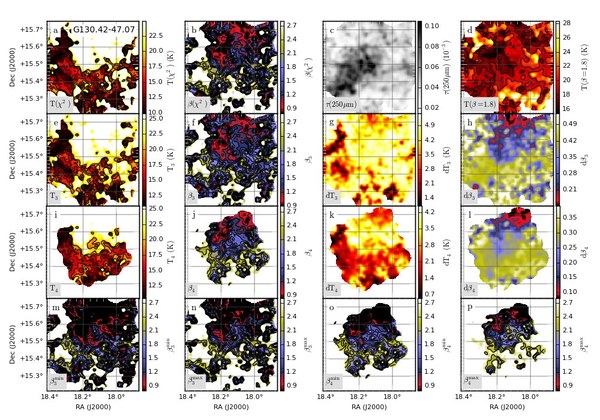}
\caption{Continued{\ldots} Field G130.42-47.07}
\end{figure*}

\clearpage

\begin{figure*}
\includegraphics[width=16cm]{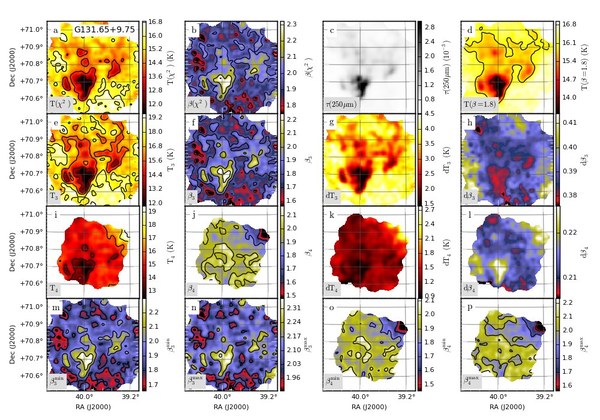}
\caption{Continued{\ldots} Field G131.65+9.75}
\end{figure*}
\begin{figure*}
\includegraphics[width=16cm]{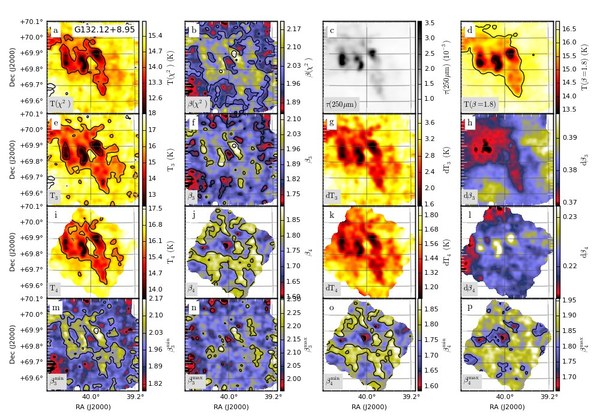}
\caption{Continued{\ldots} Field G132.12+8.95}
\end{figure*}
\begin{figure*}
\includegraphics[width=16cm]{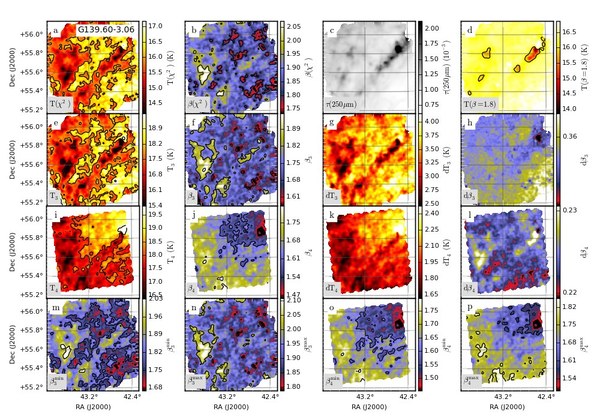}
\caption{Continued{\ldots} Field G139.60-3.06}
\end{figure*}
\begin{figure*}
\includegraphics[width=16cm]{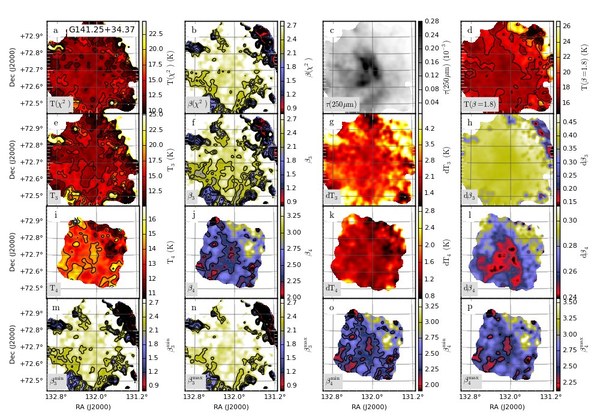}
\caption{Continued{\ldots} Field G141.25+34.37}
\end{figure*}
\begin{figure*}
\includegraphics[width=16cm]{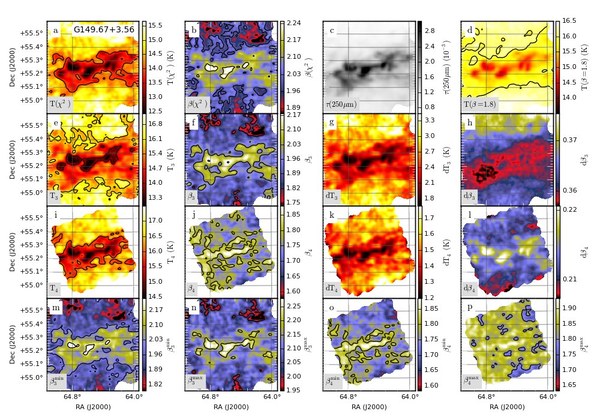}
\caption{Continued{\ldots} Field G149.67+3.56}
\end{figure*}
\begin{figure*}
\includegraphics[width=16cm]{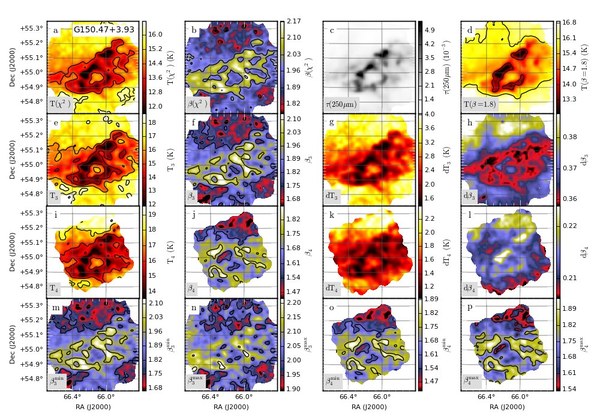}
\caption{Continued{\ldots} Field G150.47+3.93}
\end{figure*}
\begin{figure*}
\includegraphics[width=16cm]{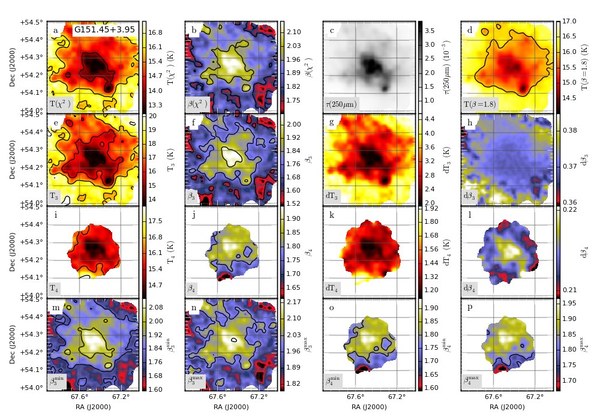}
\caption{Continued{\ldots} Field G151.45+3.95}
\end{figure*}
\begin{figure*}
\includegraphics[width=16cm]{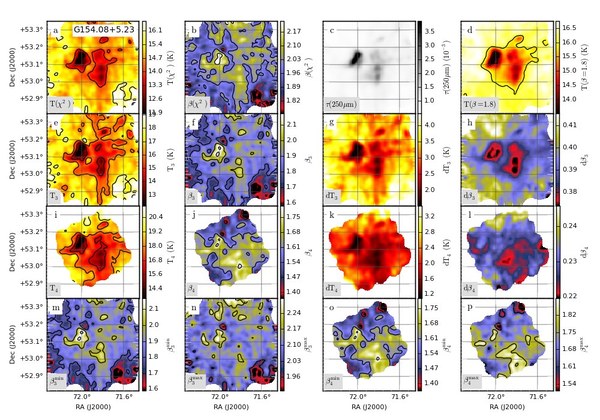}
\caption{Continued{\ldots} Field G154.08+5.23}
\end{figure*}
\begin{figure*}
\includegraphics[width=16cm]{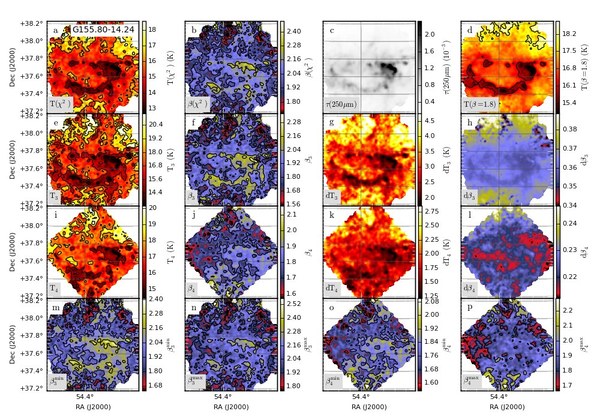}
\caption{Continued{\ldots} Field G155.80-14.24}
\end{figure*}
\begin{figure*}
\includegraphics[width=16cm]{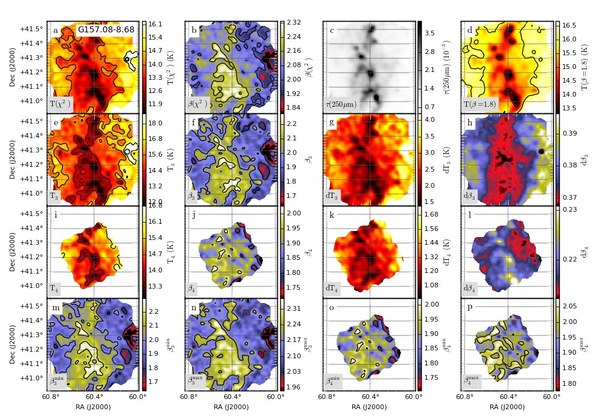}
\caption{Continued{\ldots} Field G157.08-8.68}
\end{figure*}

\clearpage

\begin{figure*}
\includegraphics[width=16cm]{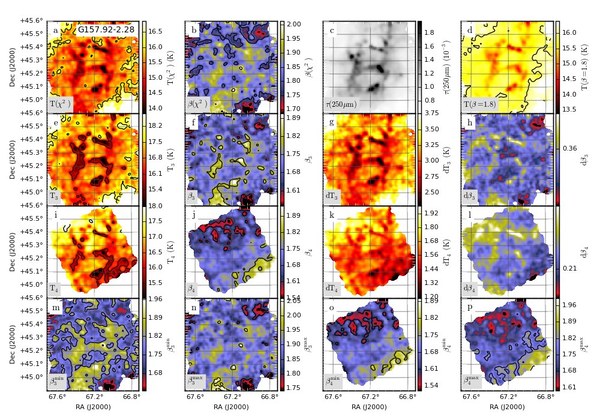}
\caption{Continued{\ldots} Field G157.92-2.28}
\end{figure*}
\begin{figure*}
\includegraphics[width=16cm]{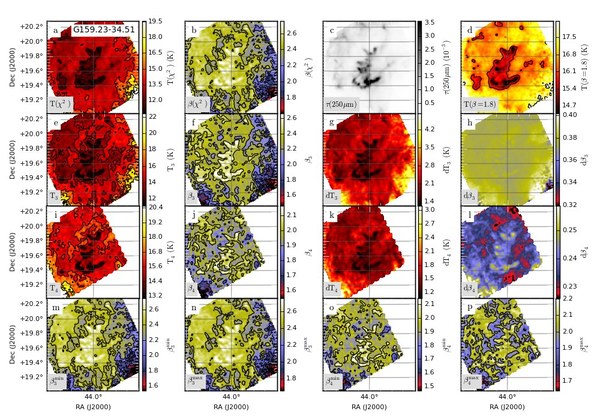}
\caption{Continued{\ldots} Field G159.23-34.51}
\end{figure*}
\begin{figure*}
\includegraphics[width=16cm]{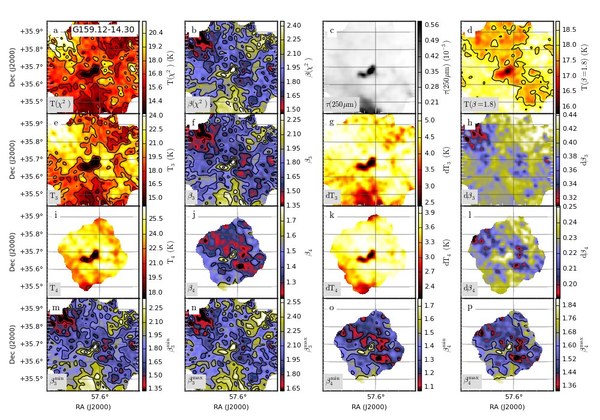}
\caption{Continued{\ldots} Field G159.12-14.30}
\end{figure*}
\begin{figure*}
\includegraphics[width=16cm]{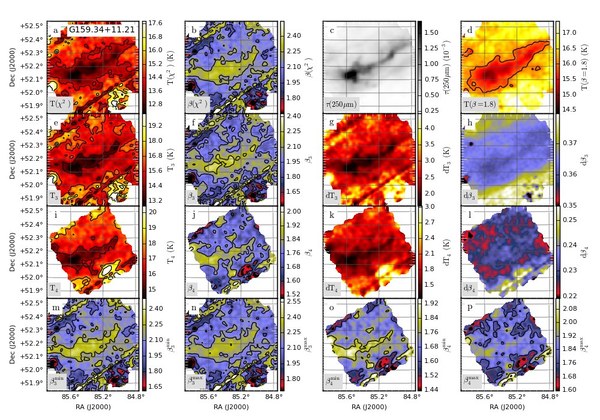}
\caption{Continued{\ldots} Field G159.34+11.21}
\end{figure*}
\begin{figure*}
\includegraphics[width=16cm]{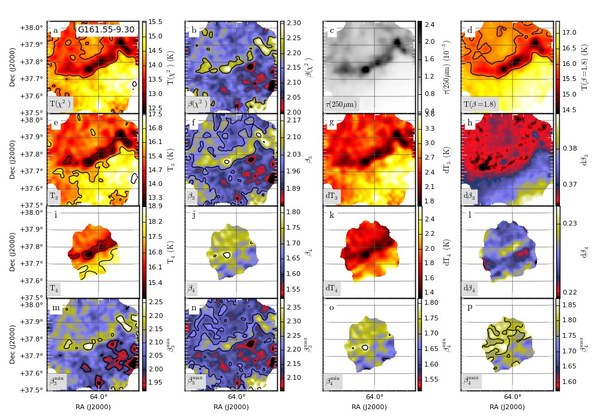}
\caption{Continued{\ldots} Field G161.55-9.30}
\end{figure*}
\begin{figure*}
\includegraphics[width=16cm]{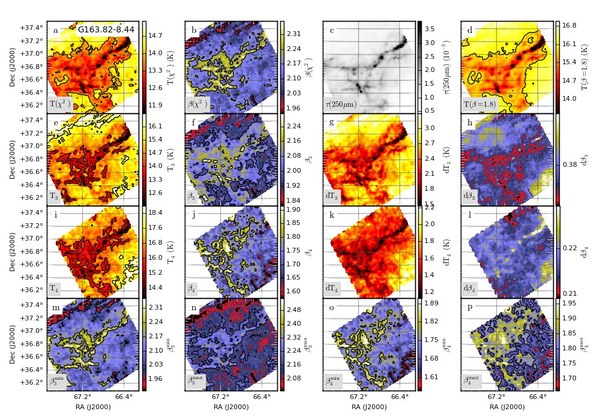}
\caption{Continued{\ldots} Field G163.82-8.44}
\end{figure*}
\begin{figure*}
\includegraphics[width=16cm]{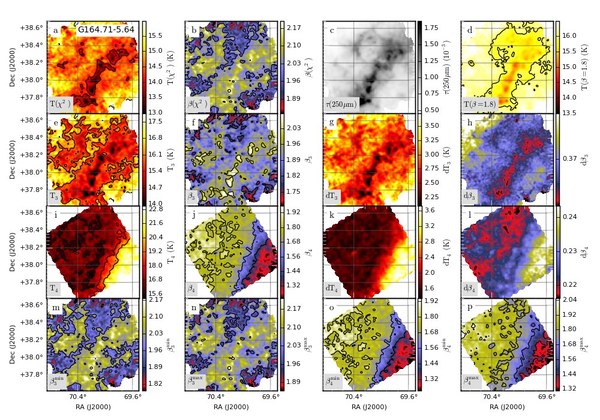}
\caption{Continued{\ldots} Field G164.71-5.64}
\end{figure*}
\begin{figure*}
\includegraphics[width=16cm]{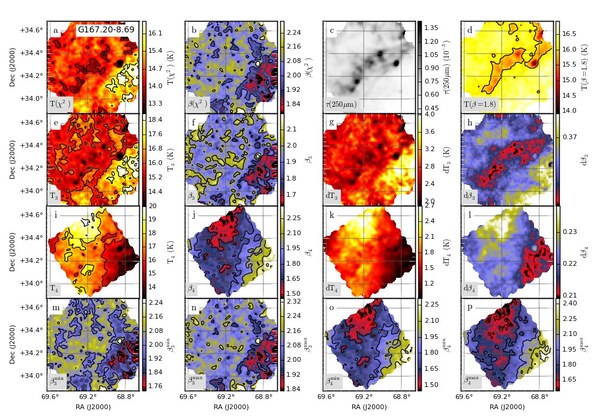}
\caption{Continued{\ldots} Field G167.20-8.69}
\end{figure*}
\begin{figure*}
\includegraphics[width=16cm]{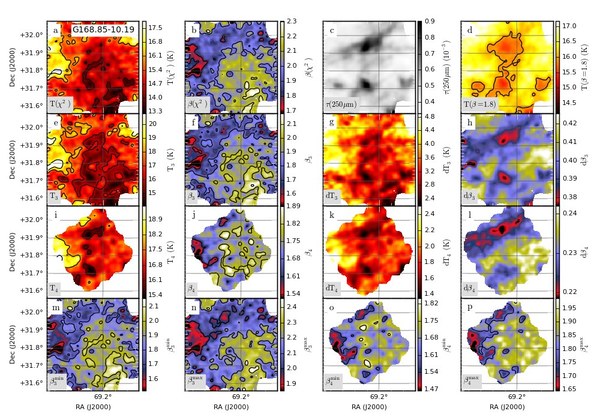}
\caption{Continued{\ldots} Field G168.85-10.19}
\end{figure*}
\begin{figure*}
\includegraphics[width=16cm]{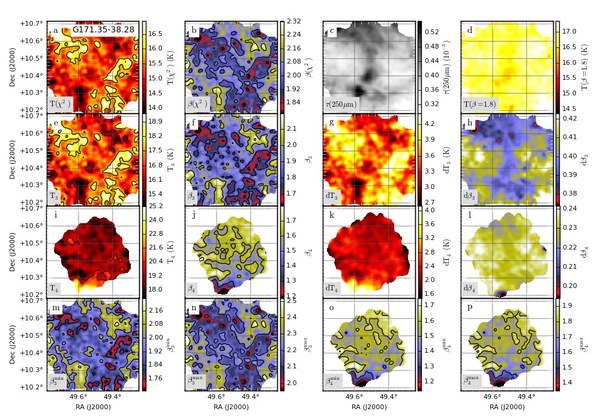}
\caption{Continued{\ldots} Field G171.35-38.28}
\end{figure*}

\clearpage

\begin{figure*}
\includegraphics[width=16cm]{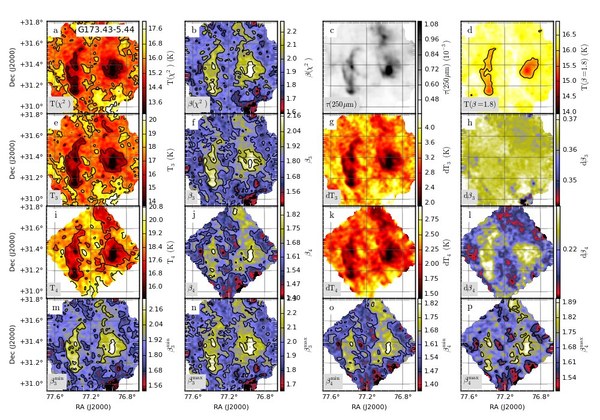}
\caption{Continued{\ldots} Field G173.43-5.44}
\end{figure*}
\begin{figure*}
\includegraphics[width=16cm]{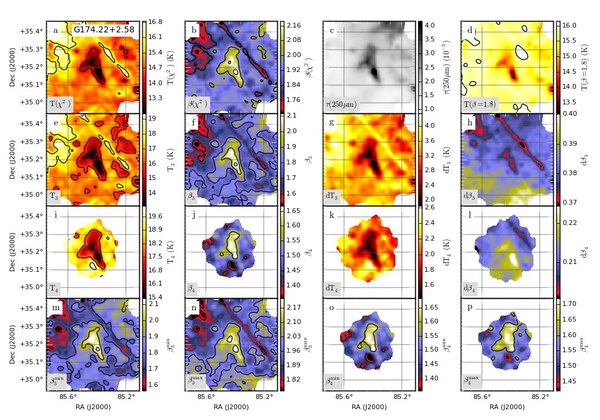}
\caption{Continued{\ldots} Field G174.22+2.58}
\end{figure*}
\begin{figure*}
\includegraphics[width=16cm]{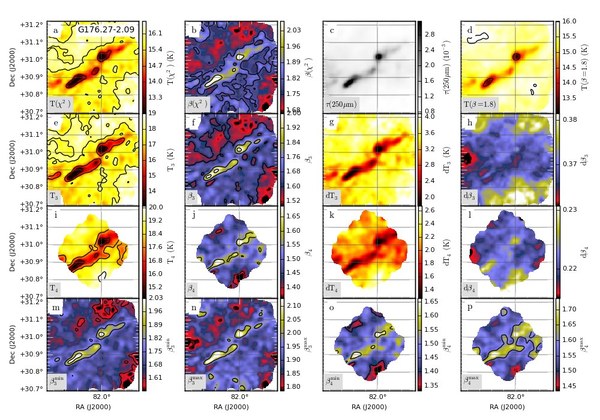}
\caption{Continued{\ldots} Field G176.27-2.09}
\end{figure*}
\begin{figure*}
\includegraphics[width=16cm]{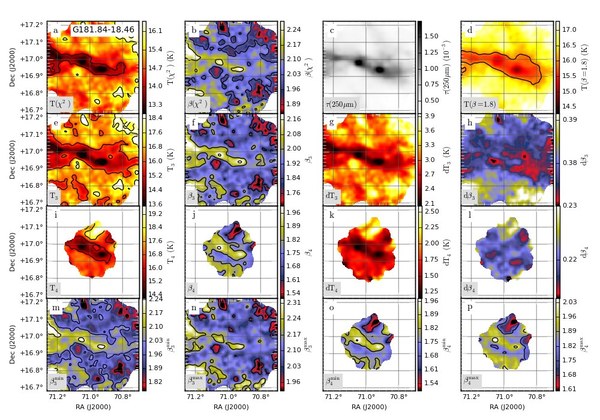}
\caption{Continued{\ldots} Field G181.84-18.46}
\end{figure*}
\begin{figure*}
\includegraphics[width=16cm]{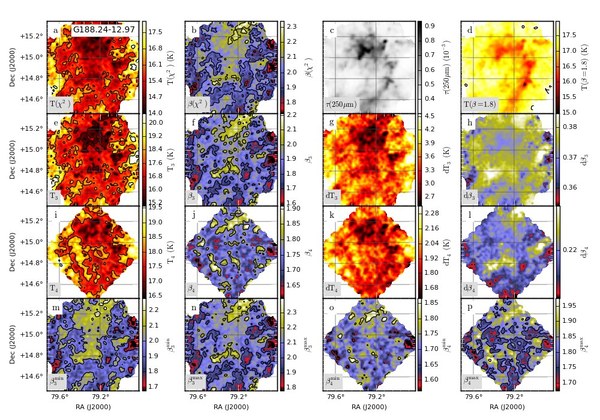}
\caption{Continued{\ldots} Field G188.24-12.97}
\end{figure*}
\begin{figure*}
\includegraphics[width=16cm]{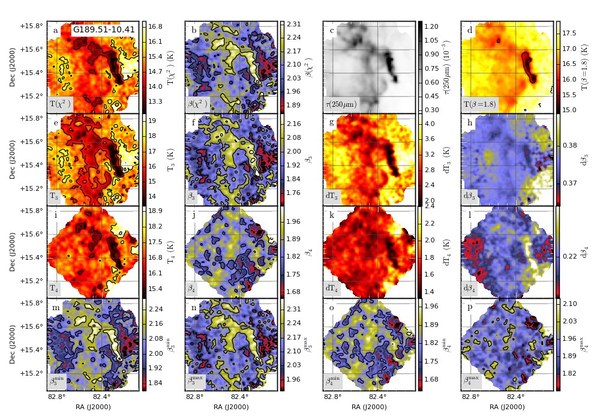}
\caption{Continued{\ldots} Field G189.51-10.41}
\end{figure*}
\begin{figure*}
\includegraphics[width=16cm]{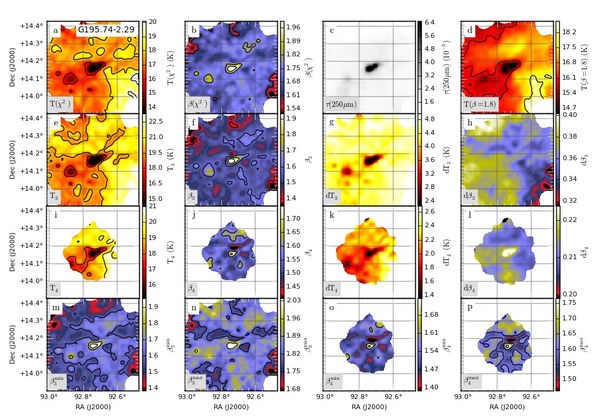}
\caption{Continued{\ldots} Field G195.74-2.29}
\end{figure*}
\begin{figure*}
\includegraphics[width=16cm]{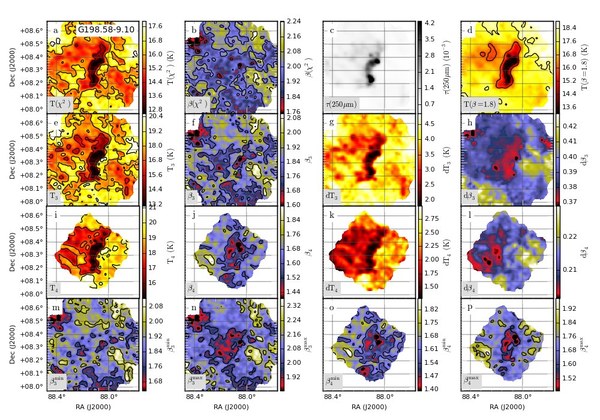}
\caption{Continued{\ldots} Field G198.58-9.10}
\end{figure*}
\begin{figure*}
\includegraphics[width=16cm]{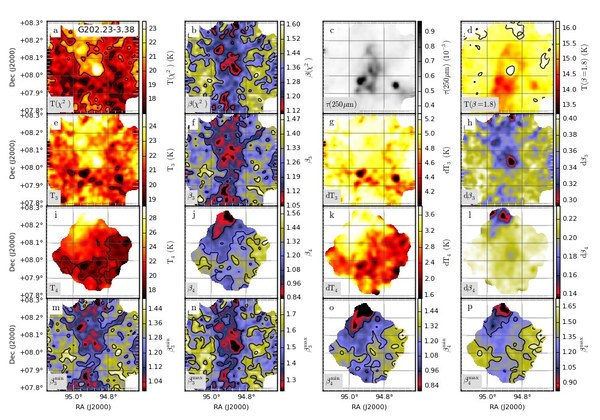}
\caption{Continued{\ldots} Field G202.23-3.38}
\end{figure*}
\begin{figure*}
\includegraphics[width=16cm]{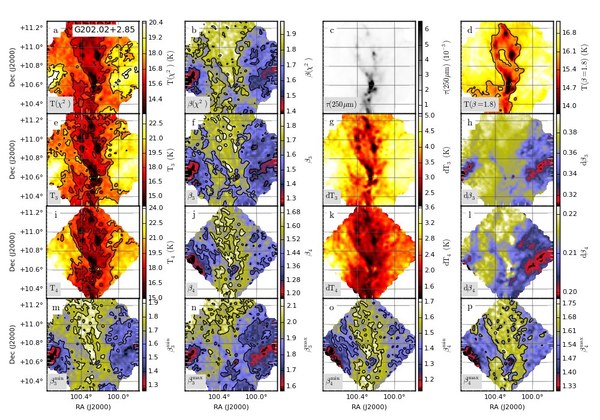}
\caption{Continued{\ldots} Field G202.02+2.85}
\end{figure*}

\clearpage

\begin{figure*}
\includegraphics[width=16cm]{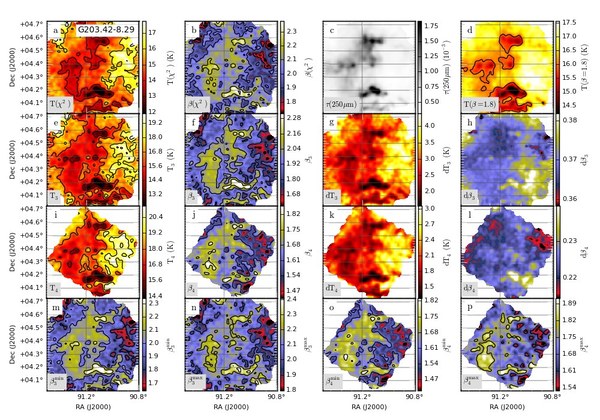}
\caption{Continued{\ldots} Field G203.42-8.29}
\end{figure*}
\begin{figure*}
\includegraphics[width=16cm]{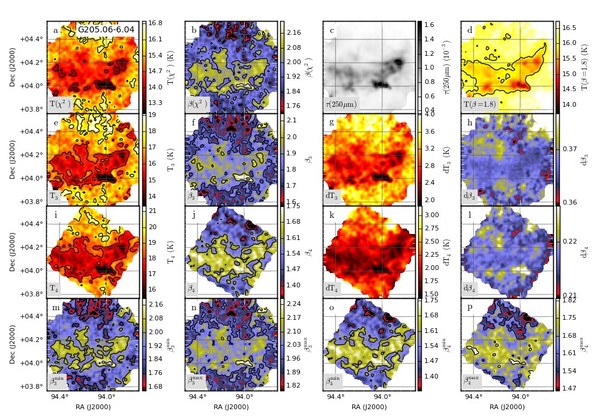}
\caption{Continued{\ldots} Field G205.06-6.04}
\end{figure*}
\begin{figure*}
\includegraphics[width=16cm]{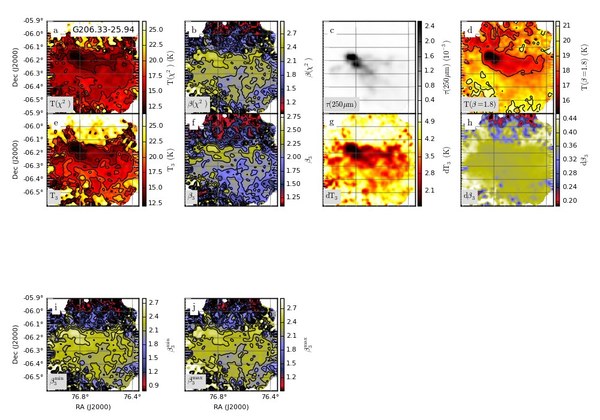}
\caption{Continued{\ldots} Field G206.33-25.94}
\end{figure*}
\begin{figure*}
\includegraphics[width=16cm]{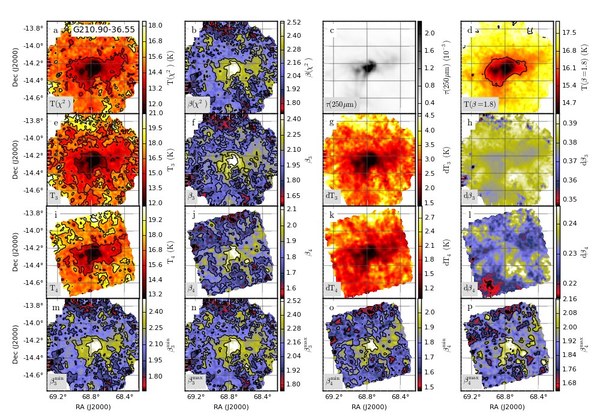}
\caption{Continued{\ldots} Field G210.90-36.55}
\end{figure*}
\begin{figure*}
\includegraphics[width=16cm]{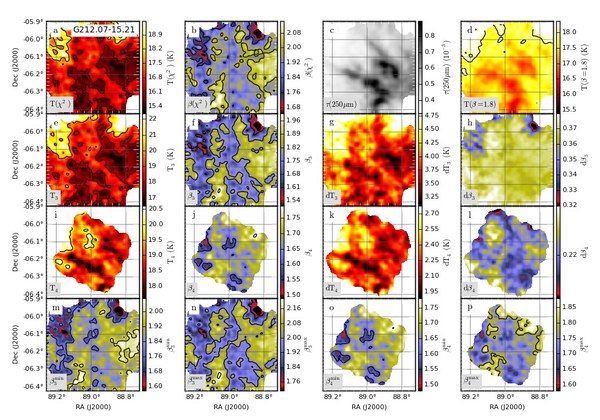}
\caption{Continued{\ldots} Field G212.07-15.21}
\end{figure*}
\begin{figure*}
\includegraphics[width=16cm]{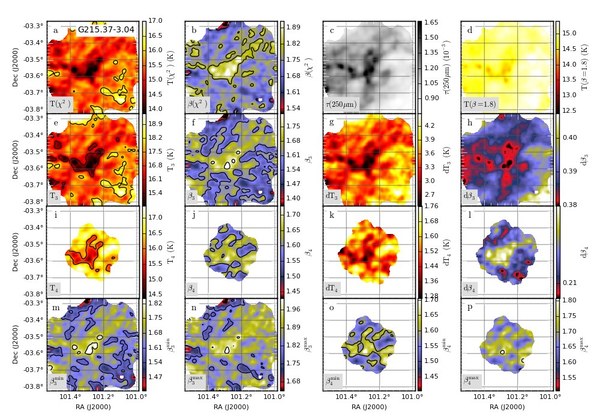}
\caption{Continued{\ldots} Field G215.37-3.04}
\end{figure*}
\begin{figure*}
\includegraphics[width=16cm]{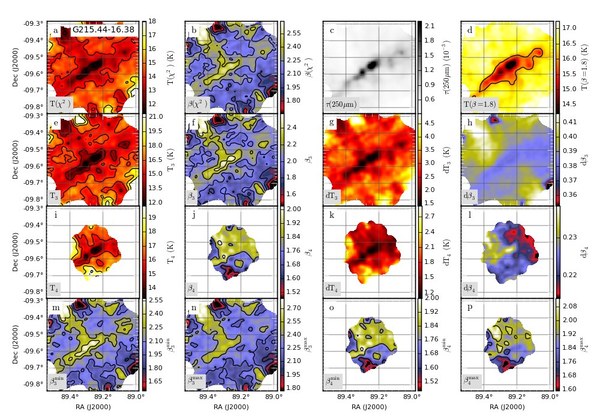}
\caption{Continued{\ldots} Field G215.44-16.38}
\end{figure*}
\begin{figure*}
\includegraphics[width=16cm]{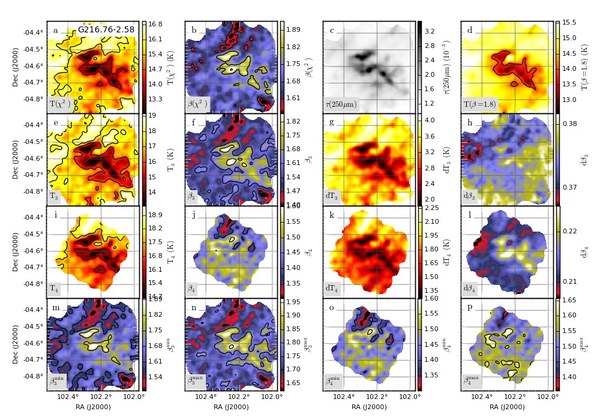}
\caption{Continued{\ldots} Field G216.76-2.58}
\end{figure*}
\begin{figure*}
\includegraphics[width=16cm]{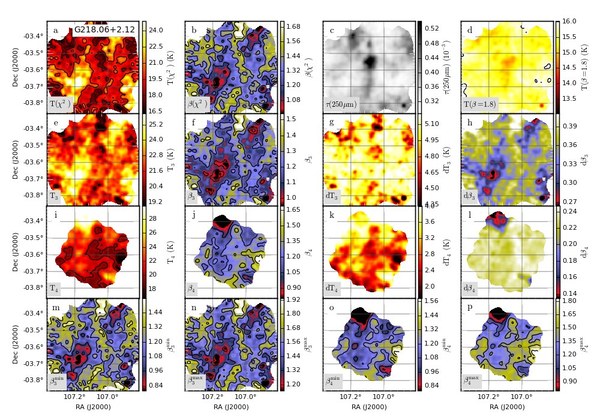}
\caption{Continued{\ldots} Field G218.06+2.12}
\end{figure*}
\begin{figure*}
\includegraphics[width=16cm]{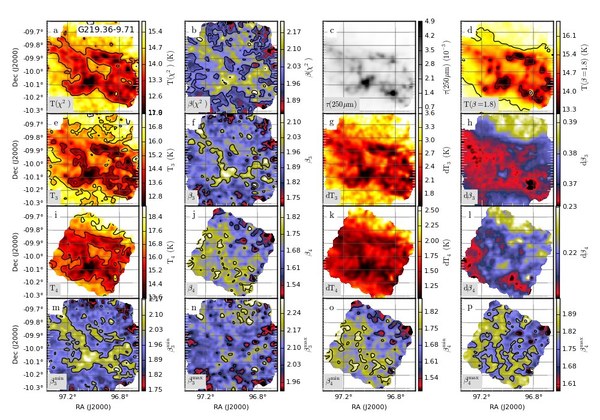}
\caption{Continued{\ldots} Field G219.36-9.71}
\end{figure*}

\clearpage

\begin{figure*}
\includegraphics[width=16cm]{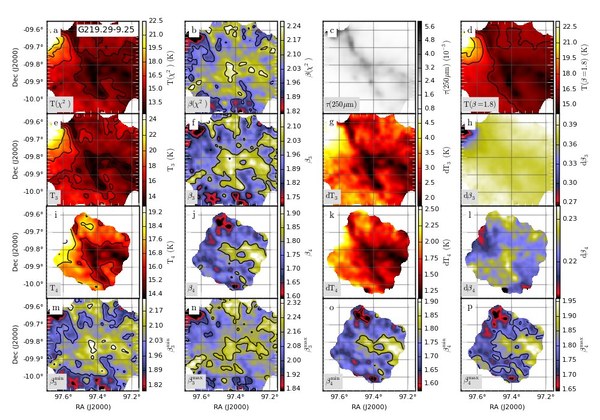}
\caption{Continued{\ldots} Field G219.29-9.25}
\end{figure*}
\begin{figure*}
\includegraphics[width=16cm]{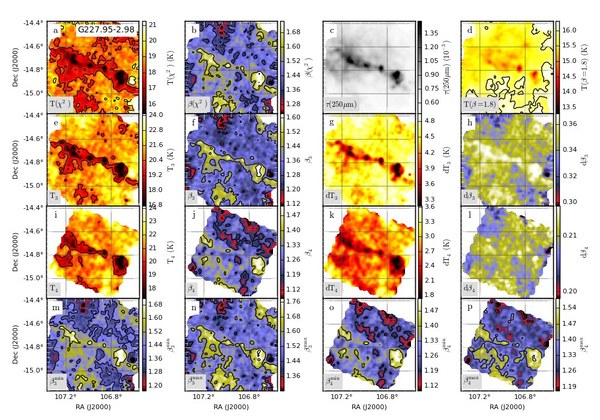}
\caption{Continued{\ldots} Field G227.95-2.98}
\end{figure*}
\begin{figure*}
\includegraphics[width=16cm]{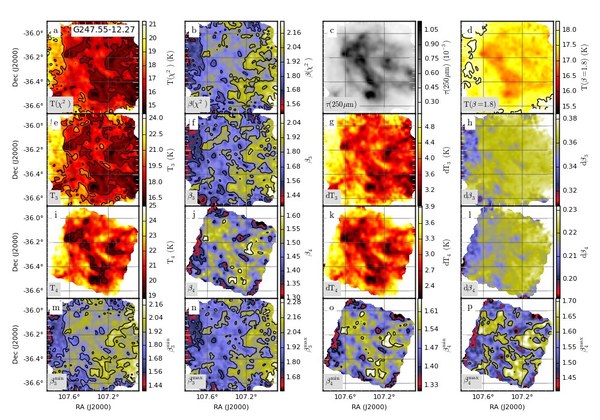}
\caption{Continued{\ldots} Field G247.55-12.27}
\end{figure*}
\begin{figure*}
\includegraphics[width=16cm]{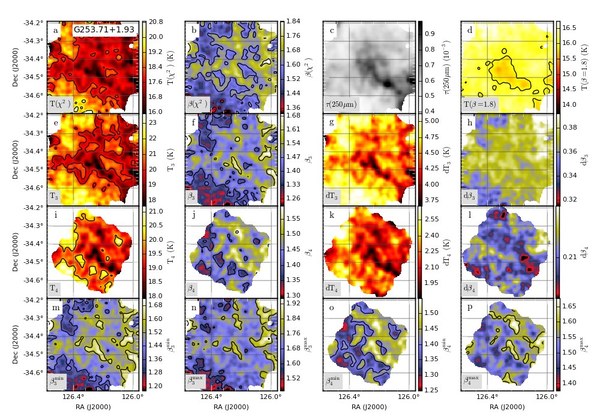}
\caption{Continued{\ldots} Field G253.71+1.93}
\end{figure*}
\begin{figure*}
\includegraphics[width=16cm]{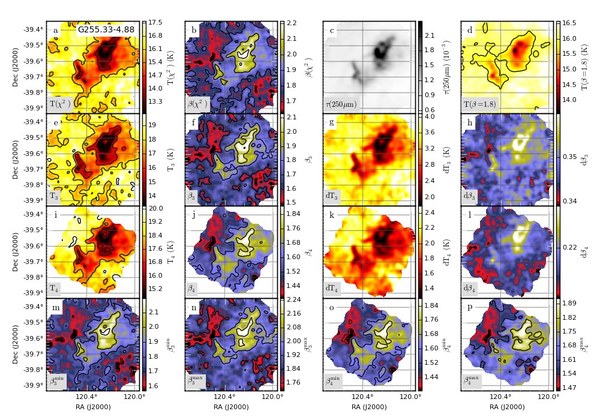}
\caption{Continued{\ldots} Field G255.33-4.88}
\end{figure*}
\begin{figure*}
\includegraphics[width=16cm]{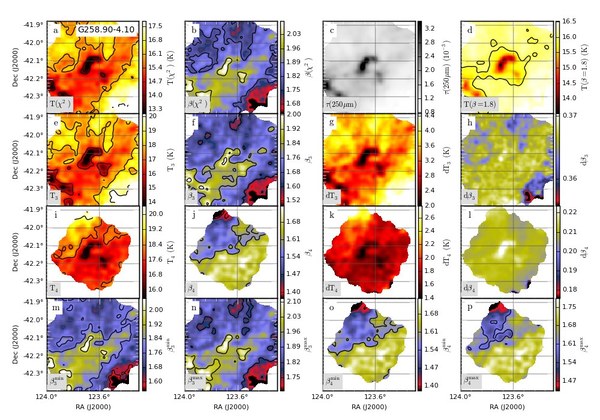}
\caption{Continued{\ldots} Field G258.90-4.10}
\end{figure*}
\begin{figure*}
\includegraphics[width=16cm]{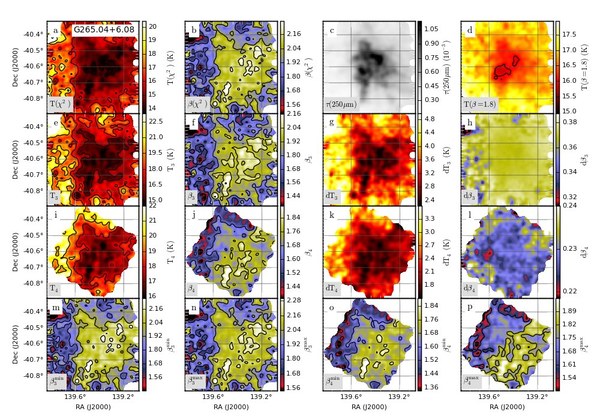}
\caption{Continued{\ldots} Field G265.04+6.08}
\end{figure*}
\begin{figure*}
\includegraphics[width=16cm]{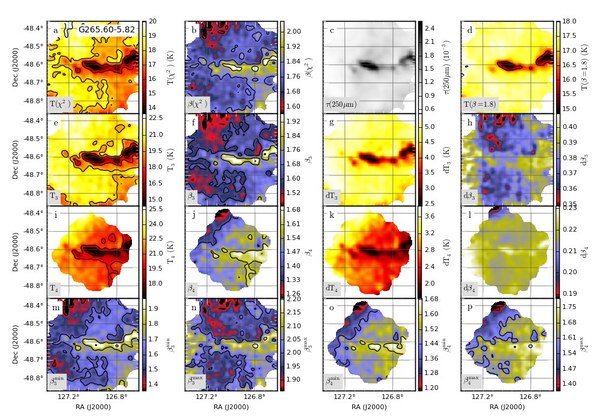}
\caption{Continued{\ldots} Field G265.60-5.82}
\end{figure*}
\begin{figure*}
\includegraphics[width=16cm]{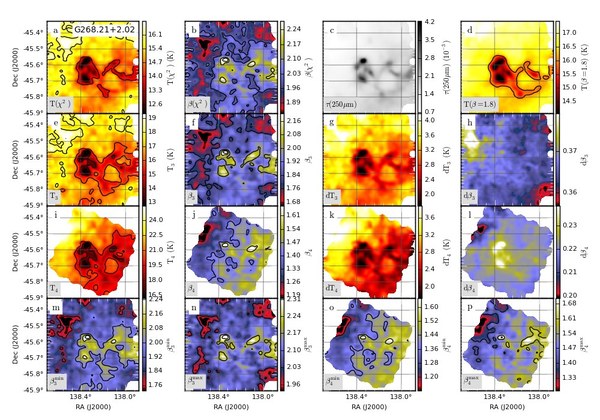}
\caption{Continued{\ldots} Field G268.21+2.02}
\end{figure*}
\begin{figure*}
\includegraphics[width=16cm]{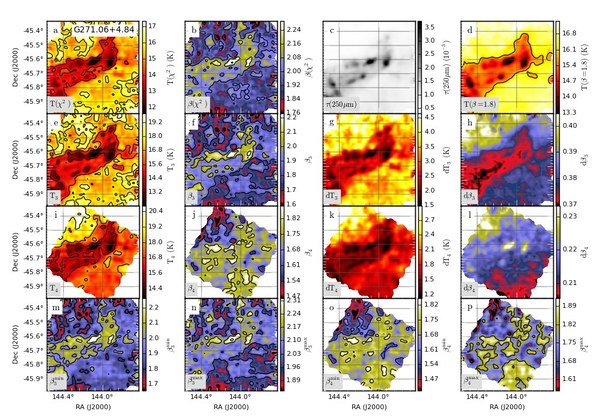}
\caption{Continued{\ldots} Field G271.06+4.84}
\end{figure*}

\clearpage

\begin{figure*}
\includegraphics[width=16cm]{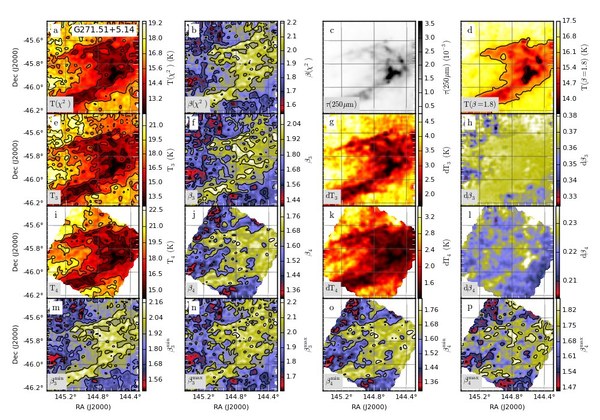}
\caption{Continued{\ldots} Field G271.51+5.14}
\end{figure*}
\begin{figure*}
\includegraphics[width=16cm]{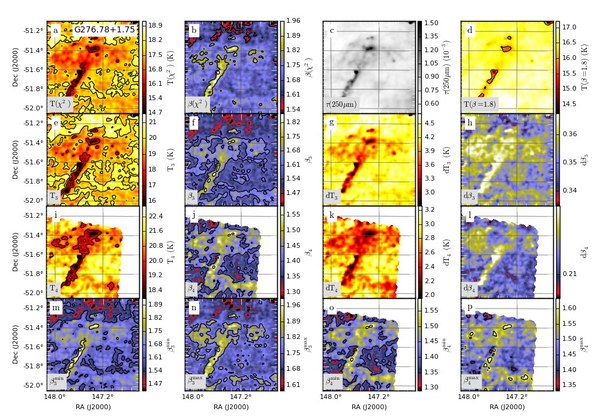}
\caption{Continued{\ldots} Field G276.78+1.75}
\end{figure*}
\begin{figure*}
\includegraphics[width=16cm]{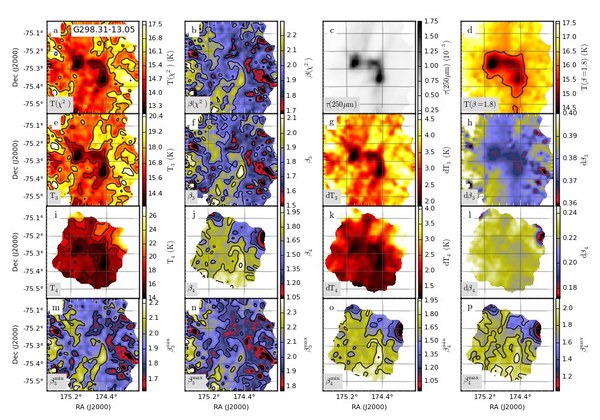}
\caption{Continued{\ldots} Field G298.31-13.05}
\end{figure*}
\begin{figure*}
\includegraphics[width=16cm]{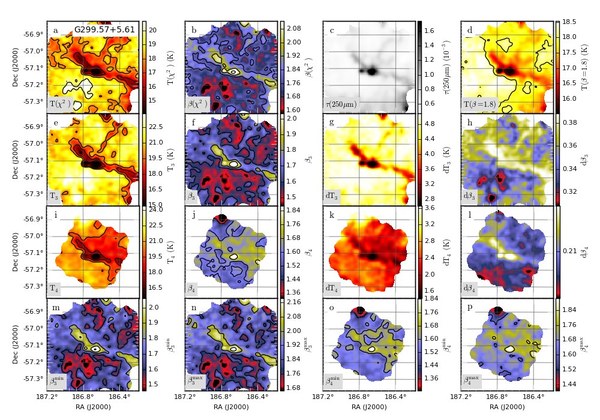}
\caption{Continued{\ldots} Field G299.57+5.61}
\end{figure*}
\begin{figure*}
\includegraphics[width=16cm]{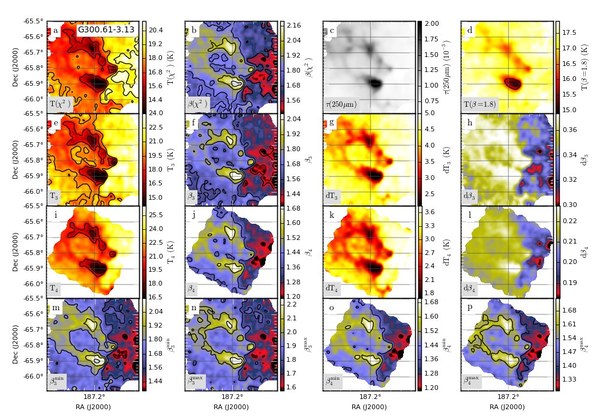}
\caption{Continued{\ldots} Field G300.61-3.13}
\end{figure*}
\begin{figure*}
\includegraphics[width=16cm]{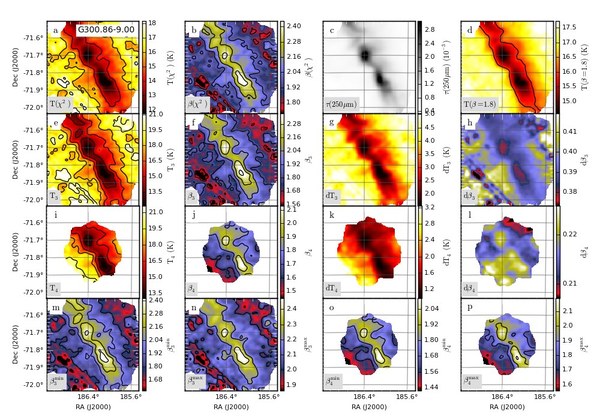}
\caption{Continued{\ldots} Field G300.86-9.00}
\end{figure*}
\begin{figure*}
\includegraphics[width=16cm]{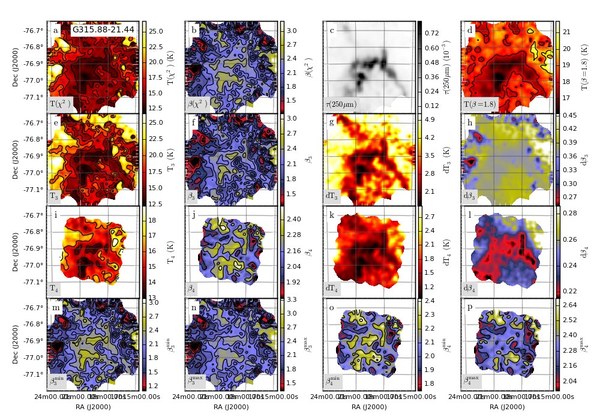}
\caption{Continued{\ldots} Field G315.88-21.44}
\end{figure*}
\begin{figure*}
\includegraphics[width=16cm]{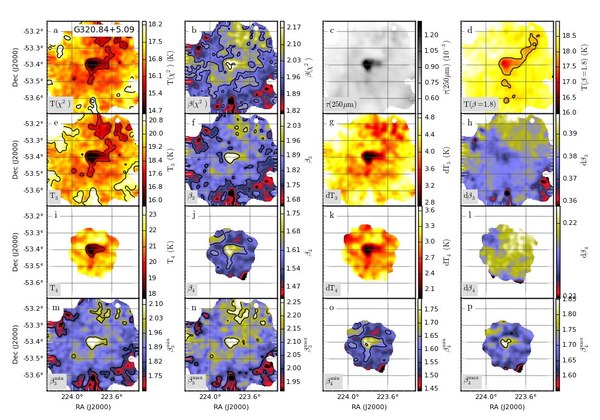}
\caption{Continued{\ldots} Field G320.84+5.09}
\end{figure*}
\begin{figure*}
\includegraphics[width=16cm]{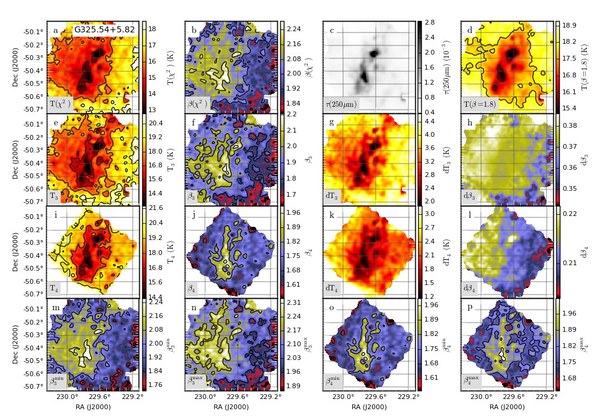}
\caption{Continued{\ldots} Field G325.54+5.82}
\end{figure*}
\begin{figure*}
\includegraphics[width=16cm]{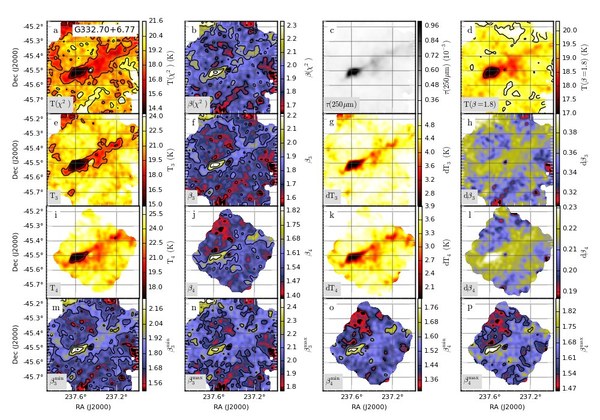}
\caption{Continued{\ldots} Field G332.70+6.77}
\end{figure*}

\clearpage

\begin{figure*}
\includegraphics[width=16cm]{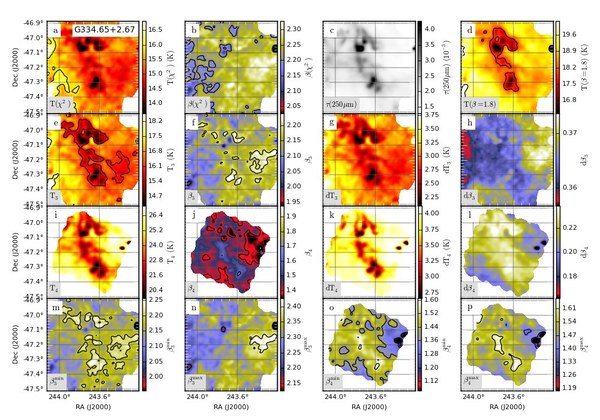}
\caption{Continued{\ldots} Field G334.65+2.67}
\end{figure*}
\begin{figure*}
\includegraphics[width=16cm]{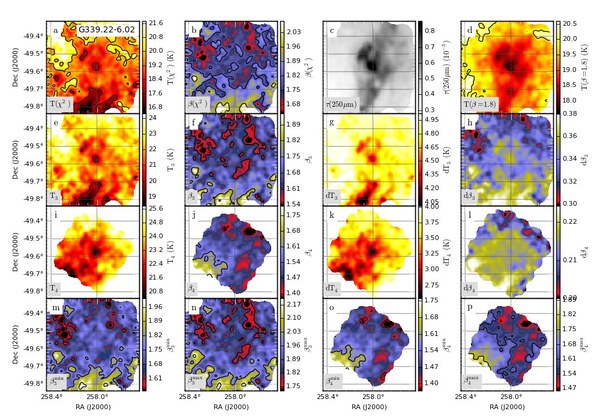}
\caption{Continued{\ldots} Field G339.22-6.02}
\end{figure*}
\begin{figure*}
\includegraphics[width=16cm]{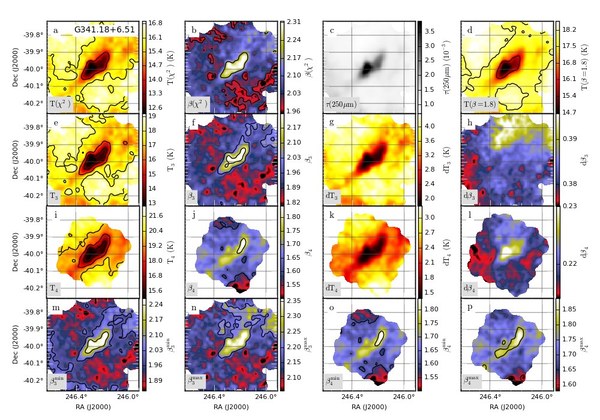}
\caption{Continued{\ldots} Field G341.18+6.51}
\end{figure*}
\begin{figure*}
\includegraphics[width=16cm]{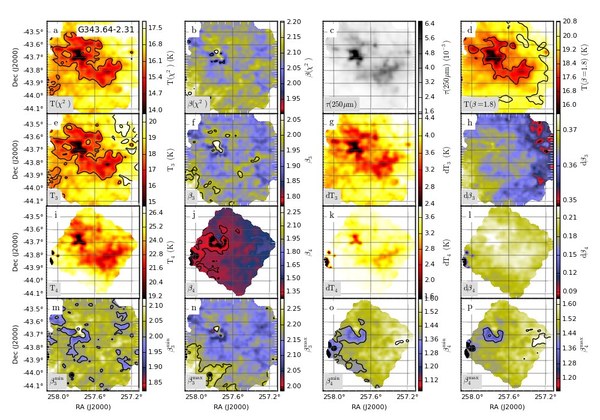}
\caption{Continued{\ldots} Field G343.64-2.31}
\end{figure*}
\begin{figure*}
\includegraphics[width=16cm]{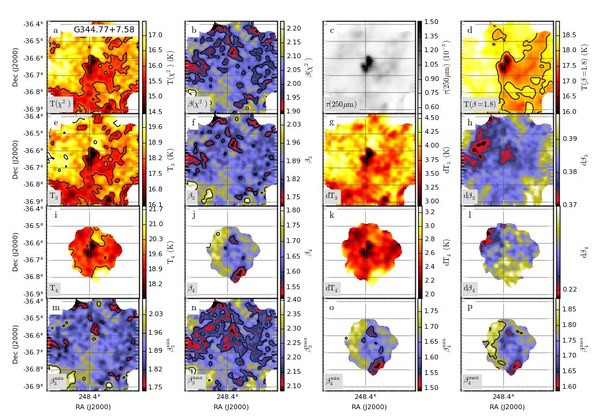}
\caption{Continued{\ldots} Field G344.77+7.58}
\end{figure*}
\begin{figure*}
\includegraphics[width=16cm]{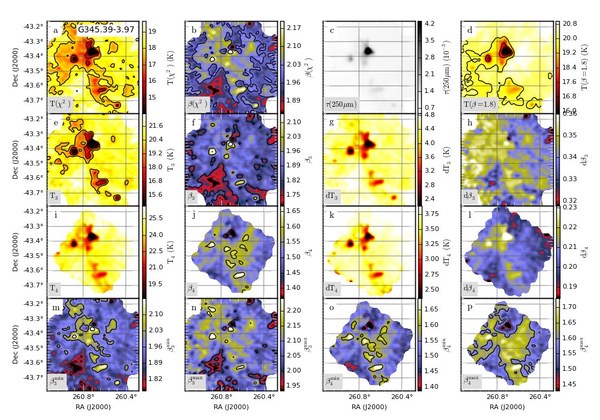}
\caption{Continued{\ldots} Field G345.39-3.97}
\label{fig:TB_PLWBEAM_115}
\end{figure*}
\begin{figure*}
\includegraphics[width=16cm]{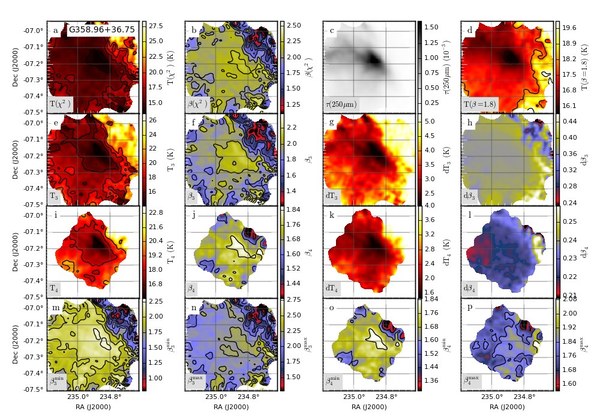}
\caption{Continued{\ldots} Field G358.96+36.75}
\label{fig:TB_PLWBEAM_116}
\end{figure*}

\clearpage

\section{Combined ($T$, $\beta$) fits to {\it Planck} and {\it Herschel} data} \label{app:hpfit}

The joint fit of {\it Herschel} (250\,$\mu$m--500\,$\mu$m) and {\it Planck}
(857\,GHz-217\,GHz) data was carried out only for the 53 fields with estimated distances
$d<\le$400\,pc (see Sect.~\ref{sect:combined}). For comparison, the figures also
include 353\,GHz and 217\,GHz residuals from fits where we assume twice the default CO
correction (see Sect.~\ref{sect:Planck_data}). The larger correction could in some cases
eliminate or even reverse the sign of the 217\,GHz residual. However, for example in the
case of field G6.03+36.73, ground-based observations confirm that the line ratios
assumed in the default CO correction are of the correct magnitude (see
Sect.~\ref{sect:Planck_data}). 

\begin{figure*}
\includegraphics[width=16cm]{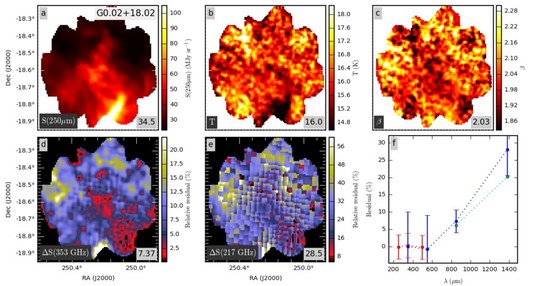}
\caption{
Modified blackbody fits in field G0.02+18.02 using the combination of {\it Herschel} and
{\it Planck} data. The uppermost frames show the fitted intensity at 250\,$\mu$m, and
the colour temperature and spectral index maps. The relative residuals (observation
minus model, divided by model prediction) are shown in frames $d$ and $f$ for the {\it
Planck} bands of 353\,GHz and 217\,GHz. Frame $f$ shows the median values of the
residuals for the three {\em Herschel} bands (red symbols) and the four {\it Planck} bands
(blue symbols). The error bars correspond to the median error estimate of the surface
brightness over the map (mainly the assumed calibration errors and the uncertainties of
the CO corrections). The lower data points at 850\,$\mu$m and 1380\,$\mu$m ({\em Planck}
bands 353\,GHz and 217\,GHz; cyan symbols) correspond to twice the default CO
correction.
}
\label{fig:BRUTE_001}
\end{figure*}

\begin{figure*}
\includegraphics[width=16cm]{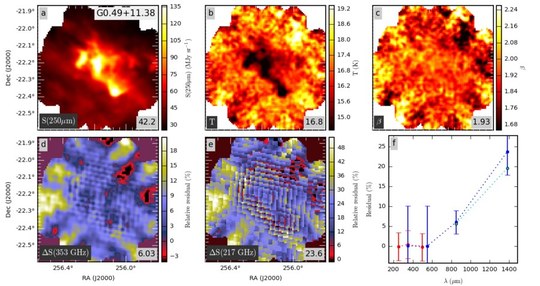}
\caption{Continued{\ldots} Field G0.49+11.38}
\end{figure*}
\begin{figure*}
\includegraphics[width=16cm]{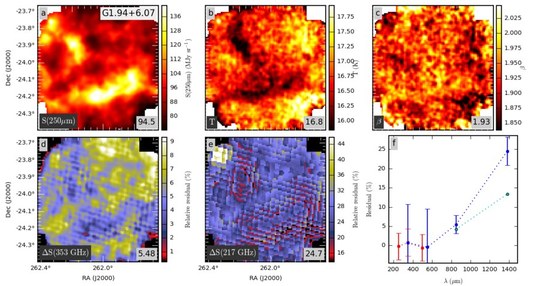}
\caption{Continued{\ldots} Field G1.94+6.07}
\end{figure*}
\begin{figure*}
\includegraphics[width=16cm]{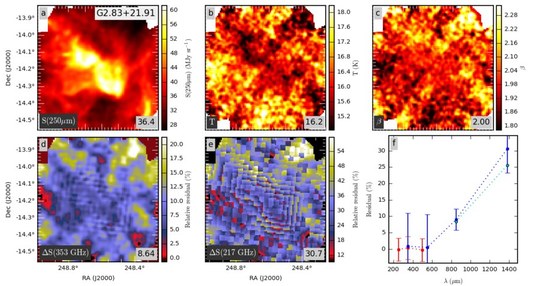}
\caption{Continued{\ldots} Field G2.83+21.91}
\end{figure*}
\begin{figure*}
\includegraphics[width=16cm]{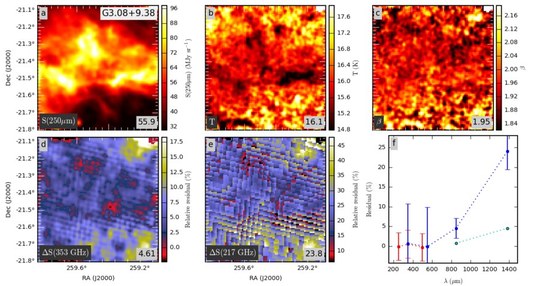}
\caption{Continued{\ldots} Field G3.08+9.38}
\end{figure*}
\begin{figure*}
\includegraphics[width=16cm]{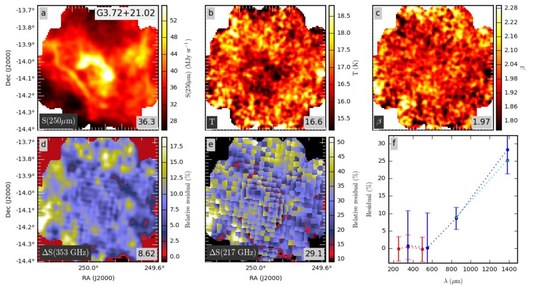}
\caption{Continued{\ldots} Field G3.72+21.02}
\end{figure*}
\begin{figure*}
\includegraphics[width=16cm]{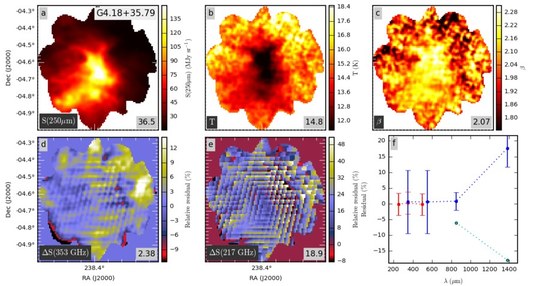}
\caption{Continued{\ldots} Field G4.18+35.79}
\end{figure*}
\begin{figure*}
\includegraphics[width=16cm]{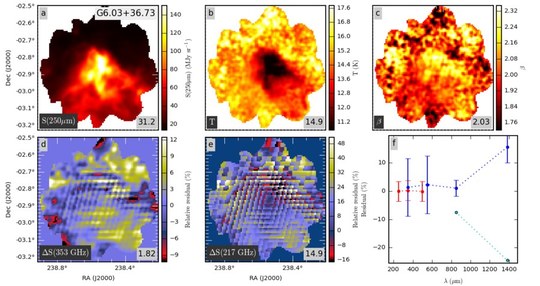}
\caption{Continued{\ldots} Field G6.03+36.73}
\end{figure*}
\begin{figure*}
\includegraphics[width=16cm]{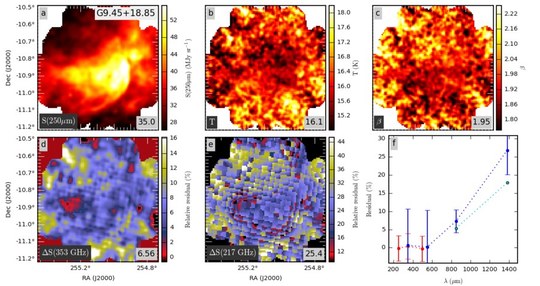}
\caption{Continued{\ldots} Field G9.45+18.85}
\end{figure*}
\begin{figure*}
\includegraphics[width=16cm]{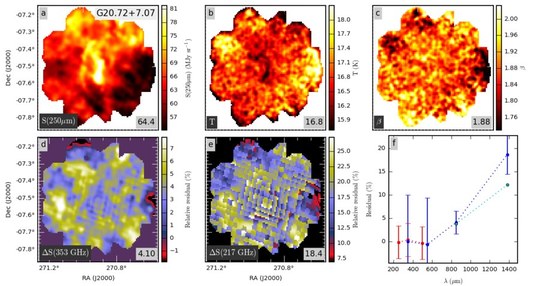}
\caption{Continued{\ldots} Field G20.72+7.07}
\end{figure*}
\begin{figure*}
\includegraphics[width=16cm]{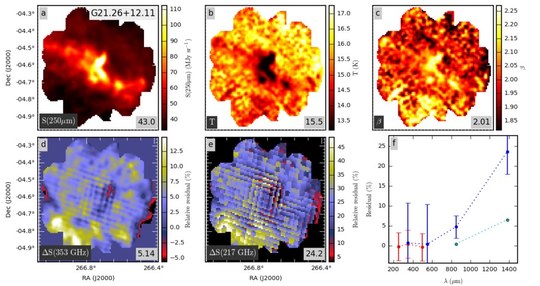}
\caption{Continued{\ldots} Field G21.26+12.11}
\end{figure*}
\begin{figure*}
\includegraphics[width=16cm]{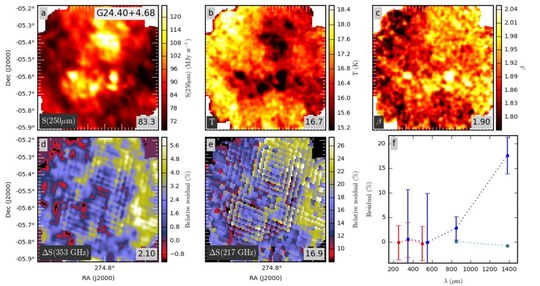}
\caption{Continued{\ldots} Field G24.40+4.68}
\end{figure*}
\clearpage
\begin{figure*}
\includegraphics[width=16cm]{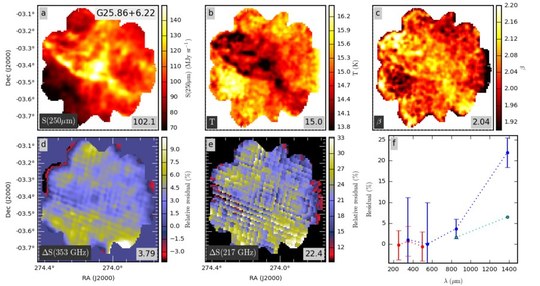}
\caption{Continued{\ldots} Field G25.86+6.22}
\end{figure*}
\begin{figure*}
\includegraphics[width=16cm]{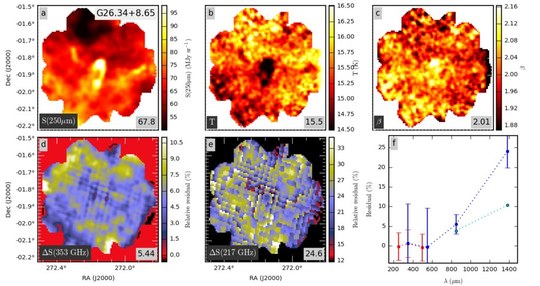}
\caption{Continued{\ldots} Field G26.34+8.65}
\end{figure*}
\begin{figure*}
\includegraphics[width=16cm]{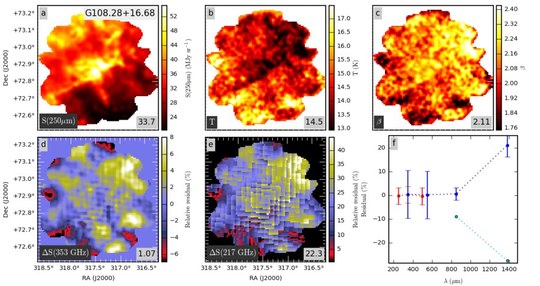}
\caption{Continued{\ldots} Field G108.28+16.68}
\end{figure*}
\begin{figure*}
\includegraphics[width=16cm]{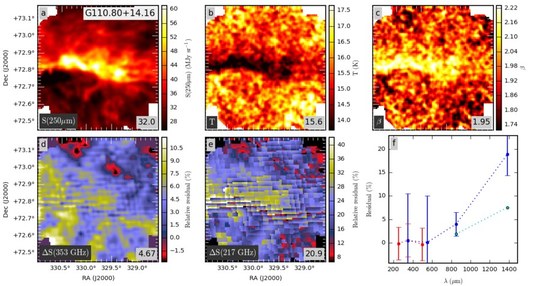}
\caption{Continued{\ldots} Field G110.80+14.16}
\end{figure*}
\begin{figure*}
\includegraphics[width=16cm]{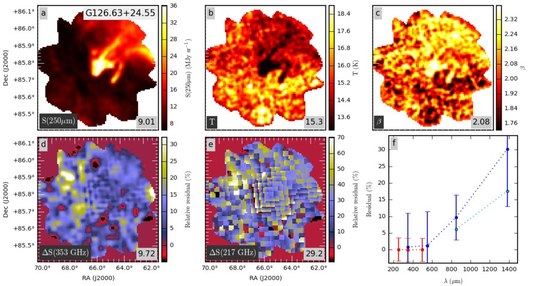}
\caption{Continued{\ldots} Field G126.63+24.55}
\end{figure*}
\begin{figure*}
\includegraphics[width=16cm]{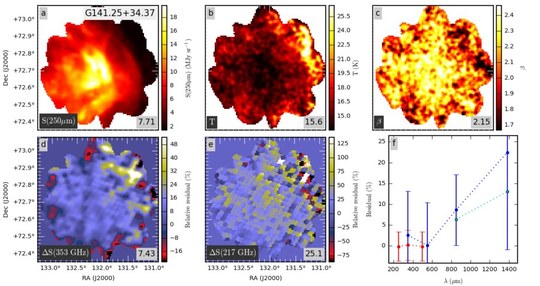}
\caption{Continued{\ldots} Field G141.25+34.37}
\end{figure*}
\begin{figure*}
\includegraphics[width=16cm]{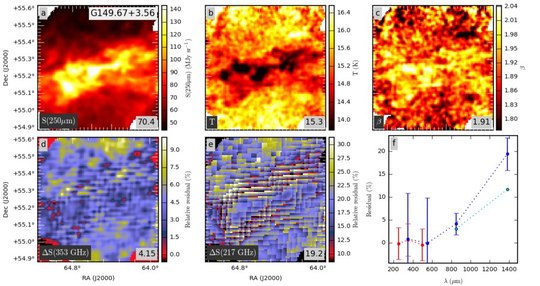}
\caption{Continued{\ldots} Field G149.67+3.56}
\end{figure*}
\begin{figure*}
\includegraphics[width=16cm]{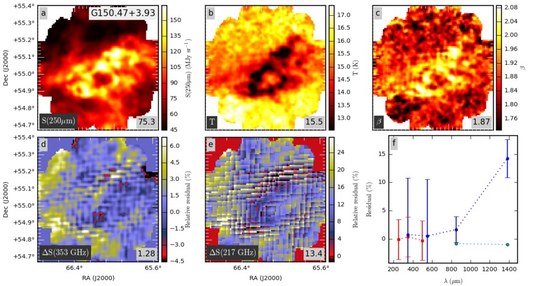}
\caption{Continued{\ldots} Field G150.47+3.93}
\end{figure*}
\begin{figure*}
\includegraphics[width=16cm]{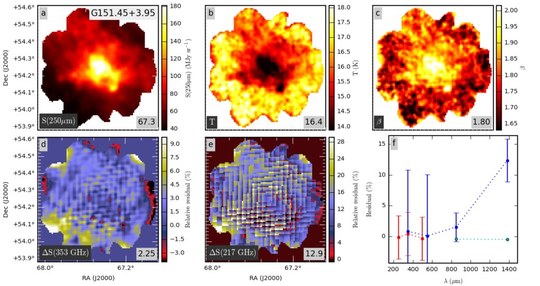}
\caption{Continued{\ldots} Field G151.45+3.95}
\end{figure*}
\begin{figure*}
\includegraphics[width=16cm]{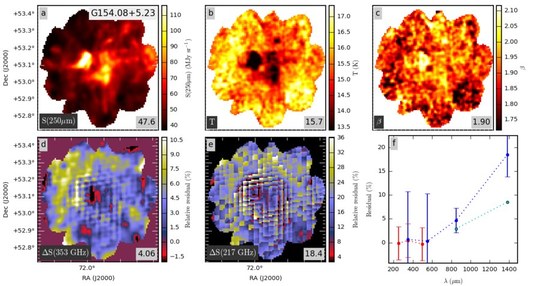}
\caption{Continued{\ldots} Field G154.08+5.23}
\end{figure*}
\begin{figure*}
\includegraphics[width=16cm]{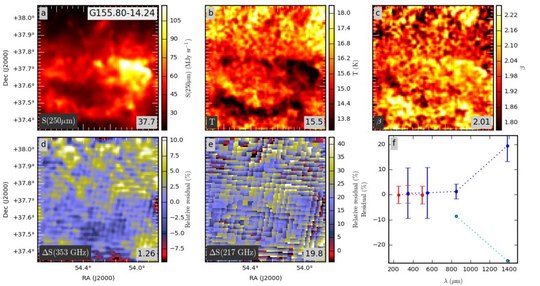}
\caption{Continued{\ldots} Field G155.80-14.24}
\end{figure*}
\begin{figure*}
\includegraphics[width=16cm]{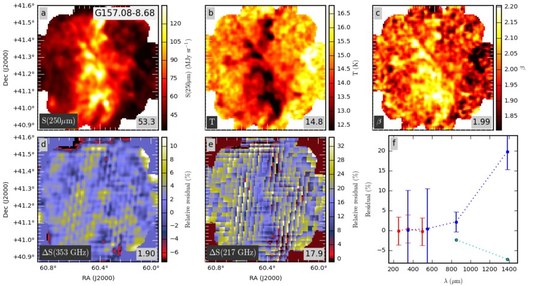}
\caption{Continued{\ldots} Field G157.08-8.68}
\end{figure*}
\clearpage
\begin{figure*}
\includegraphics[width=16cm]{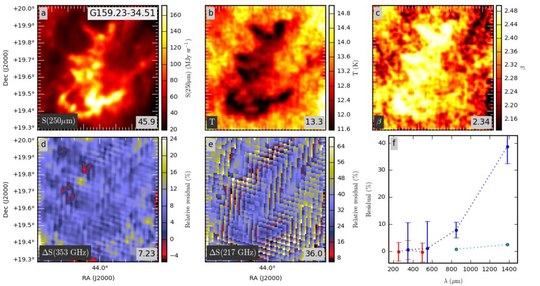}
\caption{Continued{\ldots} Field G159.23-34.51}
\end{figure*}
\begin{figure*}
\includegraphics[width=16cm]{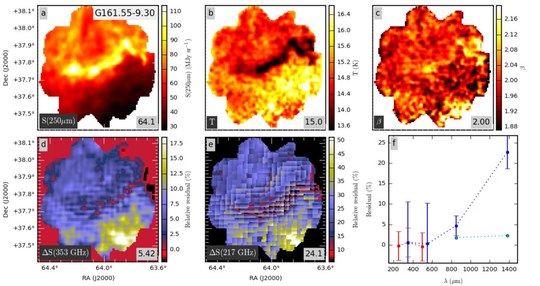}
\caption{Continued{\ldots} Field G161.55-9.30}
\end{figure*}
\begin{figure*}
\includegraphics[width=16cm]{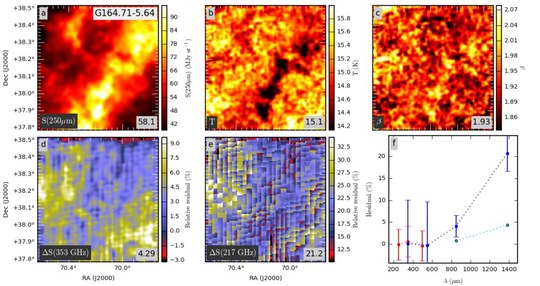}
\caption{Continued{\ldots} Field G164.71-5.64}
\end{figure*}
\begin{figure*}
\includegraphics[width=16cm]{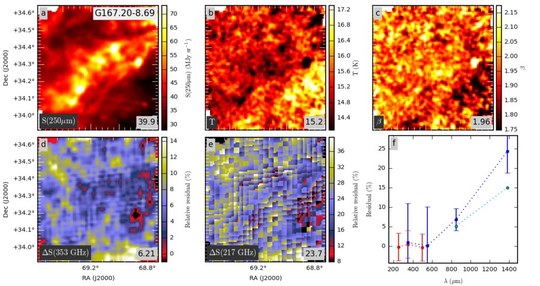}
\caption{Continued{\ldots} Field G167.20-8.69}
\end{figure*}
\begin{figure*}
\includegraphics[width=16cm]{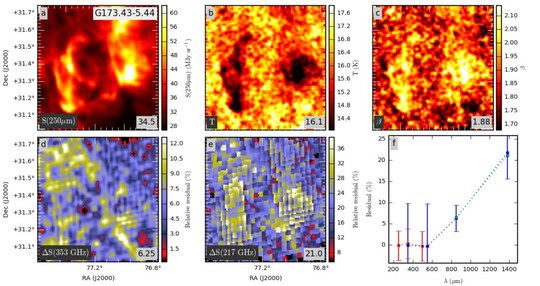}
\caption{Continued{\ldots} Field G173.43-5.44}
\end{figure*}
\begin{figure*}
\includegraphics[width=16cm]{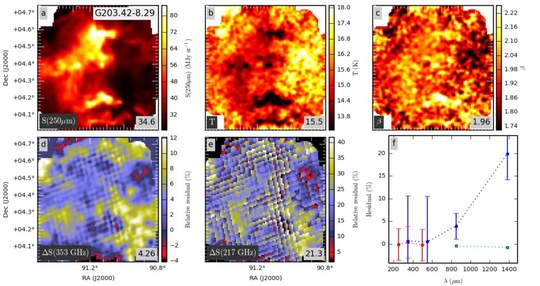}
\caption{Continued{\ldots} Field G203.42-8.29}
\end{figure*}
\begin{figure*}
\includegraphics[width=16cm]{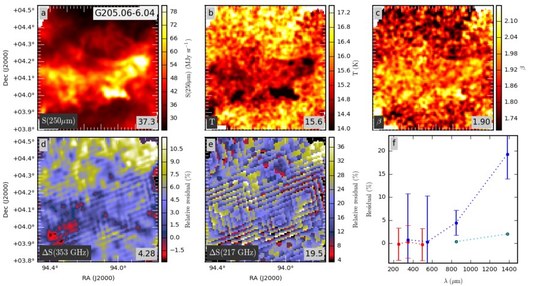}
\caption{Continued{\ldots} Field G205.06-6.04}
\end{figure*}
\begin{figure*}
\includegraphics[width=16cm]{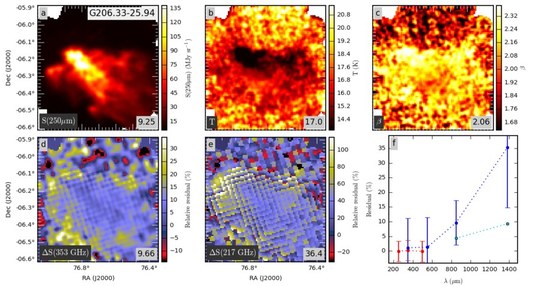}
\caption{Continued{\ldots} Field G206.33-25.94}
\end{figure*}
\begin{figure*}
\includegraphics[width=16cm]{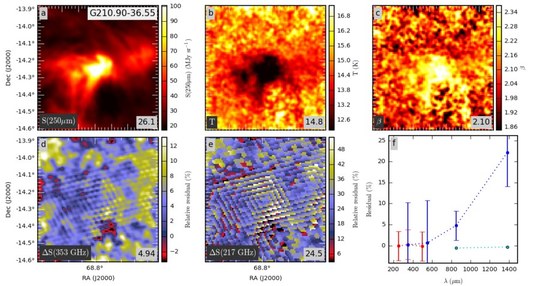}
\caption{Continued{\ldots} Field G210.90-36.55}
\end{figure*}
\begin{figure*}
\includegraphics[width=16cm]{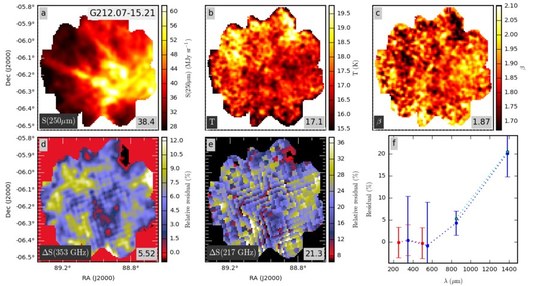}
\caption{Continued{\ldots} Field G212.07-15.21}
\end{figure*}
\begin{figure*}
\includegraphics[width=16cm]{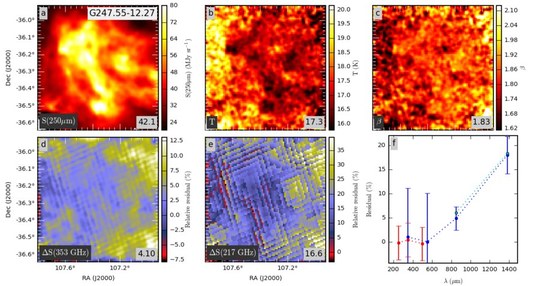}
\caption{Continued{\ldots} Field G247.55-12.27}
\end{figure*}
\begin{figure*}
\includegraphics[width=16cm]{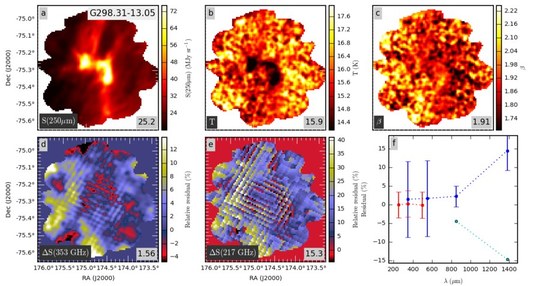}
\caption{Continued{\ldots} Field G298.31-13.05}
\end{figure*}
\clearpage
\begin{figure*}
\includegraphics[width=16cm]{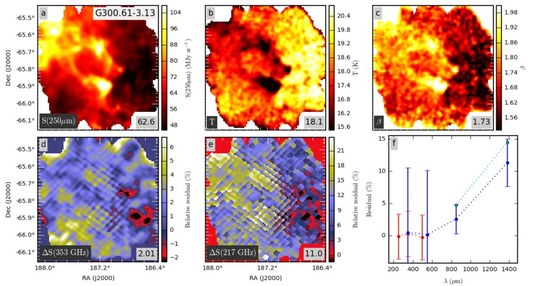}
\caption{Continued{\ldots} Field G300.61-3.13}
\end{figure*}
\begin{figure*}
\includegraphics[width=16cm]{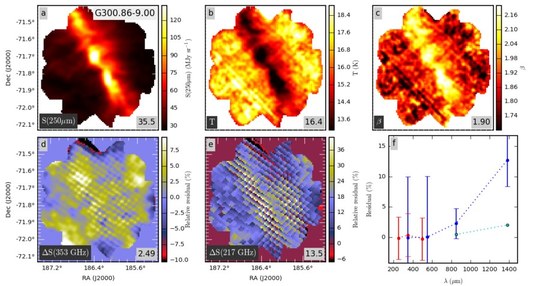}
\caption{Continued{\ldots} Field G300.86-9.00}
\end{figure*}
\begin{figure*}
\includegraphics[width=16cm]{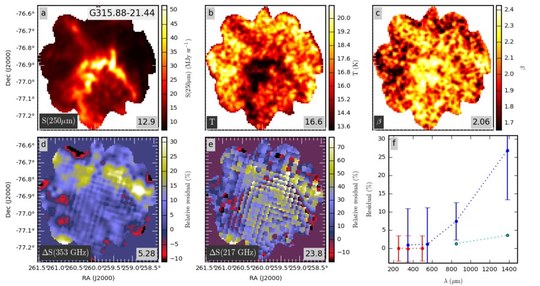}
\caption{Continued{\ldots} Field G315.88-21.44}
\end{figure*}
\begin{figure*}
\includegraphics[width=16cm]{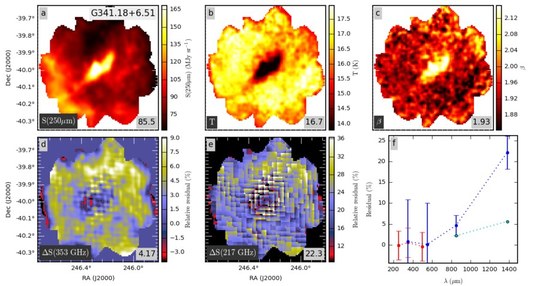}
\caption{Continued{\ldots} Field G341.18+6.51}
\end{figure*}
\begin{figure*}
\includegraphics[width=16cm]{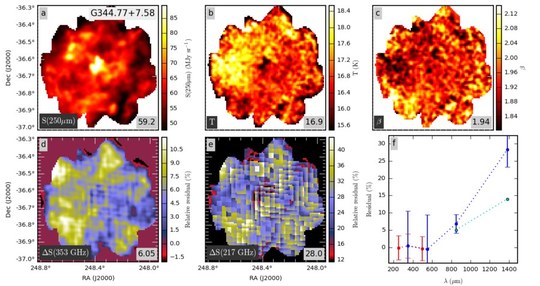}
\caption{Continued{\ldots} Field G344.77+7.58}
\end{figure*}
\begin{figure*}
\includegraphics[width=16cm]{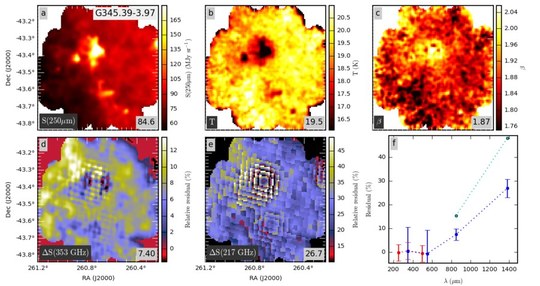}
\caption{Continued{\ldots} Field G345.39-3.97}
\end{figure*}
\begin{figure*}
\includegraphics[width=16cm]{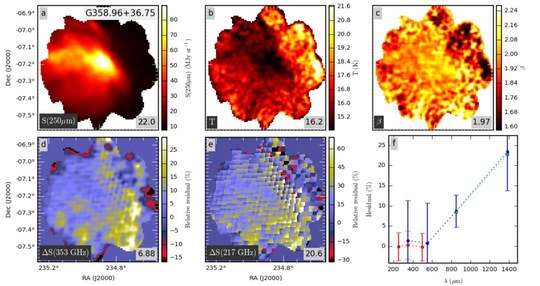}
\caption{Continued{\ldots} Field G358.96+36.75}
\label{fig:BRUTE_043}
\end{figure*}

\end{document}